\documentclass[aps,prb,twocolumn,showpacs,10pt]{revtex4-1}
\usepackage{graphics,graphicx,amsmath,amssymb}
\usepackage{tabularx,float}
\usepackage{epsfig}
\usepackage{subfigure}
\usepackage{bm}
\usepackage{wasysym}
\usepackage{bm}
\usepackage{color}
\usepackage{verbatim}

\newcommand{\PRB}{Phys. Rev. B }

\newcommand{\PRD}{Phys. Rev. D }

\newcommand{\PRL}{Phys. Rev. Lett. }
\newcommand{\RMP}{Rev. Mod. Phys. }
\newcommand{\nJ}{Nature }
\newcommand{\npJ}{Nature Physics }
\newcommand{\ntJ}{Nature Nanotechnology }
\newcommand{\nmJ}{Nature Materials }
\newcommand{\ncJ}{Nature Communications }
\newcommand{\SJ}{Science }
\def \tit#1#2#3#4{{#1 {\bf #2}, #3 (#4)}}

\newcommand{\Nmath}{\mathcal{N}}
\newcommand{\Qmath}{\mathcal{Q}}

\newcommand{\bmm}{{\bf m}}
\newcommand{\bM}{{\bf M}}

\newcommand{\bD}{{\bf D}}

\newcommand{\br}{{\bf r}}

\newcommand{\ua}{\uparrow}
\newcommand{\da}{\downarrow}
\newcommand{\nn}{\nonumber}
\def \U#1{{\rm{U}(#1)}}
\def \SU#1{{\rm{SU}(#1)}}

\def \SO#1{{\rm{SO}(#1)}}

\def \CP#1{{\rm{CP}^{#1}}}

\begin{document}
\title{Spin-valley skyrmions in graphene at filling factor $\nu=-1$}
\date{\today}
\author{Yunlong Lian and Mark O. Goerbig}
\affiliation{
\centerline{Laboratoire de Physique des Solides CNRS-UMR-8502,}
\centerline{Universit\'e Paris-Sud $\subset$ Universit\'e Paris-Saclay, F-91405 Orsay Cedex, France}
}

\begin{abstract}
We model quantum Hall skyrmions in graphene monolayer at quarter filling by a theory of $\CP{3}$ fields and study the energy minimizing skyrmions in presence of valley pseudospin anisotropy and Zeeman coupling. We present a diagram of all types of skyrmions in a wide range of the anisotropy parameters. 
For each type of skyrmion, we visualize it on three Bloch spheres, and present the profiles of its texture on the graphene honeycomb lattice, thus providing references for the STM/STS imaging of spin-pseudospin textures in graphene monolayer in quantum Hall regime. Besides the spin and pseudospin skyrmions for the corresponding degrees of freedom of an electron in the $N=0$ Landau level, we discuss two unusual types -- the ``entanglement skyrmion'' whose texture lies in the space of the entanglement of spin and pseudospin, as well as the ``deflated pseudospin skyrmion'' with partial entanglement.  
For all skyrmion types, we study the dependence of the energy and the size of a skyrmion on the anisotropy parameters and perpendicular magnetic field. 
We also propose three ways to modify the anisotropy energy, namely the sample tilting, the substrate anisotropy and the valley pseudospin analogue of Zeeman coupling. 
\end{abstract}
\maketitle

\section{Introduction}
\label{sec:Introduction}
A skyrmion, first introduced in high-energy physics in the description of baryons,\cite{Skyrme1961} is a topologically non-trivial configuration of a  continuous field $\phi(x)$, which is localized, in the sense that its gradient $\nabla\phi(x)$ is significant only in a finite spatial region $\Omega$. Far away from $\Omega$, the field $\phi(x)$ approaches a uniform configuration, which is conveniently described by a boundary condition at spatial infinity.\cite{textbooks}  
The Euclidean action $S_{\rm E}[\phi(x)]$ usually has the form of non-linear sigma model (NLSM), with the lower bounds of energy being guaranteed by the Bogomol'nyi--Prasad--Sommerfield inequality. \cite{Bogomolny1975} Such lower bound is proportional to $|{\cal Q}|$ -- the absolute value of an integer ``topological charge'' ${\cal Q}$, which classifies the field $\phi(x)$ as a continuous mapping. 
In the work of Skyrme, \cite{Skyrme1961} $\phi(x)$ maps the (compactified) 3+1 dimensional space-time to the manifold of Lie group $\SU{2}$ (which is isomorphic to $S^3$, the 3-sphere), and the third homotopy group $\pi_3(S^3)=\mathbb{Z}$ provides the topological charge ${\cal Q}$, which was identified as particle number in that work. 

After its introduction in high-energy physics, skyrmion physics regained interest in the context of condensed-matter physics as a topological object in a two-dimensional ferromagnetic state. They have been discussed in quantum Hall systems,\cite{Sondhi1993,Moon1995} and in chiral magnets\cite{Roessler2006,Schulz2012,Nagaosa2013,Freimuth2013} and magnetic thin films.\cite{Bogdanov2001,Sampaio2013,MoreauLuchaire2016} While skyrmions have been experimentally viewed in scanning-tunnelling microscopy and spectroscopy (STM/STS) in the latter two systems,\cite{skyrmExp1,skyrmExp2} skyrmions in quantum Hall systems have mainly been identified through their magnetic properties in nuclear-magnetic-resonance experiments,\cite{BarrettNMR} and a spectroscopic identification is yet outstanding. Furthermore, skyrmions have been used to describe special configuration of the superconducting order parameter in superconductors.\cite{Knigavko1999,Li2009} The magnetization or the superconducting order parameter, denoted as $\phi(x)$, usually maps the (compactified) two-dimensional plane to the Bloch sphere $S^2$. The second homotopy group $\pi_2(S^2)=\mathbb{Z}$ provides the topological charge ${\cal Q}$, which is the degree of the mapping between two 2-spheres. 
A more familiar case is the vortex configuration of the complex scalar field $\phi(x)$ on two-dimensional plane, which is punctured by the vortex core. The phase of the complex scalar field lives in a ring $S^1$. Then the homotopy group $\pi_1(S^1)=\mathbb{Z}$ provides us the topological charge ${\cal Q}$, known as winding number. 
Some more exotic cases appeal to the second homotopy group of\cite{dadda1978} $\CP{\Nmath-1}$ that allows for establish of skyrmions in $\Nmath$-component systems, such as in Refs.~[\onlinecite{Garaud2013}] for $\Nmath=3$ and [\onlinecite{Ezawa1999,Ghosh2000,Cote2007}] for $\Nmath=4$, or, to the homotopy group $\pi_3(\rm{RP}^2)=\mathbb{Z}$ in Ref.~[\onlinecite{Smalyukh2010,Ackerman2015}]. In the present study, we choose a $\CP{3}$ field on two-dimensional plane to model the static configuration of QH skyrmions in graphene monolayer. 

The distinguished feature of quantum Hall (QH) skyrmion is that it carries \emph{electric} charge of $q={\cal Q}\,e$, where ${\cal Q}$ is its \emph{topological} charge and $e$ is the charge of an electron. Moreover, such equality holds locally, i.e. the electric charge is smeared in the texture of the QH skyrmion, The excess charge density $\delta \rho_{\rm el}(\br)$ in a QH skyrmion is equal\cite{Sondhi1993,Moon1995} to $e \rho_{\rm topo}(\br)$, where $\rho_{\rm topo}(\br)$ characterizes the spatial variation of the texture in the QH skyrmion, and the integral $\int\rho_{\rm topo}(\br)d^2r$ reproduces its topological charge ${\cal Q}$. As a consequence of the smeared charge density, a QH skyrmion has a lower energy in the $N=0$ Landau level (LL) than the quasiparticle of the same charge, therefore reduces the charge transport gap. A lattice of QH skyrmions also has lower energy than the corresponding Wigner crystal.\cite{Cote2008} We finally notice that, as a consequence of the intimate relation between the topological and the electric charge, a skyrmion is a fermionic excitation in QH systems, in contrast to most of the aforementioned magnetic systems.

In this work, we study the QH skyrmions in the $N=0$ LL in graphene monolayer. Besides the robust quantum Hall effects in graphene in a wide range of temperature\cite{Novoselov2007} and magnetic field,\cite{Lafont2015} there are several advantages to choose graphene as the host system for QH skyrmions. 
First, the Dirac valley of electron states can be described by a pseudospin degree of freedom, and in the presence of strong perpendicular magnetic field the Coulomb interaction is approximately symmetric\cite{goerbigRev} under global $\SU{4}$ transformations in the spin$\otimes$valley space. In contrast, the layer pseudospin in double-layer QH system does not enjoy such a symmetric Coulomb interaction in the spin$\otimes$valley space, because the layer separation always breaks the pseudospin symmetry of the Coulomb interaction. In our approach, the $\SU{4}$ spin-valley symmetry is explicitly broken by terms at lower energy scales (compared to the Coulomb energy), namely the Zeeman coupling that couples directly to the spin, and the valley pseudospin anisotropy. 
As a second advantage of the  graphene monolayer, one may point out that the valley degree of freedom coincides with the sublattice degree of freedom for the electron states in the $N=0$ LL. Such a coincidence provides a convenient way to probe the $z$-component of the valley pseudospin directly -- it can be read off from the sublattice occupation, thus allowing a direct imaging of the pseudospin texture via STM/STS experiments. 
Moreover, the spin texture in graphene under QH conditions can also be imaged by spin-resolved STM/STS experiments. The combination of the lattice-resolved images for spin and pseudospin textures thus allows for an identification of the various skyrmion types in graphene. 
The fact that graphene is naturally a surface system renders these states spectroscopically more accessible than QH systems in semiconductor heterostructures, where the two-dimensional electron system appears at the interface between two types of semiconductors. For STM/STS experiments in semiconductor heterostructures, two-dimensional electron systems have been realized on an n-doped InSb surface\cite{hashimoto2008} albeit with a mobility that does not reach that of the more common GaAs hetero-structures or graphene. 

Our work completes our previously obtained phase diagram for QH skyrmions in graphene\cite{lian2016} in several respects. Apart from a full description of the FM states one encounters at $\nu=-1$ (and $\nu=+1$ by particle-hole symmetry), we provide here a detailed characterization of skyrmions with entanglement. In addition to a pure entanglement skyrmion, one obtains an exotic type of \textit{deflated} pseudospin skyrmions with partial entanglement. We characterize all encountered skyrmion types with the help of three different Bloch spheres that describe spin, valley pseudospin and entanglement, respectively. Furthermore, all ferromagnetic and skyrmion states are analysed from their appearance in lattice-resolved density plots that may use as a guideline in an experimental STM/STS investigation. Finally, we discuss the scaling relations of the skyrmion size and energy close to the transitions between different underlying FM states -- while a critical behavior is obtained in the case of a symmetry restoration at the transition, we find that the scaling is truncated in the other cases where a (subordinate) symmetry is not fully restored. These scaling relations are then discussed in the context of further modifications of the anisotropy parameters that we vary to obtain the skyrmion phase diagram, e.g. in the case of a tilted sample. 

The paper is organized as follows. In Sec.~\ref{sec:Quantum-Hall-Ferromagnetic-State} we propose an anisotropic energy for the ferromagnetic state in the $N=0$ LL at quarter filling, and discuss the $\CP{3}$ representation of the QH ferromagnetic states. There are four types of ferromagnetic ground states for the anisotropic Hamiltonian. Each type is described by a $\CP{3}$ spinor that contains six angular variables and visualized on the spin, pseudospin and entanglement Bloch spheres. We also visualize them on the honeycomb lattice. 
In Sec.~\ref{sec:CP3-Skyrmion-of-charge-Q-1} we discuss the energy functional and ansatz for the $\CP{3}$ skyrmion on top of the QH ferromagnetic states, and present the skyrmion type diagram, which is obtained by minimization of the energy functional with the ansatz and under various input parameters of the model. 
The detailed analysis of the minimization result is presented in Sec.~\ref{sec:4-types-skyrmion}. We visualize the skyrmion $\CP{3}$-field on three Bloch spheres in a same manner as for the QH ferromagnetic states. These visualizations explicitly demonstrate that the skyrmions can be viewed as a wrapping of the $xy$-plane on Bloch spheres. Besides, the $\CP{3}$-skyrmion is also visualized on the honeycomb lattice, in order to show the difference between two types of skyrmions. The lattice-resolved profiles of the $\CP{3}$-skyrmion serve as references for the STM/STS imaging of spin-pseudospin textures in graphene monolayer under QH conditions. 
The qualitative discussion on the $\CP{3}$-skyrmions are followed by quantitative discussions in Sec.~\ref{sec:Size_and_energy}, where we present the dependence of size and energy of the $\CP{3}$-skyrmions on the input parameters of the model. 
Finally in Sec.~\ref{sec:Modification-of-the-anisotropy-energy}, we discuss three ways to modify the anisotropic energy of the $\CP{3}$-skyrmions. We demonstrate how the type of a $\CP{3}$-skyrmion is altered by these modifications. These modifications are relevant in certain experimental regimes.

\section{Quantum Hall Ferromagnetic State}
\label{sec:Quantum-Hall-Ferromagnetic-State}
A multicomponent quantum Hall system is characterized by the $n$-fold degeneracy of the Landau levels which originates from internal degrees of freedom of 2D electrons in a strong magnetic field. For example, when the Zeeman energy is much smaller than the Coulomb interaction, one may first neglect the Zeeman coupling of the electron spin to the applied magnetic field and consider spin as an internal degree of freedom which yields a two-fold degeneracy of the Landau levels. At half-filling of such LL (i.e. integer filling of Landau sublevel), the major cause for the complete spin polarization in the ground state is the Coulomb interaction, which is larger than the Zeeman energy and prefers a totally asymmetric orbital part of the many-particle wave function, so that the electrons are separated as far as possible, and the spinor part is totally symmetric. The Zeeman coupling at a smaller energy scale then conspires with the Coulomb energy and orients the polarized spins along the applied magnetic field. In this way, the electrons in a single LL form a \emph{quantum Hall ferromagnet} (QHFM).

In graphene monolayer, the internal degrees of freedom of the Landau levels are doubled by the valley degeneracy, which is described by a \emph{pseudospin} analogous to the electron spin. The Coulomb interaction has approximate $\SU{4}$ symmetry,\cite{goerbigRev} which is broken by spin/pseudospin anisotropic interactions at smaller energy scale.  

At quarter filling, i.e. when one of the four spin-valley sublevels is completely filled ($\nu=-1$ for electron filling) or completely empty ($\nu=1$ for hole filling), the ground state $\left|F\right\rangle $ for the electron system is the $\SU{4}$ quantum Hall ferromagnetic (QHFM) state\cite{Yang2006,Nomura2006}
\begin{equation}
\left|F\right\rangle =\prod_{X}(f_1 C_{X\uparrow K}^{\dagger}+f_2 C_{X\downarrow K}^{\dagger}+f_3 C_{X\uparrow K'}^{\dagger}+f_4 C_{X\downarrow K'}^{\dagger})\left|0\right\rangle\label{eq:QHFM}
\end{equation}
where $C_{X\sigma\xi}^{\dagger}$ creates an electron 
in the $N=0$ LL with spin $\sigma$ ($=\uparrow,\downarrow$) and pseudospin $\xi$ ($=K,K'$). The quantum number $X$ is the Landau orbit, which characterizes the LL degeneracy. It is related to the underlying translation invariance -- while this invariance is, strictly speaking, broken by the position-dependent vector potential, the physical magnetic field is constant in the entire plane. The kinetic energy of an electron in this field does therefore not depend on the center of its quantized cyclotron motion. The Landau orbit $X$ is precisely the quantum number associated with the position of this center.
In the case of one electron per Landau orbit, which we consider here, the coefficients $f_1,f_2,f_3,f_4\in \mathbb{C}$ satisfy $\sum^{4}_{i=1}\left|f_i\right|^{2}=1$. To model electron states with spatially homogeneous spin-pseudospin magnetization at a length-scale larger than $l_B$, we assume that $F_i$ do not carry Landau orbit index $X$. 
Since $\left|F\right\rangle$ and $e^{i\theta}\left|F\right\rangle$ ($\theta$ is a real constant) correspond to the same quantum state, the coefficients in Eq.~(\ref{eq:QHFM}) can be uniquely represented by a CP$^{3}$-spinor $F=(f_1,f_2,f_3,f_4)^{T}$, which is normalized $F^{\dagger}F=1$ and is equivalent to $e^{i\theta}F$. (See Appendix.~\ref{subsec:CP3-spinor-and-NLSM}) 
The meaning of the four components in the CP$^{3}$-spinor is inherited from $\left|F\right\rangle$.

In this section, based on a general proposal of valley pseudospin anisotropy, we discuss different types of QHFM ground states at quarter filling ($\nu=-1$ for electron filling and $\nu=1$ for hole filling) of the four-fold degenerate $N=0$ LL in graphene monolayer. We propose the anisotropic energy in Sec.~\ref{subsec:Anisotropic-energy}, and then discuss the parametrization of CP$^{3}$-spinor in Sec.~\ref{subsec:Parametrization}. The four types of QHFM ground state are discussed in Sec.~\ref{subsec:QHFM-ground-states}. We schematically visualize the QHFM states on the honeycomb lattice of graphene in Sec.~\ref{subsec:QHFM-visualization}.

\subsection{Anisotropic energy}
\label{subsec:Anisotropic-energy}

The electrons are restricted to the $N=0$ LL in our problem and hence the kinetic energy is quenched and set to a constant. Due to the $\SU{4}$ symmetry of the interaction Hamiltonian\cite{goerbigRev}
\begin{equation}
H_{\rm C}=\int\overline{\rho}(r)V(r-r')\overline{\rho}(r')d^{2}rd^{2}r',\label{eq:CoulombH}
\end{equation}
the Coulomb energy $E_{C}[F]=\left\langle F\right|H_{\rm C}\left|F\right\rangle$ does not depend on the QHFM state $\left|F\right\rangle$. 
The $\SU{4}$ symmetry is explicitly broken by the spin-pseudospin anisotropies \cite{Nomura2009}
\begin{equation}
H_{\rm A} = \int d^{2}r \left\{ U_{\perp}(P_{\rm x}^{2}+P_{\rm y}^{2})+ U_{\rm z}P_{\rm z}^{2} + U_{0}\left|\mathbf{S}\right|^{2}-\frac{1}{2}\Delta_{\rm Z}S_{z}\right\}\label{eq:AnisotropicH}
\end{equation}
where $S_i$ and $P_i$ are spin and pseudospin density operators. The first terms reflect a pseudospin anisotropy that can be generated, e.g. by short-range interactions of Hubbard type\cite{Alicea2006} 
($U_0$ and $U_{\rm z}$)
or an out-of-plane\cite{Fuchs2007} (also contributing to $U_{\rm z}$) and in-plane Kekul\'e-type \cite{Nomura2009} lattice deformation due to electron-phonon couplings. The coefficients $U_{0},U_{\perp},U_{\rm z},\Delta_{\rm Z}$ have the dimension of energy, and $U_{0},U_{\perp},U_{\rm z}$ are proportional to the perpendicular component $B_\perp$ of the applied magnetic field, whereas $\Delta_{\rm Z}$ is proportional to the total applied magnetic field $B_{\rm T}$. 
Their numerical values are estimated as $1.0\times B_\perp[T]K$, $2.0\times B_\perp[T]K$, $0.5\times B_\perp[T]K$ and $1.3\times B_{\rm T}[T]K$ respectively in Ref.~[\onlinecite{Nomura2009}]. 
$H_{\rm A}$ explicitly breaks the $\SU{4}$ symmetry down to U$(1)_S\times$U$(1)_{\rm P}\times$U$(1)_E\times\mathbb{Z}_2$. While the U$(1)_S$ and the U$(1)_P$ symmetries reflect rotations around the $z$-quantization axes
of the spin and pseudospin, respectively, $\mathbb{Z}_2$ indicates that the two orientations $z$ and $-z$ of the pseudospin are equivalent. They are not equivalent in the spin channel due to the Zeeman coupling. The U$(1)_E$ symmetry is more subtle and related to spin-pseudospin entanglement,\cite{Doucot2008}
as we discuss in more detail below. 

The system energy is determined by minimization of the anisotropic energy $E_{\rm A}[F]=\left\langle F\right|H_{\rm A}\left|F\right\rangle$, which we propose to be the following form:
\begin{equation}
\label{eq:EnA}
E_{\rm A}[F] = A_{0}\frac{\Delta_{\rm Z}}{2}\left\{ u_{\perp}\left(M_{\rm Px}^{2}+M_{\rm Py}^{2}\right)+u_{\rm z}M_{\rm Pz}^{2}-M_{\rm Sz} \right\}.
\end{equation}
where $A_{0}$ is a dimensionless area to ensure an extensive energy, $\Delta_{\rm Z}\ge0$, and the dimensionless parameters $u_{\perp}$ and $u_{\rm z}$ characterize the pseudospin anisotropy. In the present work, $u_{\perp}$ and $u_{\rm z}$ are treated as control parameters that generate the different phases even if they may be more difficult to vary in a typical situation.
In the above equation, we denote the spin and pseudospin magnetizations as
\begin{eqnarray}
\bM_{\rm S} &=F^{\dagger}(1\otimes\boldsymbol{\sigma})F \label{eq:Si_F},\\
{\rm and} \qquad \bM_{\rm P} &=F^{\dagger}(\boldsymbol{\sigma}\otimes 1)F, \label{eq:Ti_F}
\end{eqnarray}
respectively.
They can be expanded explicitly in the components of the $CP^{3}$-spinor $F$. For instance, the $z$-component of the spin magnetization $S_{z}(\br)$ is 
\begin{equation}
S_{z} = F^{\dagger}(1\otimes\sigma_{z})F = \left|f_{K\uparrow}\right|^{2}-\left|f_{K\downarrow}\right|^{2}+\left|f_{K'\uparrow}\right|^{2}-\left|f_{K'\downarrow}\right|^{2},
\end{equation}
and the $z$-component of the pseudospin magnetization $P_{z}\left(r\right)$ is
\begin{equation}
P_{z} = F^{\dagger}(\sigma_{z}\otimes1)F = \left|f_{K\uparrow}\right|^{2}+\left|f_{K\downarrow}\right|^{2}-\left|f_{K'\uparrow}\right|^{2}-\left|f_{K'\downarrow}\right|^{2},
\end{equation}
where we use the labels $K\ua$, $K\da$, $K'\ua$ and $K'\da$ for the index $1,2,3,4$ of the components of $F$ to emphasize their meaning. 

\subsection{Parametrization of CP$^{3}$-spinor}
\label{subsec:Parametrization}
It is essential for later discussions to use the Schmidt decomposition to parametrize a general CP$^{3}$-spinor $Y$ with six real parameters, namely $\theta_{\rm S},\theta_{\rm P},\alpha\in[0,\pi]$ and $\phi_{\rm S},\phi_{\rm P},\beta\in[0,2\pi)$:\cite{Doucot2008}
\begin{eqnarray}
Y &=& \cos\frac{\alpha}{2}\psi^{\rm P}\otimes\psi^{\rm S}+ e^{i\beta}\sin\frac{\alpha}{2}\chi^{\rm P}\otimes\chi^{\rm S}\label{eq:parametrizationZ}\\
\psi^{\rm J} &=& (\cos\frac{\theta_{\rm J}}{2},\sin\frac{\theta_{\rm J}}{2}e^{i\phi_{\rm J}})^{T}\label{eq:psi_basis}\\
\chi^{\rm J} &=& (-\sin\frac{\theta_{\rm J}}{2}e^{-i\phi_{\rm J}},\cos\frac{\theta_{\rm J}}{2})^{T}\label{eq:chi_basis}
\end{eqnarray}
where $(\bmm_{\rm J}\cdot\boldsymbol{\sigma})\psi^{\rm J}=+\psi^{\rm J}$ and $(\bmm_{\rm J}\cdot\boldsymbol{\sigma})\chi^{\rm J}=-\chi^{\rm J}$, $\boldsymbol{\sigma}=(\sigma_{1},\sigma_{2},\sigma_{3})$ are the Pauli matrices, $\bmm_{\rm J}=(\sin\theta_{\rm J}\cos\phi_{\rm J},\sin\theta_{\rm J}\sin\phi_{\rm J},\cos\theta_{\rm J})$ is the unit vector for the direction of magnetization. The subscript $\rm{J}=\rm{S}$ ($\rm{P}$) stands for spin (pseudospin). The Pauli matrices for spin and pseudospin are designed as $1\otimes\boldsymbol{\sigma}$ and $\boldsymbol{\sigma}\otimes 1$ respectively, in agreement with the earlier convention for the CP$^{3}$-spinor $F$. In the present section, where we discuss QHFM states, the real parameters $\theta_{\rm S},\phi_{\rm S}\theta_{\rm P},\phi_{\rm P},\alpha$, and $\beta$ are constant in space, i.e. independent of the Landau orbit $X$, while we consider an explicit position dependence in the following section on skyrmions. To simplify notations, we therefore omit here to write explicitly the argument $(\br)$ in the spinors and parameters. 
Under the parametrization Eq.~(\ref{eq:parametrizationZ}), the spin and the pseudospin magnetization for the CP$^{3}$-spinor $Y$ are 
\begin{eqnarray}
\bM_{\rm S} &=& Y^{\dagger}(1\otimes\boldsymbol{\sigma})Y= \bmm_{\rm S}\cos\alpha, \label{eq:Si_Y}\\
{\rm and} \qquad \bM_{\rm P} &=& Y^{\dagger}(\boldsymbol{\sigma}\otimes 1)Y= \bmm_{\rm P}\cos\alpha ,\label{eq:Ti_Y}
\end{eqnarray}
respectively.
The meaning of $\theta_{\rm S},\phi_{\rm S}$ and $\theta_{\rm P},\phi_{\rm P}$ in the parametrization Eq.~(\ref{eq:parametrizationZ}) is evident from the above equations -- they are the polar and azimuthal angle of the spin and pseudospin magnetizations. 

As we discuss in detail later, the parameter $\alpha$ enriches the types of FM ground state of the system. It can be understood in terms of ``entanglement''.\cite{Doucot2008} It is possible to rewrite the state $Y$ into a direct product of spinors for electron spin and and valley pseudospin \emph{only} when $\alpha=0$ or $\pi$. In this case the spin and pseudospin are \emph{unentangled}. Otherwise, $Y$ is a superposition of two ``product states'' carrying opposite spin and pseudospin. Consequently, as indicated in Eq.~(\ref{eq:Si_Y}) and Eq.~(\ref{eq:Ti_Y}), the spin and pseudospin magnetizations have magnitudes smaller than $1$. 
Remarkably, these magnitudes are equal to $\cos\alpha$, because in a $\CP{3}$-spinor the spin and pseudospin are on equal footing, analogous to the two entangled $1/2$-spins in the Bell state.\cite{NielsenChuang} 
Since we have specified the forms of $\psi_{\rm S/P}$ and $\chi_{\rm S/P}$ in Eq.~(\ref{eq:psi_basis}) and Eq.~(\ref{eq:chi_basis}), the coefficients in Eq.~(\ref{eq:parametrizationZ}) are complex numbers in general and the parameter $\beta$ is just the relative phase of these complex coefficients. 
Throughout this paper, we use the term ``entanglement'' as a synonym of the magnitude of the spin or pseudospin magnetizations: ``maximal entanglement'' means $\alpha=\pi/2$; ``unentangled'' refers to $\alpha=0$ or $\pi$. 

\subsection{Four types of QHFM ground states}
\label{subsec:QHFM-ground-states}

\begin{figure}[t]
\includegraphics[width=0.8\columnwidth]{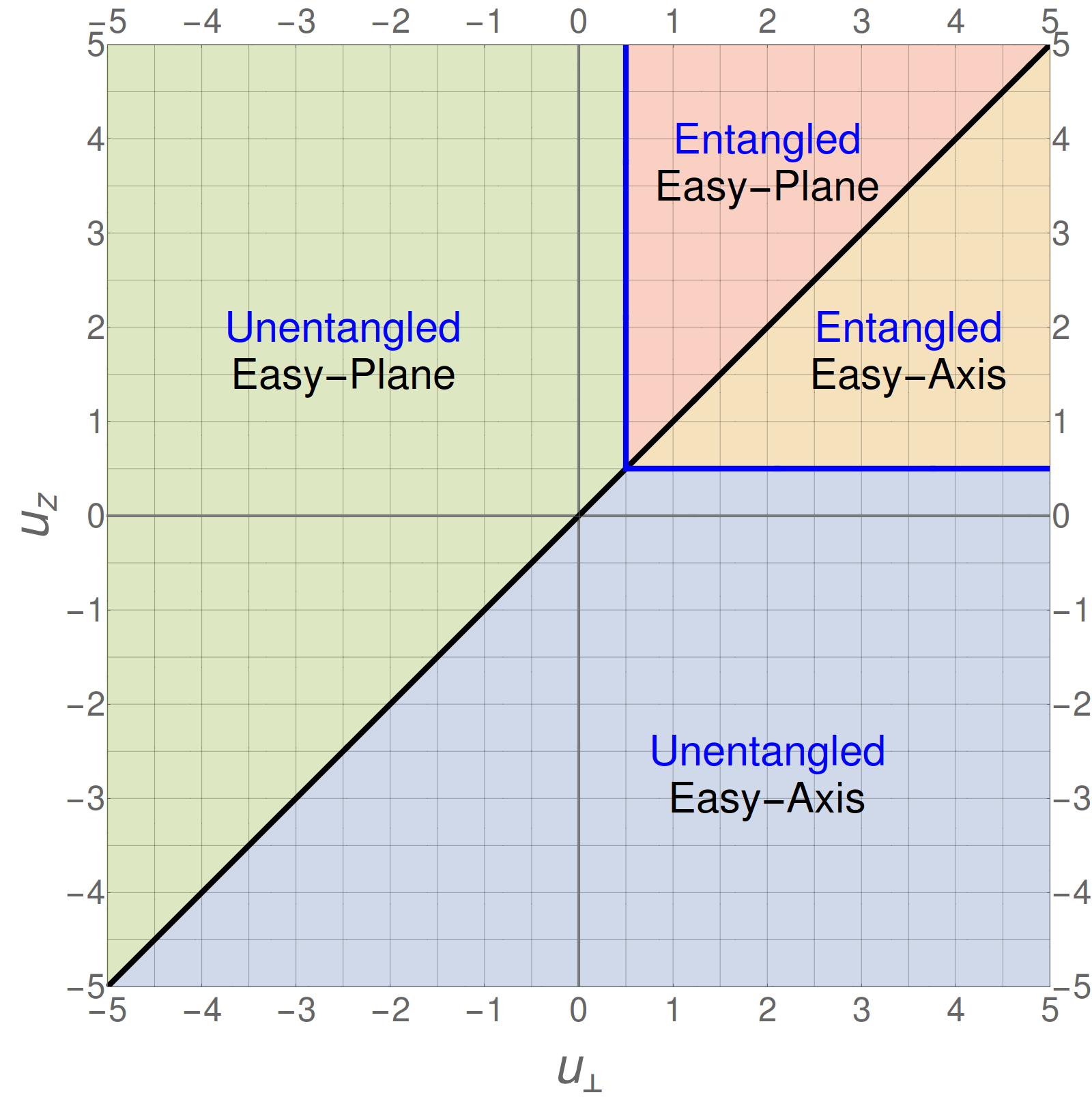}
\caption{\label{fig:FM-region}Four types of QHFM ground states on the $u_{\perp}-u_{\rm z}$ plane. They are listed in Tab.~\ref{tab:FM-tab}.}
\end{figure}

Applying the parametrization in Eq.~(\ref{eq:parametrizationZ}) to the CP$^{3}$-spinor $F$ representing the ferromagnetic state (\ref{eq:QHFM}), we minimize $E_{\rm A}[F]$ with respect to the six parameters of $F$ and obtain four types of QHFM ground states, which are displayed in Fig.~\ref{fig:FM-region} and Tab. \ref{tab:FM-tab}. 
Due to the Zeeman coupling, a CP$^{3}$-spinor $F$ with polarized spin along the $z$-axis is always energetically favourable, independent of the pseudospin and entanglement. It follows that the optimal value is $\theta_{\rm S}=0$ in \textit{all regions} throughout the $u_{\perp}-u_{\rm z}$ plane, if we assume $\alpha\in[0,\pi/2]$. 
Moreover, when $u_{\perp}>u_{\rm z}$ (region 1, 3), it is energetically favourable to orient the pseudospin along the $z$ (or $-z$) axis and one thus obtains 
an easy-axis FM state with optimal value $\theta_{\rm P}=0$ or $\pi$. On the contrary, when $u_{\perp}<u_{\rm z}$ (region 2, 4), a pseudospin magnetization in the $xy$-plane is favoured, and one obtains
an easy-plane FM state with optimal value $\theta_{\rm P}=\pi/2$ and arbitrary optimal value for $\phi_{\rm P}$. On the line $u_{\perp}=u_{\rm z}$, the $\SU{2}$ symmetry of pseudospin is restored and the optimal values for $\theta_{\rm P}$ and $\phi_{\rm P}$ are arbitrary. 

\begin{table}[b]
\begin{tabular}{|c|c|}
\hline 
QHFM type & Region\tabularnewline
\hline 
\multicolumn{2}{|c|}{CP$^{3}$-spinor $F$}\tabularnewline
\hline 
\hline 
Unentangled easy-axis pspin FM & (1) $u_{\perp}>u_{\rm z}$, $u_{\rm z}\le\frac{1}{2}$\tabularnewline
\hline 
\multicolumn{2}{|c|}{$F=(1,0,0,0)^T$ or $(0,0,1,0)^T$}\tabularnewline
\hline 
\hline 
Unentangled easy-plane pspin FM & (2) $u_{\perp}<u_{\rm z}$, $u_{\perp}\le\frac{1}{2}$\tabularnewline
\hline 
\multicolumn{2}{|c|}{$F=\frac{1}{\sqrt{2}}(1,0,e^{i\phi_{\rm P}},0)^T$ with arbitrary $\phi_{\rm P}$}\tabularnewline
\hline 
\hline 
Entangled FM with  & (3) $u_{\perp}>u_{\rm z}>\frac{1}{2}$\tabularnewline
preferential easy-axis pspin & \tabularnewline
\hline 
\multicolumn{2}{|c|}{$F=(\cos\frac{\alpha}{2},0,0,e^{i\beta}\sin\frac{\alpha}{2})^T$ or $(0,e^{i\beta}\sin\frac{\alpha}{2},\cos\frac{\alpha}{2},0)^T$}\tabularnewline
\multicolumn{2}{|c|}{$\alpha=\sec^{-1}(2u_{\rm z})$}\tabularnewline
\hline 
\hline 
Entangled FM with  & (4) $u_{\rm z}>u_{\perp}>\frac{1}{2}$\tabularnewline
preferential easy-plane pspin & \tabularnewline
\hline 
\multicolumn{2}{|c|}{$F=\frac{1}{\sqrt{2}}\left(\cos\frac{\alpha}{2},-e^{i(\beta-\phi_{\rm P})} \sin\frac{\alpha}{2},e^{i\phi_{\rm P}}\cos\frac{\alpha}{2},e^{i\beta}\sin\frac{\alpha}{2}\right)^T$}\tabularnewline
\multicolumn{2}{|c|}{with $\alpha=\sec^{-1}(2u_{\perp})$ and arbitrary $\phi_{\rm P},\,\beta$}\tabularnewline
\hline 
\end{tabular}
\caption{\label{tab:FM-tab}Four types of QHFM ground states correspond to four distinct regions on the $u_{\perp}-u_{\rm z}$ plane in Fig.~(\ref{fig:FM-region}). 
}
\end{table}

Entanglement enriches the types of FM states because tuning the parameter $\alpha$ allows the magnitudes of spin and pseudospin magnetization to decrease and thus to lower the anisotropic energy $E_{\rm A}$. When $\min(u_{\perp},u_{\rm z})\le0$, the optimal value of $\alpha$ is always 0. This is easily understood: since at least one of the parameters $u_\perp$ or $u_{\rm z}$ is negative, $E_A$ is minimized when the spin and pseudospin magnetizations are maximal, i.e. when $|\bM_{\rm S}|=|\bM_{\rm P}|=1$. For $0\le\min(u_{\perp},u_{\rm z})\le 1/2$, the Zeeman energy is crucial for the presence of unentangled FM states, which are of the same type as in the previous case. Were there no term proportional to $M_{\rm Sz}$ in the expression of $E_{\rm A}$, all the other terms related to pseudospin would be non-negative, and the minimization would result in a spinor with $|\bM_{\rm P}|=0$ (maximal entanglement or $\alpha=\pi/2$) for all possible $(u_{\perp},u_{\rm z})$ in the 1-quadrant. The term proportional to $M_{\rm Sz}$ prefers larger magnitudes of $\bM_{\rm S}$ along the applied magnetic field. 
Since $|\bM_{\rm S}|=|\bM_{\rm P}|$ always holds, there is a competition between the tendency towards small magnitude of pseudospin magnetization due to the pseudospin contribution to $E_{\rm A}$, and the tendency towards large magnitude of the spin magnetization due to the spin contribution. Here in the case of $0\le\min(u_{\perp},u_{\rm z})\le 1/2$, spin contributes more, so that a maximal value of $|M_{\rm Sz}|$ is favoured, very much as for negative values of $u_\perp$ and $u_{\rm z}$, and spin and pseudospin remain unentangled. The above-mentioned energy competition yields different results when $\frac{1}{2}<\min(u_{\perp},u_{\rm z})$. In this case the pseudospin contribution to $E_{\rm A}$ is positive, which can be lowered not only by proper choice of the direction ($\theta_{\rm P},\phi_{\rm P}$) of pseudospin magnetization, but also by shrinking the magnitude $|\bM_{\rm P}|$ of pseudospin magnetization. Although the later yield a less negative value of the spin contribution to $E_{\rm A}$ due to the identity $|\bM_{\rm S}|=|\bM_{\rm P}|$, the reduction of energy contribution from pseudospin overcomes the increase of the contribution from spin, so that the overall minimum of $E_{\rm A}$ is reached when
\begin{equation}
|\bM_{\rm S}|=|\bM_{\rm P}| = \cos\alpha = \frac{1}{2\min(u_{\perp},u_{\rm z})}.\label{eq:entanglement_value}
\end{equation}


\subsection{Visualization of the QHFM states}
\label{subsec:QHFM-visualization}

\begin{figure}[t]
\begin{tabular}{ccc}
\includegraphics[width=0.3\columnwidth]{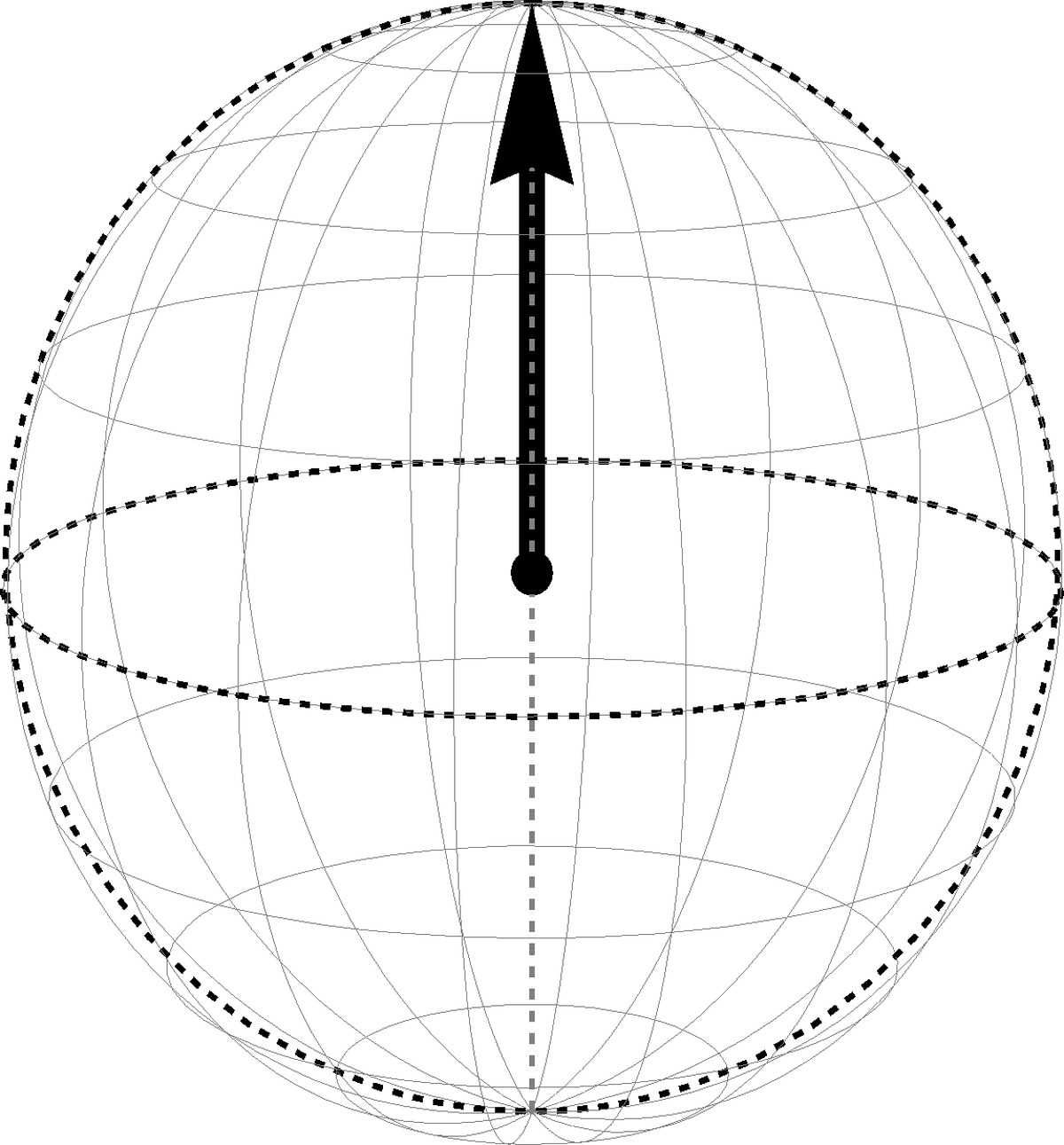} & \includegraphics[width=0.3\columnwidth]{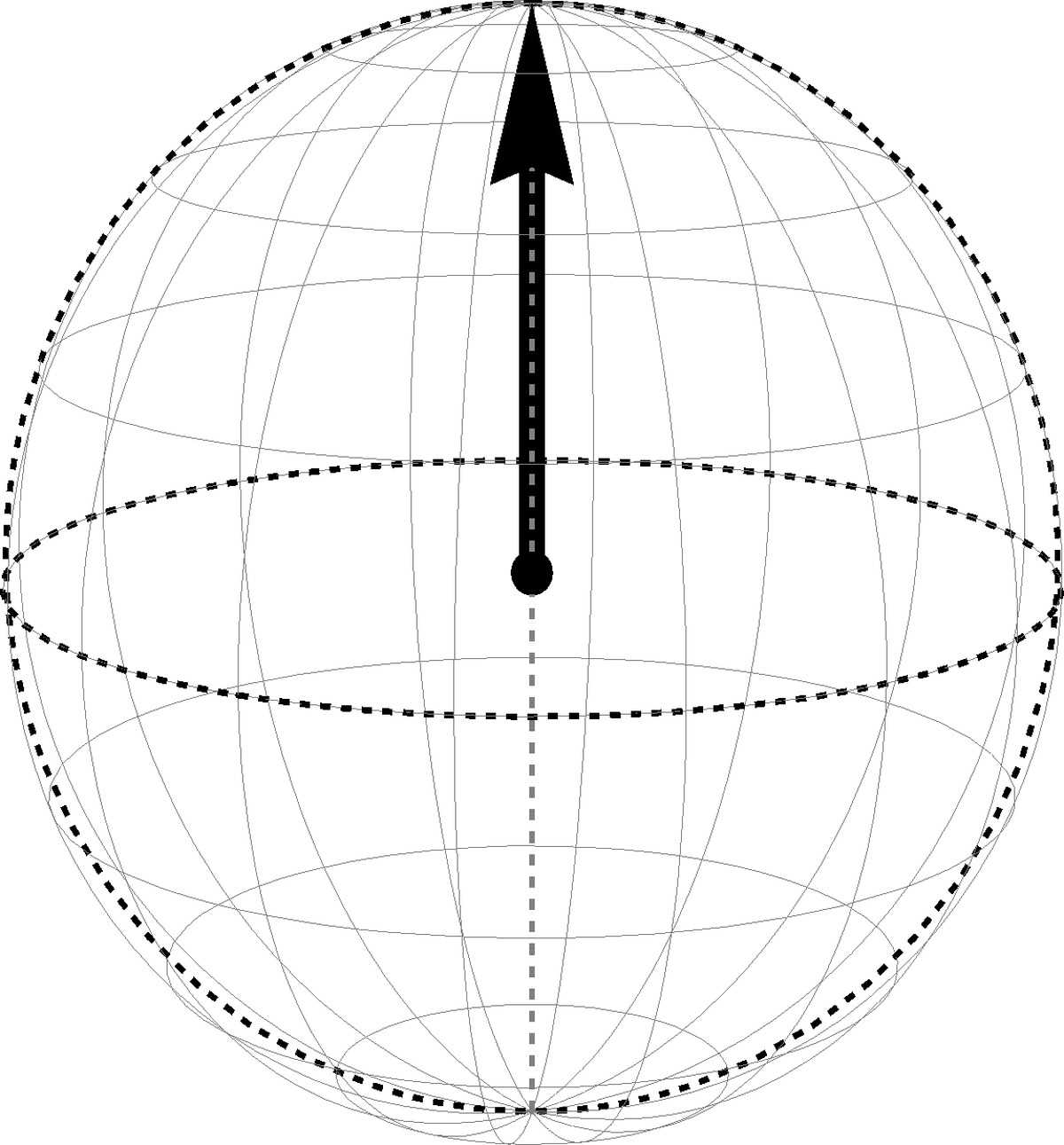} & \includegraphics[width=0.3\columnwidth]{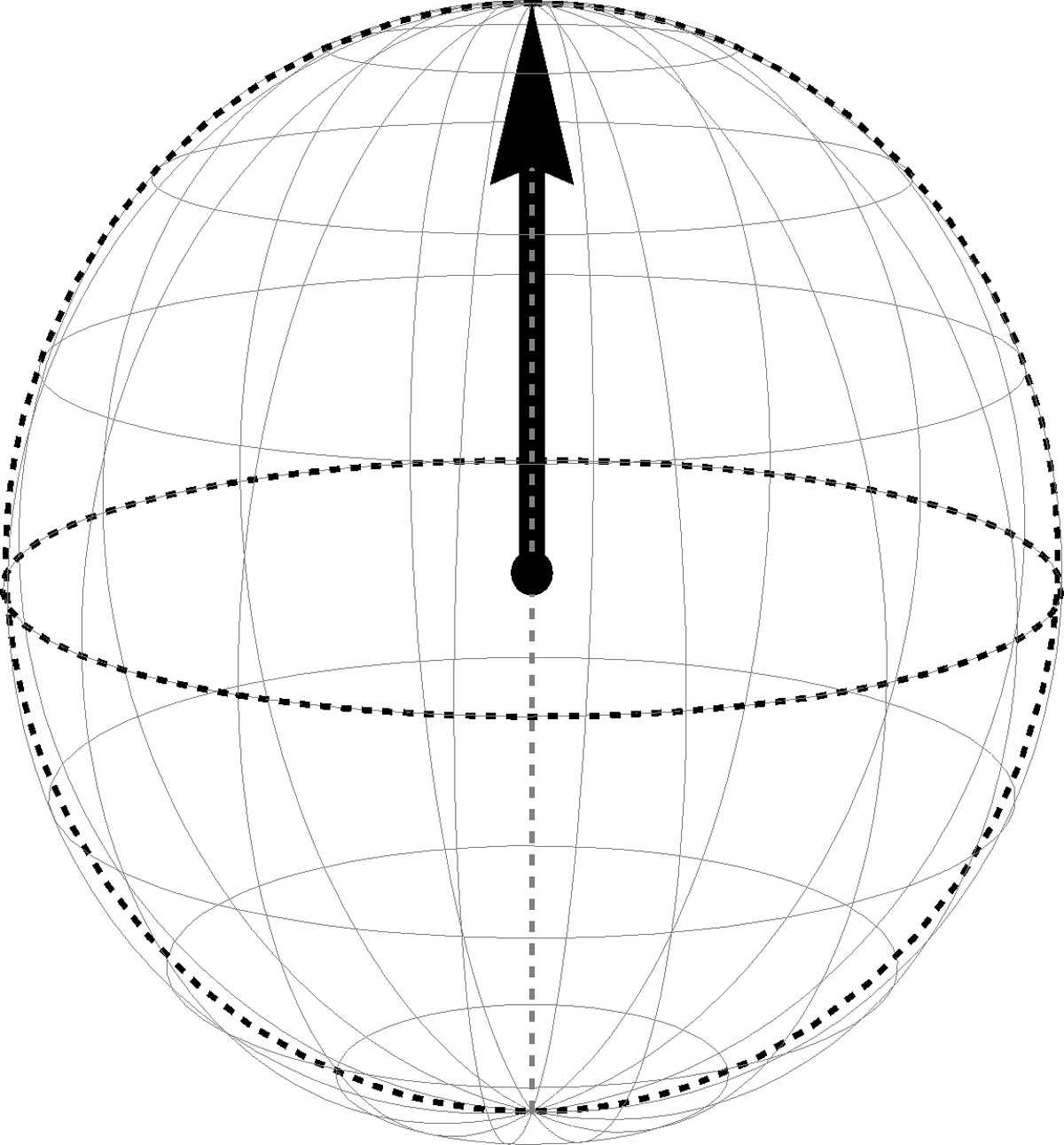}\tabularnewline
(a) Spin BS & (b) Pspin BS & (c) Entanglement BS\tabularnewline
\end{tabular}
\includegraphics[width=\columnwidth]{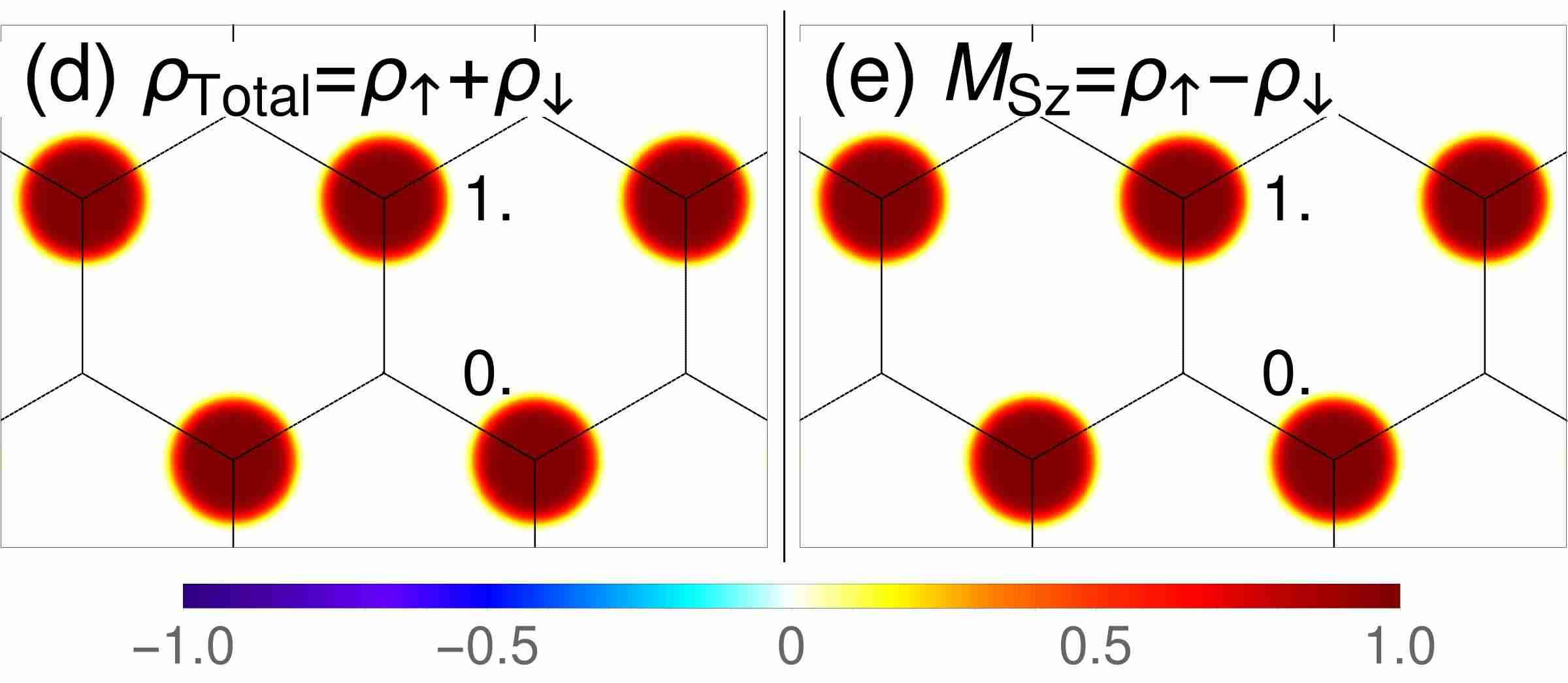}
\caption{\label{fig:QHFM_UEA} Visualization of unentangled easy-axis QH ferromagnetic state on the spin (a), pseudospin (b) and entanglement (c) Bloch spheres, as well as the lattice-scale profiles of the electron density $\rho_{\rm Total}(\br)$ (d) and the $z$-component of spin magnetization $M_{\rm Sz}(\br)$ (e).
}
\end{figure}

\begin{figure}[t]
\begin{tabular}{ccc}
\includegraphics[width=0.3\columnwidth]{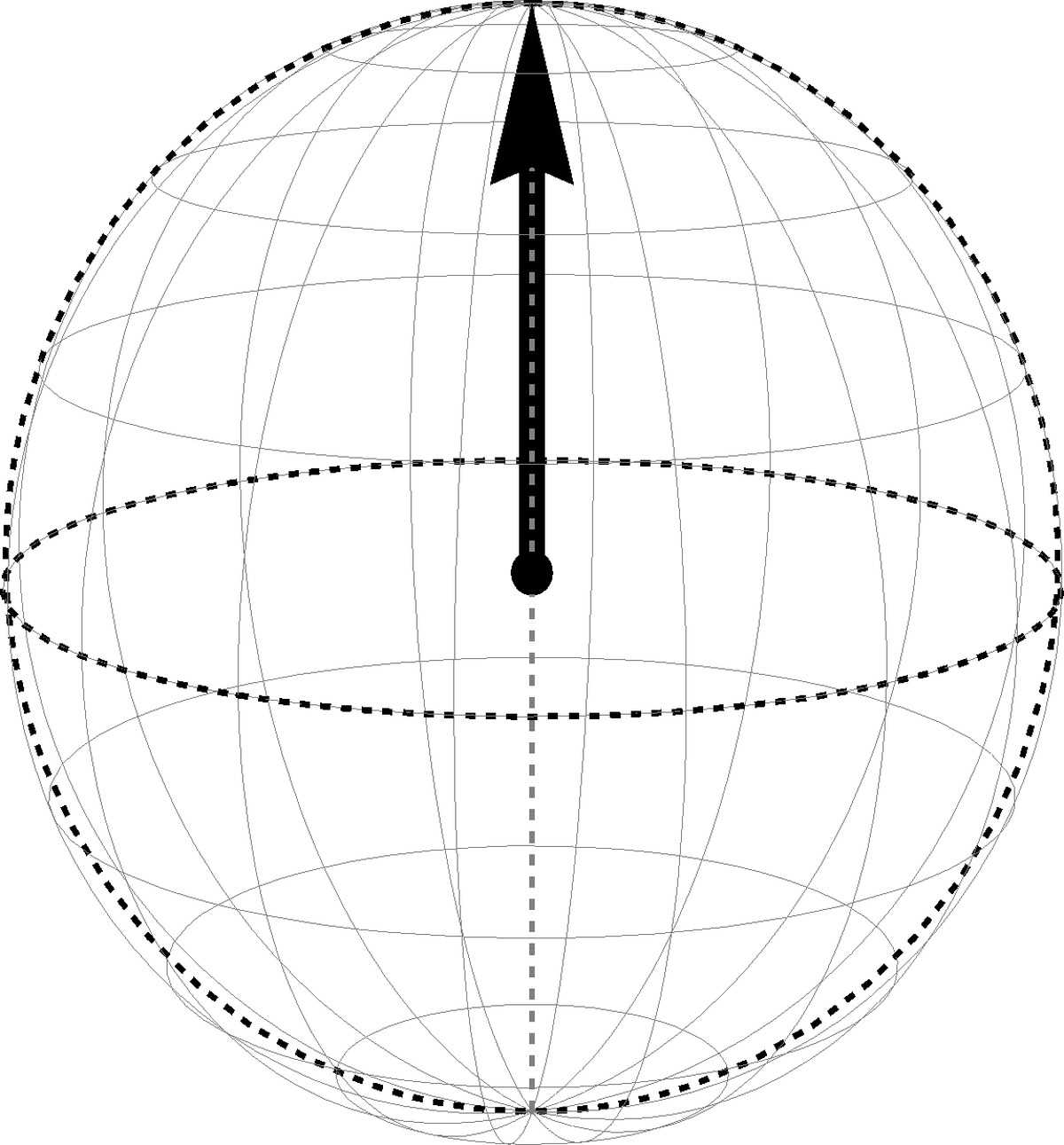} & \includegraphics[width=0.3\columnwidth]{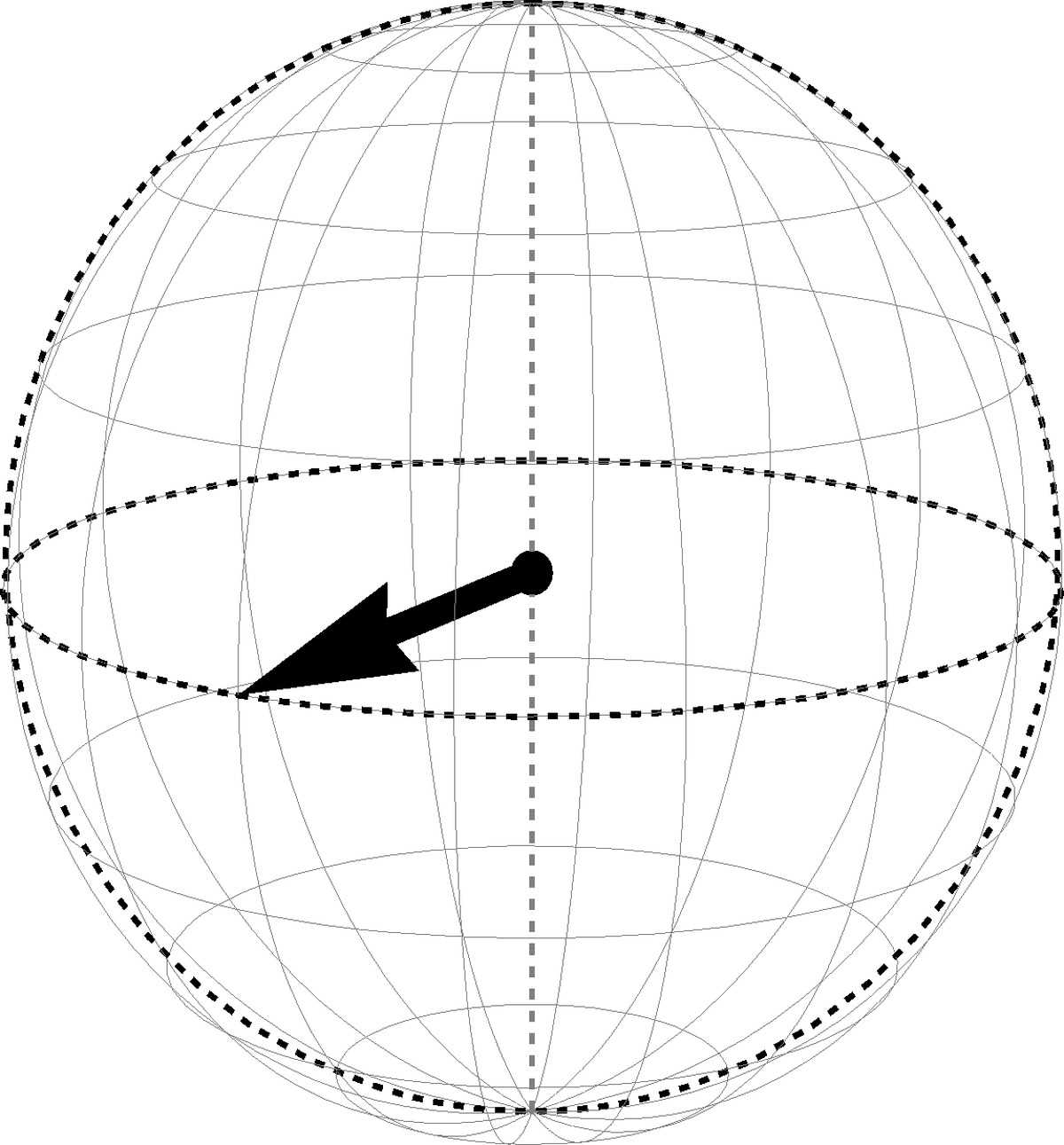} & \includegraphics[width=0.3\columnwidth]{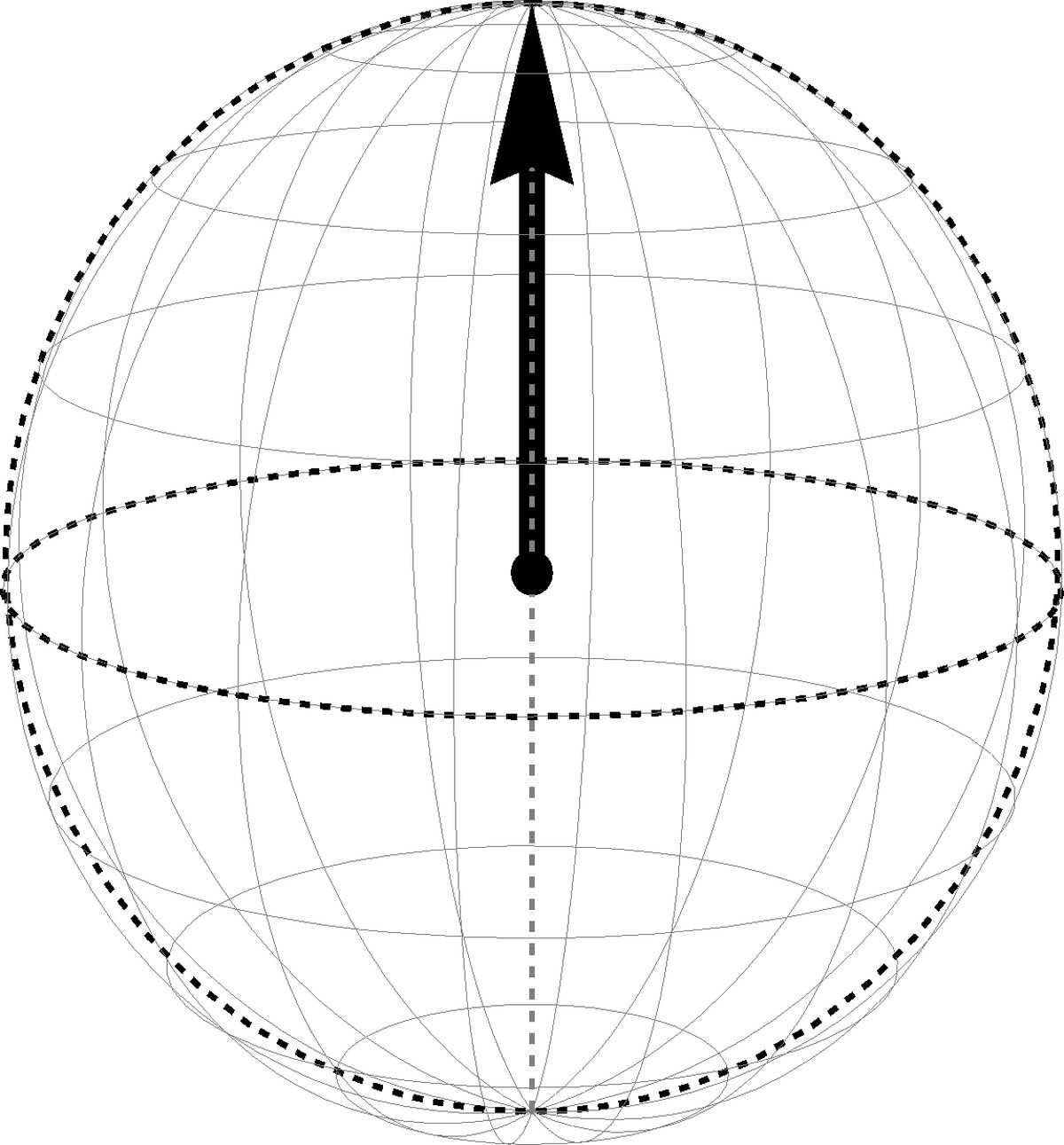}\tabularnewline
(a) Spin BS & (b) Pspin BS & (c) Entanglement BS\tabularnewline
\end{tabular}
\includegraphics[width=\columnwidth]{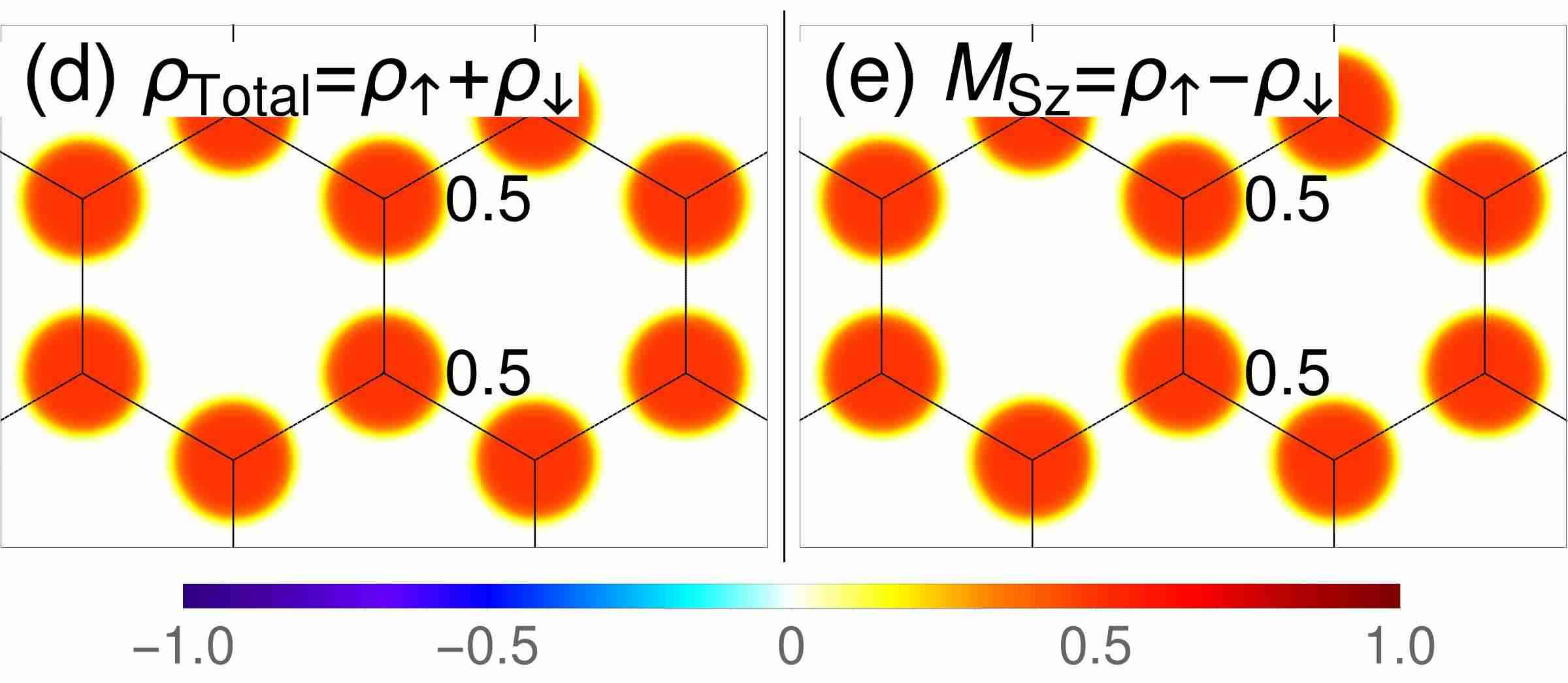}
\caption{\label{fig:QHFM_UEP} Visualization of unentangled easy-plane QHFM state in the same ways as Fig.~\ref{fig:QHFM_UEA}.
}
\end{figure}

\begin{figure}[t]
\begin{tabular}{ccc}
\includegraphics[width=0.3\columnwidth]{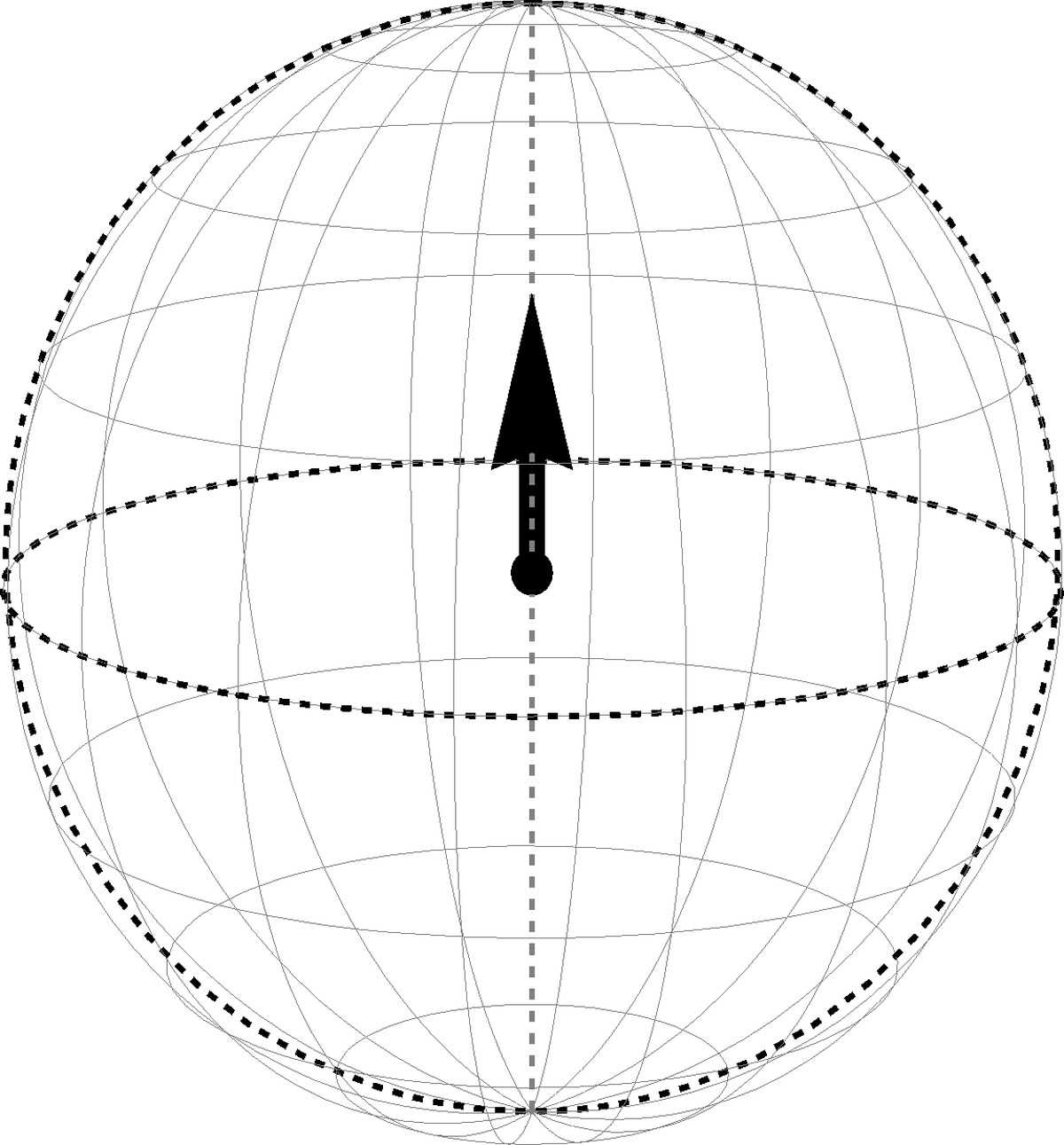} & \includegraphics[width=0.3\columnwidth]{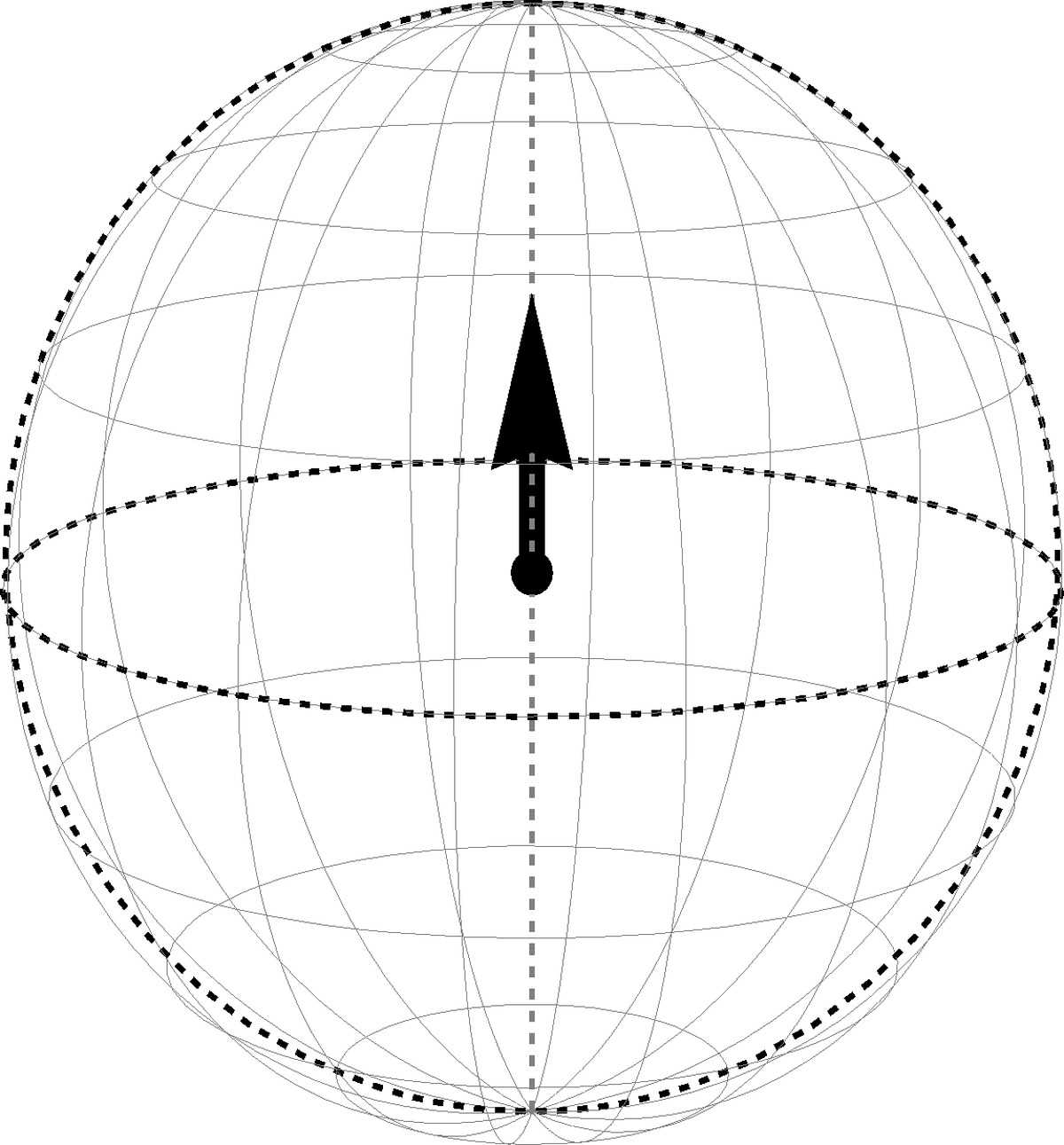} & \includegraphics[width=0.3\columnwidth]{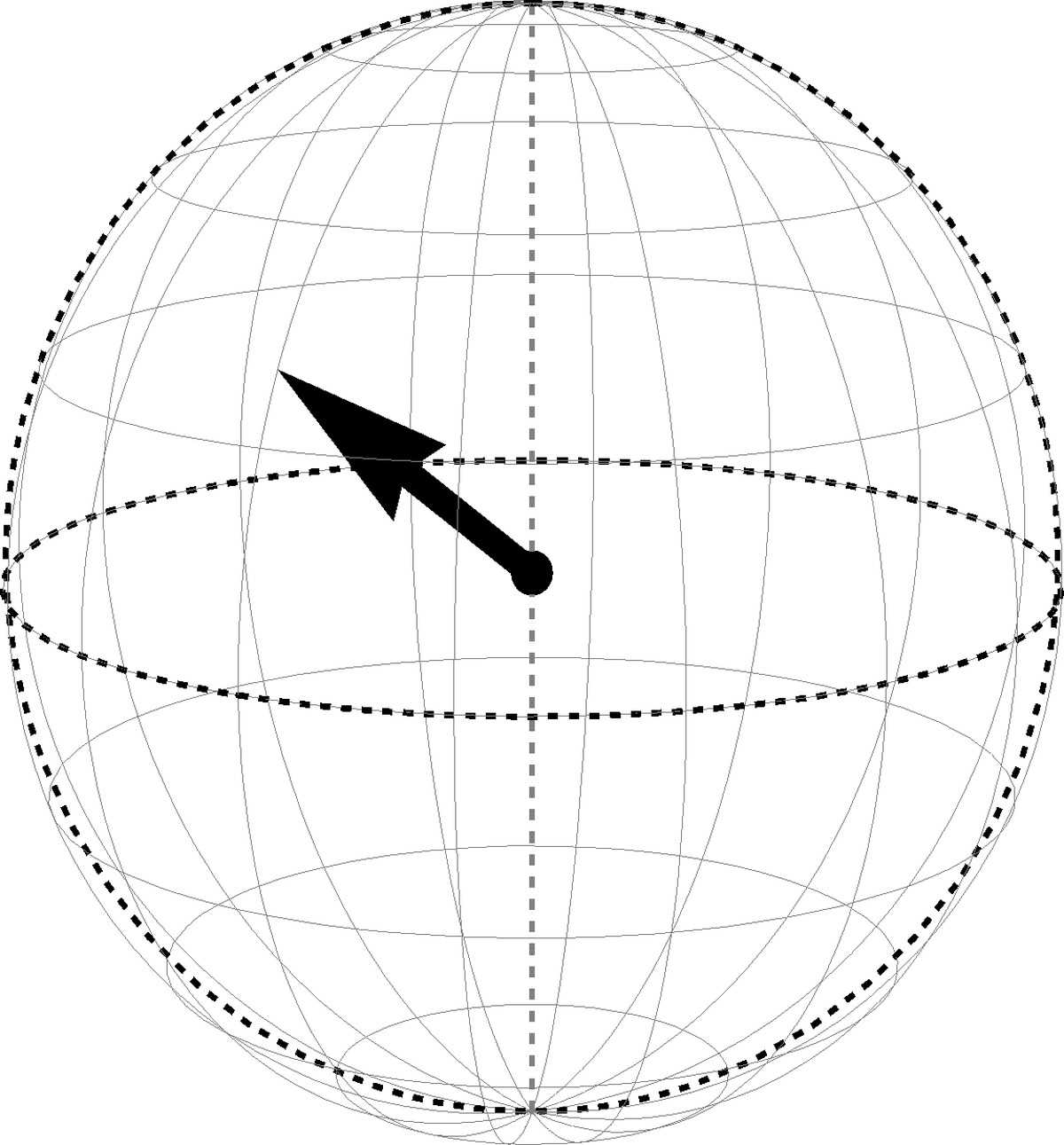}\tabularnewline
(a) Spin BS & (b) Pspin BS & (c) Entanglement BS\tabularnewline
\end{tabular}
\includegraphics[width=\columnwidth]{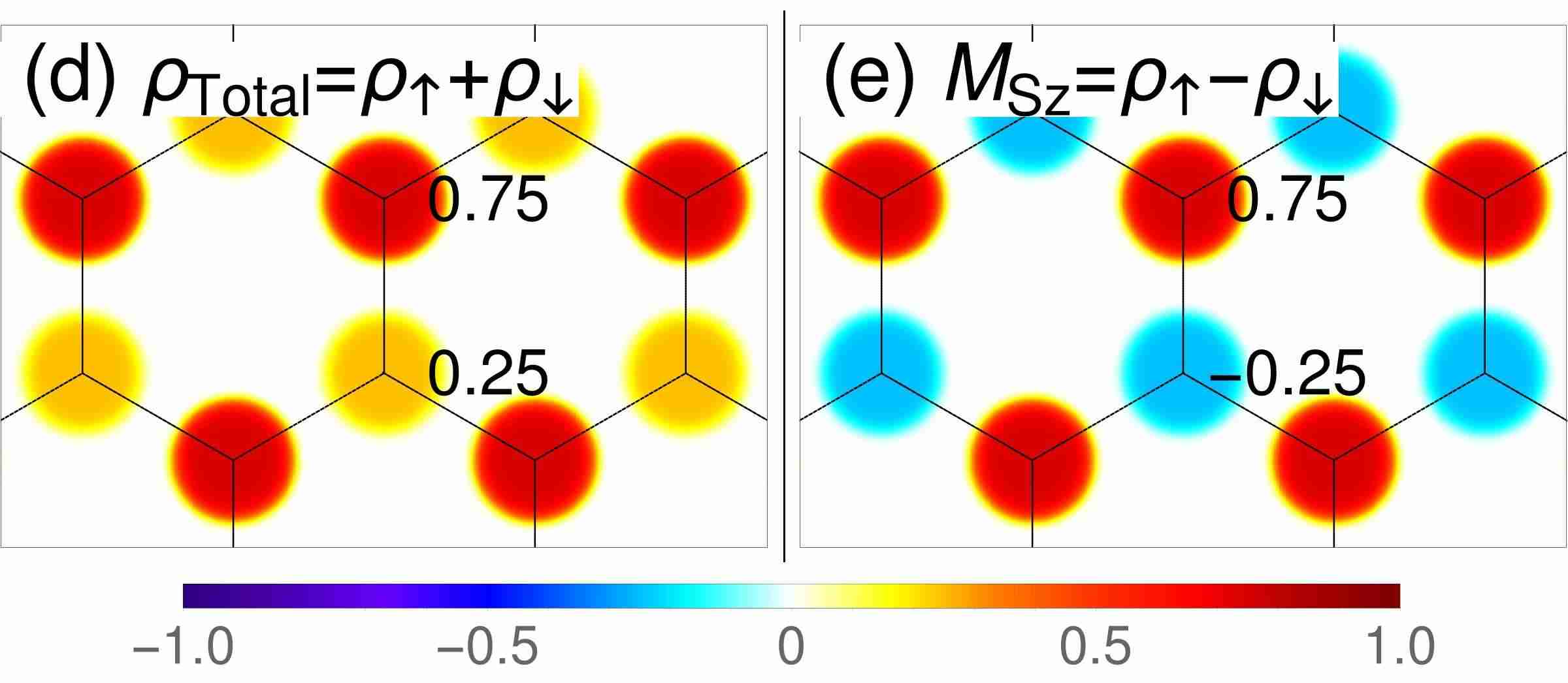}
\caption{\label{fig:QHFM_EEA} Visualization of entangled easy-axis QHFM state in the same ways as Fig.~\ref{fig:QHFM_UEA}.
}
\end{figure}

\begin{figure}[t]
\begin{tabular}{ccc}
\includegraphics[width=0.3\columnwidth]{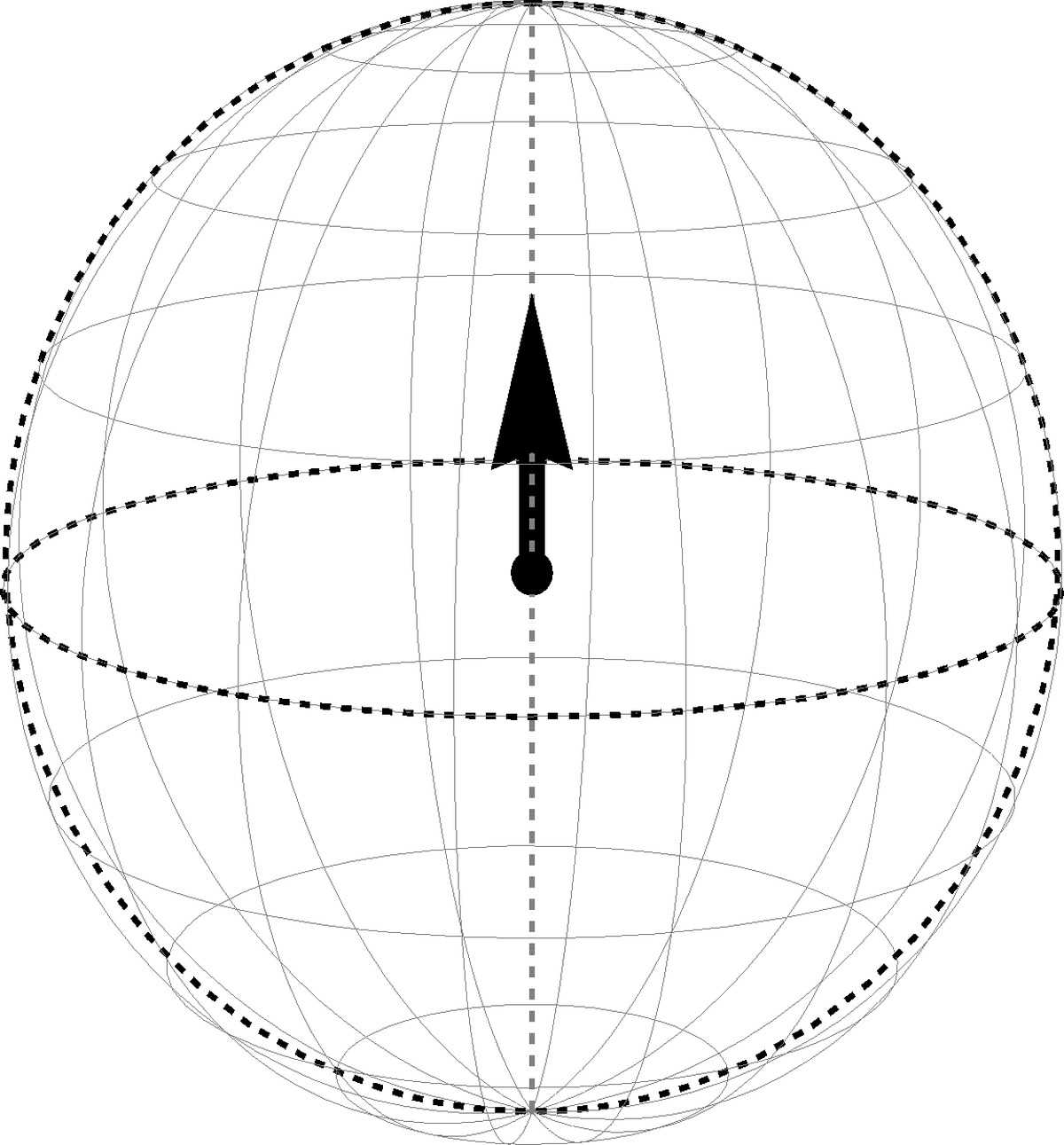} & \includegraphics[width=0.3\columnwidth]{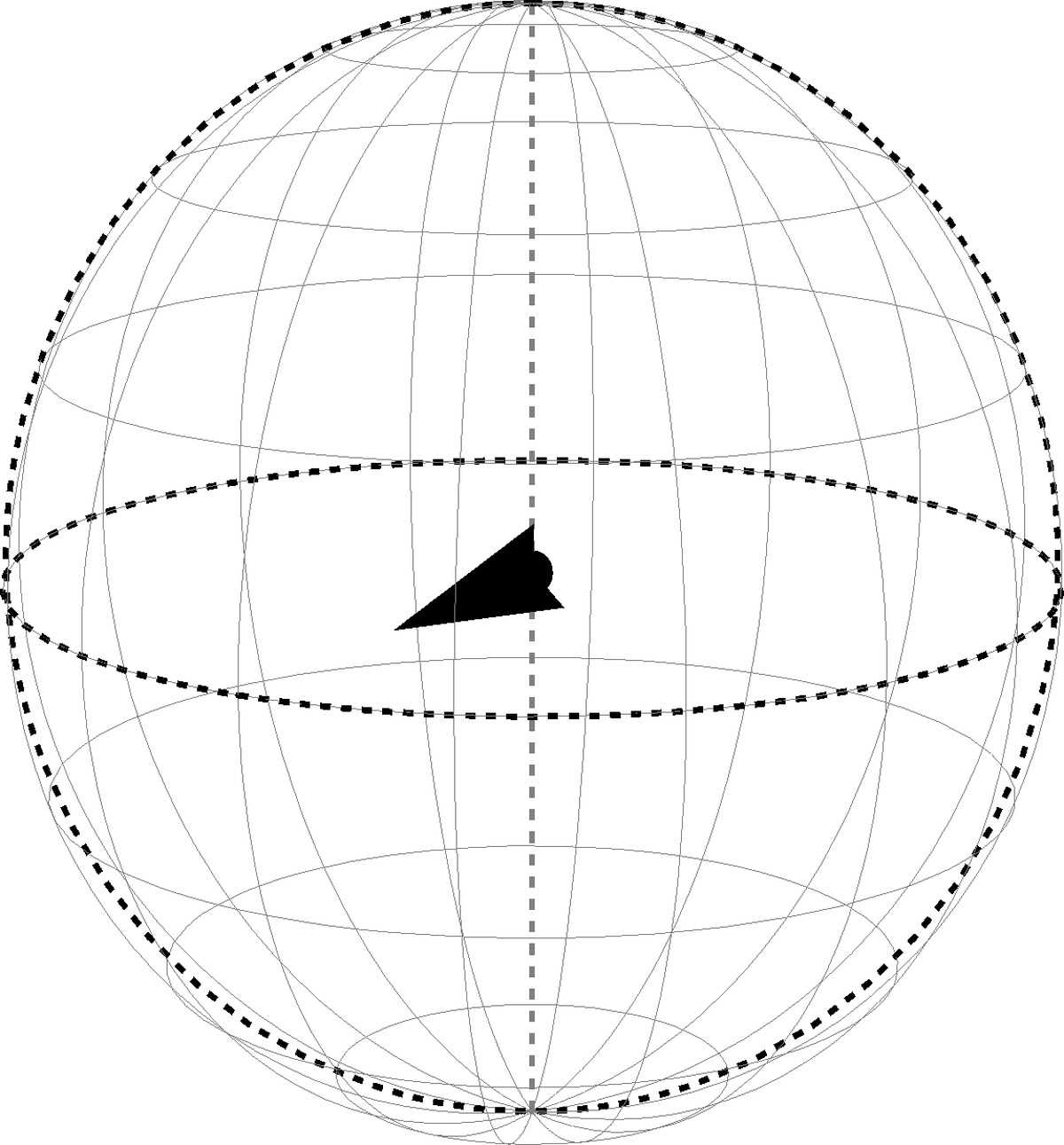} & \includegraphics[width=0.3\columnwidth]{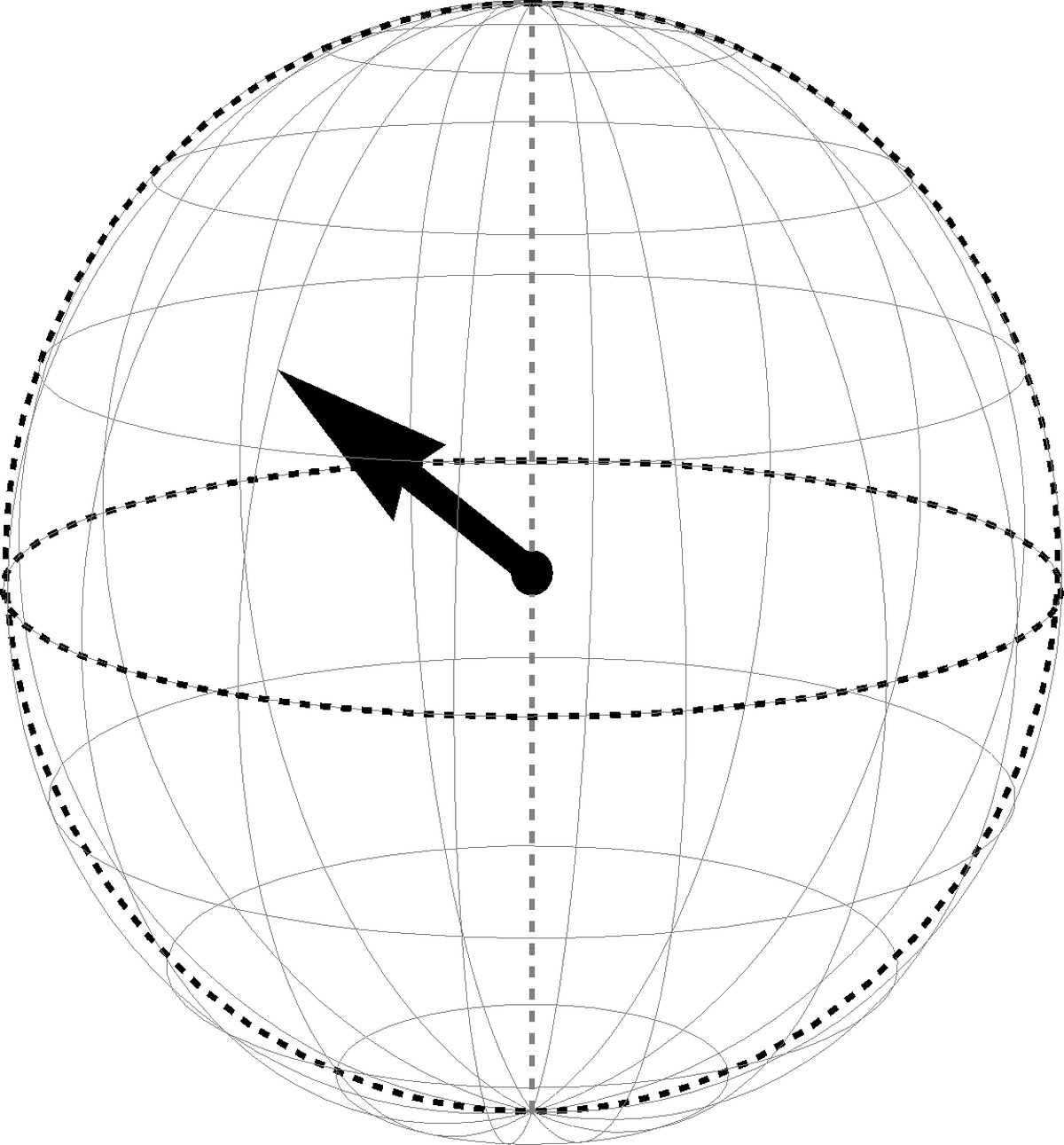}\tabularnewline
(a) Spin BS & (b) Pspin BS & (c) Entanglement BS\tabularnewline
\end{tabular}
\includegraphics[width=\columnwidth]{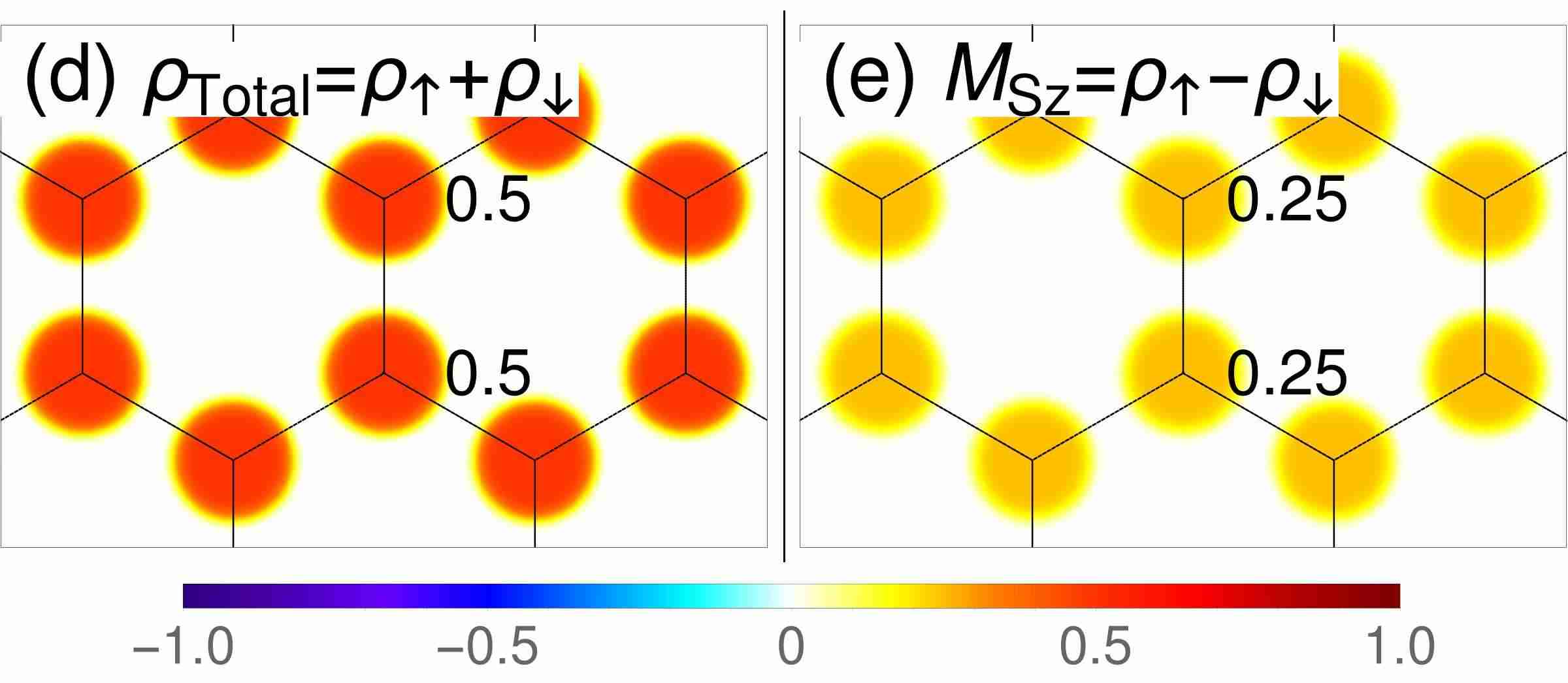}
\caption{\label{fig:QHFM_EEP} Visualization of entangled easy-plane QHFM state in the same ways as Fig.~\ref{fig:QHFM_UEA}.
}
\end{figure}

As a consequence of the identity between valley pseudospin and sublattice index in $N=0$, the different QHFM bear a clear fingerprint in the spin-polarized electronic occupation of the two graphene sublattices.
The four types of QHFM ground states are visualized in Figs.~\ref{fig:QHFM_UEA}, \ref{fig:QHFM_UEP}, \ref{fig:QHFM_EEA} and \ref{fig:QHFM_EEP}. The upper parts (panels a, b and c) in each figure are the Bloch sphere representations for the CP$^{3}$-spinor $F$ of the QHFM ground state. 
In particular, panel (a) shows the spin magnetization $\bM_{\rm S}=F^{\dagger}(1\otimes\boldsymbol{\sigma})F$ (Eq. \ref{eq:Si_Y}) of the CP$^{3}$-spinor $F$ in the spin Bloch sphere. 
For the CP$^{3}$-spinor $F$ corresponding to the unentangled QHFM states, with an easy-plane (Fig.~\ref{fig:QHFM_UEP}) and an easy-axis magnetization (Fig.~\ref{fig:QHFM_UEA}), $\bM_{\rm S}$ is a unit vector and its arrowhead is located \emph{on} the spin Bloch sphere. In contrast, for the CP$^{3}$-spinor $F$ corresponding to the entangled QHFM states, again with an easy-plane (Fig.~\ref{fig:QHFM_EEP}) and an easy-axis magnetization (Fig.~\ref{fig:QHFM_EEA}), the vector $\bM_{\rm S}$ has a magnitude $|\cos\alpha|$ smaller than $1$, and explores the \emph{interior} of the spin Bloch sphere. 
The appearance of the pseudospin magnetization $\bM_{\rm P}=F^{\dagger}(\boldsymbol{\sigma}\otimes 1)F$ in the pseudospin Bloch sphere in panel (b) can be understood similarly. In the case of an easy-axis pseudospin ferromagnetic state that corresponds to the electronic occupation of a single sublattice, the arrow points along the $z$-axis, while a balanced occupation of both sublattices goes along with pseudospin magnetization pointing to the equator of the Bloch sphere. Again, in the presence of spin-pseudospin entanglement, the arrow starts to exploit the inside of the sphere.
Panel (c) shows the \emph{entanglement vector} 
\begin{equation}
\boldsymbol{m}_{\rm E}(\alpha,\beta) = (\sin\alpha\cos\beta,\sin\alpha\sin\beta,\cos\alpha)\label{eq:Entanglement-vector}
\end{equation}
for a CP$^{3}$-spinor $F$ of the QHFM state, where $\alpha$ and $\beta$ are obtained by the parametrization of $F$ with Eq. \ref{eq:parametrizationZ}. The bounding sphere of the entanglement vector is called the \emph{entanglement Bloch sphere} by analogy. For the CP$^{3}$-spinor $F$ corresponding to the unentangled QHFM states, both with easy-plane (Fig.~\ref{fig:QHFM_UEP}) and easy-axis magnetization (Fig.~\ref{fig:QHFM_UEA}), the entanglement vector $\boldsymbol{m}_{\rm E}$ points to the north pole of the entanglement Bloch sphere. Meanwhile, for the CP$^{3}$-spinor $F$ corresponding to the entangled QHFM states (Figs.~\ref{fig:QHFM_EEP} and \ref{fig:QHFM_EEA}), the direction of $\boldsymbol{m}_{\rm E}$ is generic due to a non-vanishing value of $\alpha$. 

The Bloch sphere representation of the $\CP{3}$-spinor is helpful in the later discussions about the $\CP{3}$-field. For a CP$^{3}$-field $Z(\br)$, the collection of the endpoints for each of the three vectors $\bM_{\rm S}(\br)$, $\bM_{\rm P}(\br)$ and $\bmm_{\rm E}(\br)$ at different $\br$ on the $xy$-plane forms a closed surface in/on the corresponding Bloch sphere. In Sec.~\ref{sec:4-types-skyrmion} we will visualize the skyrmions as special configuration of the CP$^{3}$-field on the Bloch spheres by these closed surfaces. Notice that the Bloch-sphere representation, which we have introduced above, is not unique also due to redundancies in the spinor representation. However, it is a natural one in that it keeps track as much as possible of the spin and pseudospin magnetizations, which are experimentally accessible quantities.

In addition to the Bloch sphere representations, we also show the lattice-resolved profiles of the total electron density in the $N=0$ LL $\rho_{\rm Total}$ and the $z$-component of spin magnetization $M_{\rm Sz}$ in the lower parts (panels d and e) of Figs.~\ref{fig:QHFM_UEA}, \ref{fig:QHFM_UEP}, \ref{fig:QHFM_EEA} and \ref{fig:QHFM_EEP}. For a concrete CP$^{3}$-spinor $F$, the total electron density and the $z$-component of the spin magnetization at sublattice $\lambda=A,B$ can be computed as
\begin{eqnarray}
\rho_{\rm Total}(\lambda) &=\rho_{\uparrow}(\lambda)+\rho_{\downarrow}(\lambda)\\
 M_{\rm Sz}(\lambda) &=\rho_{\uparrow}(\lambda)-\rho_{\downarrow}(\lambda)
\end{eqnarray}
where 
\begin{eqnarray}
\rho_{\uparrow}(\lambda=A)=f_1^{*}f_1 ,\,& \rho_{\downarrow}(\lambda=A)=f_2^{*}f_2,\\
\rho_{\uparrow}(\lambda=B)=f_3^{*}f_3 ,\,& \rho_{\downarrow}(\lambda=B)=f_4^{*}f_4.
\end{eqnarray}
We have used the fact that in the $N=0$ LL, the eigenstate for electrons at K(K') valley occupies only A(B) sublattice. The electron density profiles at lattice scale is rendered by convolution of the $\CP{3}$-field with a form factor, which is a superposition of Gaussian functions peaked at different lattice sites. A detailed description of the rendering method is provided in Appendix.~\ref{subsec:Visualization-CP3-skyrmion-on-honeycomb-lattice}.

The unentangled QHFM states are visualized in panels (d) and (e) of Figs.~\ref{fig:QHFM_UEA} and \ref{fig:QHFM_UEP} for unentangled easy-axis and easy-plane pseudospin magnetizations, respectively. In these two cases, the profiles of $\rho_{\rm Total}$ and $M_{\rm Sz}$ are identical, i.e. $\rho_{\downarrow}(A)=\rho_{\downarrow}(B)=0$ and $\rho_{\rm Total}(\lambda)= M_{\rm Sz}(\lambda)=\rho_{\uparrow}(\lambda)$
because the completely polarized spin can be factored out from the CP$^{3}$-spinor $F$. 
Since the pseudospin magnetization $\bM_{\rm P}$ points along the $z$-axis of the pseudospin space, the electrons occupy only one of the A/B sublattices in the $\rho_{\rm Total}$ profile for an unentangled easy-axis QHFM state (Fig.~\ref{fig:QHFM_UEA}(d)). Meanwhile, in the $\rho_{\rm Total}$ profile for an unentangled easy-plane QHFM state (Fig.~\ref{fig:QHFM_UEP}(d)), both sublattices are equally occupied. 

The entangled QHFM states are visualized in panel (d) and (e) of Figs.~\ref{fig:QHFM_EEA} and \ref{fig:QHFM_EEP} for an easy-axis and an easy-plane pseudospin magnetization, respectively. In these two cases, the profiles of $\rho_{\rm Total}$ and $M_{\rm Sz}$ are different in general. For instance, in Fig.~\ref{fig:QHFM_EEA}(d) for the entangled easy-axis QHFM state, one can see the unequal sublattice occupations. Indeed, we have $\rho_{\rm Total}(A)=\cos^2(\alpha/2)$ and $\rho_{\rm Total}(B)=\sin^2(\alpha/2)$, as one obtains directly from the spinor $F=(\cos\alpha/2,0,0,\exp(i\beta)\sin\alpha/2)^T$ in the third block of Tab.~(\ref{tab:FM-tab}). Such sublattice occupation pattern also appears for a CP$^{3}$-spinor $F$ with some pseudospin magnetization of $\theta_{\rm P}\in(0,\pi/2)$. 
To distinguish them, we notice that for the entangled easy-axis QHFM state, the spin magnetizations on two sublattices have opposite directions and different magnitudes, which resembles an \textit{anti-ferrimagnetic} pattern and is shown in Fig.~\ref{fig:QHFM_EEA}(e). The pattern would be fully anti-ferromagnetic in the absence of the Zeeman coupling, but due to the latter, there remains a non-zero spin polarization, whence the term ``anti-ferrimagnet'' -- it is similar to the canted anti-ferromagnetic states discussed in the framework of QHFM at $\nu=0$.\cite{Kharitonov2012} 
This is due to the superposition of two the basis states 
$\psi^{\rm P}\otimes \psi^{\rm S}$ and $\chi^{\rm P}\otimes \chi^{\rm S}$
with opposite spin and pseudospin in Eq.~(\ref{eq:parametrizationZ}) at generic values of $\alpha$. Such anti-ferrimagnetic patterns appear also in entanglement $\CP{3}$-skyrmions as discussed in Sec.~\ref{subsec:entanglement-skyrmion}. 
The $\rho_{\rm Total}$ and $M_{\rm Sz}$ profiles are also different in the case of the entangled easy-plane QHFM [Fig.~\ref{fig:QHFM_EEP}(d)(e)]. The $\rho_{\rm Total}$ profile has the same appearance as the unentangled easy-plane QHFM -- both sublattices are equally occupied. However, the $M_{\rm Sz}$ profile shows the equal but diminished magnitudes at the two sublattices. As we have discussed earlier, the spin magnetization diminishes its magnitude in order to lower the pseudospin contribution of the anisotropic energy and thus achieve an overall minimization of both parts so that we do no longer have $\rho_{\rm Total}=M_{\rm Sz}$ as in the unentangled case. 

The above discussions show that the four types of QHFM ground states are clearly distinguished by the lattice profiles of $\rho_{\rm Total}$ and $M_{\rm Sz}$. Besides, the profiles have common features among the four types of QHFM ground states. First, due to the normalization of the CP$^{3}$-spinor $F$, in the the $\rho_{\rm Total}$ profile of $F$ the relation $\rho_{\rm Total}(A)+\rho_{\rm Total}(B)=1$ always holds. Second, in the $M_{\rm Sz}$ profile of $F$ representing an unentangled CP$^{3}$-spinor, the signs of $M_{\rm Sz}(A)$ and $M_{\rm Sz}(B)$ are always the same. However, the reverse is not true (for instance, consider the $M_{\rm Sz}$ profile for the entangled easy-plane QHFM in Fig.~\ref{fig:QHFM_EEP}). 

The $\rho_{\rm Total}$ and $M_{\rm Sz}$ profiles of the QHFM states are helpful also in the following discussion of the various $\CP{3}$-skyrmions represented by the position-dependent field $Z(\br)$. Since these fields vary slowly on the lattice scale, the lattice-resolved profiles of $\rho_{\rm Total}$ and $M_{\rm Sz}$ are recovered in the vicinity of a point $\br=\br_0$ in the $xy$-plane, i.e. the state represented by $Z(r)$ is locally ferromagnetic. In Sec.~\ref{sec:4-types-skyrmion} we will examine the local structures of the $\rho_{\rm Total}$ and $M_{\rm Sz}$ profiles for a CP$^{3}$-field $Z(\br)$ for skyrmions.

\section{CP$^{3}$-Skyrmion of charge ${\cal Q}=1$}
\label{sec:CP3-Skyrmion-of-charge-Q-1}
In this section we study the smoothly varying CP$^{3}$-field $Z(\br)$, which describes a skyrmion, i.e. a ``locally ferromagnetic'', charge-carrying spin-pseudospin texture.\cite{Sondhi1993,Moon1995,Arovas1999,Yang2006,Ezawa2003,Ezawa2005} Because the second homotopy group of $\CP{3}$ is $\pi_2(\CP{3})=\mathbb{Z}$ (see App.~\ref{subsec:second-homotopy}), these textures are described by an integer topological charge. 
For the field $Z(\br)$, we use the same parametrization as for the QHFM states in Eq.~(\ref{eq:parametrizationZ}), but now in terms of position-dependent angles $\theta_{\rm S}, \phi_{\rm S}, \theta_{\rm P}, \phi_{\rm P}, \alpha$, and $\beta$. Again, we omit the position dependence in the symbols to simplify the notations. 
The energy functional for the CP$^{3}$-field proposed in Refs.~[\onlinecite{Arovas1999}, \onlinecite{Ezawa2005}] contains the non-linear sigma model energy, which supports the skyrmionic configurations of $Z(\br)$, and the Coulomb energy of topological charge density, which tends to enlarge the skyrmion. The Coulomb energy is counterbalanced by the anisotropic energy, which is locally identical to the anisotropic energy proposed in Sec.~\ref{subsec:Anisotropic-energy}. 

The energy functional for $Z(\br)$ is presented in Sec.~\ref{subsec:Energy-functional}. With the help of the skyrmion Ansatz discussed in Sec.~\ref{subsec:Skyrmion-Ansatz} and Sec.~\ref{subsec:Radial-deformation}, we minimize the energy of a skyrmion and present the result in Sec.~\ref{subsec:Minimization-results}.

\subsection{Energy functional}
\label{subsec:Energy-functional}
The CP$^{3}$-field $Z(\br)$ describes the 
skyrmions in the $N=0$ LL of graphene.\cite{Moon1995,Arovas1999,Yang2006} 
The wave function ansatz $\left|\psi[Z(\br)]\right\rangle $ is constructed following Ref.~[\onlinecite{Arovas1999}]. 
Then the static energy functional $$E[Z]=\left\langle \psi[Z]\right|H_{\rm C}+H_{\rm A}\left|\psi[Z]\right\rangle $$ is the average of wave function ansatz on the Hamiltonian of the system. 

The gradient expansion of $E[Z]$ to leading order yields an expression in the form of the non-linear sigma model\cite{Moon1995,Arovas1999}
\begin{equation}
E_{\rm NLSM}[Z] =  2\rho_s\int d^2r \bD Z^{\dagger}(\br)\cdot \bD Z(\br),\label{eq:EnNLSM}
\end{equation}
where we defined the covariant derivative
$\bD Z = \nabla Z(\br) - [Z^{\dagger}(\br)\nabla Z(\br)] Z(\br)$. The spin stiffness $\rho_{\rm s}=e^{2}/16\sqrt{2\pi}\epsilon l_{B}$ measures the coupling strength of the neighbouring local magnetic moments, and it is consistent with the single-mode approximation analysis of spin waves at small wave vectors.\cite{Yang2006}

The next order in the gradient expansion of $E[Z]$ is the interaction energy of a topological charge density\cite{Moon1995,Arovas1999} 
\begin{equation}
E_{\rm C}[Z]=\frac{1}{2}\int d^{2}r\int d^{2}r'\rho_{\rm topo}(\br)V(\br-\br')\rho_{\rm topo}(\br'),\label{eq:EnInteraction}
\end{equation}
where the Coulomb potential is $V(r)=e^{2}/4\pi\epsilon r$, and the topological charge density $\rho_{\rm topo}$ is defined as
\begin{equation}
\rho_{\rm topo}(\br)= \frac{1}{2\pi i}[\bD Z(\br)^{\dagger} \times \bD Z(\br)]_z . \label{eq:TopologicalDensity}
\end{equation}

The quantity $\rho_{\rm topo}$ is essential for the understanding of quantum Hall skyrmions. 
On the one hand, the configuration space for the finite energy CP$^{3}$-fields $Z(\br)$ is a disjoint union of topological sectors ${\cal C}_{{\cal Q}}$ labelled by the topological charge ${\cal Q}=\int d^{2}r\rho_{\rm topo}(\br)$ (see App.~\ref{subsec:CP3-spinor-and-NLSM}.) 
On the other hand, the topological charge density $\rho_{\rm topo}$ is equal to the excess electron density $\delta\rho_{\rm el}$ at integer filling of the Landau sublevels.\cite{Sondhi1993,Moon1995,Arovas1999} The spin-pseudospin texture state is charged, and $E_{\rm C}$ can be understood as its Coulomb energy.

In contrast to the gradient expansion terms $E_{\rm NLSM}$ and $E_{\rm C}$, which are invariant under a global $\SU{4}$ transformation of the $\CP{3}$-field $Z$, the anisotropic energy $E_{\rm A}[Z]$ does not contain any gradients of $Z(\br)$ and prefers particular directions of the spin and pseudospin magnetizations. We assume that the system is locally ferromagnetic at large length scale $\Lambda=\left\Vert \nabla Z\right\Vert ^{-1}\gg l_{B}$, and use the same form for the anisotropic energy of smooth fields $Z(\br)$ as for the homogeneous FM states in Eq.~(\ref{eq:EnA}),
\begin{equation}
E_{\rm A}[Z]=\frac{\Delta_{\rm Z}}{2}\int\frac{d^{2}r}{2\pi l_{B}^{2}}\left[u_{\perp}(M_{\rm Px}^{2}+M_{\rm Py}^{2})+u_{\rm z}M_{\rm Pz}^{2}-M_{\rm Sz}\right].\label{eq:EnAnisotropic}
\end{equation}
Here, the spin and pseudospin magnetizations are now local quantities
\begin{eqnarray}
\bM_S(\br) &= Z^{\dagger}(\br)(1\otimes\boldsymbol{\sigma})Z(\br)\label{eq:Si_Z}\\
\bM_{\rm P}(\br) &= Z^{\dagger}(\br)(\boldsymbol{\sigma}\otimes1)Z(\br)\label{eq:Ti_Z}
\end{eqnarray}
expressed inn terms  of the $\CP{3}$-field $Z(\br)$ and
replace $\left\langle S_{z}\right\rangle$ and $\left\langle P_{\rm z}\right\rangle$ in Eq.~(\ref{eq:EnA}). The coefficients $\Delta_{\rm Z}$, $u_{\perp}$, $u_{\rm z}$ remain unchanged.

In this work, we use the magnetic length $l_{B}$ as the unit of length. In the rest of the article, the coordinate $\br$ in the $xy$-plane, the radius $r$, and the $\lambda_0$ parameter are thus dimensionless, and we use $e^{2}/4\pi\epsilon l_{B}$ as the unit of energy. 
For graphene on h-BN substrate, we have $\epsilon=\varepsilon_0 \varepsilon_r$ with $\varepsilon_r=5.5$.\cite{goerbigRev}
After rescaling of length and energy in the above units respectively, we have $\rho_{\rm s}=\sqrt{\pi/32}$ in Eq.~(\ref{eq:EnNLSM}), $V(r)=1/r$ in Eq.~(\ref{eq:EnInteraction}), and Eq.~(\ref{eq:EnAnisotropic}) becomes $\delta_{\rm Z}\int d^{2}r(...)$ with 
\begin{eqnarray}
\delta_{\rm Z} &=\frac{1}{2\pi}\frac{\Delta_{\rm Z}/2}{e^{2}/4\pi\epsilon l_{B}} = 9.04\times 10^{-4} B_{\rm T}\mbox{[T]}/\sqrt{B_{\perp}\mbox{[T]}}\nonumber\\
&=9.04\times 10^{-4} \sqrt{B_{\perp}\mbox{[T]}}/\cos\varphi.\label{eq:delta_Z-B}
\end{eqnarray}
where $B_{\perp}=B_{\rm T} \cos\varphi$ is the perpendicular component of the total magnetic field $B_{\rm T}$ when the sample is tiled away from an upright position by angle $\varphi$. 
The above numerical value indicates that the anisotropic energy $\left\langle H_{\rm A}\right\rangle$ is much smaller than the Coulomb interaction energy $\left\langle H_{\rm C}\right\rangle$ under strong magnetic field. The relative energy scales of $E_{\rm NLSM}$ and $E_{\rm C}$ are characterized by the order of gradient expansion, i.e. $E_{\rm NLSM}\propto(l_{B}/\Lambda)^{2}$ and $E_{\rm C}\propto(l_{B}/\Lambda)^{4}$ where $\Lambda=\left\Vert \nabla Z\right\Vert^{-1}\gg l_{B}$ characterizes the length scale of the spatial variation of the field $Z$. Therefore we first discuss the minimization of the largest part $E_{\rm NLSM}[Z]$. 

\subsection{Skyrmion Ansatz}
\label{subsec:Skyrmion-Ansatz}
The CP$^{3}$-non-linear sigma model is briefly reviewed in Appendix.~\ref{subsec:CP3-spinor-and-NLSM}. As mentioned there, each field configuration $Z(r)$ is associated with a unique topological index ${\cal Q}$ and falls into the corresponding topological sector ${\cal C}_{{\cal Q}}$. Two configurations from different topological sectors cannot be connected by continuous deformation. Therefore, in search of the minimal field configuration of $E[Z]$, the variational analysis is performed only within each topological sector. We are particularly interested in the configuration $Z_{\rm sk}(\br)\in{\cal C}_{1}$ which minimizes $E_{\rm NLSM}[Z]$ to the value $4\pi\rho_{\rm S}$ in the topological sector ${\cal C}_{1}$. The (scale-invariant) CP$^{3}$-skyrmion of charge ${\cal Q}=1$ is a holomorphic polynomial of degree one (see App.~\ref{subsec:Energy-minimizing-solutions-in-each-topological-sector}),
\begin{equation}
Z_{\rm sk}(x,y) = {\cal N}(r)^{-1} [ (x+iy)F-\lambda_0 C ]\label{eq:Z1},
\end{equation}
with $r=|\br|$. Here, $F$ and $C\in\mathbb{C}^{4}$ are normalized
CP$^{3}$-spinors, and one has 
\begin{equation}\label{eq:spinorOrth}
F^{\dagger}C=0.
\end{equation}
The parameter $\lambda_0$ is chosen to be real because a phase factor could be combined with $C$. The normalization factor is denoted as ${\cal N} = \sqrt{x^{2}+y^{2}+\lambda_0^2}$. In the rest of this paper, the term CP$^{3}$-skyrmion always refers to the case of a CP$^{3}$-skyrmion carrying topological charge ${\cal Q}=1$. By making such an ansatz, we reduce the functional minimization to a minimization problem of several real parameters.

We make several technical comments on the above field configuration in order to display how we determine the parameters for a skyrmion from energy minimization.

\paragraph*{(i)} $Z_{\rm sk}$ is an interpolation between $F$ and $C$ along the radius $r$. Far from the center of a CP$^{3}$-skyrmion, i.e. $r=\sqrt{x^{2}+y^{2}}\gg\lambda_0$, $Z_{\rm sk}$ approaches $e^{i\mbox{arg}(x+iy)}F\sim F$, which is the ground state of the host QHFM.
$F$ may thus be viewed as the \textit{FM background spinor}. 
 The phase factor $e^{i\mbox{arg}(x+iy)}$ of $Z_{\rm sk}$ at large radius has negligible contribution to the NLSM energy density $\bD Z_{\rm sk}^{\dagger}(\br)\cdot \bD Z_{\rm sk}(\br)$. In this sense, a skyrmion described by $Z_{\rm sk}$ can be viewed as a localized texture, which is embedded into the ferromagnetic background.

\paragraph*{(ii)} The \emph{center spinor} $C=-Z_{\rm sk}(\br=0)$ is parametrized with respect to the same basis as in the parametrization of $F$. The meaning of the condition (\ref{eq:spinorOrth}) is now apparent : the center spinor is orthogonal to the FM background spinor identified by the above energy minimization which leads to the FM phase diagram in Fig. \ref{fig:FM-region}. Within our variational approach, Eq.~(\ref{eq:spinorOrth}) represents a constraint reducing the set of free parameters, with respect to which we minimize the skyrmion energy, and we may thus write
\begin{eqnarray}\label{eq:Para_C}
C = & \cos\theta_C\left[ -e^{-i\beta}\sin\frac{\alpha}{2}\psi_{\rm P}\otimes\psi_{\rm S} +\cos\frac{\alpha}{2}\chi_{\rm P}\otimes\chi_{\rm S} \right]\nonumber\\
+ & \sin\theta_C\left[ e^{i \varphi_A}\cos\phi_C\psi_{\rm P}\otimes\chi_{\rm S} +e^{i \varphi_B}\sin\phi_C\chi_{\rm P}\otimes\psi_{\rm S} \right].\nonumber\\
\end{eqnarray}
The entanglement parameters $\alpha$ and $\beta$ are inherited from $F$, but $C$ contains 4 more parameters $\phi_C$, $\theta_C$, $\varphi_A$ and $\varphi_B$ in addition to the six fixed angles in $F$. The center spinor $C$ and FM background spinor $F$ determine the various types of skyrmions, which are discussed in Sec.~\ref{sec:4-types-skyrmion}. 

\paragraph*{(iii)} The scaling invariance $E_{\rm NLSM}[Z]=E_{\rm NLSM}[Z(\alpha\br)]$ is manifest for $Z_{\rm sk}$, since we have $E_{\rm NLSM}[Z_{\rm sk}]=4\pi\rho_{\rm S}$ independent of $\lambda_0$, and $Z_{\rm sk}(\br)$ with parameter $\lambda_0$ is equivalent to $Z_{\rm sk}(\gamma\br)$ with parameter $\gamma\lambda_0$. To emphasize such property, we call $Z_{\rm sk}$ the \emph{scale-invariant} skyrmion. 

\paragraph*{(iv)} The parameter $\lambda_0$ is the width of the bell-shaped topological charge density
\begin{equation}
\rho_{\rm topo}(\br)=\frac{\lambda_0^{2}}{\pi(r^{2}+\lambda_0^{2})^{2}}\label{eq:rho_Topo_Z1}
\end{equation}
for a skyrmion $Z_{\rm sk}$ and can be viewed as the \emph{size} of the skyrmion. Notice that the length unit is $l_B$, and $\lambda_0$, $x$, $y$, $r$ are dimensionless. Here we perform another rescaling $\br\rightarrow \lambda_0 \br$ to extract the $\lambda_0$-dependence in the three components of $E[Z_{\rm sk}]$ :
\begin{equation}
E'[Z_{\rm sk}]=E_{\rm NLSM}[\tilde{Z}_{\rm sk}]+\frac{1}{\lambda_0} E_{\rm C}[\tilde{Z}_{\rm sk}] + \lambda_0^2 E'_{\rm A}[\tilde{Z}_{\rm sk}]\label{eq:lambda_dep}
\end{equation}
where $\tilde{Z}_{\rm sk}:=\left.Z_{\rm sk}\right|_{\lambda_0=1}$, and the prime on $E$ and $E_{\rm A}$ will be discussed in the next paragraph. The CP$^{3}$-skyrmion is stabilized by the competition between the second and the third terms in the above equation. 

The skyrmion ansatz Eq.~(\ref{eq:Z1}) cannot be used directly for energy minimization. 
Prior to each minimization for a given set of parameters $\delta_{\rm Z}$, $u_{\perp}$ and $u_{\rm z}$, the corresponding FM background spinor $F$ is obtained by minimizing solely the anisotropic energy $E_{\rm A}[F]$, which is then subtracted from the energy of the system. 
This is because the skyrmions are embedded into the ferromagnetic background, and a large portion of the $xy$-plane far from the skyrmion center is still in the QHFM ground state in presence of a single skyrmion. 
In fact, we are minimizing the \emph{excess energy} 
\begin{equation}
E'[Z_{\rm sk}] = E[Z_{\rm sk}] - E_{\rm A}[F]\label{eq:sub-EAF}
\end{equation}
of a skyrmion state in the vicinity of quarter filling of the $N=0$ LL, with respect to the corresponding QHFM ground state in which the skyrmion is embedded. The term $E'[Z_{\rm sk}]$ and  $E'_{\rm A}[\tilde{Z}_{\rm sk}]$ in Eq.~(\ref{eq:lambda_dep}) are shorthand notations for the excess energy $E[Z_{\rm sk}] - E_{\rm A}[F]$ and $E_{\rm A}[\tilde{Z}_{\rm sk}]-E_{\rm A}[F]$ respectively. 
Such energy difference is bounded from below because a scale-invariant skyrmion of topological charge ${\cal Q}=1$ comes along with a minimal energy cost of $4\pi\rho_s$, and it is increased by the amount of $E[Z_{\rm sk}]-E_{\rm A}[F]-4\pi\rho_s$ in the presence of the Coulomb interaction energy $E_{\rm C}$ and anisotropic energy $E_{\rm A}$. Minimization of the energy difference results in stable skyrmions. 

\subsection{Radial deformation}
\label{subsec:Radial-deformation}
Before presenting our results of the aforementioned energy minimization, we discuss here a subtle issue concerning a necessary deformation of the scale-invariant skyrmion $Z_{\rm sk}$ in Eq.~(\ref{eq:Z1}). 
In practice, it has a \emph{logarithmically divergent} anisotropic energy $E'_{\rm A}[Z_{\rm sk}]=E_{\rm A}[Z_{\rm sk}]-E_{\rm A}[F]$ with respect to the FM background. 
However, a skyrmion is expected to have finite energy because the texture is localized and non-singular. 
The apparent contradiction is solved by allowing the skyrmion Ansatz to slightly deviate from the scale-invariant skyrmion $Z_{\rm sk}$, since it is the optimal configuration only for the NLSM part $E_{\rm NLSM}$ of the energy of skyrmion. 
The radial deformation is achieved by replacing the the size parameter $\lambda_0$ in the scale-invariant skyrmion in Eq.~(\ref{eq:Z1}) by the following $r$-dependent function $\lambda(r)$ to ``shrink'' the skyrmion:
\begin{equation}
\lambda(r)=\lambda_0 \exp(-\frac{r^2}{\kappa \lambda_0^2}),\label{eq:lambda_r}
\end{equation}
so that the divergence in $E_{\rm A}$ is controlled by the parameter $\kappa$ (see App.~\ref{subsec:kappa-dependence-of-energy-for-deformed-skyrmions}). This leads to the radially deformed skyrmion Ansatz 
\begin{equation}
\check{Z}_{\rm sk}(x,y) = {\cal N}(r)^{-1} [ (x+iy)F-\lambda(r) C ]\label{eq:Zdeformed},
\end{equation}
which is obtained by replacing $\lambda_0$ by $\lambda(r)$ in Eq.~(\ref{eq:Z1}). 
We put the factor $1/\lambda^2_0$ to make it convenient to extract $\lambda_0$ from the expression of $E_{\rm C}[\check{Z}_{\rm sk}]$ and $E_{\rm A}[\check{Z}_{\rm sk}]$ by a redefinition $\br/\lambda_0\rightarrow\br$, which is in the same spirit of Eq.~(\ref{eq:lambda_dep}). Notice that we still use $l_B$ as the length unit during the whole discussion, and $\lambda_0$, $\br$ are dimensionless.

At $\lambda_0 = 1$ and $\kappa\rightarrow\infty$, $E_{\rm NSLM}[\check{Z}_{\rm sk}]$ and $E_{\rm C}[\check{Z}_{\rm sk}]$ decrease monotonically to the value $(4\pi\sqrt{\pi/32})(e^{2}/4\pi\epsilon l_{B})$ and $(3\pi^2/64)(e^{2}/4\pi\epsilon l_{B})$ respectively, which are consistent with the corresponding energy values for the scale-invariant skyrmion. Meanwhile, the anisotropic energy has a linear dependence on $\log\kappa$. At $\lambda_0=1$ and finite $\kappa$, the price to pay for the control of divergence in the anisotropic energy is a slight increase in the NLSM energy and the interaction energy from the $\kappa\rightarrow\infty$ limit. More details can be found in App.~\ref{subsec:kappa-dependence-of-energy-for-deformed-skyrmions}. 

The radial deformation makes the components of the CP$^{3}$-field deviate from holomorphic functions of $z=x+i y$. Notice that this deviation from holomorphic functions is not in contradiction with the lowest-LL condition, which requires the electronic wave functions to be analytic. Indeed, the CP$^{3}$-field $\check{Z}_{\rm sk}(\br)$ is an envelope function over these LL wave-functions $\phi_m(z)$ and does not need itself to be analytic. The holomorphic solution from minimization of $E_{\rm NLSM}[Z]$ in Eq.~(\ref{eq:EnNLSM}) or Eq.~(\ref{eq:NLSM-general}) and the holomorphic Landau wave function $\phi_m(z)$ have different origins. 

We are now armed to minimize the skyrmion energy $E[\check{Z}_{\rm sk}]-E_{\rm A}[F]$ for a  given set of parameters $u_{\perp}$, $u_{\rm z}$ and $\delta_{\rm Z}$. Using the radially deformed Ansatz $\check{Z}_{\rm sk}$ in Eq.~(\ref{eq:Zdeformed}) for the CP$^{3}$-skyrmion, which contains the size parameter $\lambda_0$, the deformation parameter $\kappa$, and $\phi_C$, $\theta_C$, $\varphi_A$ and $\varphi_B$ in the center spinor $C$. The FM background spinor $F$ is determined by minimization of $E_{\rm A}[F]$. 

\subsection{Minimization results}
\label{subsec:Minimization-results}
\begin{table}[b]
\caption{\label{tab:skyrmion-types}Summary of various types of optimal CP$^{3}$-skyrmions. From the parametrization of the center spinor $C$ in Eq.~(\ref{eq:Para_C}), we notice that when $\theta_C=0$ or $\pi$ (the case for ${\cal X}=0$), the value of $\phi_C$ is irrelevant.}
\begin{tabular}{|c||c|c|c|}
\hline 
Color & Blue & Yellow & Red\tabularnewline
\hline 
${\cal X}$ & $-2$ & $0$ & $+2$\tabularnewline
\hline 
$\theta_{C}$ & $\frac{\pi}{2}$ & $0$ & $\frac{\pi}{2}$\tabularnewline
\hline 
$\phi_{C}$ & $\frac{\pi}{2}$ & $0$ & $0$\tabularnewline
 & $(0,\frac{\pi}{2})$ & (irrelevant) & \tabularnewline
\hline 
Skyrmion & pspin SK & Entanglement SK & Spin SK\tabularnewline
(SK) types & Deflated pspin SK &   &  \tabularnewline
\hline 
\end{tabular}
\end{table}

\begin{figure}[t]
\includegraphics[width=0.95\columnwidth]{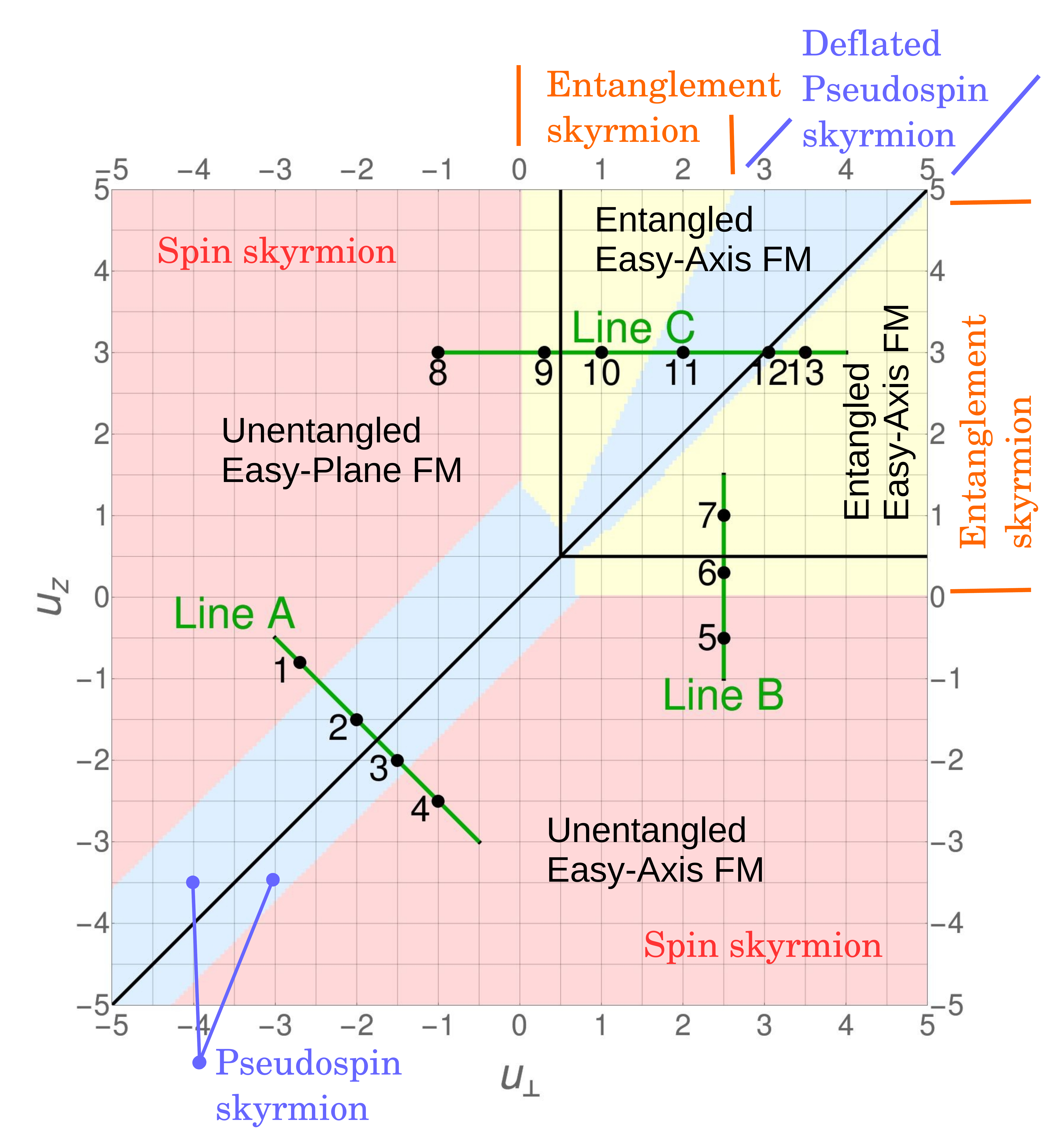}
\caption{\label{fig:Skyrmion_Type_Diagram}The color plot for the skyrmion type indicator $\cal{X}$, which by design has three possible values $-2,0,2$, marked by blue, yellow and red respectively. Black lines highlight the boundaries of four regions for the four types of FM background of a skyrmion. 
Three cuts, as well as 13 representative points on these lines, are selected for detailed discussion. 
The equations of the green lines: (A) $u_{\perp}+u_{\rm z}=-3.5$, $u_{\perp}\in[-3,-0.5]$; (B) $u_{\perp}=2.5$, $u_{\rm z}\in[-1,1.5]$; (C) $u_{\rm z}=3$, $u_{\perp}\in[-1,4]$. 
The $(u_{\perp},u_{\rm z})$ coordinates of the 13 points are: 
on line (A) $(-2.7, -0.8)$, $(-2, -1.5)$, $(-1.5, -2)$, $(-1, -2.5)$ labelled by 1, 2, 3, 4; 
on line (B) $(2.5, -0.5)$, $(2.5, 0.3)$, $(2.5,1)$, labelled by 5, 6, 7; 
on line (C) $(-1, 3)$, $(0.3, 3)$, $(1, 3)$, $(2, 3)$, $(3.05, 3)$, $(3.5, 3)$, labelled by numbers 8-13.
The minimization is done at $\delta_{\rm Z}=0.0005$ for each $(u_{\perp},u_{\rm z})$ point.}
\end{figure}

We set $\delta_{\rm Z}=0.0005$ (which corresponds to rather high but experimentally achievable value $B\sim 50\mbox{T}$ for the magnetic field) and minimize the system energy in the presence of one skyrmion at various combinations of $u_{\perp}$ and $u_{\rm z}$ in the interval $[-5,5]$.

Our result is presented in Fig.~\ref{fig:Skyrmion_Type_Diagram} with the help of the skyrmion-type indicator 
\begin{equation}
{\cal X}=\frac{\bM_{\rm P}(0)\cdot\bM_{\rm P}(\infty)}{\left|\bM_{\rm P}(0)\right|\left|\bM_{\rm P}(\infty)\right|}-\frac{\bM_{\rm S}(0)\cdot\bM_{\rm S}(\infty)}{\left|\bM_{\rm S}(0)\right|\left|\bM_{\rm S}(\infty)\right|},\label{eq:SkX-type-indicator}
\end{equation}
which is a useful quantity to characterize the relative orientations of spin and pseudospin magnetization at the skyrmion center 
\begin{eqnarray}
\bM_{\rm S}(\br=0) &= C^{\dagger}\sigma_{0}\otimes\boldsymbol{\sigma}C\\
\bM_{\rm P}(\br=0) &= C^{\dagger}\boldsymbol{\sigma}\otimes\sigma_{0}C
\end{eqnarray} 
with respect to the FM background far from the center 
\begin{eqnarray}
\bM_{\rm S}(r\rightarrow\infty) &= F^{\dagger}\sigma_{0}\otimes\boldsymbol{\sigma}F\\
\bM_{\rm P}(r\rightarrow\infty) &= F^{\dagger}\boldsymbol{\sigma}\otimes\sigma_{0}F.
\end{eqnarray} 
By checking the explicit expressions of $\bM_{S,T}(0)$ and $\bM_{S,T}(\infty)$ for a skyrmion state $Z_{\rm sk}$, one finds that ${\cal X}$ takes three values $-2,0,+2$. In the skyrmion type diagram Fig.~\ref{fig:Skyrmion_Type_Diagram}, they are colored in blue, yellow and red respectively. For the spin skyrmion and pseudospin skyrmion, we have ${\cal X}=2$ and ${\cal X}=-2$, respectively, because the pseudospin magnetization in a \emph{pseudospin skyrmion} is reversed from the skyrmion center to infinity, whereas for a \emph{spin skyrmion} the spin magnetization is flipped. The case ${\cal X}=0$ means both spin and pseudospin magnetization are reversed and it corresponds to an \emph{entanglement skyrmion}. Moreover, with the help of the the $(u_{\perp},u_{\rm z})$-dependence of the optimal value of $\phi_{C}$, we have identified a fourth type of CP$^{3}$-skyrmions, namely the \emph{deflated pseudospin skyrmion}. In Tab.~\ref{tab:skyrmion-types}, we list the skyrmion types and corresponding optimal values of relevant parameters. 

Since a skyrmion is a localized texture state and it is smoothly connected to the ferromagnetic background, in Fig.~(\ref{fig:Skyrmion_Type_Diagram}) we highlight the borders of the four types of background FM states studied in Sec.~\ref{subsec:QHFM-ground-states}, in order to stress the fact that there are different types of CP$^{3}$-skyrmions embedded into the same ferromagnetic background. 
Guided by this observation, we have selected three lines (labelled A, B, C, colored in green) and 13 representative points (labelled by number 1 to 13, colored in black) for a more detailed analysis. 
In the following sections, the qualitative and quantitative features of the different skyrmion types are investigated in more detail. 

\section{Four types of CP$^{3}$-skyrmions}
\label{sec:4-types-skyrmion}

In this section, we present a detailed discussion of the four types of CP$^{3}$-skyrmions which we obtain and summarize in Tab.~\ref{tab:skyrmion-types}. We choose the 13 representative points among the minimization results presented in Fig.~\ref{fig:Skyrmion_Type_Diagram}. 
For each skyrmion type, we visualize the minimization result at selected points on the three Bloch spheres which are introduced in Sec.~\ref{subsec:QHFM-visualization} for the discussion of the FM states. The methodology is described in more detail in App.~\ref{subsec:Visualization-CP3-skyrmion-on-Bloch-spheres}.
We also display the profiles of the electron density $\rho_{\rm Total}(\br)$ and the $z$-component of the spin magnetization $M_{\rm Sz}(\br)$ on the honeycomb lattice associated with the different skyrmion types in the same manner as for the QHFM states (technical details of the lattice-scale representation can be found in App.~\ref{subsec:Visualization-CP3-skyrmion-on-honeycomb-lattice}). 

The aim of visualization is twofold -- on the one hand, the visualization of skyrmions on Bloch spheres makes it transparent that ``a skyrmion is a wrapping of Bloch sphere''; on the other hand, the lattice-scale profiles provide hints for direct imaging via spin-resolved STM/STS experiments. Notice that for mere illustration purposes, we have chosen a rather large magnetic field of $B\sim1000{\rm T}$ to visualize the lattice-scale profiles. 
While these fields are not achieved in a typical experimental situation, more realistic fields would yield skyrmion sizes on the order of some hundred lattice spacings and the profiles presented in Fig.~\ref{fig:Pt4_spin_skyrmion} to Fig.~\ref{fig:Pt12_deflated_skyrmion} simply need to be upscaled. 
A more quantitative analysis on CP$^{3}$-skyrmions is given in Sec.~\ref{sec:Size_and_energy}. 

In Sec.~\ref{subsec:spin-skyrmion} and Sec.~\ref{subsec:pseudospin-skyrmion}, we discuss the CP$^{3}$-spin skyrmion and the CP$^{3}$-pseudospin skyrmion. For these types, the CP$^{3}$-field $\check{Z}_{\rm sk}(x,y)$ factorizes into a direct product of two $CP^1$-spinors for spin and pseudospin throughout the entire $xy$-plane. We call them \emph{factorizable} CP$^{3}$-skyrmions. 
In Sec.~\ref{subsec:entanglement-skyrmion}, we discuss the CP$^{3}$-\emph{entanglement skyrmion} with the help of the Schmidt-decomposition Eq.~(\ref{eq:parametrizationZ}) of the CP$^{3}$-spinors. For this type of skyrmion, our parametrization Eq.~(\ref{eq:parametrizationZ}) provides a one-to-one mapping between the $xy$-plane and the ``entanglement Bloch sphere'' for the angles $\alpha$ and $\beta$, precisely in the same manner as the spin magnetization of a CP$^{3}$-spin skyrmion, and the pseudospin magnetization of a CP$^{3}$-pseudospin skyrmion. 
Finally in Sec.~\ref{subsec:deflated-pseudospin-skyrmion}, we discuss a more subtle type of skyrmion. Although the skyrmion-type indicator ${\cal X}$ does not help one to distinguish this type from the (factorizable) CP$^{3}$-pseudospin skyrmion, the visualizations on the Bloch spheres and on the lattice indicate a certain degree of entanglement. According to the appearance on the Bloch spheres, we call this type the \emph{deflated CP$^{3}$-pseudospin skyrmion}.

\subsection{Spin skyrmion}
\label{subsec:spin-skyrmion}

The spin skyrmion is probably the most studied texture state in the literature.\cite{DasSarmaPinczuk} 
Our energy minimization shows that the  CP$^{3}$-skyrmion appears as a spin skyrmion when the parameters $(u_{\perp},u_{\rm z})$ fall in the red region of Fig.~\ref{fig:Skyrmion_Type_Diagram}. In this case, we have $\alpha=0$ throughout the $xy$-plane, so that the CP$^{3}$-field factorizes into a direct product of spin and pseudospin:
\begin{equation}
Z_{\rm spin}(x,y)={\cal N}(r)^{-1}\psi^{\rm P}\otimes\left[ (x+iy)\psi^{\rm S} -\lambda(r)\chi^{\rm S} \right].
\end{equation}

The constant $CP^1$-spinor $\psi^{\rm P}$ is defined in Eq.~(\ref{eq:psi_basis}). It is determined by the pseudospin ferromagnetic background in which the skyrmion is embedded and that is not affected by the formation of the spin texture. Therefore the coordinate dependence of the spin skyrmion lies entirely in the spin part, where we have $\psi^{\rm S}=(1,0)^T$ and $\chi^{\rm S}=(0,1)^T$. Regardless of the radial deformation, the spin part of a spin skyrmion is identical to an O$(3)$-skyrmion written in $\CP{1}$-spinor form.\cite{Han2010}

\begin{figure}[t]
\begin{tabular}{ccc}
\includegraphics[width=0.3\columnwidth]{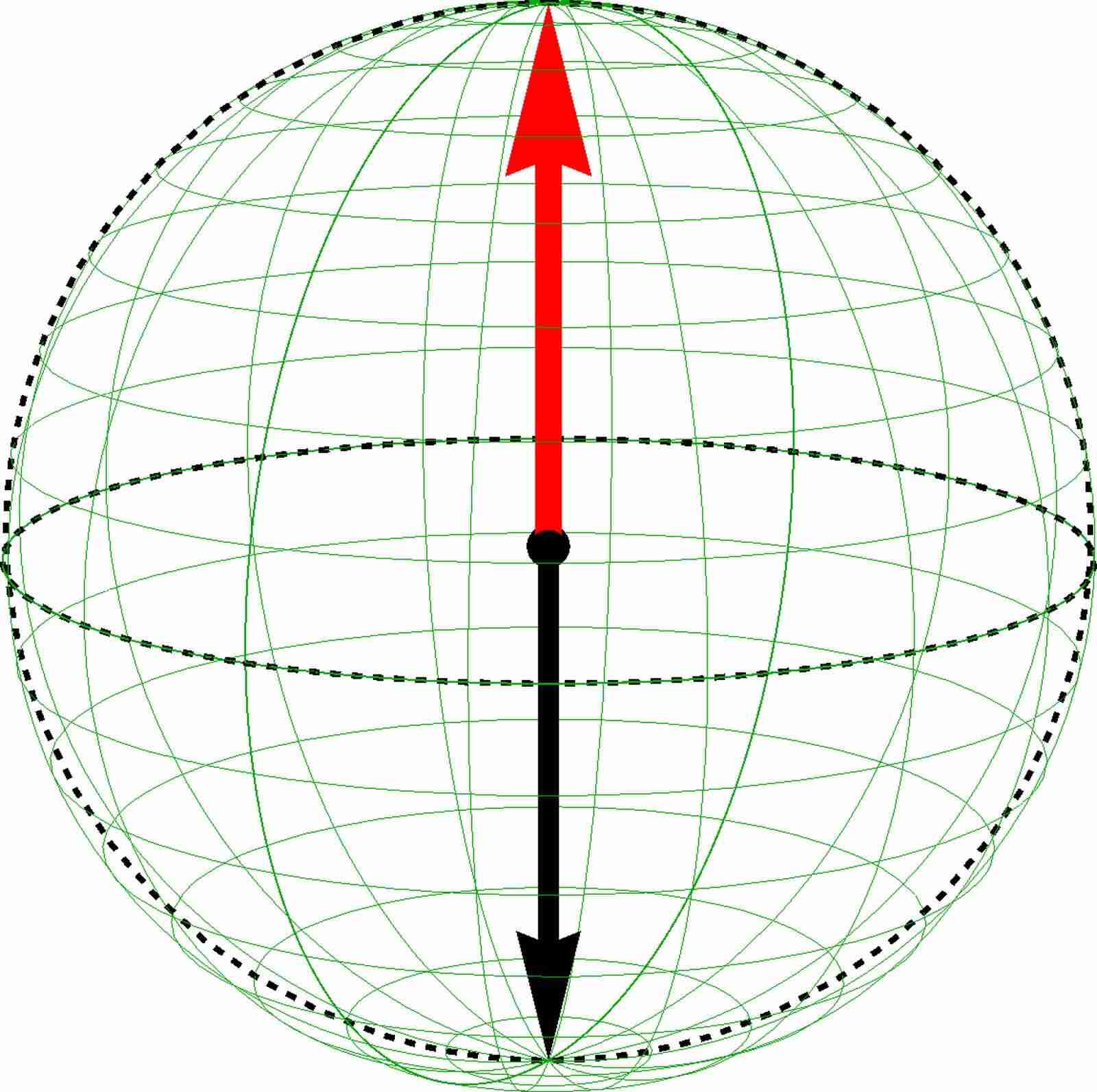} & \includegraphics[width=0.3\columnwidth]{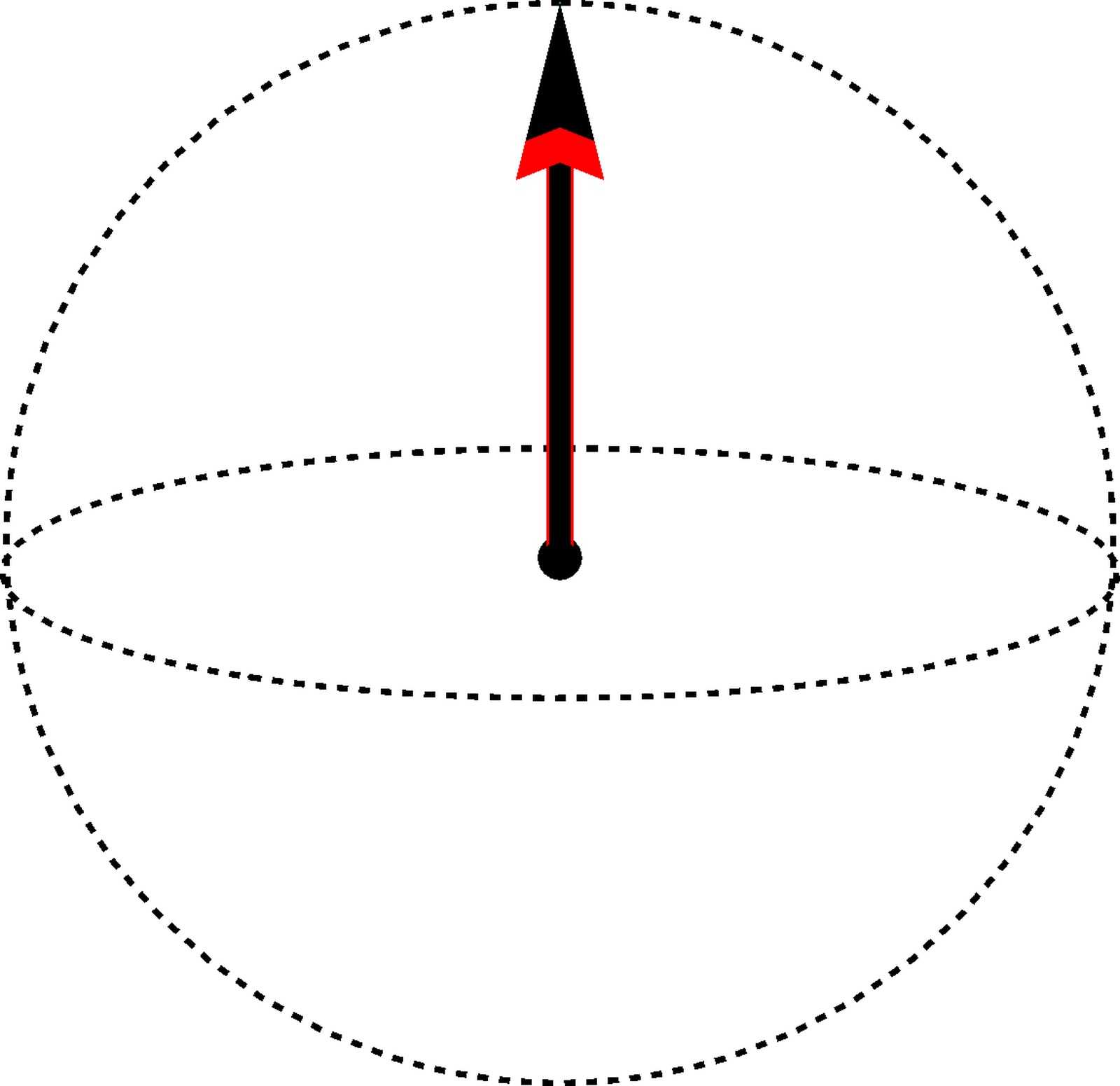} & \includegraphics[width=0.3\columnwidth]{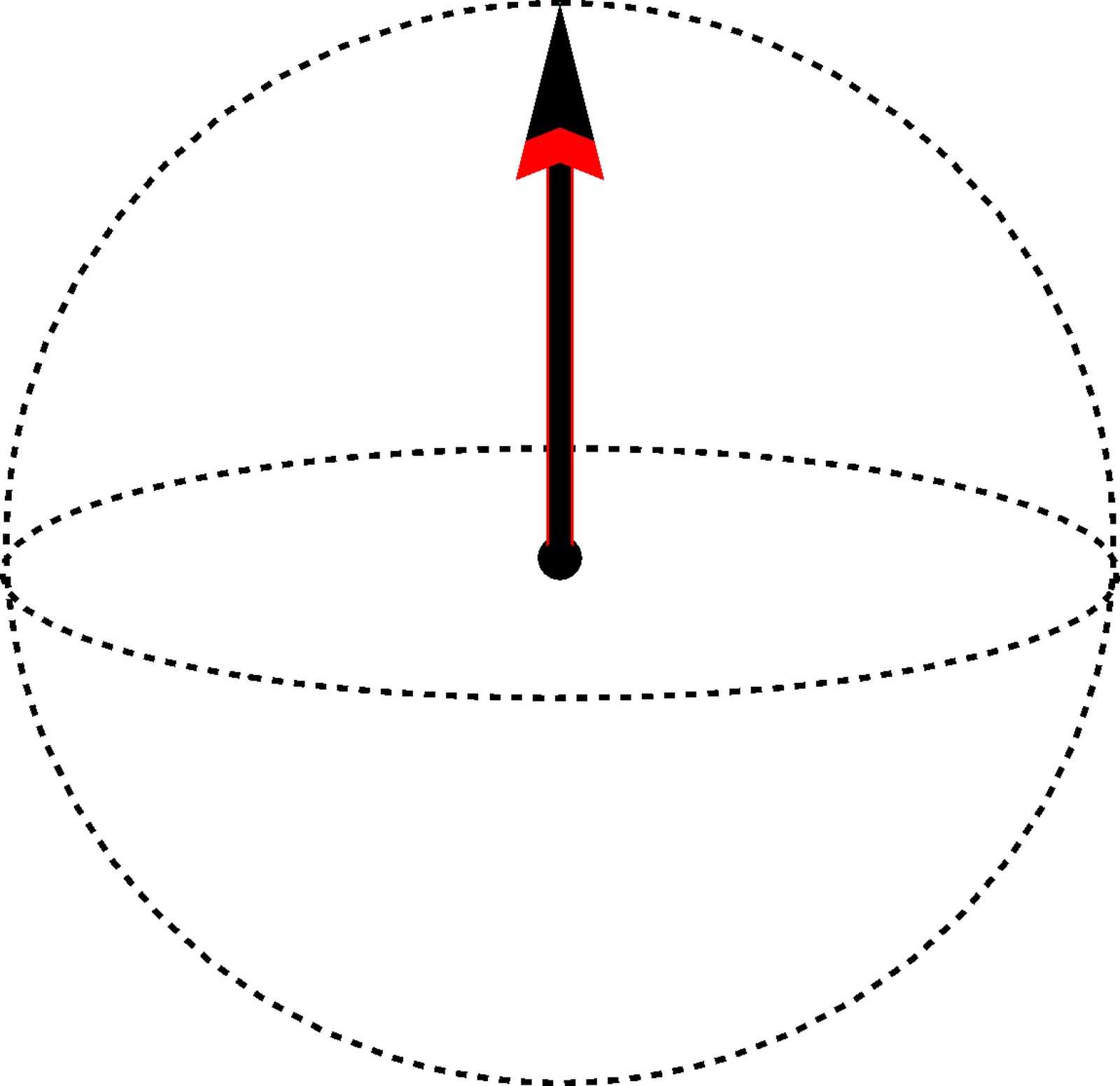}\tabularnewline
(a) Spin BS & (b) Pspin BS & (c) Entanglement BS\tabularnewline
\end{tabular}
\includegraphics[width=\columnwidth]{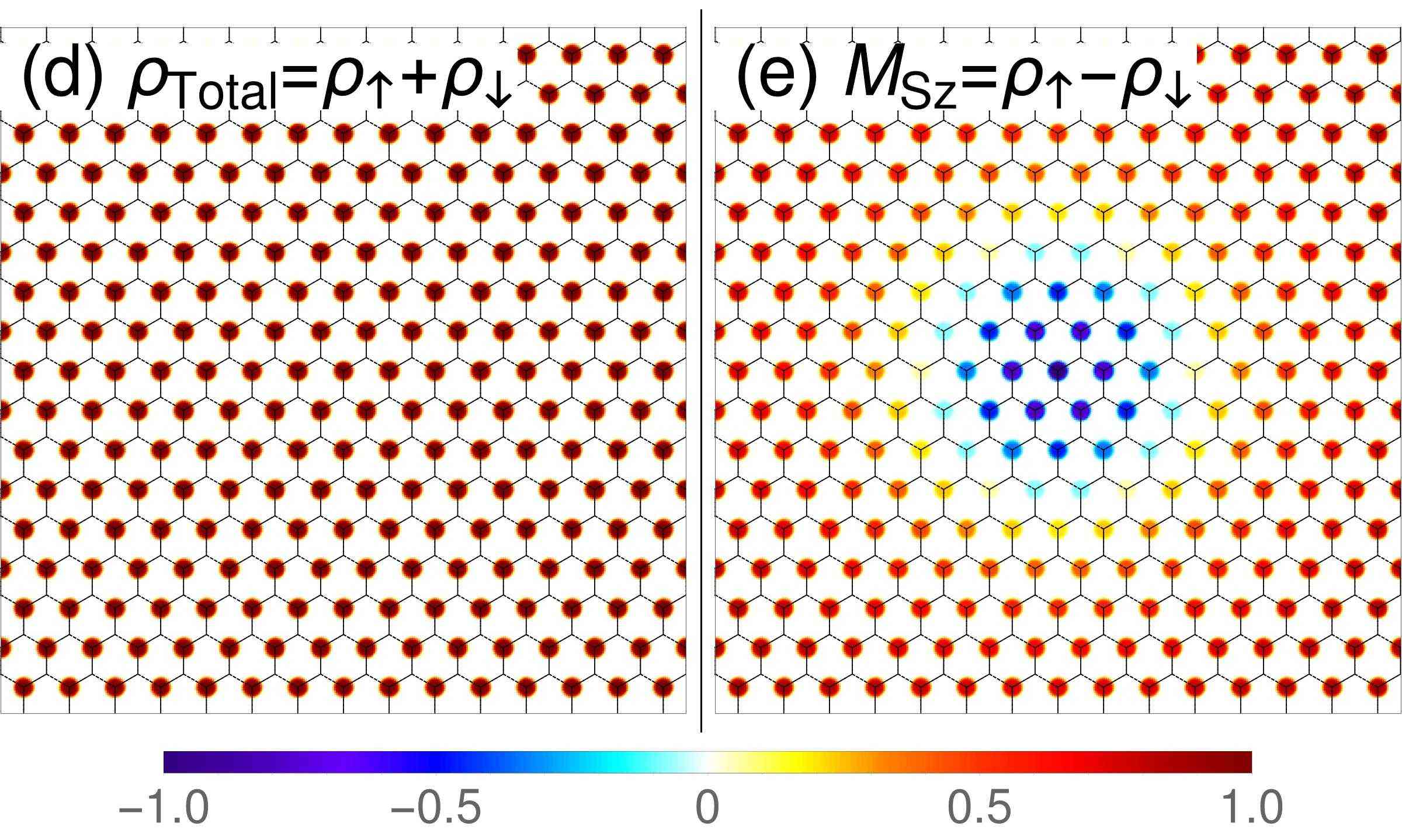}
\caption{\label{fig:Pt4_spin_skyrmion}Visualization of the CP$^{3}$-spin skyrmion in the unentangled easy-axis FM background corresponding to the optimal configurations at point 4 in the skyrmion type diagram Fig. \ref{fig:Skyrmion_Type_Diagram}. Panels (a), (b) and (c) show the texture visualized on the spin, pseudospin and entanglement Bloch spheres respectively. Red arrows indicate the polarizations for $r\rightarrow\infty$, while the black arrows indicate the polarizations at the center of the texture. (d) and (e) show the lattice-resolved profiles of the electron density $\rho_{\rm Total}(\br)$ and the $z$-component of spin magnetization $M_{\rm Sz}(\br)$. The center of the texture is always chosen to be at the center of the figure. Here and in the following figures, we have chosen a field $B\sim1000{\rm T}$ simply to illustrate the profiles. For more realistic $B$-fields, the patterns need to be upscaled. 
}
\end{figure}

\begin{figure}[t]
\begin{tabular}{ccc}
\includegraphics[width=0.3\columnwidth]{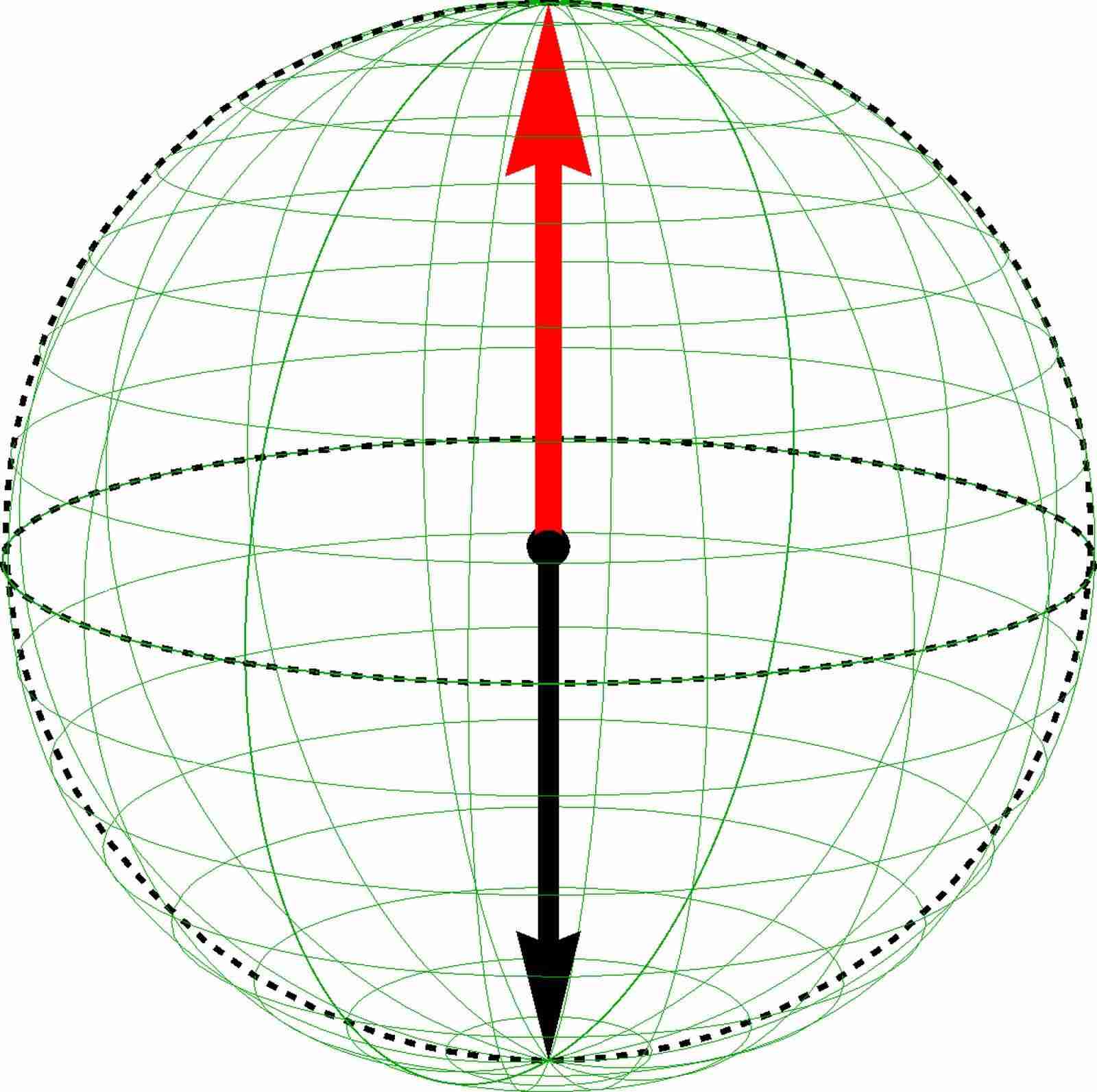} & \includegraphics[width=0.3\columnwidth]{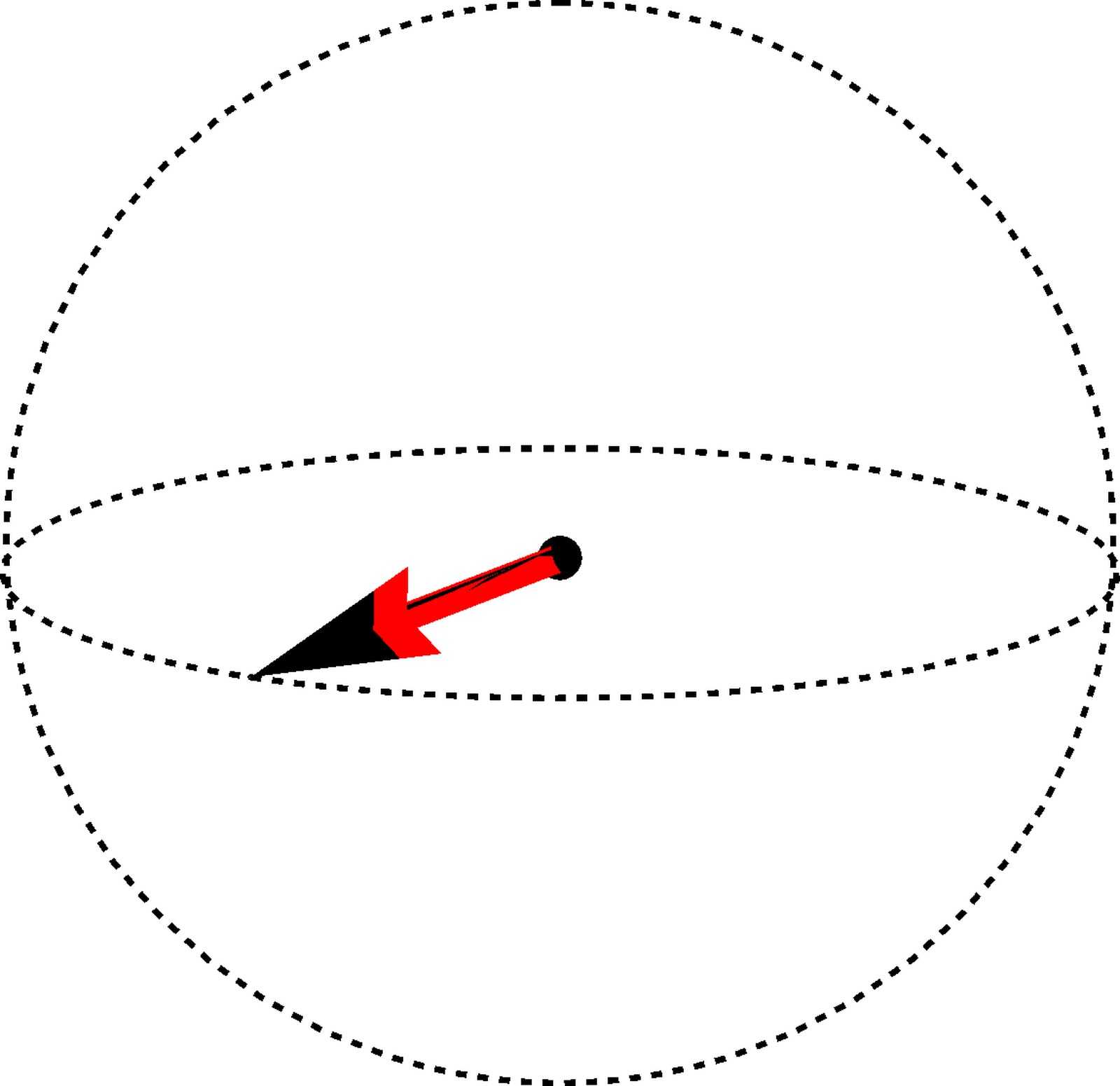} & \includegraphics[width=0.3\columnwidth]{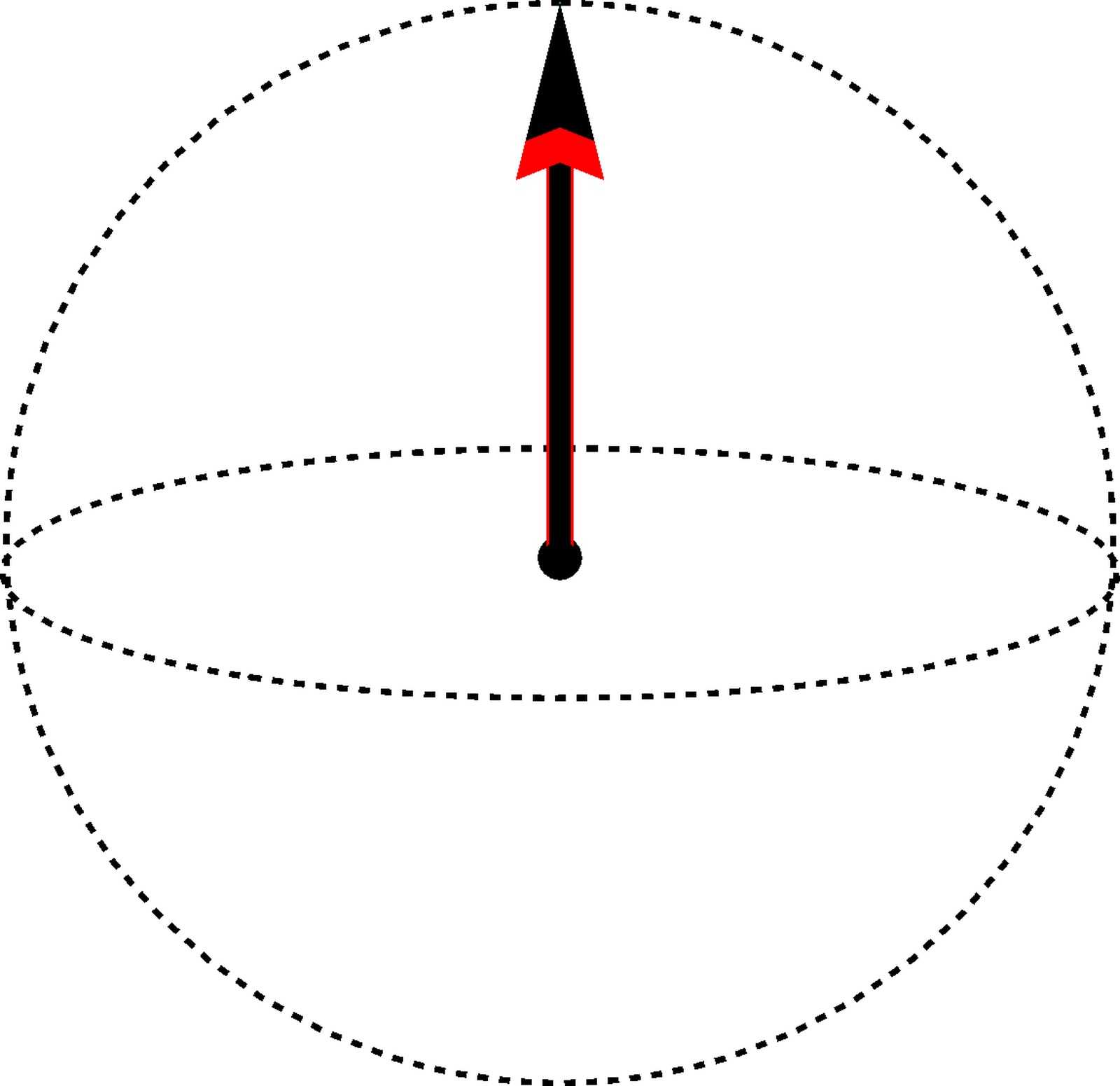}\tabularnewline
(a) Spin BS & (b) Pspin BS & (c) Entanglement BS\tabularnewline
\end{tabular}
\includegraphics[width=\columnwidth]{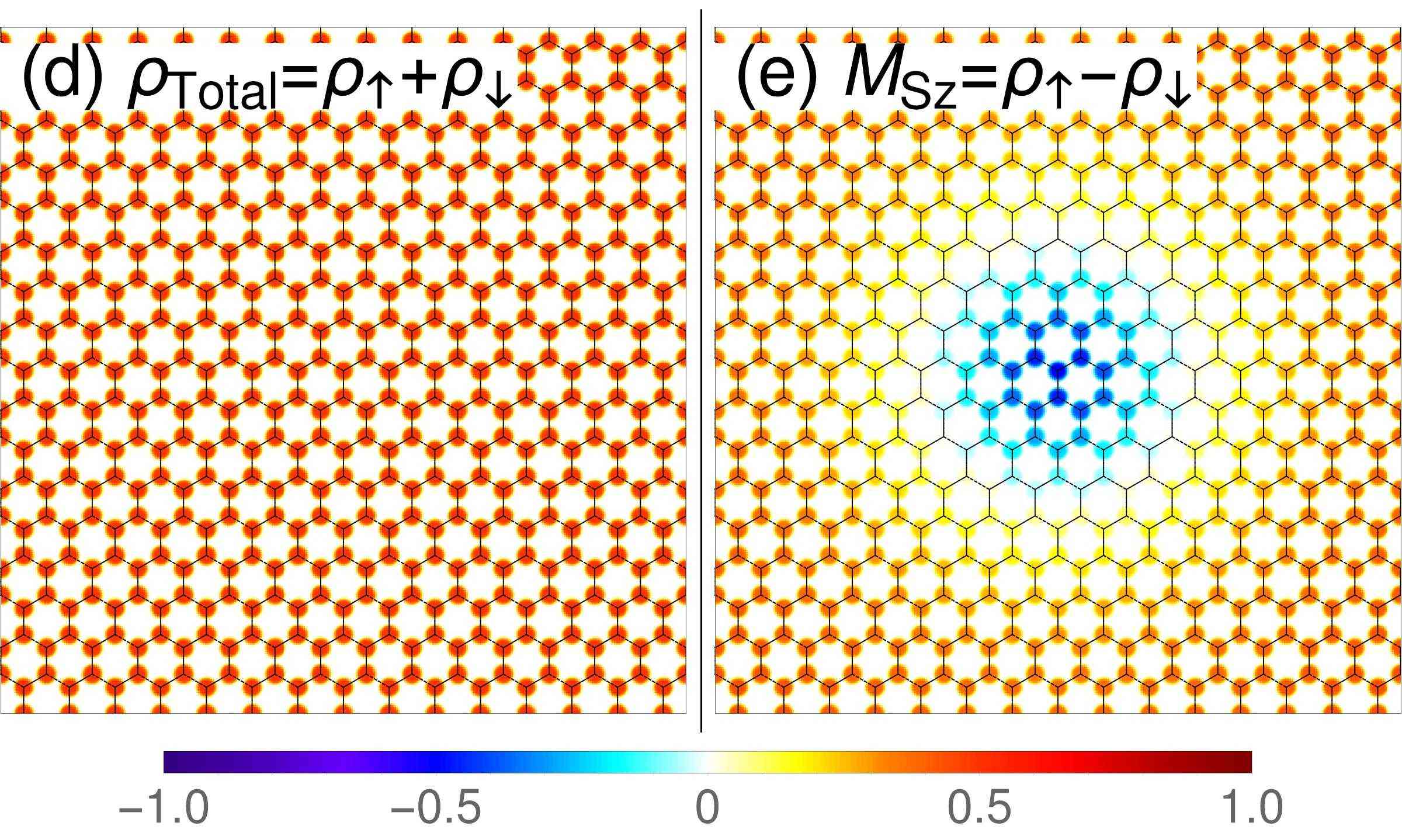}
\caption{\label{fig:Pt1_spin_skyrmion}Visualization of the CP$^{3}$-spin skyrmion embedded in unentangled easy-plane FM background (point 1 in Fig.\ref{fig:Skyrmion_Type_Diagram}). Conventions are the same as in Fig.~\ref{fig:Pt4_spin_skyrmion}.
}
\end{figure}

In Fig.~\ref{fig:Pt4_spin_skyrmion}(a),(b),(c) and Fig.~\ref{fig:Pt1_spin_skyrmion}(a),(b),(c) e visualize the CP$^{3}$-spin skyrmions on the spin, pseudospin and entanglement Bloch spheres. In order to obtain this representation, we first compactify the $xy$-plane to the Riemann sphere by adding the infinity point (see Appendix.~\ref{subsec:Visualization-CP3-skyrmion-on-Bloch-spheres}), and then use a ``Gauss map'' to map the Riemann sphere on the Bloch sphere. 
Consider the spin magnetization vector field $\bM_{\rm S}(\br)$ for example. At a specific point $\br_0$ in the $xy$-plane, the spin magnetization for a CP$^{3}$-field $Z(\br_0)$ corresponds to a point on the spin Bloch sphere specified by $\bM_{\rm S}(\br_0)$. Collecting all such points with coordinate $\br_0$ throughout the compactified $xy$-plane (the Riemann sphere), we obtain a surface on the spin Bloch sphere.  Such surface is continuous because a CP$^{3}$-skyrmion can be viewed as an interpolation between its center spinor $C$ and its FM background spinor $F$. It is also closed because the boundary conditions for a CP$^{3}$-field $Z(\br)$ require that all $Z(\br_0)$ are equivalent when $|\br_0|\rightarrow\infty$ and thus give the same spin magnetization vector $\bM_{\rm S}(\infty)$. In this sense, the spin magnetization vector field $\bM_{\rm S}(\br)$, as well as the pseudospin magnetization vector field $\bM_{\rm P}(\br)$ and the entanglement vector field $\boldsymbol{m}_{\rm E}(\br)$ by analogy, can be understood as mappings from the Riemann sphere to the corresponding Bloch spheres.    
The surface on the pseudospin Bloch sphere can be obtained similarly by collecting the end points of the pseudospin magnetization vector field $\bM_{\rm P}(\br_0)$ for all $\br_0$ in the compactified $xy$-plane. 
To keep track of the continuous change of spin magnetization on the surface of the spin Bloch sphere, we draw the image of longitudinal and latitudinal lines of the Riemann sphere on the spin Bloch sphere in green.

In the case of CP$^{3}$-spin skyrmion embedded into an unentangled easy-axis FM background, the spin Bloch sphere [Fig.~\ref{fig:Pt4_spin_skyrmion}(a)] is completely covered by the image of Riemann sphere, while on the pseudospin Bloch sphere [Fig.~\ref{fig:Pt4_spin_skyrmion}(b)] the image is a single point at the north pole, which indicates that all electrons remain on a single sublattice. 
In the entanglement Bloch sphere the image of the Riemann sphere is also a point and the entanglement vector $\boldsymbol{m}_{\rm E}(\alpha,\beta)$ is pointing upward to the north pole, because the factorizable CP$^{3}$-skyrmion means $\alpha=0$ throughout the $xy$-plane. 
On the spin Bloch spheres one also notices the opposite direction of the spin magnetization at the skyrmion center (black arrow pointing towards the south pole) and infinity (red arrow pointing towards the north pole).

The Bloch sphere representation of the CP$^{3}$-spin skyrmion embedded into an unentangled easy-plane FM background [Figs.~\ref{fig:Pt1_spin_skyrmion} (a)-(c)] is essentially the same as that of the previous case, apart from the in-plane direction of pseudospin magnetization in Fig.~\ref{fig:Pt1_spin_skyrmion}(b).

In addition to the visualization of the CP$^{3}$-spin skyrmions on Bloch spheres, we display the profiles of electron density $\rho_{\rm Total}(\br)$ and the $z$-component of spin magnetization $M_{\rm Sz}(\br)$ in lattice resolution. 
Fig.~\ref{fig:Pt4_spin_skyrmion}(d) and Fig.~\ref{fig:Pt1_spin_skyrmion}(d) show that there are no textures in the electron density profiles, i.e. the sublattice occupation pattern is constant throughout the $xy$-plane. In particular, in a CP$^{3}$-spin skyrmion with unentangled easy-axis FM background, the sublattice A is fully filled and sublattice B is empty -- we have  
$\rho_{\rm Total}(B)=\rho_{\uparrow}(B)+\rho_{\downarrow}(B)=0$ 
and 
$\rho_{\rm Total}(A)=\rho_{\uparrow}(A)+\rho_{\downarrow}(A)=1$ everywhere. In contrast to this, 
a spin skyrmion in unentangled easy-plane FM background has equal occupation of the two sublattices -- we have 
$\rho_{\rm Total}(B)=\rho_{\uparrow}(B)+\rho_{\downarrow}(B)=1/2$ 
and 
$\rho_{\rm Total}(A)=\rho_{\uparrow}(A)+\rho_{\downarrow}(A)=1/2$ everywhere.

As expected, the spin texture of a CP$^{3}$-spin skyrmion becomes apparent in the lattice-scale profiles of the spin magnetization $M_{\rm Sz}(\br)$ in Fig.~\ref{fig:Pt4_spin_skyrmion}(e) and Fig.~\ref{fig:Pt1_spin_skyrmion}(e). 

\subsection{Pseudospin skyrmion}
\label{subsec:pseudospin-skyrmion}

\begin{figure}[t]
\begin{tabular}{ccc}
\includegraphics[width=0.3\columnwidth]{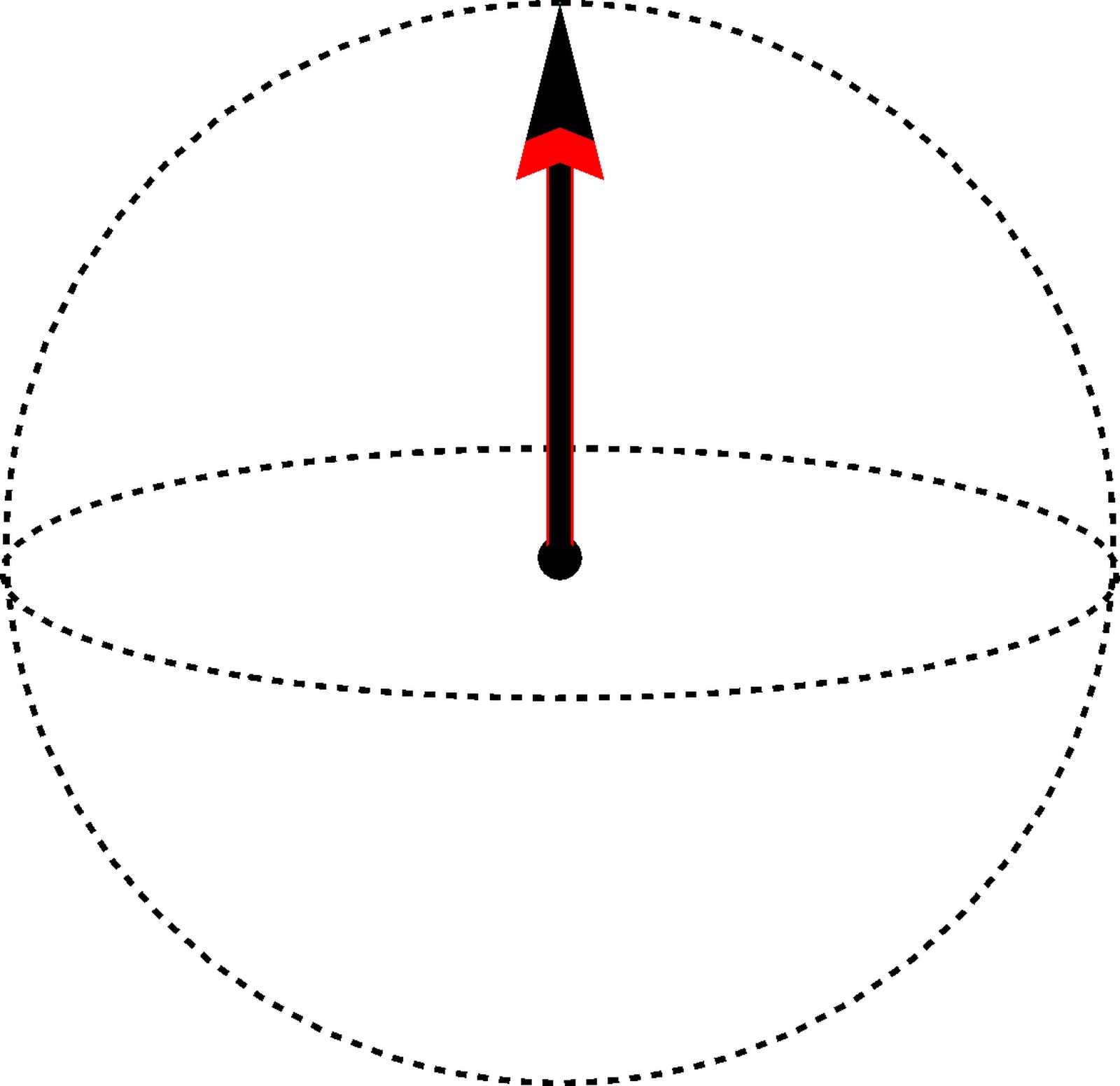} & \includegraphics[width=0.3\columnwidth]{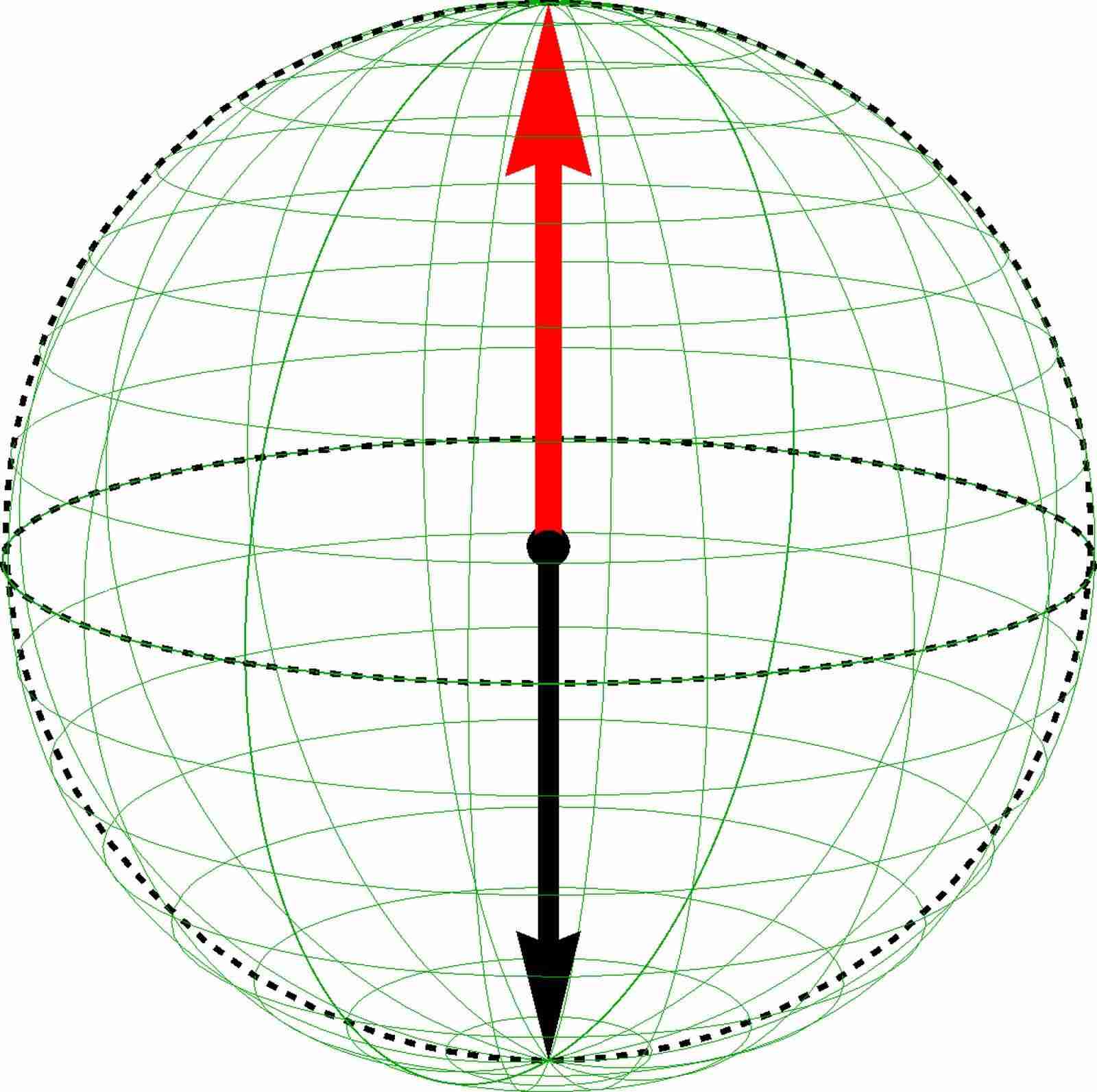} & \includegraphics[width=0.3\columnwidth]{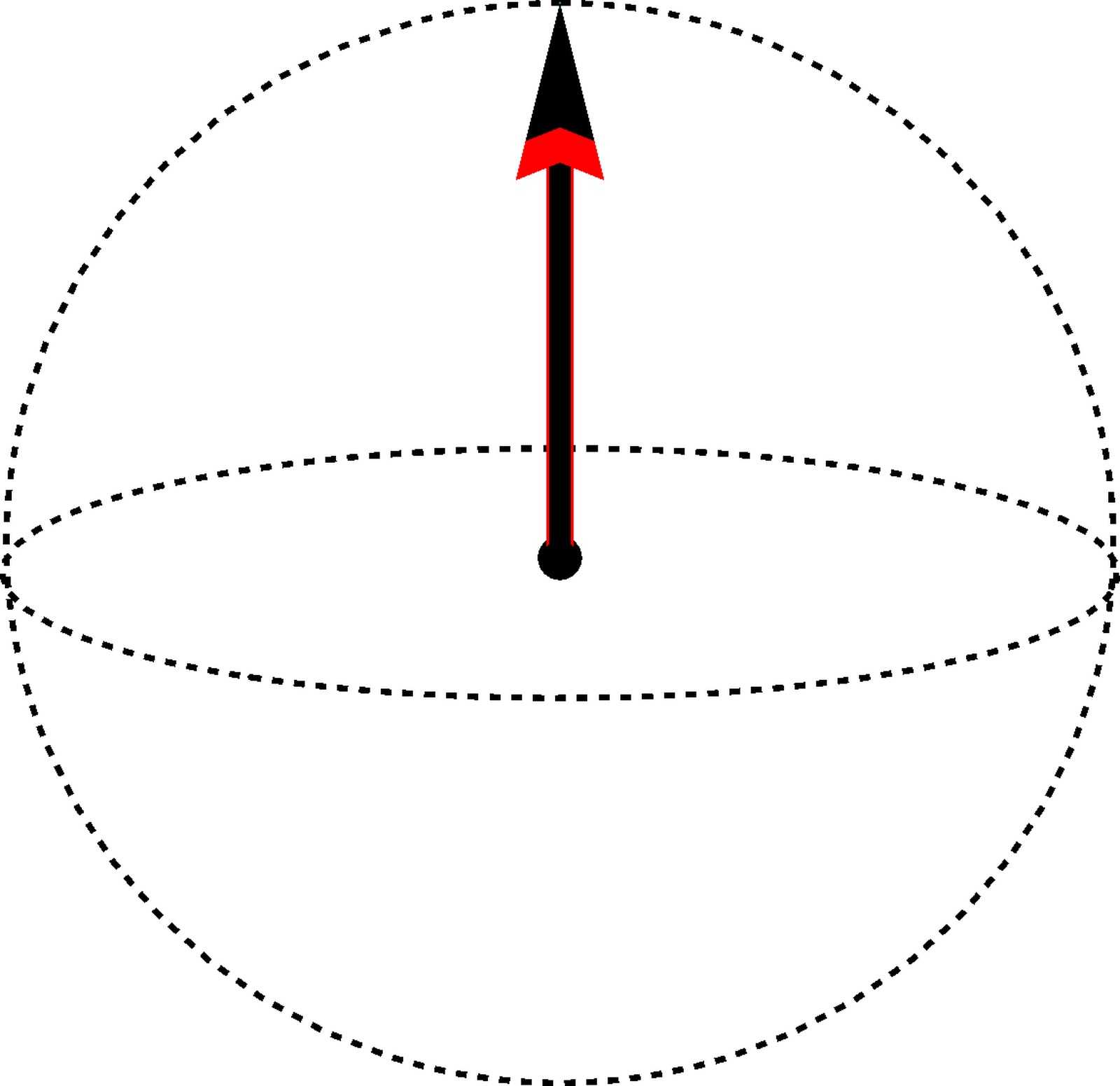}\tabularnewline
(a) Spin BS & (b) Pspin BS & (c) Entanglement BS\tabularnewline
\end{tabular}
\includegraphics[width=\columnwidth]{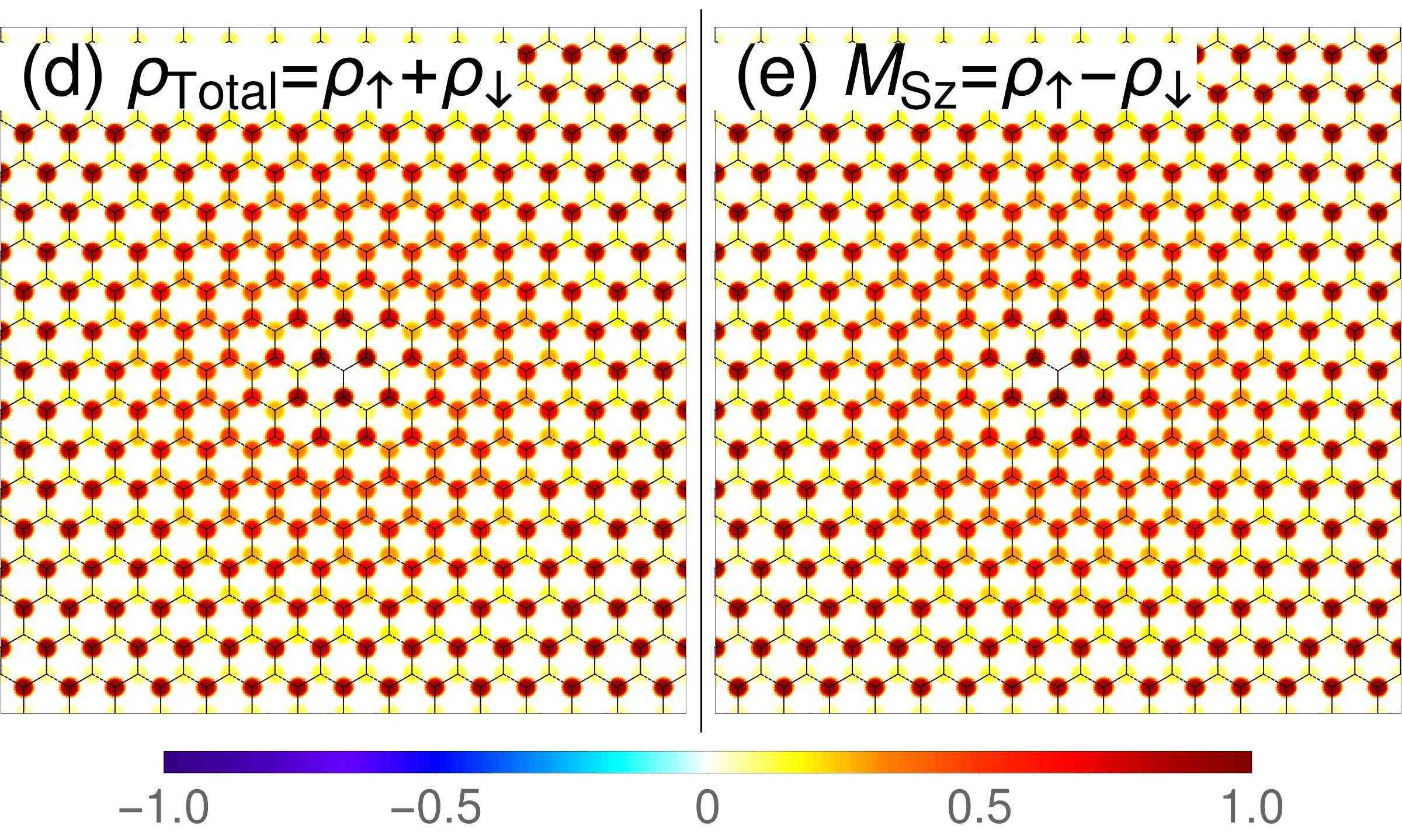}
\caption{\label{fig:Pt3_ppin_skyrmion}Visualization of the CP$^{3}$-pseudospin skyrmion embedded in an unentangled easy-axis FM background (point 3 in Fig.\ref{fig:Skyrmion_Type_Diagram}). Conventions are the same as in Fig.~\ref{fig:Pt4_spin_skyrmion}.
}
\end{figure}
 
Interchanging the role of spin and pseudospin in the CP$^{3}$-spin skyrmion, we obtain the CP$^{3}$-pseudospin skyrmion : 
\begin{equation}
Z_{\rm pspin}(x,y)={\cal N}(r)^{-1}\left[(x+iy)\psi^{\rm P} - \lambda(r)\chi^{\rm P} \right]\otimes\psi^{\rm S},
\end{equation}
where the $\CP{1}$-spinor basis $\psi^{\rm P}$ and $\chi^{\rm P}$  for pseudospin are defined in Eq.~(\ref{eq:psi_basis}) and Eq.~(\ref{eq:chi_basis}). 
Dual to the case of the spin skyrmion, the CP$^{3}$-pseudospin skyrmion is factorized into two parts: a constant $\CP{1}$-spinor $\psi^{\rm S}=(1,0)^T$ for spin throughout the $xy$-plane, tensor product to a $\CP{1}$-pseudospin skyrmion embedded in an unentangled easy-axis FM background, or a $\CP{1}$-pseudospin bimeron embedded in an unentangled easy-plane FM background. 

\begin{figure}[t]
\begin{tabular}{ccc}
\includegraphics[width=0.3\columnwidth]{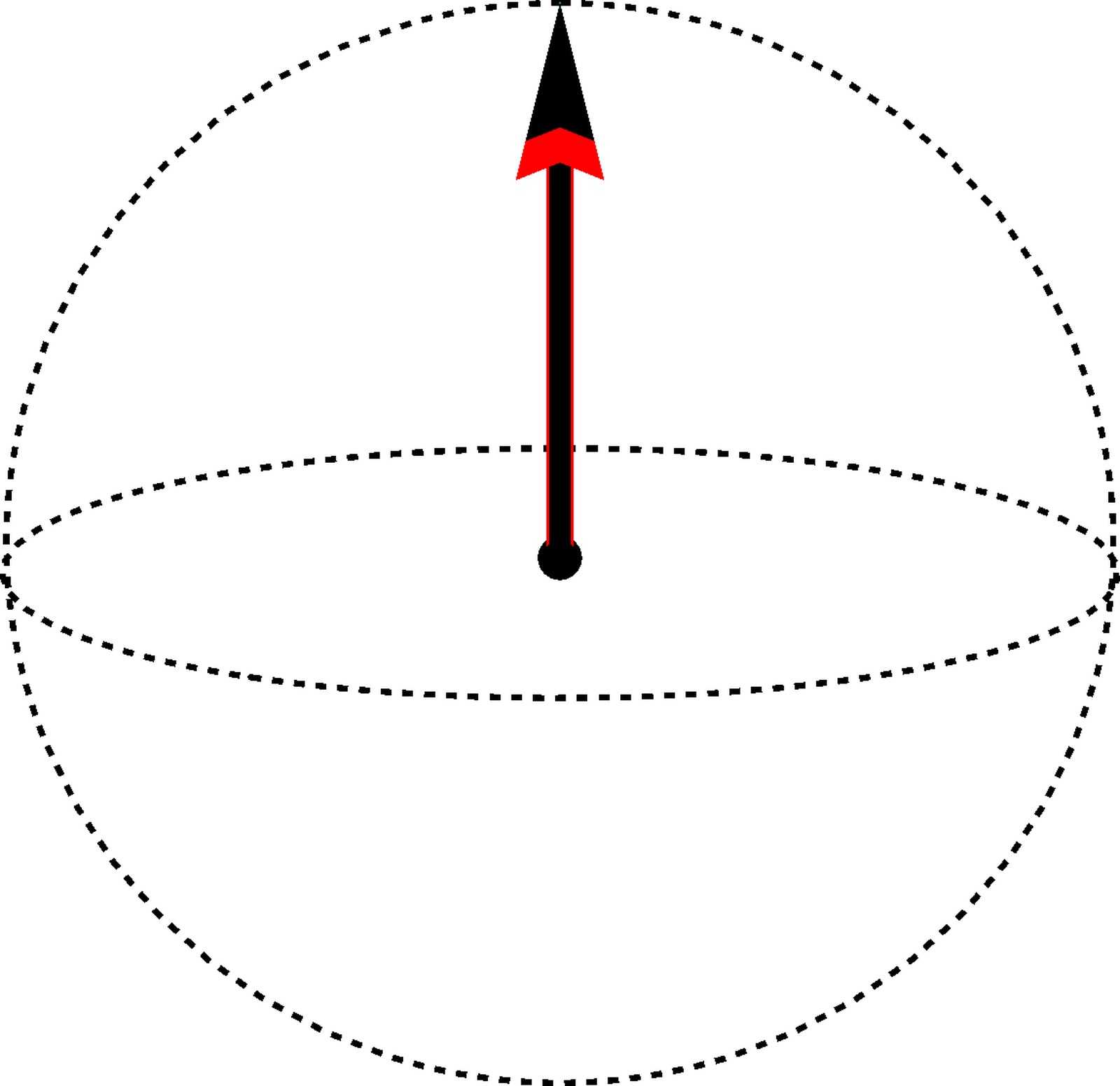} & \includegraphics[width=0.3\columnwidth]{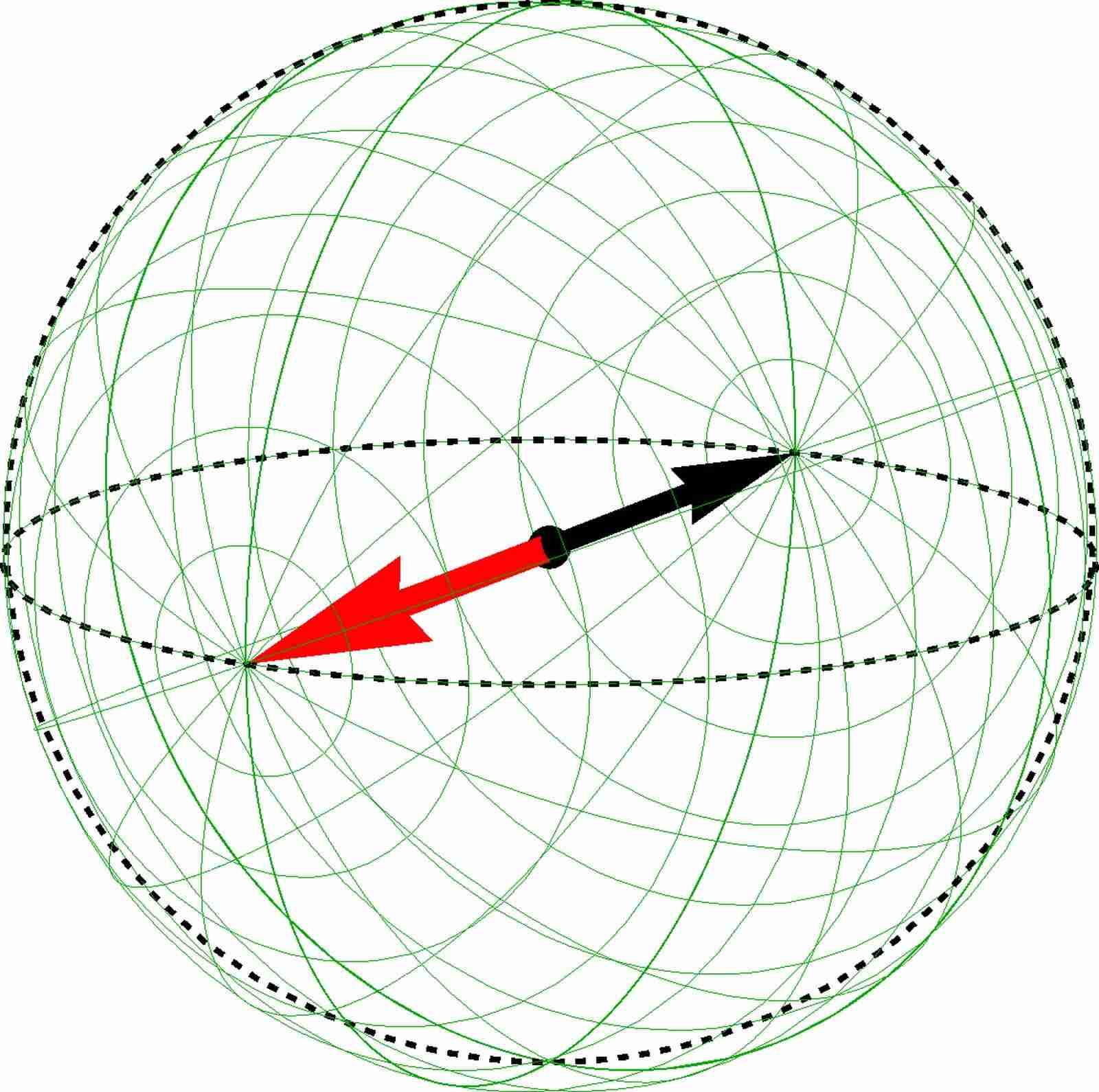} & \includegraphics[width=0.3\columnwidth]{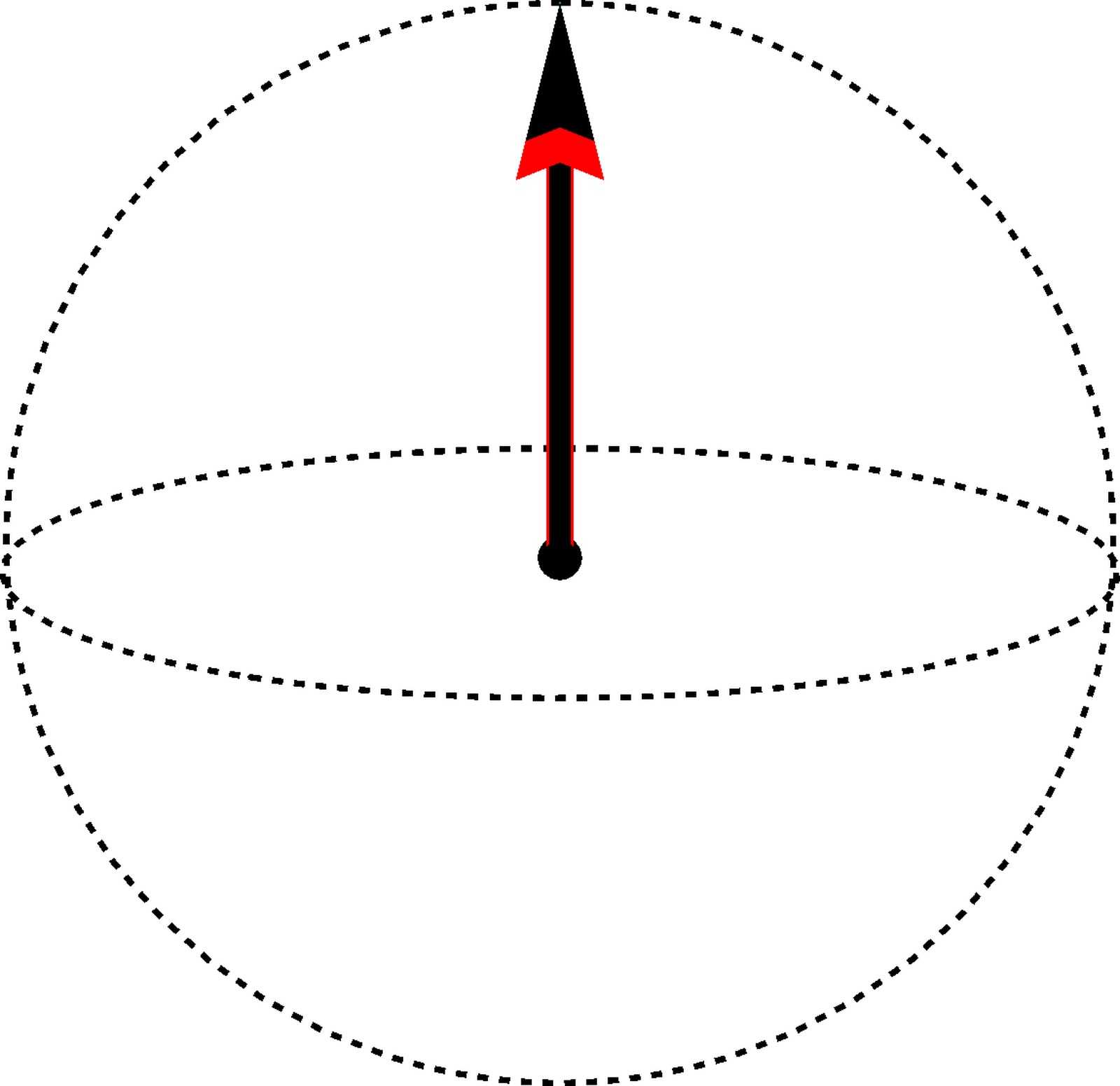}\tabularnewline
(a) Spin BS & (b) Pspin BS & (c) Entanglement BS\tabularnewline
\end{tabular}
\includegraphics[width=\columnwidth]{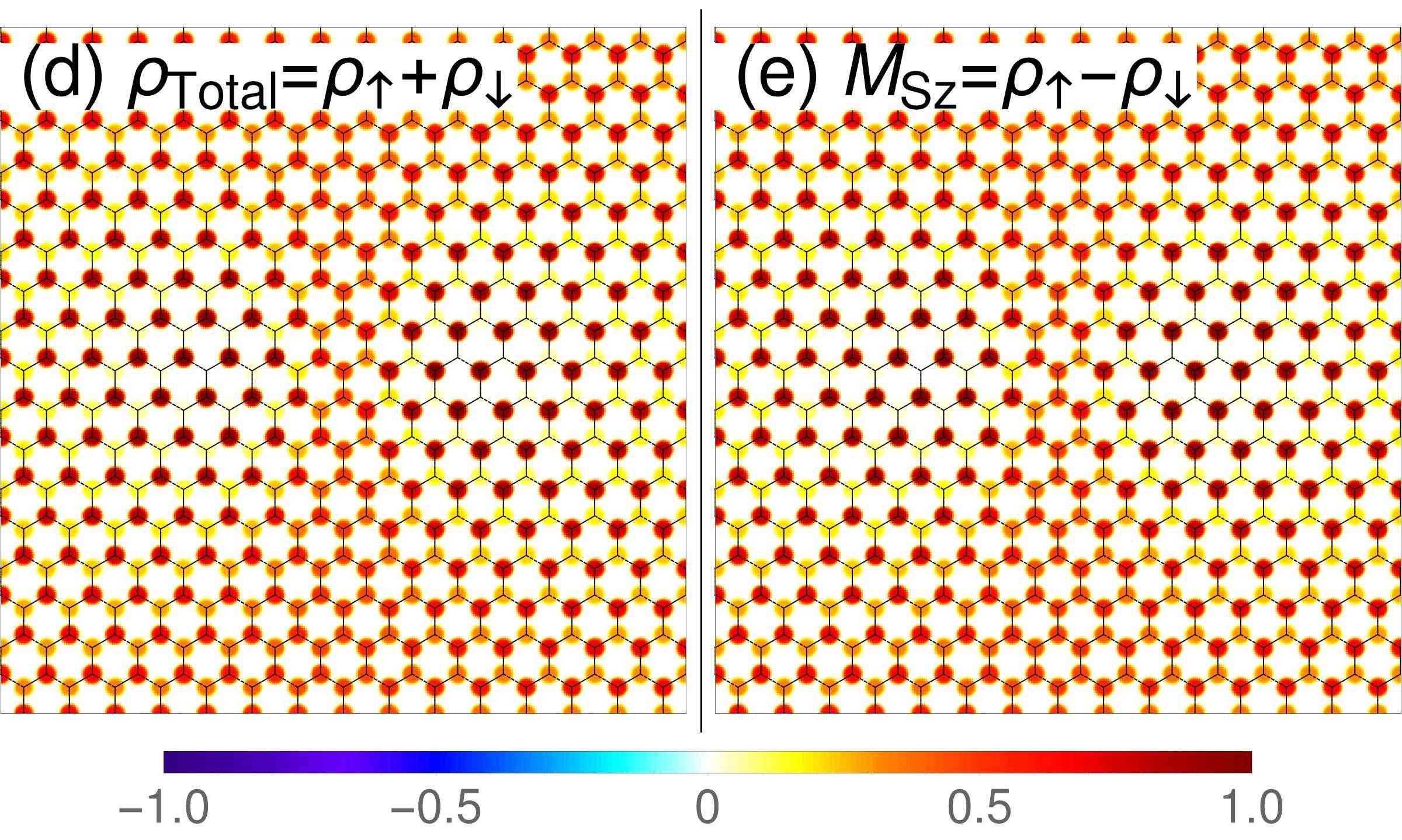}
\caption{\label{fig:Pt2_ppin_skyrmion}Visualization of the CP$^{3}$-pseudospin skyrmion embedded in an unentangled easy-plane FM background (point 2 in Fig.\ref{fig:Skyrmion_Type_Diagram}). Conventions are the same as in Fig.~\ref{fig:Pt4_spin_skyrmion}.
}
\end{figure}

We visualize the CP$^{3}$-pseudospin skyrmion on the spin, pseudospin and entanglement Bloch spheres in Figs.~\ref{fig:Pt3_ppin_skyrmion}(a)-(c) and Figs.~\ref{fig:Pt2_ppin_skyrmion}(a)-(c) in the same manner as for CP$^{3}$-spin skyrmions. 
In contrast to the spin skyrmions discussed in the previous section, we have constant spin magnetization for a CP$^{3}$-pseudospin skyrmion, and it appears as a point at the north pole on the spin Bloch sphere [Figs.~\ref{fig:Pt3_ppin_skyrmion}(a) and \ref{fig:Pt2_ppin_skyrmion}(a)]. The pseudospin texture appears as the image of the Riemann sphere (i.e. the compactified $xy$-plane), which completely covers the pseudospin Bloch sphere [Fig.~\ref{fig:Pt3_ppin_skyrmion}(b) and \ref{fig:Pt2_ppin_skyrmion}(b)]. 
For the CP$^{3}$-pseudospin skyrmion in an unentangled easy-axis FM background (Fig.~\ref{fig:Pt3_ppin_skyrmion}), the pseudospin magnetization of the FM background spinor $F$ and the center spinor $C$ point to the south and north pole respectively in the pseudospin Bloch sphere, while for the CP$^{3}$-pseudospin skyrmion in an unentangled easy-plane FM background (Fig.~\ref{fig:Pt2_ppin_skyrmion}), the two magnetizations of opposite directions lie in the equatorial plane. In the same manner as for the CP$^{3}$-spin skyrmion, the CP$^{3}$-pseudospin skyrmion is factorizable (because $\alpha=0$) and the image of the Riemann sphere in the entanglement Bloch sphere is simply a point at the north pole [Figs.~\ref{fig:Pt3_ppin_skyrmion}(c) and \ref{fig:Pt2_ppin_skyrmion}(c)]. 

For the CP$^{3}$-pseudospin skyrmion, the lattice-scale profile of $\rho_{\rm Total}(r)$ is identical to that of $M_{\rm Sz}(\br)$, because the spin magnetization is constant over the $xy$-plane and it is fully polarized along the $z$-axis, i.e. we have $\rho_{\downarrow}(A)=\rho_{\downarrow}(B)=0$ and $\rho_{\rm Total}(A/B)=\rho_{\uparrow}(A/B)=M_{\rm Sz}(A/B)$. Since in the $N=0$ LL of graphene, a pseudospin $M_{\rm Pz}=+1(-1)$ means full occupation of the A(B) sublattice, we can read off the $z$-component of the pseudospin magnetization from the sublattice occupation pattern. In particular, for the CP$^{3}$-pseudospin skyrmion in an unentangled easy-axis FM background [visualized in Figs.~\ref{fig:Pt3_ppin_skyrmion}(d) and \ref{fig:Pt3_ppin_skyrmion}(e)], the occupation pattern transforms continuously along the radius from fully occupation of the B-sublattice at the skyrmion center, to full occupation of the A-sublattice in the FM background. In between, we have some region where $\rho_{\rm Total}(A)$ and $\rho_{\rm Total}(B)$ are approximately $1/2$, and the pseudospin magnetization appears to be in-plane.

In Figs.~\ref{fig:Pt2_ppin_skyrmion}(d) and \ref{fig:Pt2_ppin_skyrmion}(e) for the CP$^{3}$-pseudospin skyrmion in an unentangled easy-plane FM background, we encounter a \emph{double-core} structure, which is also called a \emph{bimeron} in the literature.\cite{bimeron} 
Strictly speaking, a bimeron is a pseudospin skyrmion in an easy-plane FM background if the distance between the two cores is locked to the size of the texture in such a manner as to obtain the most homogeneous topological charge distribution. Indeed, one notices from our skyrmion ansatz (\ref{eq:Z1}) that there is only one (eventually deformed) length scale $\lambda_0$, while in a bimeron the core size can in principle be chosen different from the core separation. However, this decoupling of the length scales yields an increase of the leading energy $E_{\rm NLSM}$. One may speculate that this increase in energy can be compensated in a disorder potential that breaks translation symmetry, but this issue is beyond the scope of the present paper, and we thus only consider states of the form (\ref{eq:Z1}) here.
Since the pseudospin magnetization of a CP$^{3}$-pseudospin skyrmion explores all spherical directions, the pseudospin magnetization in the left and right ``cores'' are actually pointing in the direction close to the south and north poles of the pseudospin Bloch sphere. The opposite directions of the pseudospin magnetization in the two cores manifest themselves as full occupation of the A/B sublattice. Notice that the direction of an in-plane polarized pseudospin magnetization cannot be observed from the sublattice occupation pattern, and one therefore has access only to $M_{\rm Pz}$ in lattice-resolved spectroscopy. 

\subsection{Entanglement skyrmion}
\label{subsec:entanglement-skyrmion}

\begin{figure}[t]
\begin{tabular}{ccc}
\includegraphics[width=0.3\columnwidth]{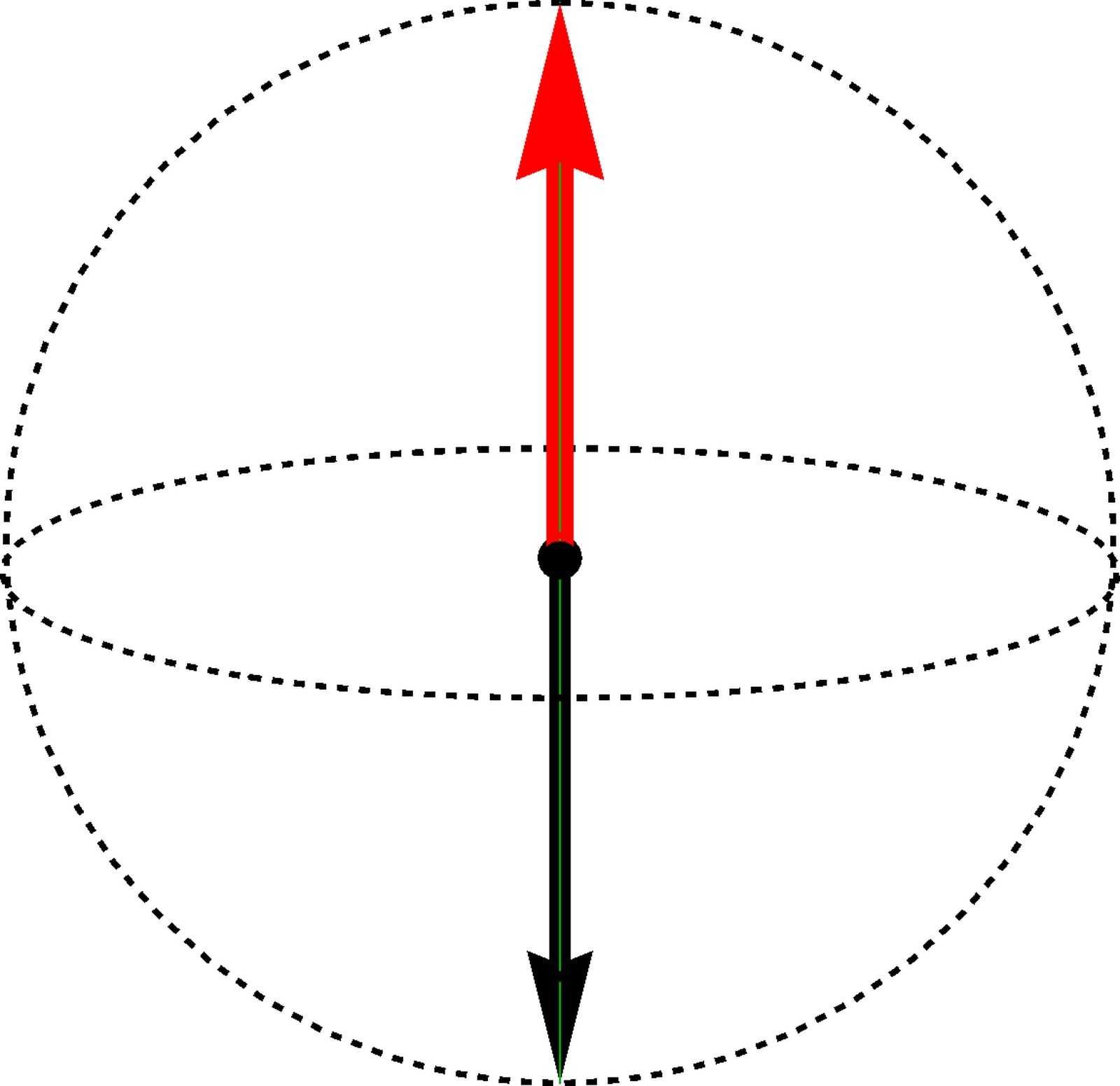} & \includegraphics[width=0.3\columnwidth]{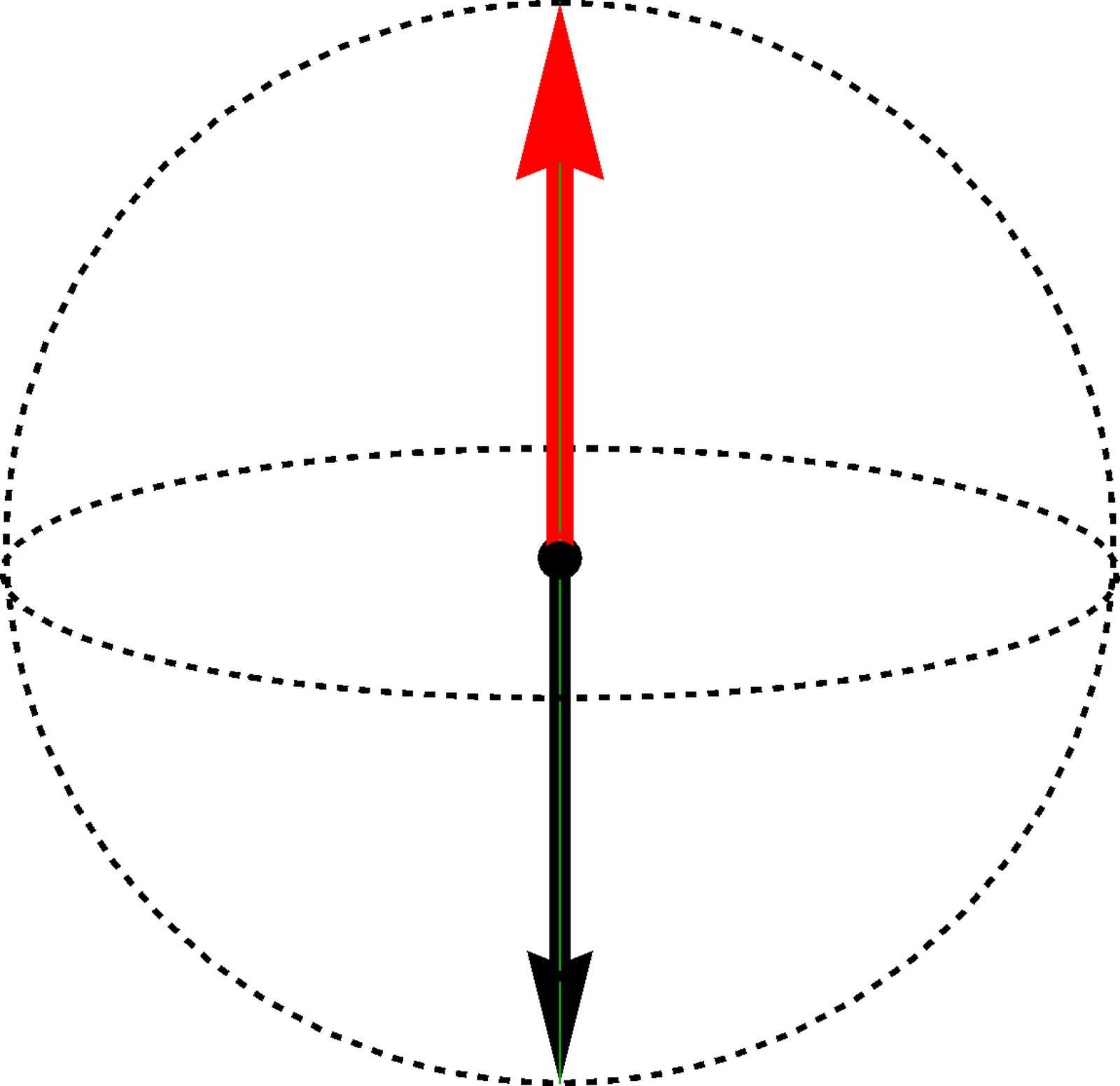} & \includegraphics[width=0.3\columnwidth]{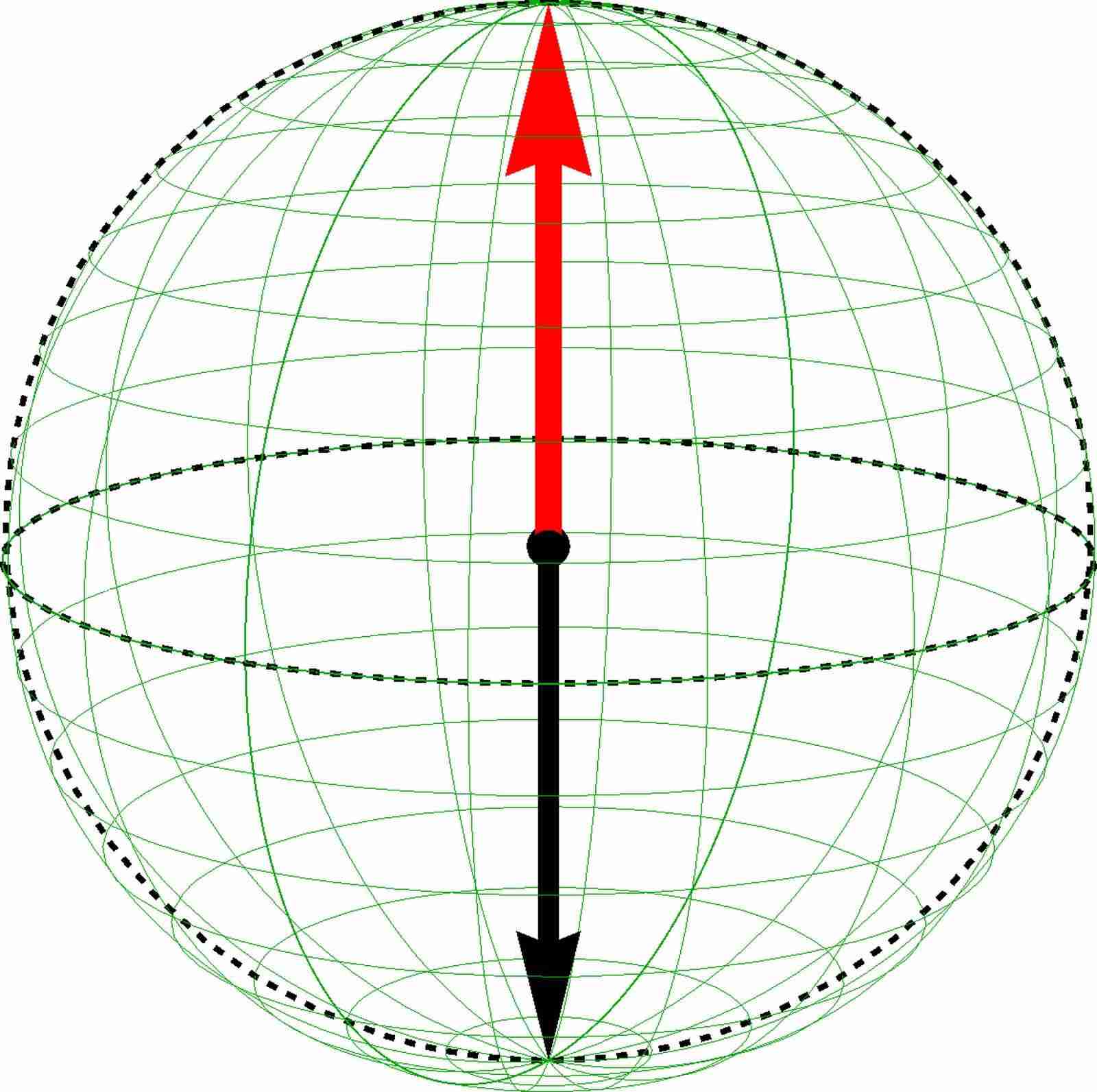}\tabularnewline
(a) Spin BS & (b) Pspin BS & (c) Entanglement BS\tabularnewline
\end{tabular}
\includegraphics[width=\columnwidth]{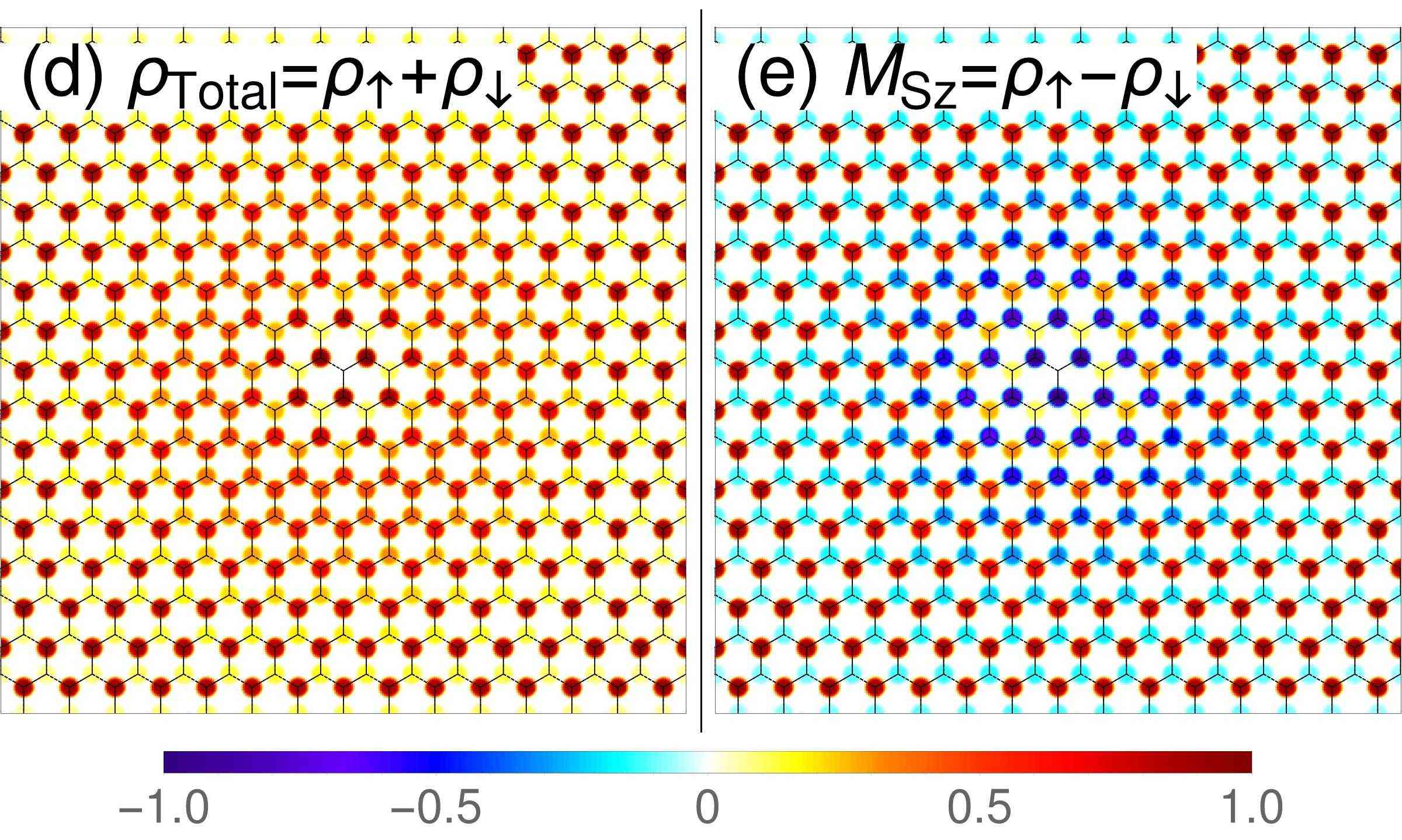}
\caption{\label{fig:Pt6_ent_skyrmion}Visualization of the CP$^{3}$-entanglement skyrmion embedded in an unentangled easy-axis FM background (point 6 in Fig.\ref{fig:Skyrmion_Type_Diagram}). Conventions are the same as in Fig.~\ref{fig:Pt4_spin_skyrmion}.
}
\end{figure}

\begin{figure}[t]
\begin{tabular}{ccc}
\includegraphics[width=0.3\columnwidth]{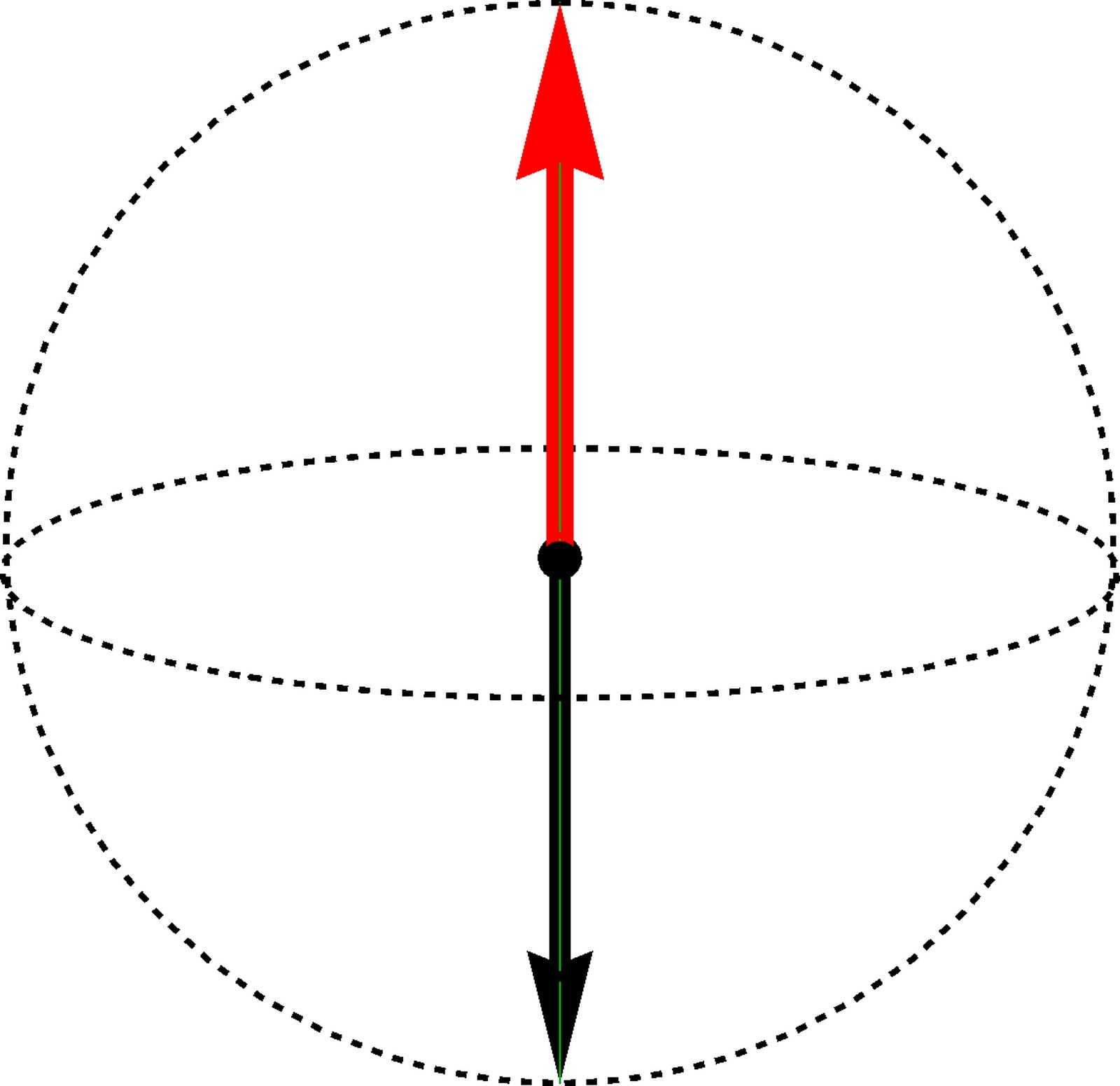} & \includegraphics[width=0.3\columnwidth]{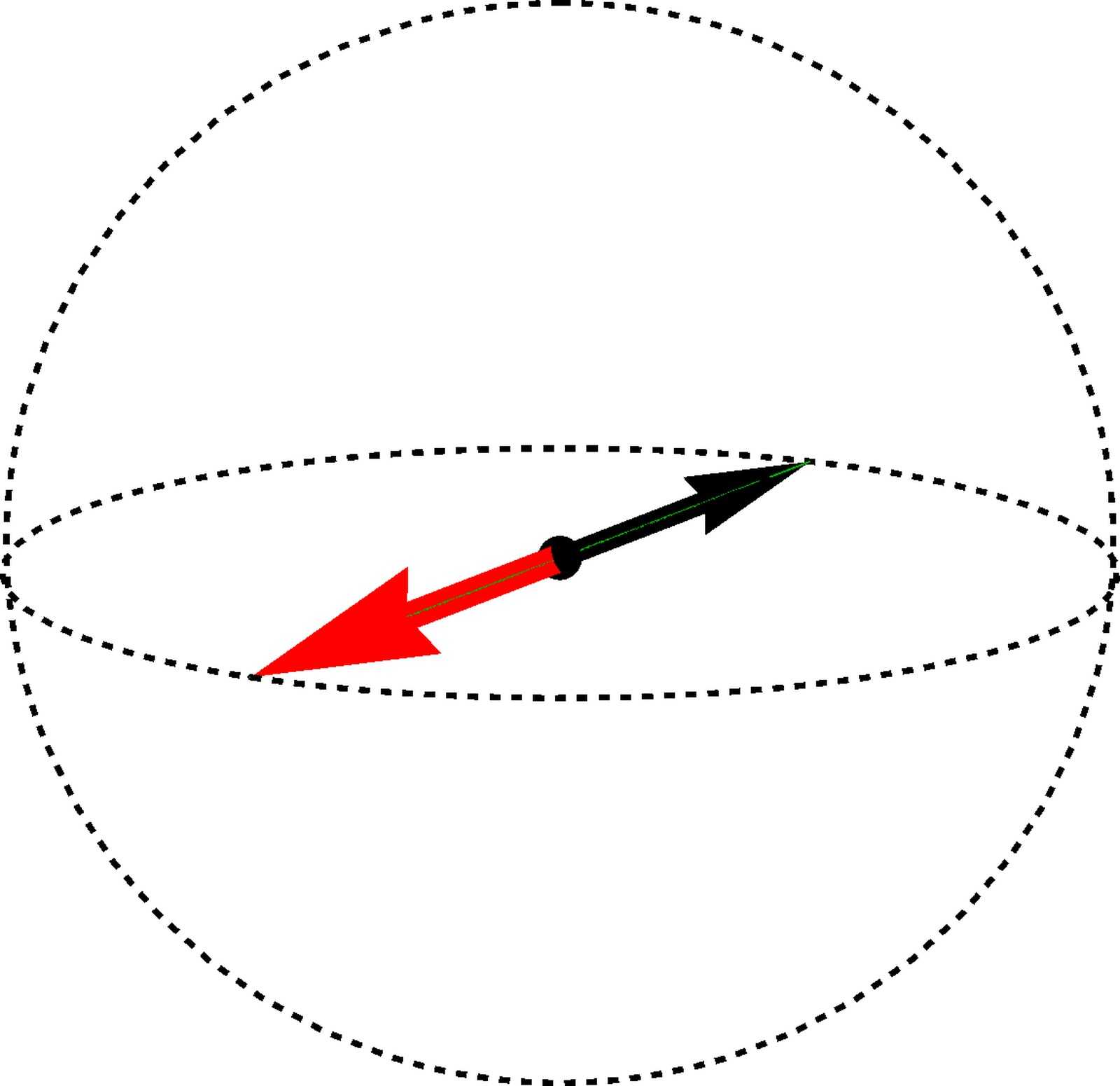} & \includegraphics[width=0.3\columnwidth]{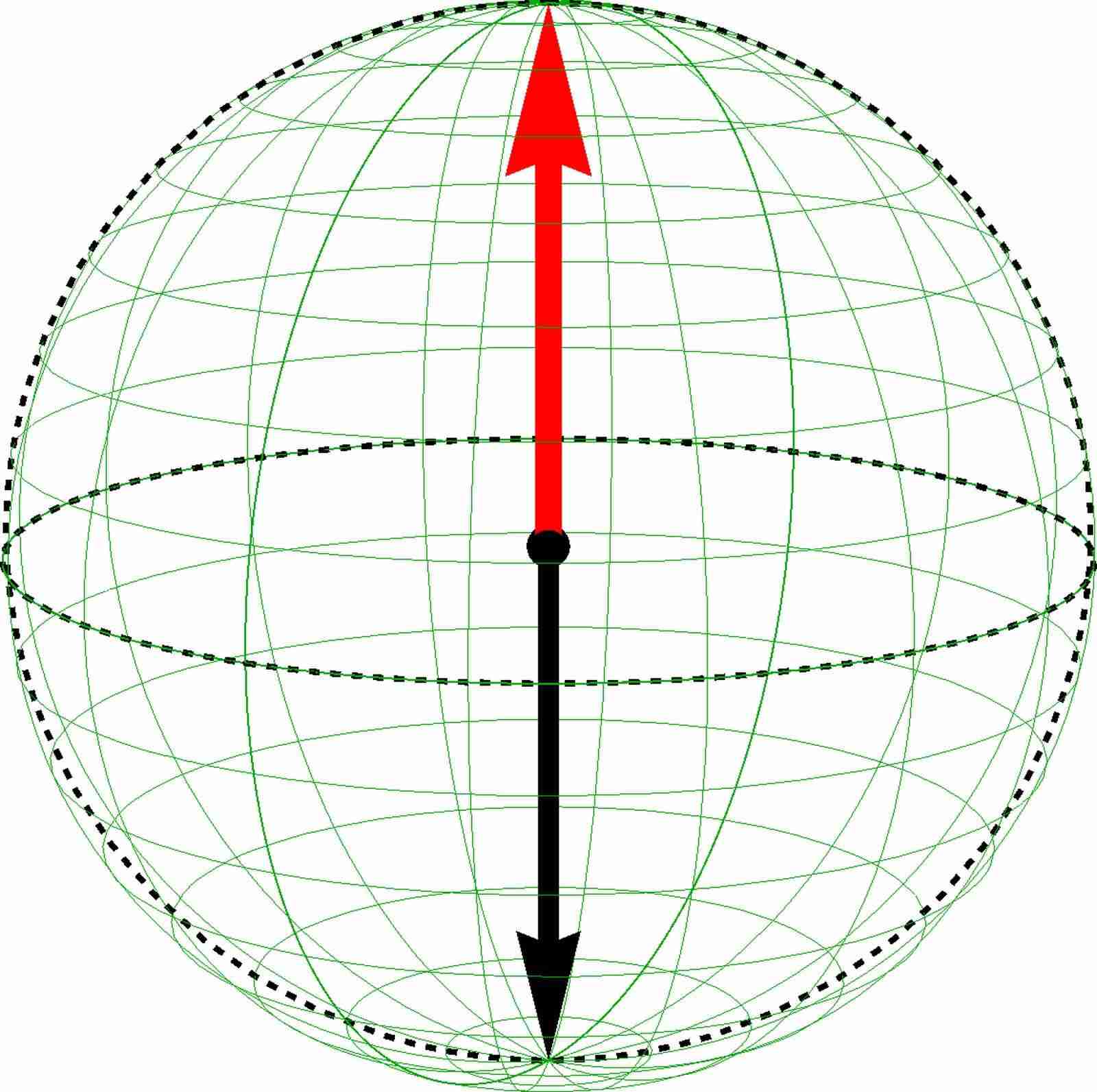}\tabularnewline
(a) Spin BS & (b) Pspin BS & (c) Entanglement BS\tabularnewline
\end{tabular}
\includegraphics[width=\columnwidth]{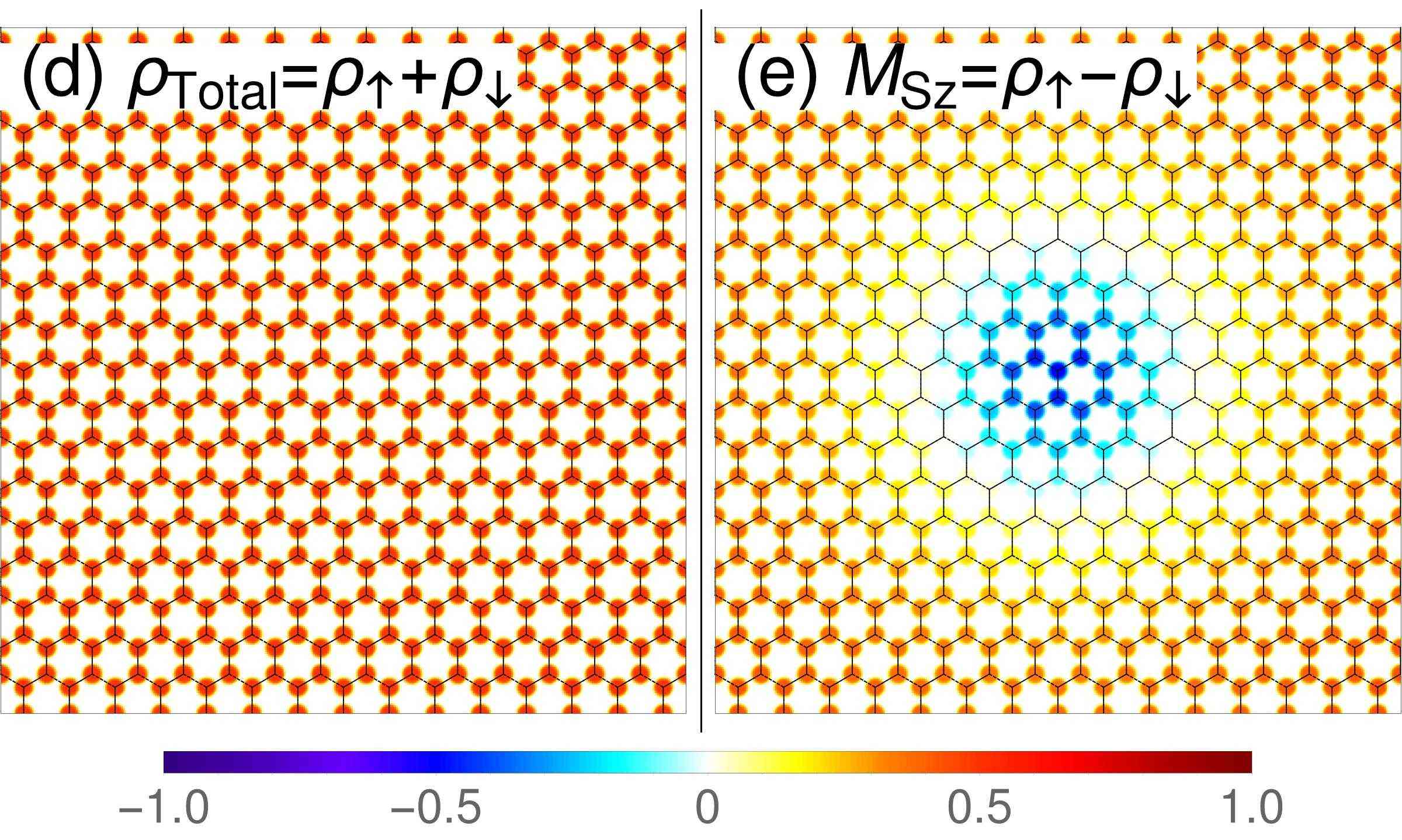}
\caption{\label{fig:Pt9_ent_skyrmion}Visualization of the CP$^{3}$-entanglement skyrmion embedded in an unentangled easy-plane FM background (point 9 in Fig.\ref{fig:Skyrmion_Type_Diagram}). Conventions are the same as in Fig.~\ref{fig:Pt4_spin_skyrmion}.
}
\end{figure}

\begin{figure}[t]
\begin{tabular}{ccc}
\includegraphics[width=0.3\columnwidth]{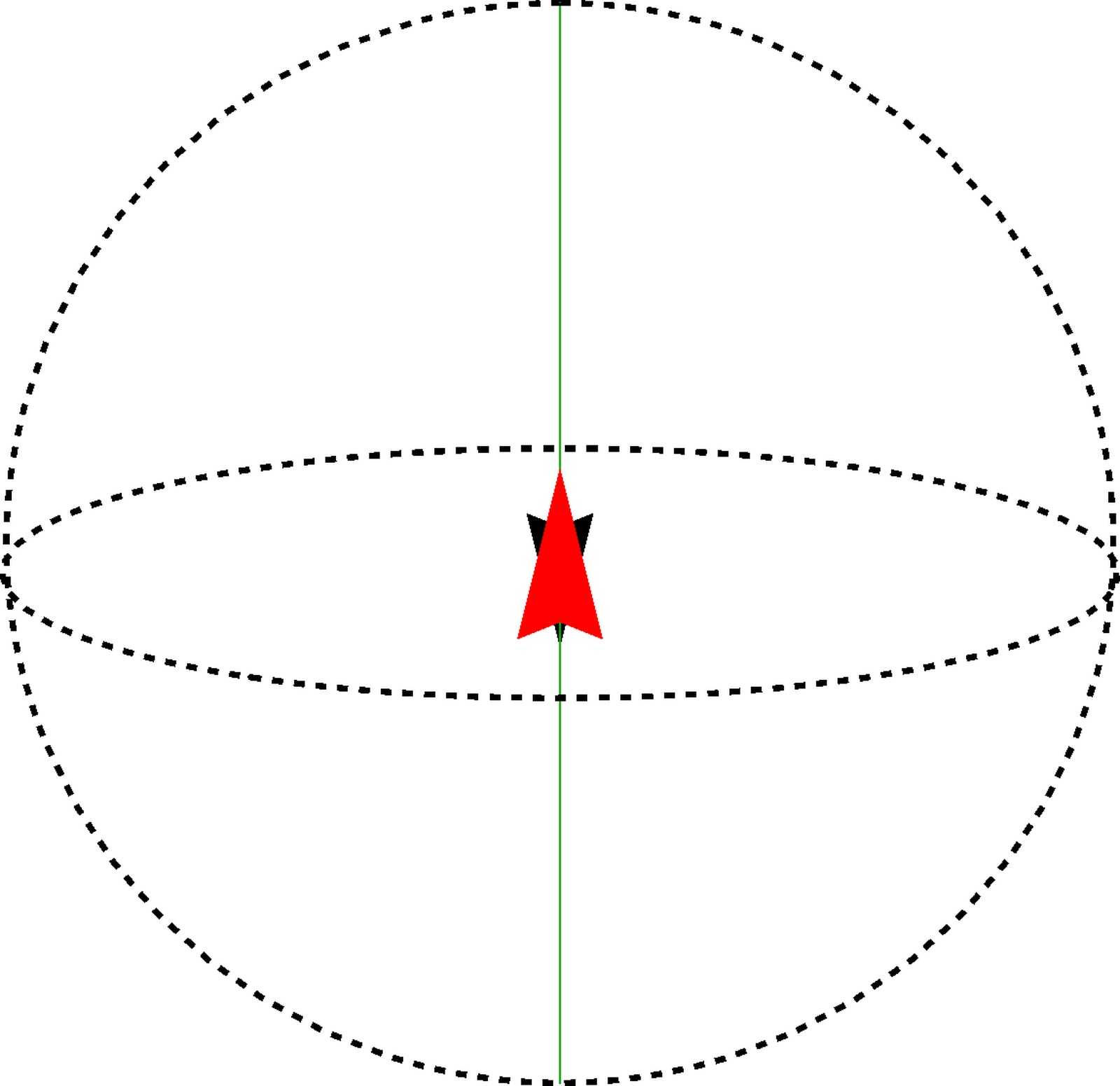} & \includegraphics[width=0.3\columnwidth]{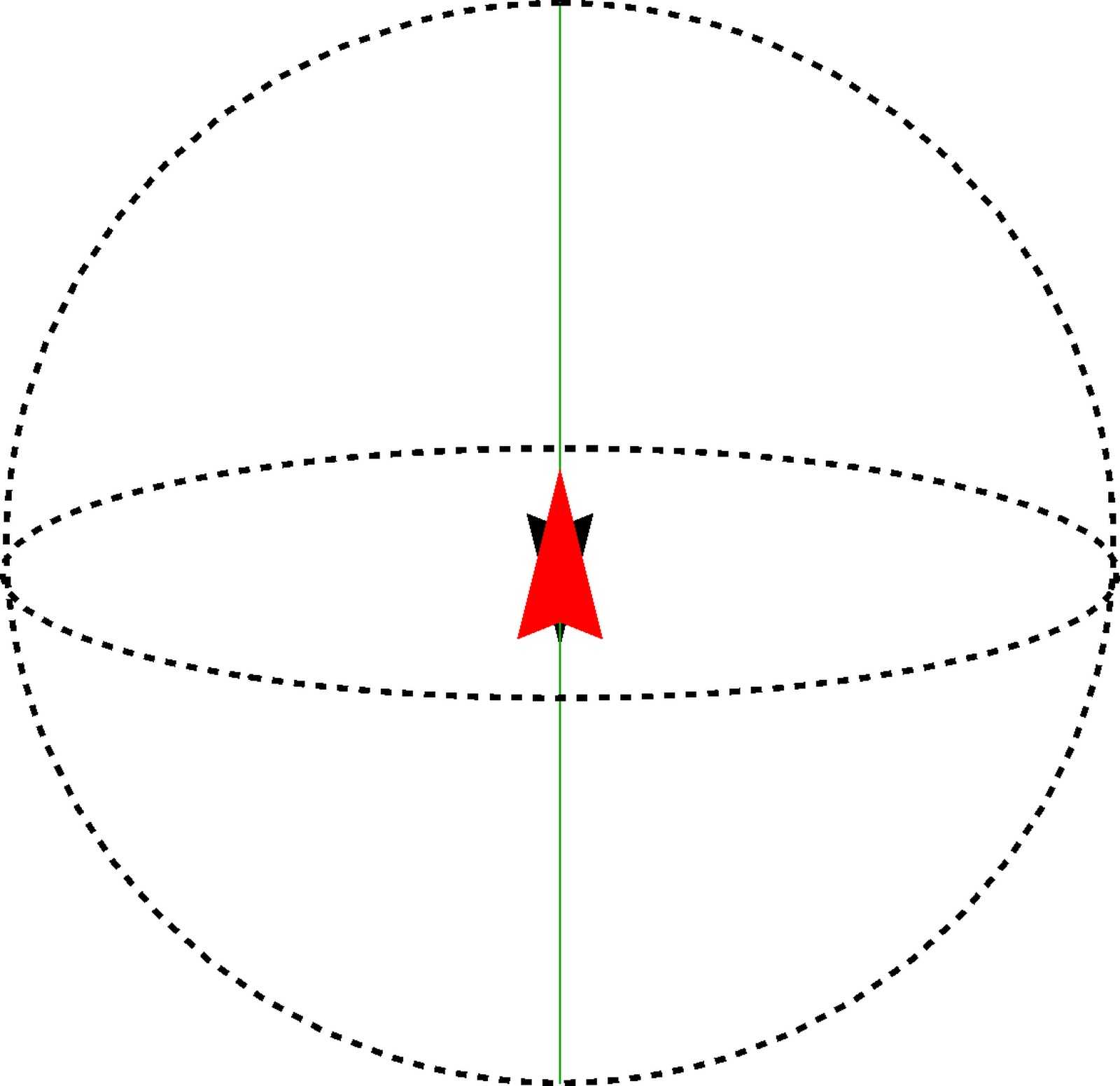} & \includegraphics[width=0.3\columnwidth]{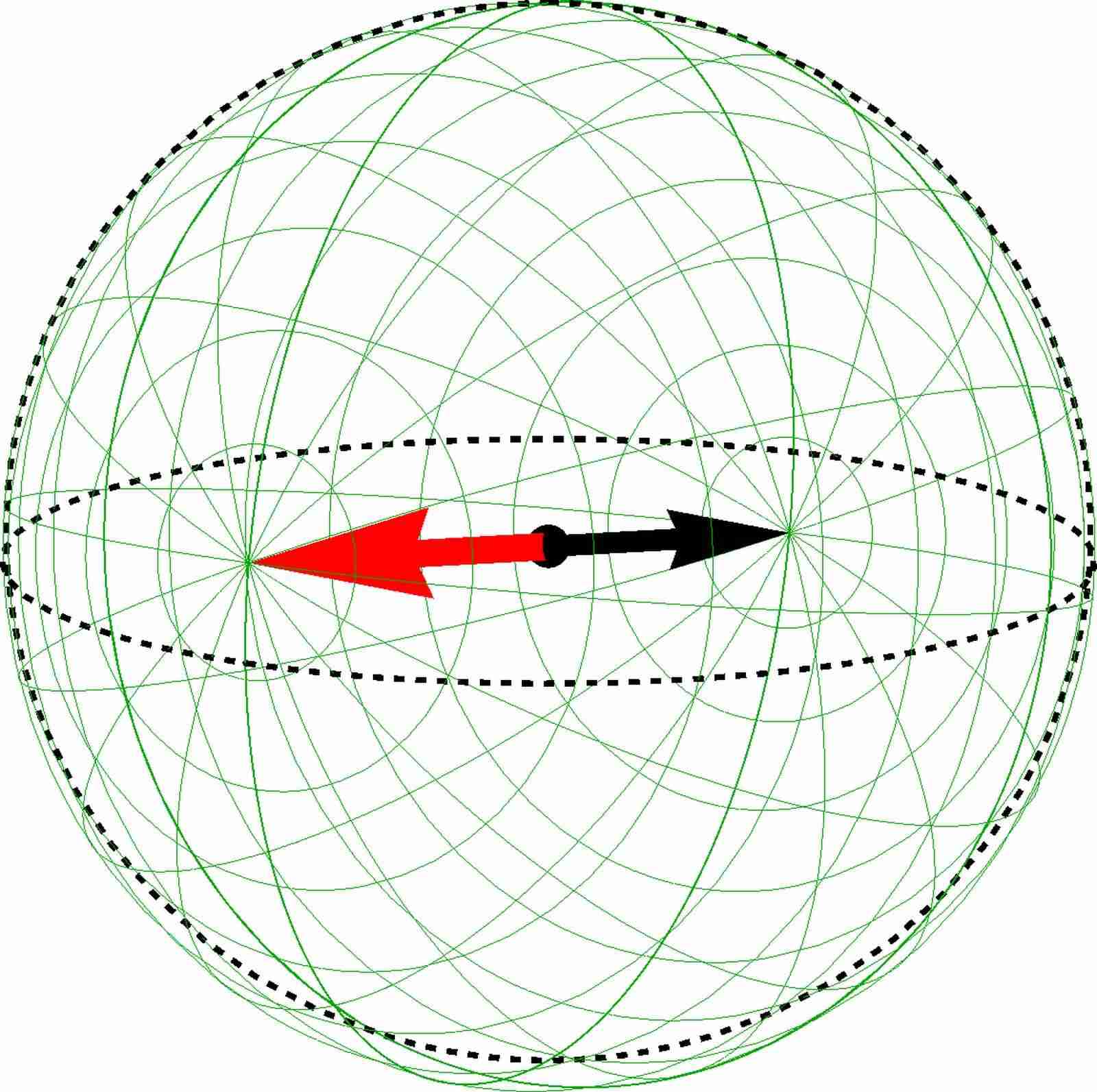}\tabularnewline
(a) Spin BS & (b) Pspin BS & (c) Entanglement BS\tabularnewline
\end{tabular}
\includegraphics[width=\columnwidth]{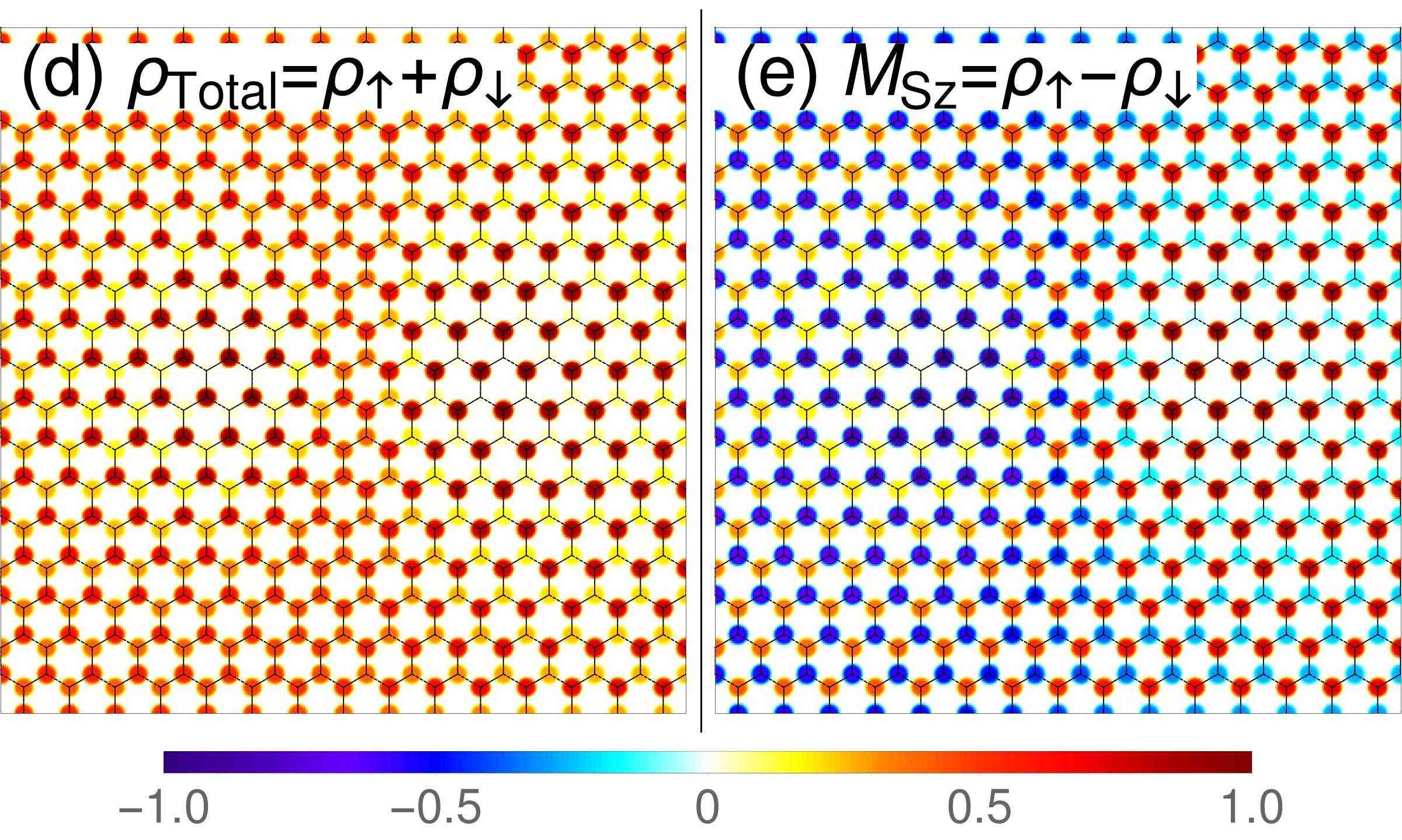}
\caption{\label{fig:Pt13_ent_skyrmion}Visualization of the CP$^{3}$-entanglement skyrmion embedded in an entangled easy-axis FM background (point 13 in Fig.\ref{fig:Skyrmion_Type_Diagram}). Conventions are the same as in Fig.~\ref{fig:Pt4_spin_skyrmion}.
}
\end{figure}

\begin{figure}[t]
\begin{tabular}{ccc}
\includegraphics[width=0.3\columnwidth]{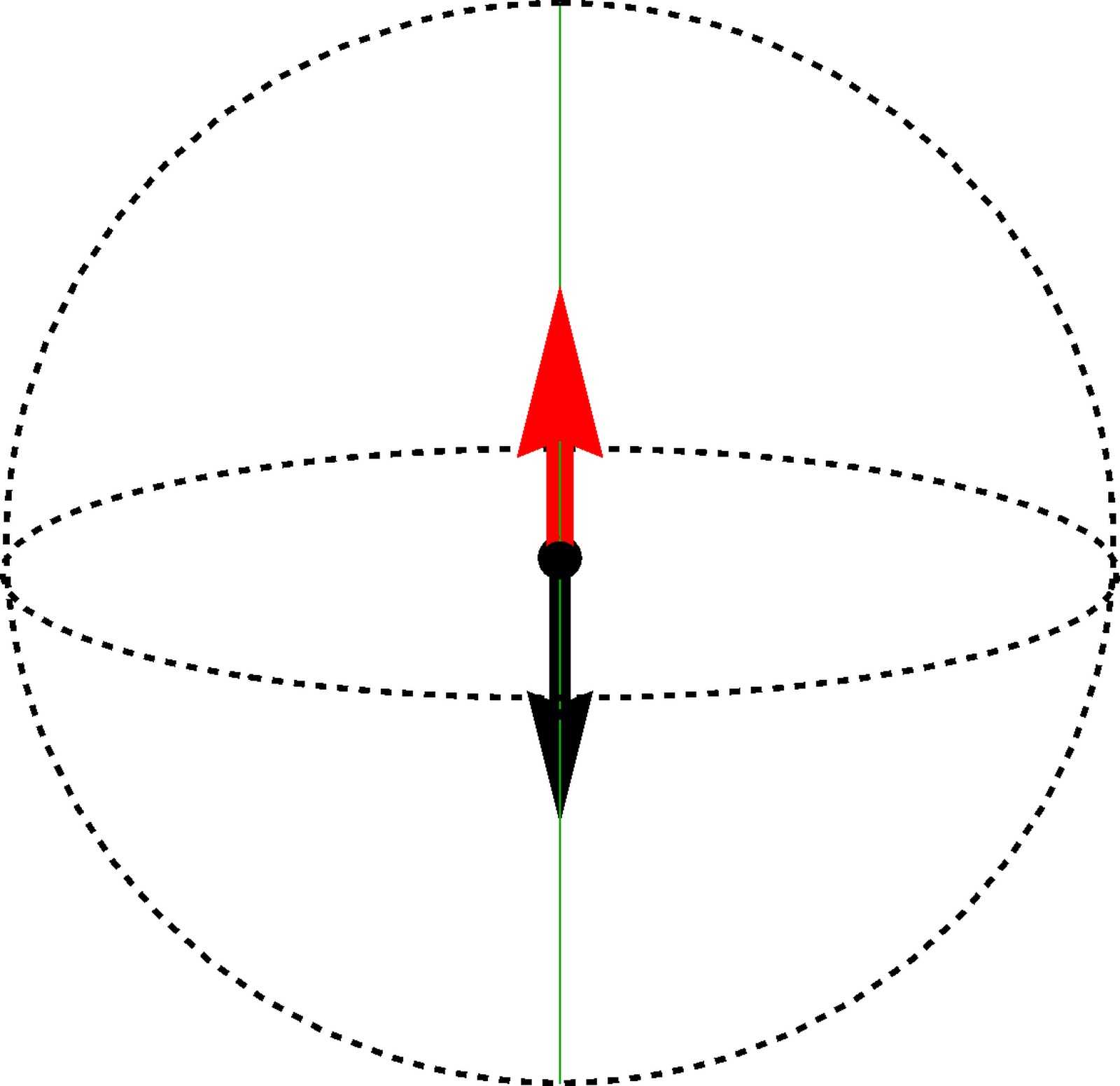} & \includegraphics[width=0.3\columnwidth]{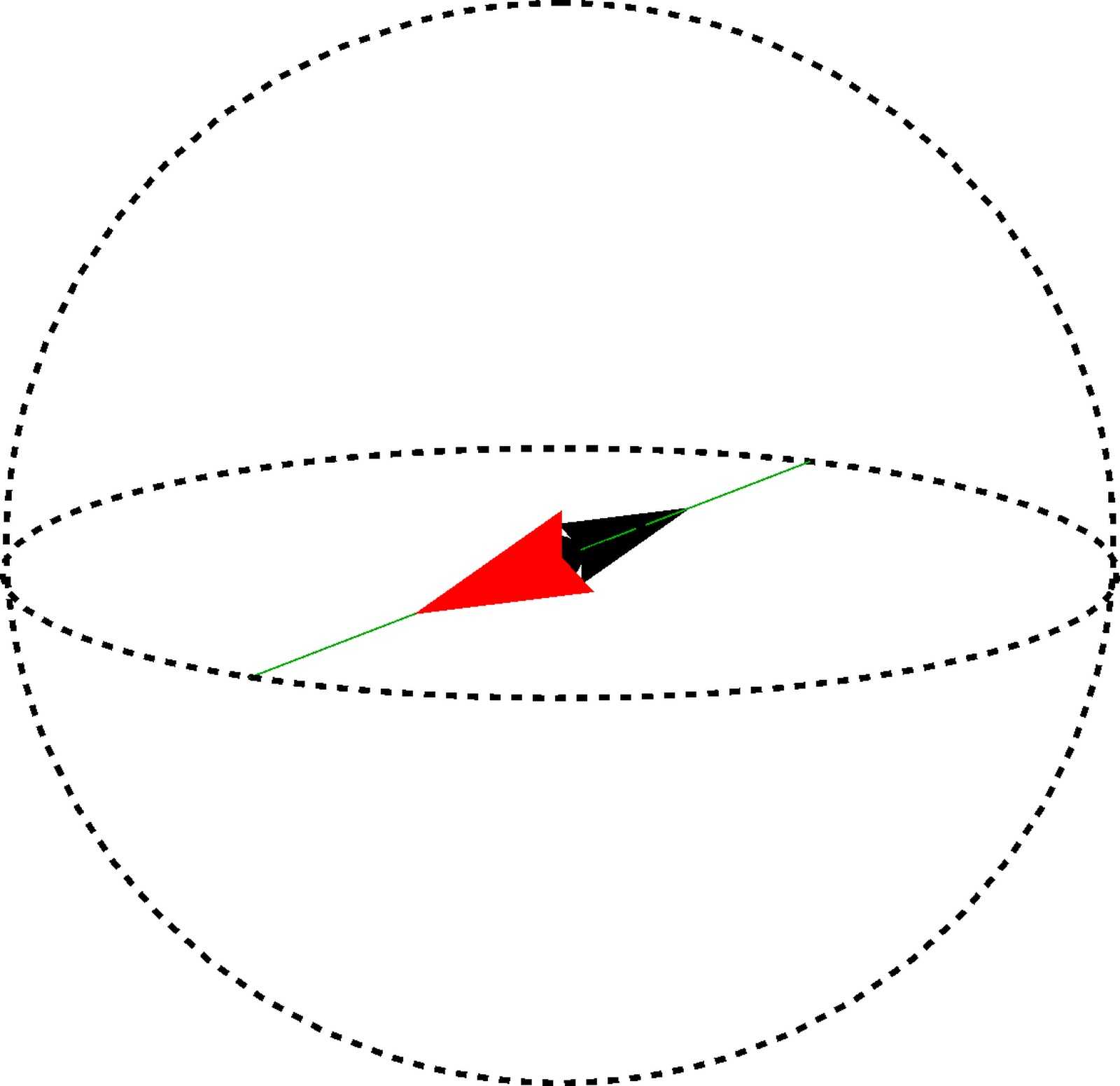} & \includegraphics[width=0.3\columnwidth]{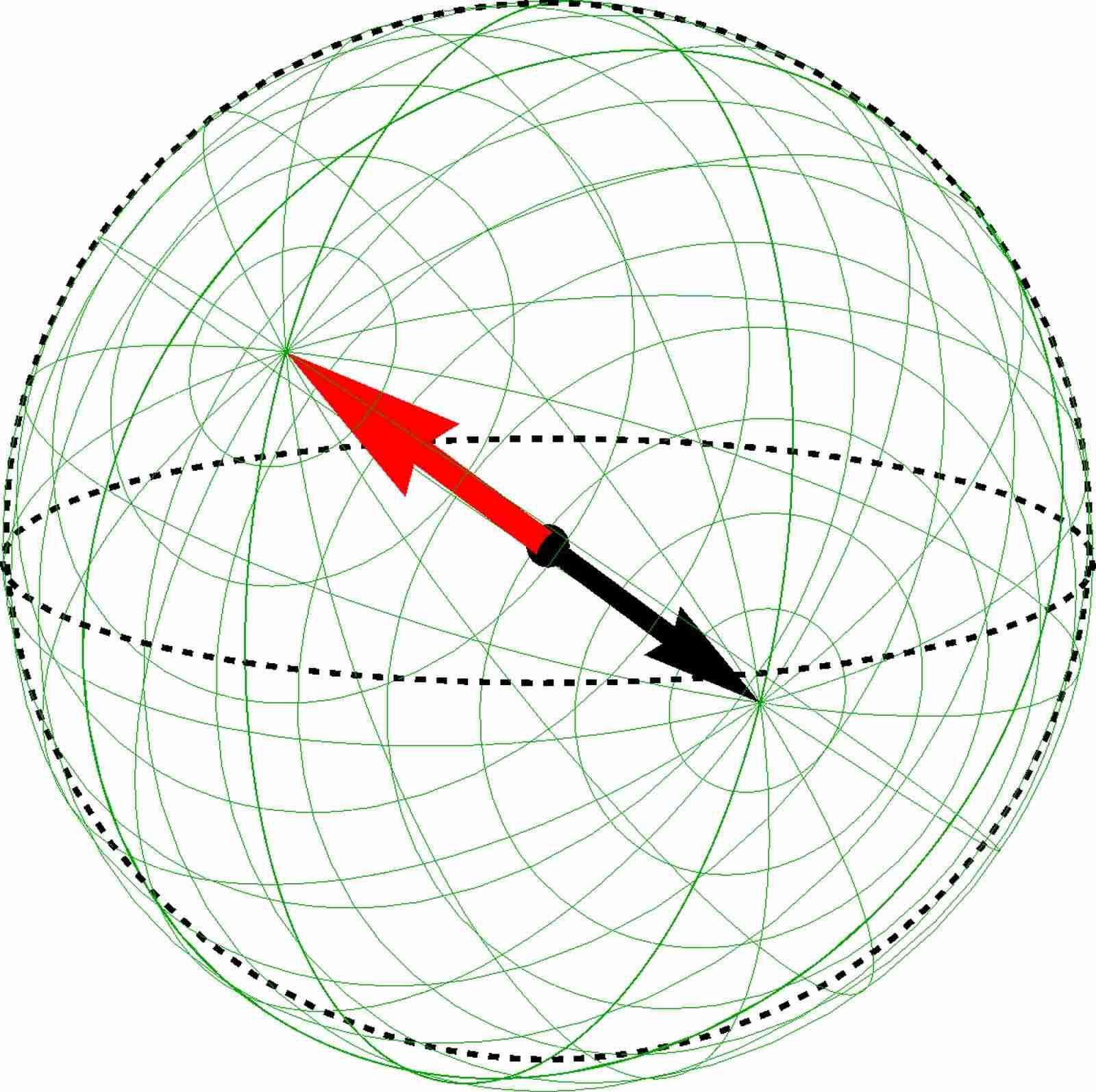}\tabularnewline
(a) Spin BS & (b) Pspin BS & (c) Entanglement BS\tabularnewline
\end{tabular}
\includegraphics[width=\columnwidth]{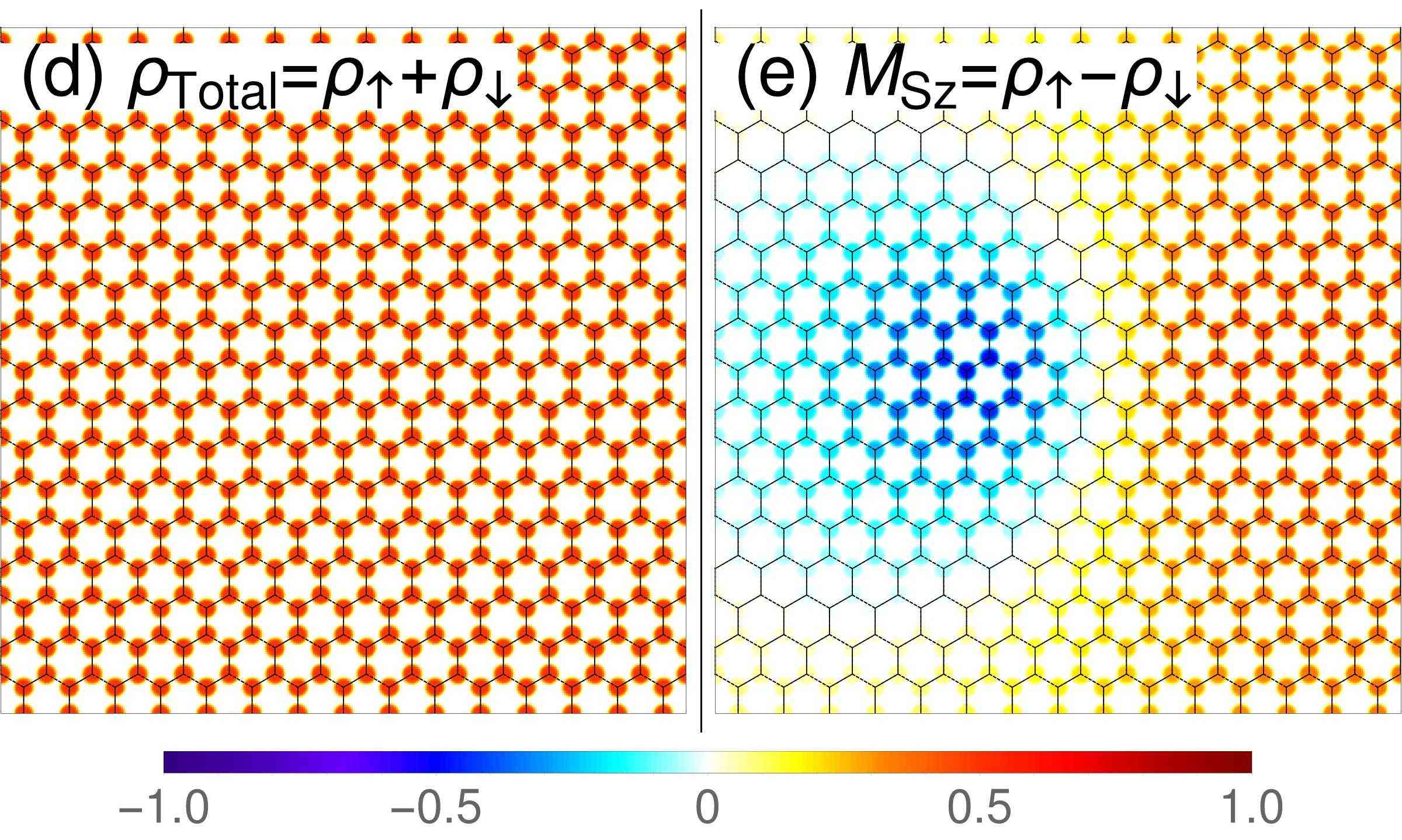}
\caption{\label{fig:Pt10_ent_skyrmion}Visualization of the CP$^{3}$-entanglement skyrmion embedded in an entangled easy-plane FM background (point 10 in Fig.\ref{fig:Skyrmion_Type_Diagram}). Conventions are the same as in Fig.~\ref{fig:Pt4_spin_skyrmion}.
}
\end{figure}

When $(u_{\rm z},u_{\perp})$ lies in the yellow region in the skyrmion-type diagram, the optimal value of the parameter $\theta_C$ of the center spinor $C$ is zero, and the center spinor $C$ Eq.~(\ref{eq:Para_C}) is simplified: 
\begin{equation}
C=-e^{-i\beta}\sin\frac{\alpha}{2}\psi^{\rm P}\otimes\psi^{\rm S} +\cos\frac{\alpha}{2}\chi^{\rm P}\otimes\chi^{\rm S}
\end{equation}
Meanwhile the FM background spinor $F$ is parametrized as 
\begin{equation}
F=\cos\frac{\alpha}{2}\psi^{\rm P}\otimes\psi^{\rm S} + e^{i\beta}\sin\frac{\alpha}{2}\chi^{\rm P}\otimes\chi^{\rm S}
\end{equation}
Recall that the parameters $\alpha$ and $\beta$ in both $F$ and $C$ are determined by the FM background spinor $F$. Therefore the CP$^{3}$-skyrmion $Z_{\rm ent}={\cal N}^{-1} \left\{(x+i y)F - \lambda C \right\}$ can be written as 
\begin{equation}
Z_{\rm ent}={\cal N}^{-1}\left\lbrace(x+iy)\left(\begin{array}{c}
\cos\frac{\alpha}{2}\\
e^{i\beta}\sin\frac{\alpha}{2}
\end{array}\right)-\lambda\left(\begin{array}{c}
-e^{-i\beta}\sin\frac{\alpha}{2}\\
\cos\frac{\alpha}{2}
\end{array}\right)\right\rbrace\label{eq:Z_ent}
\end{equation}
with respect to the fixed basis $(\psi^{\rm P}\otimes\psi^{\rm S},\chi^{\rm P}\otimes\chi^{\rm S})$, which are determined from the FM background CP$^{3}$-spinor $F$ and independent of the $x,y$ coordinates. The above form of $Z_{\rm ent}$ resembles an O(3) skyrmion written in $CP^1$ form: the entanglement vector for $Z_{\rm ent}$ defined in Eq.~(\ref{eq:Entanglement-vector}) $$\boldsymbol{m}_{\rm E}(\alpha,\beta) = (\sin\alpha\cos\beta,\sin\alpha\sin\beta,\cos\alpha)$$ is the ``magnetization'', which explores all directions of the \emph{entanglement Bloch sphere}, as shown in Fig.~\ref{fig:Pt6_ent_skyrmion}(c), \ref{fig:Pt9_ent_skyrmion}(c), \ref{fig:Pt13_ent_skyrmion}(c), \ref{fig:Pt10_ent_skyrmion}(c). Thus we call this type CP$^{3}$-\emph{entanglement skyrmion}.

The visualizations of this type of skyrmion in three Bloch spheres are special. While the spin and pseudospin polarizations continue to be aligned in the same directions in the spin and pseudospin Bloch spheres, respectively, the images of Riemann sphere shrink to a \emph{line} (instead of points in previous cases) and explore the \emph{interior} of the two Bloch spheres (see panels (a) and (b) in Figs.~\ref{fig:Pt6_ent_skyrmion}, \ref{fig:Pt9_ent_skyrmion}, \ref{fig:Pt13_ent_skyrmion} and \ref{fig:Pt10_ent_skyrmion}.)  
This is a consequence of the varying entanglement in the CP$^{3}$-spinor.
To understand this in more detail, let us first recall that the image of the Riemann sphere is the collection of the arrowheads of the vectors representing the spin and pseudospin magnetization. For the CP$^{3}$-entanglement skyrmion, when the parameter $\alpha$ explores all the possible values in $(0,\pi)$, the magnitude of the spin and pseudospin magnetizations will be less than $1$ and eventually vanish at $\alpha(\br_0)=\pi/2$ for some $\br_0$. Furthermore, it can be verified explicitly that the $x$ and $y$ components of the spin magnetization of $Z_{ent}(\br)$ are zero throughout the $xy$-plane. Therefore the spin magnetization vector is confined to the $z$-axis inside the spin Bloch sphere, and the image of the Riemann sphere shrinks to the diameter along the $z$-axis for all the four cases. To explain the similar behaviour of the pseudospin magnetization, we write down the expression of the pseudospin magnetization for $Z_{ent}(\br)$:
\begin{eqnarray}
\bM_{\rm P}(\br) &=& Z^{\dagger}_{ent}(\br)(\boldsymbol{\sigma}\otimes 1)Z_{ent}(\br)\nonumber\\
&=& {\cal N}^{-2} \left\{ \left|U\right|^2 \psi^{\dagger}_{\rm P}\boldsymbol{\sigma}\psi_{\rm P} + \left|V\right|^2 \chi^{\dagger}_{\rm P}\boldsymbol{\sigma}\chi_{\rm P}\right\}\nonumber\\
&=& {\cal N}^{-2} \left( \left|U\right|^2 - \left|V\right|^2 \right) \bmm_{\rm P}\label{eq:MT_Z_ent}
\end{eqnarray}
where 
\begin{eqnarray}
U &= (x+i y)\cos\frac{\alpha}{2} + \lambda e^{-i\beta}\sin\frac{\alpha}{2}\nonumber\\
V &= (x+ i y)e^{i\beta}\sin\frac{\alpha}{2} - \lambda \cos\frac{\alpha}{2}\nonumber
\end{eqnarray}
are the first and second components of $Z_{\rm ent}$ in Eq.~(\ref{eq:Z_ent}), and $\boldsymbol{m}_{\rm P}=\psi^{\rm{P}\dagger}\boldsymbol{\sigma}\psi^{\rm P}=-\chi^{\rm{P}\dagger}\boldsymbol{\sigma}\chi^{\rm P}$ is the unit vector of pseudospin magnetization. The above equations reveal that the pseudospin magnetization for a CP$^{3}$-entanglement skyrmion always lies in the constant direction $\boldsymbol{m}_{\rm P}$  of the pseudospin magnetization determined by the FM background, but with varying magnitude. That is to say, inside the pseudospin Bloch sphere, the shape of the Riemann sphere image collapses to a diameter line in the direction $\boldsymbol{m}_{\rm P}$. 
One notices that the orientation of the spin and pseudospin magnetization is changed when the entanglement vector crosses the equatorial plane of the entanglement Bloch sphere ($\alpha=\pi/2$). 

A CP$^{3}$-entanglement skyrmion can be embedded into all four types of FM background, as revealed by the skyrmion-type diagram Fig.~\ref{fig:Skyrmion_Type_Diagram}. 
We use the oppositely directed pair of entanglement vectors of the background spinor $F$ (colored in blue) and the center spinor $C$ (colored in red) to distinguish the four examples of the CP$^{3}$-entanglement skyrmion $Z_{\rm ent}$. For $Z_{\rm ent}$ embedded in the unentangled easy-axis and easy-plane FM backgrounds, (Fig.~\ref{fig:Pt6_ent_skyrmion} and \ref{fig:Pt9_ent_skyrmion}), respectively, the entanglement vector for the FM background spinor $F$ always points to the north pole [Figs.~\ref{fig:Pt6_ent_skyrmion}(c), \ref{fig:Pt9_ent_skyrmion}(c)] because the FM spinor $F$ is factorizable and thus $\alpha=0$ at $r\rightarrow \infty$. For $Z_{\rm ent}$ embedded in the entangled easy-axis and easy-plane FM backgrounds with generic values of $\alpha$, the entanglement vector for $F$ is tilted away from $z$-axis [Figs.~\ref{fig:Pt13_ent_skyrmion}(c), \ref{fig:Pt10_ent_skyrmion}(c)], and hence the entire image of the Riemann sphere on the entanglement Bloch sphere is \emph{tilted}. Such picture of tilting helps us to understand the relation between the CP$^{3}$-entanglement skyrmion embedded in the unentangled and entangled FM backgrounds.

Figure~\ref{fig:Pt6_ent_skyrmion} shows the case for a CP$^{3}$-entanglement skyrmion embedded in an unentangled easy-axis FM background. In this case, the $\rho_{\rm Total}(\br)$ profile is similar to that for the CP$^{3}$-pseudospin skyrmion visualized in Fig.~\ref{fig:Pt3_ppin_skyrmion}, but the $M_{\rm Sz}(\br)$ profile is different. 
Both the center spinor $C\sim Z_{\rm ent}(0)=\chi^{\rm P}\otimes\chi^{\rm S}$ and the FM background spinor $F\sim Z_{\rm ent}(\infty)=\psi^{\rm P}\otimes\psi^{\rm S}$ have no entanglement, implying that the A sublattice is occupied at the skyrmion center, whereas the B sublattice is occupied in the region far from the core, with the sublattice spin magnetization pointing into opposite directions. 
At a point $0<|\br_0|<\infty$, the lattice-scale profiles is an interpolation between $0$ and $\infty$ and has ``anti-ferrimagnetic'' appearance, which has been discussed in Sec.~\ref{subsec:QHFM-visualization} in the framework of FM with entanglement. Here, the CP$^{3}$-spinor $Z_{\rm ent}(\br_0)$ appears on the lattice as simultaneous occupation of both sublattices with different amplitudes in general, with the sublattice spin magnetization pointing along the $z$-axis with opposite directions.

Figure~\ref{fig:Pt9_ent_skyrmion} shows the case for the CP$^{3}$-entanglement skyrmion embedded in an unentangled easy-plane FM background. Most significantly, the \emph{uniform} $\rho_{\rm Total}(\br)$ profile Fig.~\ref{fig:Pt9_ent_skyrmion}(d) shows that the A/B sublattices are \emph{equally} occupied. This can be understood from the observation that the diameter line in the pseudospin Bloch sphere Fig.~\ref{fig:Pt9_ent_skyrmion}(b) lies in the equatorial plane, so the $z$-component of the pseudospin magnetization vanishes everywhere throughout the $xy$-plane. This is also true for the CP$^{3}$-spin skyrmion embedded in the unentangled easy-plane FM background (Fig.~\ref{fig:Pt1_spin_skyrmion}), as well as for the CP$^{3}$-entanglement skyrmion embedded in the entangled easy-plane FM background (Fig.~\ref{fig:Pt10_ent_skyrmion}). 
The $M_{\rm Sz}(\br)$ profile shows that the spin magnetization is opposite at the center and in regions far from the center. 
We notice that, only by the lattice-scale profiles, one cannot distinguish the CP$^{3}$-entanglement skyrmion shown in Fig.~\ref{fig:Pt9_ent_skyrmion}, from a CP$^{3}$-spin skyrmion shown in Fig.~\ref{fig:Pt1_spin_skyrmion}. Both of them are embedded in the same unentangled easy-plane FM background. 
In order to distinguish them, one can in principle use the lattice-scale profiles of the $x$ and $y$ components of spin magnetization $M_{\rm Sx}(\br)$ and $M_{\rm Sy}(\br)$, since these profiles have different appearances for the two types of skyrmions. The profiles $M_{\rm Sx}(\br)$ and $M_{\rm Sy}(\br)$ vanish for the CP$^{3}$-entanglement skyrmion, but for the CP$^{3}$-spin skyrmion they reach a maximal value in the region where $M_{\rm Sz}(\br)$ vanishes. 

For the CP$^{3}$-entanglement skyrmions embedded in entangled FM backgrounds, Fig.~\ref{fig:Pt13_ent_skyrmion} shows the case of an entangled easy-axis FM background. The double-core texture is present for both $\rho_{\rm Total}(\br)$ and $M_{\rm Sz}(\br)$ profiles. In the left core, sublattice A is occupied with the spin magnetization pointing down, while sublattice-B is empty. In the right core, sublattice B is occupied with the spin magnetization pointing up, while sublattice A is empty. Recalling the ``tilting picture'' discussed earlier in this subsection, the $\rho_{\rm Total}(\br)$ and $M_{\rm Sz}(\br)$ profiles of the two cores are in fact identical to the center and the background of a CP$^{3}$-entanglement skyrmion embedded in the unentangled easy-axis FM background (Fig.~\ref{fig:Pt6_ent_skyrmion}). It is precisely the tilting of the entanglement Bloch sphere that ``shifts'' the center and infinity in the textures of the unentangled easy-axis case to the left and right cores in the textures of the case here. 

Figure~\ref{fig:Pt10_ent_skyrmion} shows the CP$^{3}$-entanglement skyrmions embedded in the entangled easy-plane FM background. There is no texture in the $\rho_{\rm Total}(\br)$ profile, because $M_{\rm Pz}=0$ leads to equal sublattice occupation everywhere in the $xy$-plane. This is similar to the CP$^{3}$-spin (Fig.~\ref{fig:Pt1_spin_skyrmion}) and the CP$^{3}$-entanglement skyrmion (Fig.~\ref{fig:Pt9_ent_skyrmion}) when both are embedded into an unentangled easy-plane FM background. 
Meanwhile, in the $M_{\rm Sz}(\br)$ profile, the left core has $\rho_{\downarrow}(A)=\rho_{\downarrow}(B)=1/2$, whereas the right cores has $\rho_{\uparrow}(A)=\rho_{\uparrow}(B)=1/2$. The ``tilting'' of the entanglement Bloch sphere relates the textures in these two cores, to the textures at the center and infinity of the CP$^{3}$-entanglement skyrmion embedded in an unentangled easy-plane FM background (Fig.~\ref{fig:Pt9_ent_skyrmion}). 

In the $M_{\rm Sz}(\br)$ profile for a CP$^{3}$-entanglement skyrmion embedded in an unentangled (Fig.~\ref{fig:Pt9_ent_skyrmion}) and an entangled easy-plane FM background (Fig.~\ref{fig:Pt10_ent_skyrmion}), we observe the white bands, where the $z$-component of spin magnetization vanishes at both A- and B- sublattices. These regions (on the $xy$-plane) are mapped to the equator of the entanglement Bloch sphere through the entanglement vector $\bmm_{\rm E}(\alpha,\beta)$. Since $\alpha=\pi/2$ on the equator, the magnitude of the spin magnetization $\bM_{\rm S}$ and the pseudospin magnetization $\bM_{\rm P}$ of $Z_{ent}(\br)$ shrink to zero. 

\subsection{Deflated pseudospin skyrmion}
\label{subsec:deflated-pseudospin-skyrmion}

When $\min(u_{\perp},u_{\rm z})\ge 1/2$ and $(u_{\perp},u_{\rm z})$ fall in the blue region in the skyrmion-type diagram Fig.~(\ref{fig:Skyrmion_Type_Diagram}), the energy minimizing CP$^{3}$-skyrmions are of the most intricate type, because the optimal value of $\phi_C$ lies in $(0,\pi/2)$, and the interpolation between the center spinor $C$ and the FM background spinor $F$ involves all four basis spinors $\psi_{\rm P}\otimes\psi_{\rm S}$, $\psi_{\rm P}\otimes\chi_{\rm S}$, $\chi_{\rm P}\otimes\psi_{\rm S}$ and $\chi_{\rm P}\otimes\chi_{\rm S}$ introduced in the parametrization of $F$. The situation for the aforementioned three types, in contrast, is simpler, where only two basis spinors are involved in the interpolation: in the case of CP$^{3}$-spin skyrmion they are $\psi_{\rm P}\otimes\psi_{\rm S}$ and $\psi_{\rm P}\otimes\chi_{\rm S}$; for CP$^{3}$-pseudospin skyrmion $\psi_{\rm P}\otimes\psi_{\rm S}$ and $\chi_{\rm P}\otimes\psi_{\rm S}$; for CP$^{3}$-entanglement skyrmion $\psi_{\rm P}\otimes\psi_{\rm S}$ and $\chi_{\rm P}\otimes\chi_{\rm S}$. The explicit form of this type of CP$^{3}$-skyrmion $Z_{\rm defl}={\cal N}^{-1} \left\{ (x+i y) F - \lambda C \right\}$ is a lengthy expression that is not useful for further insight into this case and hence will not be displayed. 

\begin{figure}[t]
\begin{tabular}{ccc}
\includegraphics[width=0.3\columnwidth]{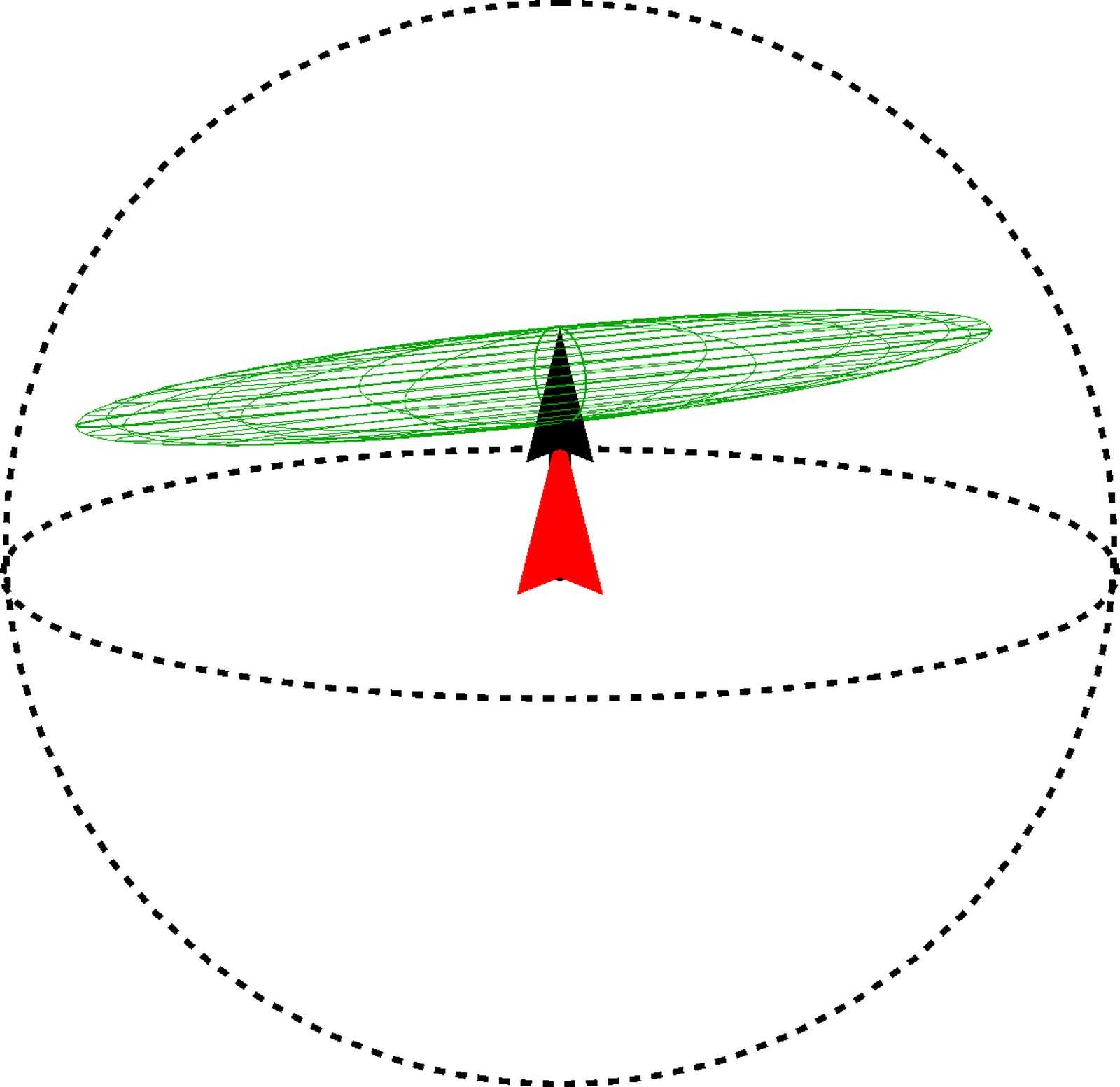} & \includegraphics[width=0.3\columnwidth]{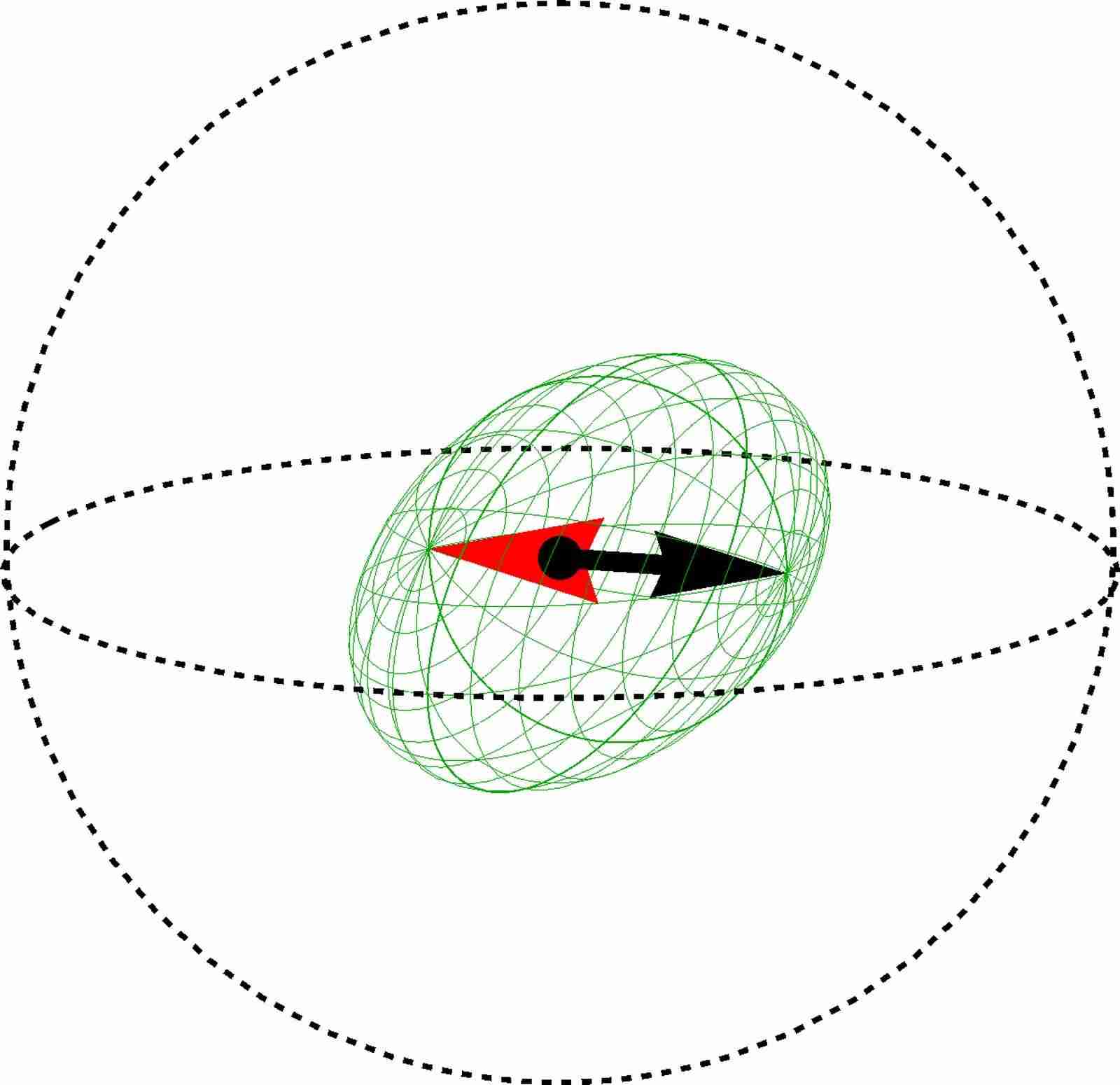} & \includegraphics[width=0.3\columnwidth]{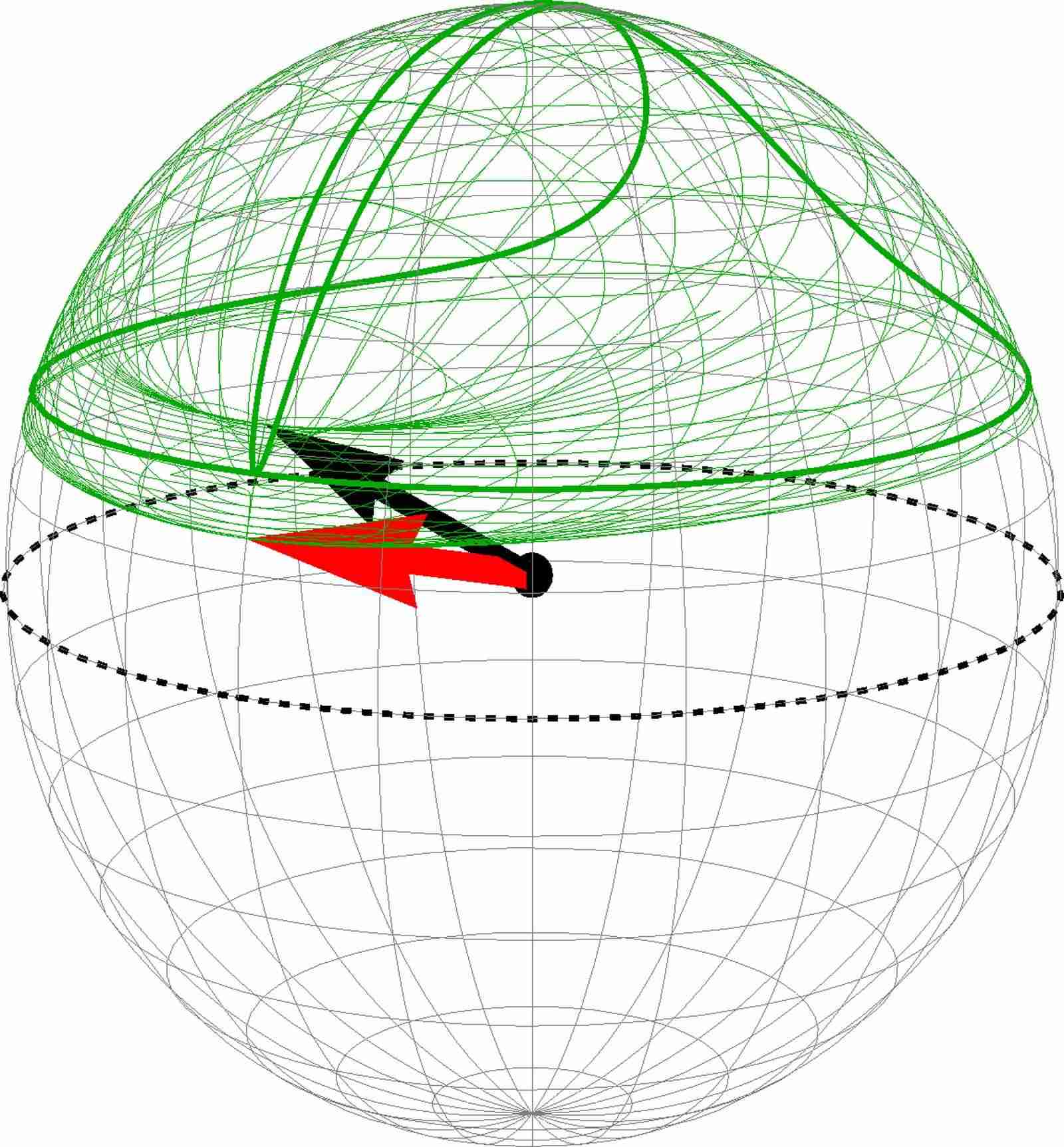}\tabularnewline
(a) Spin BS & (b) Pspin BS & (c) Entanglement BS\tabularnewline
\end{tabular}
\includegraphics[width=\columnwidth]{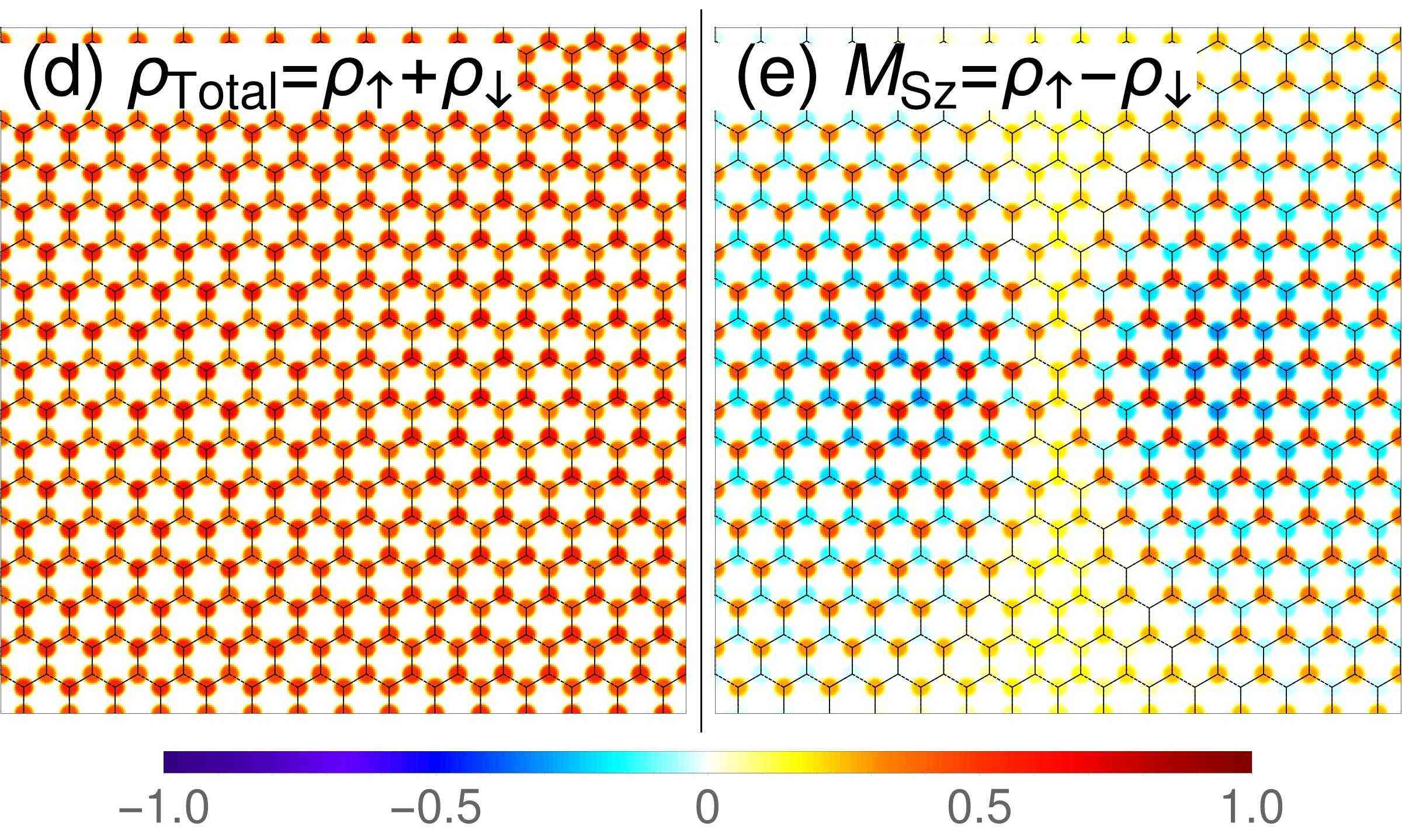}
\caption{\label{fig:Pt11_deflated_skyrmion}Visualization of the deflated CP$^{3}$-pseudospin skyrmion with entangled easy-plane FM background (point 11 in Fig.\ref{fig:Skyrmion_Type_Diagram}). Conventions are the same as in Fig.~\ref{fig:Pt4_spin_skyrmion}.
}
\end{figure}

\begin{figure}[t]
\begin{tabular}{ccc}
\includegraphics[width=0.3\columnwidth]{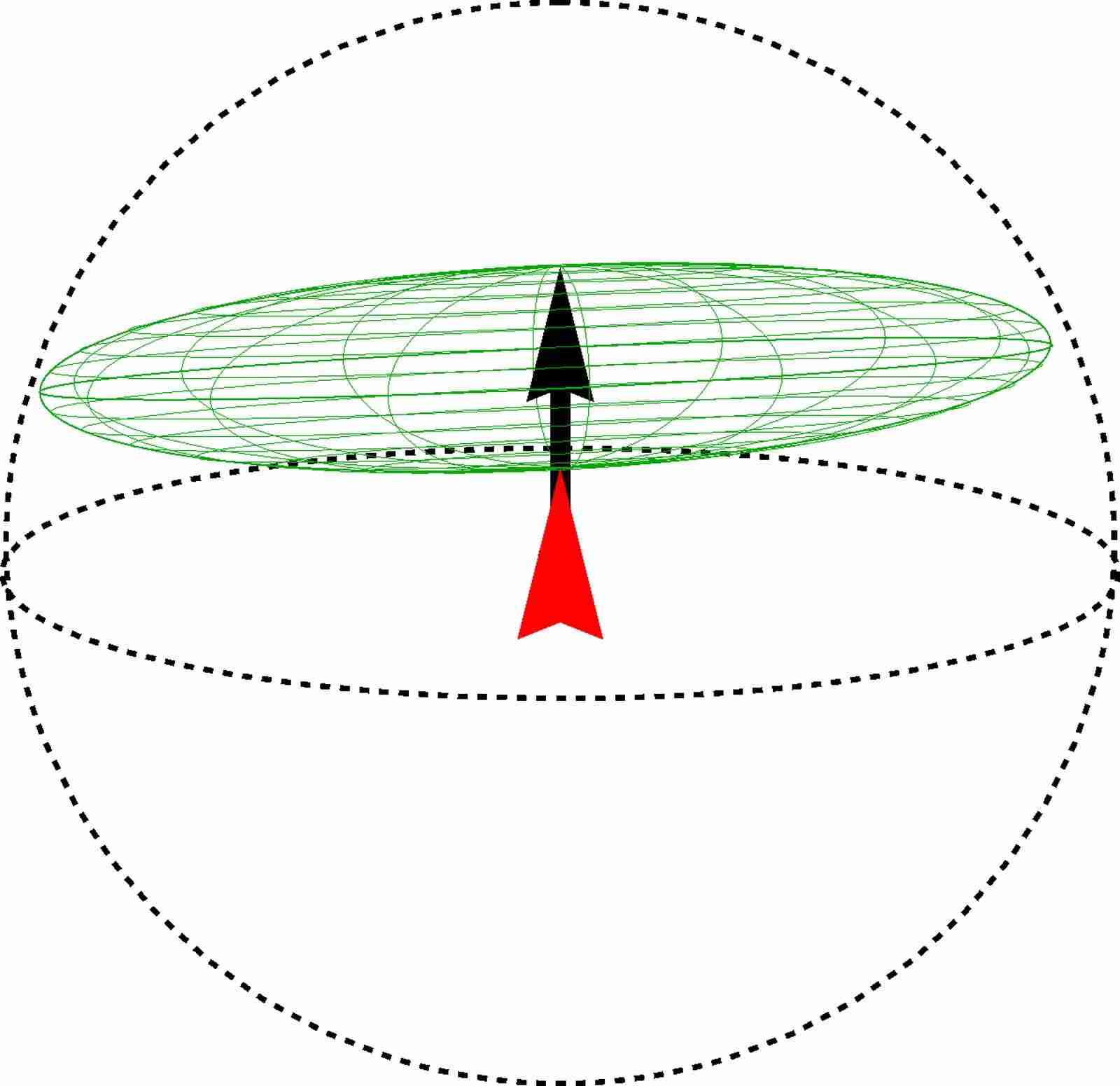} & \includegraphics[width=0.3\columnwidth]{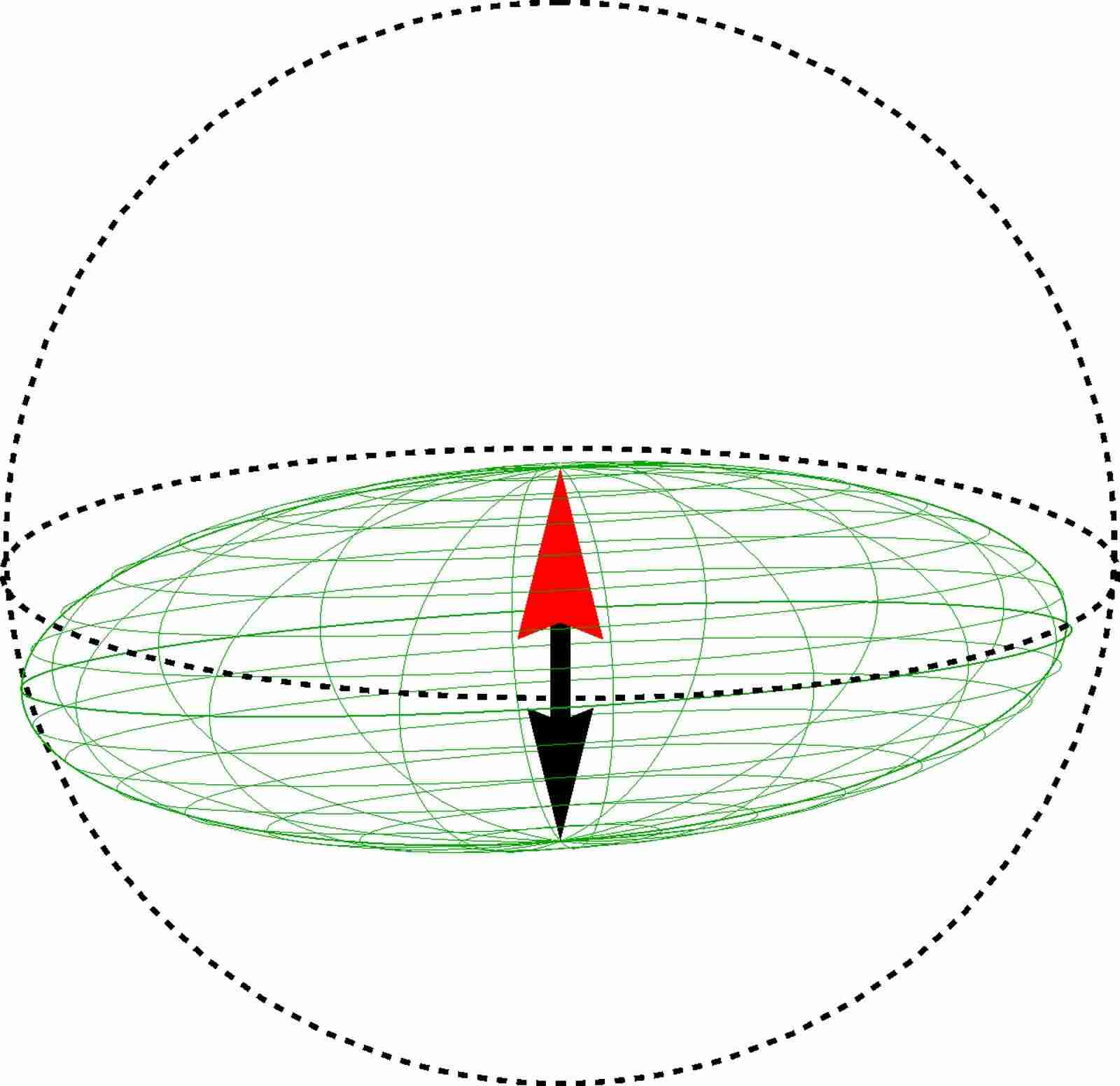} & \includegraphics[width=0.3\columnwidth]{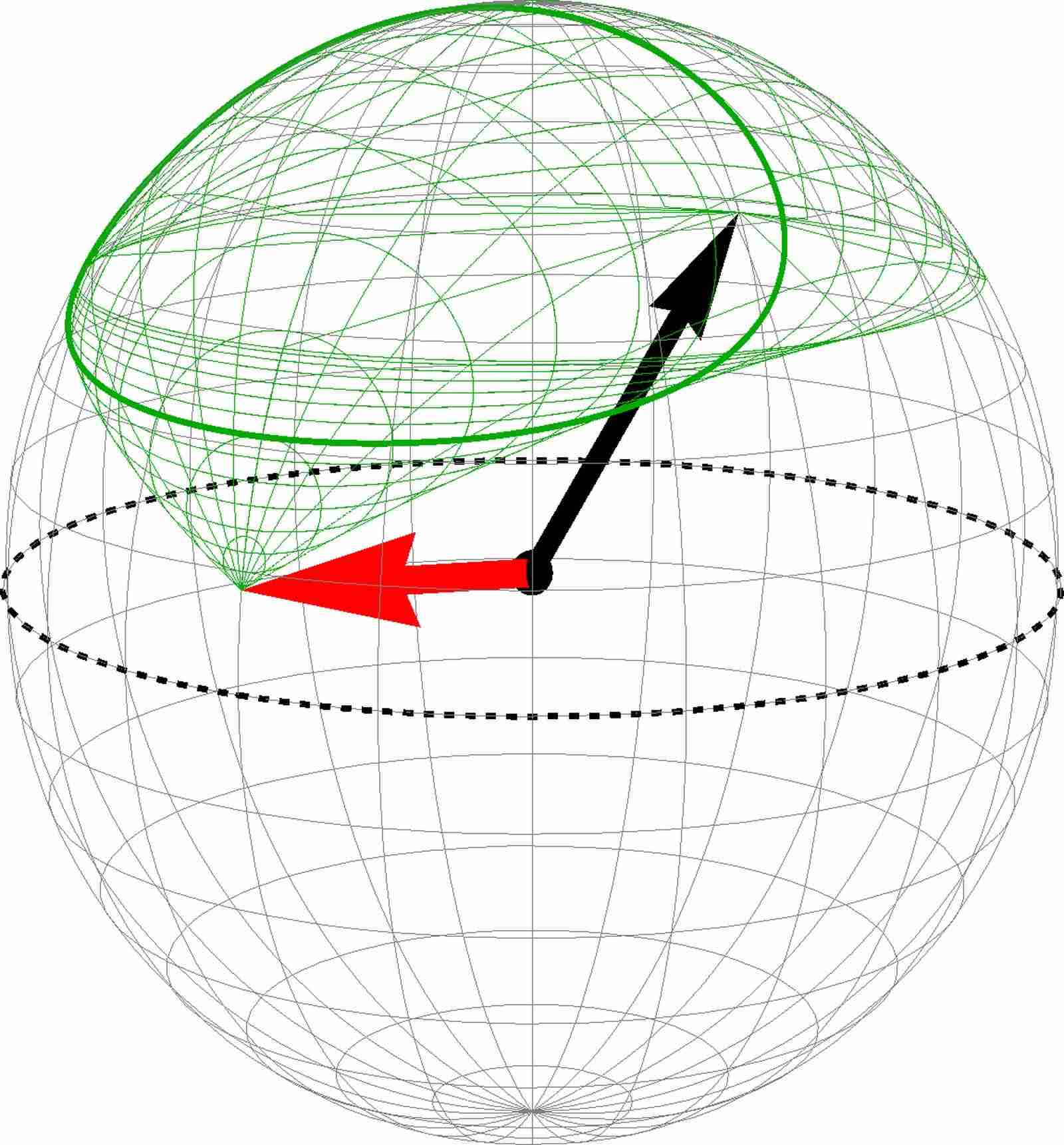}\tabularnewline
(a) Spin BS & (b) Pspin BS & (c) Entanglement BS\tabularnewline
\end{tabular}
\includegraphics[width=\columnwidth]{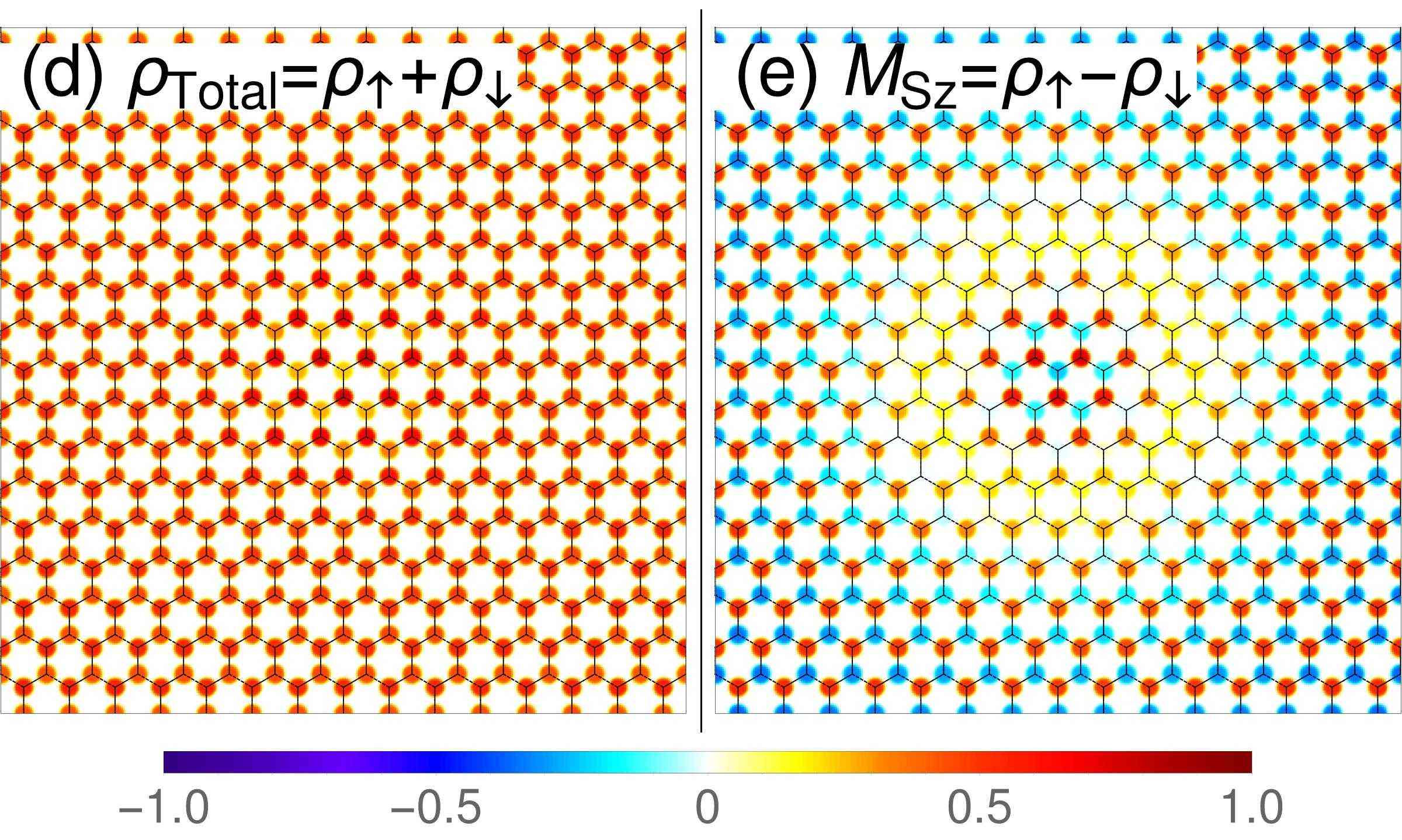}
\caption{\label{fig:Pt12_deflated_skyrmion}Visualization of the deflated CP$^{3}$-pseudospin skyrmion with entangled easy-axis FM background (point 12 in Fig.\ref{fig:Skyrmion_Type_Diagram}). Conventions are the same as in Fig.~\ref{fig:Pt4_spin_skyrmion}.
}
\end{figure}

For this type of skyrmion, the images of Riemann sphere span parts in all three Bloch spheres. They are displayed in Fig.~\ref{fig:Pt11_deflated_skyrmion}(a-c) and Fig.~\ref{fig:Pt12_deflated_skyrmion}(a-c).  Similar to the case of CP$^{3}$-entanglement skyrmion, the spin and pseudospin magnetizations explore the inside of the Bloch spheres, causing the image of Riemann sphere to ``deflate''. We therefore call this type of skyrmion the \emph{deflated CP$^{3}$-pseudospin skyrmion}. It is crucial to observe that the image of the Riemann sphere in the spin Bloch sphere \emph{never} encloses the origin point, while in the pseudospin Bloch sphere the image \emph{always} does. Hence this type of skyrmion is indeed a CP$^{3}$-pseudospin skyrmion. For the image of the Riemann sphere in the entanglement Bloch sphere, besides its incomplete covering, one also observes that it visits the north pole \emph{twice}, indicating the existence of two points $\br_1$ and $\br_2$ on the $xy$-plane, at which the CP$^{3}$-spinors $Z_{\rm defl}(\br_1)$ and $Z_{\rm defl}(\br_2)$ are \emph{unentangled}. These two points are reflected in the spin/pseudospin Bloch spheres, where the surface of the Riemann sphere image touches the surface of the Bloch sphere precisely twice, meaning that the magnitude of both the spin and the pseudospin magnetization $\cos\alpha$ reaches unity twice. The visualization here also agrees with the prediction in Ref.~[\onlinecite{Doucot2008}].

We comment on the evidence of entanglement in a CP$^{3}$-field $Z(\br)$. Although by examining the local patterns of the $\rho_{\rm Total}(\br)$ and $M_{\rm Sz}(\br)$ profiles around a generic point $\br_0$ in the $xy$-plane, it is not possible to tell whether a CP$^{3}$-field $Z(\br_0)$ carries entanglement, it is possible to do so by examine those profiles at large scale over the entire texture. As we have discussed in detail, the factorizable skyrmions do not carry entanglement at any point in the texture, and the evidence is clear in the $\rho_{\rm Total}(\br)$ and $M_{\rm Sz}(\br)$ profiles -- for a CP$^{3}$-spin skyrmion, the pattern of sublattice occupation is uniform over the $\rho_{\rm Total}(\br)$ profile because of the constant pseudospin component, whereas for a CP$^{3}$-pseudospin skyrmion, the two profiles are identical because of the constant spin component with polarization along the $z$-direction in the space of spin magnetization. 
In contrast to the factorizable skyrmions, the entanglement skyrmions (CP$^{3}$-entanglement skyrmions, deflated CP$^{3}$-pseudospin skyrmions) embedded in the easy-axis pseudospin FM background usually have an ``anti-ferrimagnetic'' pattern in the $M_{\rm Sz}(\br)$ profile, i.e. the spin magnetizations on two sublattices point in opposite directions. The CP$^{3}$-entanglement skyrmion and the CP$^{3}$-spin skyrmion which are embedded in the same easy-plane pseudospin FM background cannot be distinguished by their $\rho_{\rm Total}(\br)$ and $M_{\rm Sz}(\br)$ profiles. But we can still make a difference by comparing their $M_{\rm Sx}(\br)$ profiles.

\section{Size and energy of CP$^{3}$-skyrmion}
\label{sec:Size_and_energy}

In the previous section, we concentrated on a classification of the different skyrmion types one encounters in graphene QHFM in the $N=0$ LL. The present section is devoted to a more quantitative analysis of the skyrmions' size and energy, namely in the vicinity of phase transitions between the different FM background states.

\begin{figure}[t]
\includegraphics[width=0.75\columnwidth]{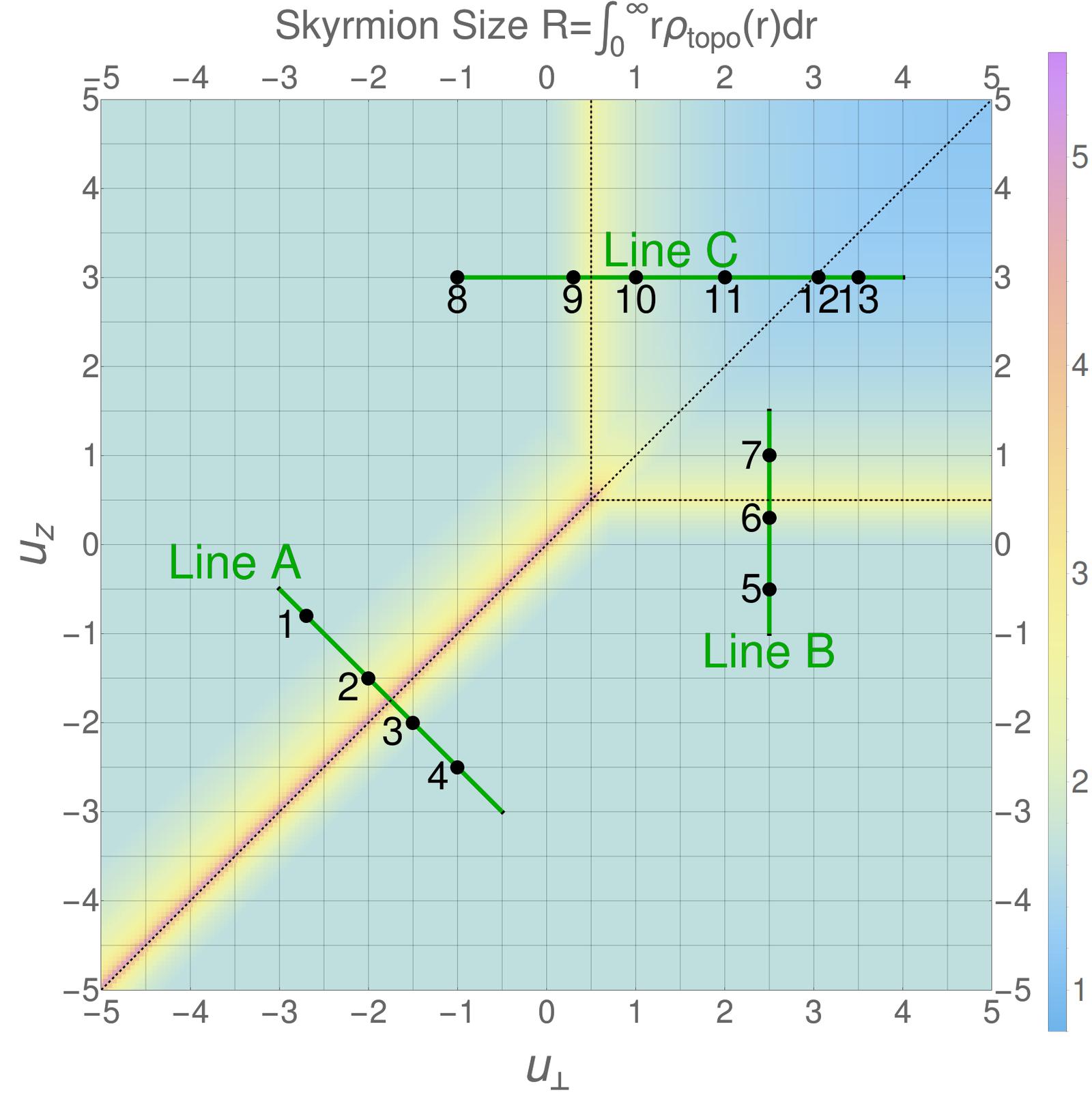}
\caption{\label{fig:Skyrmion_Size_All} Size of optimal skyrmions at $u_{\perp},u_{\rm z}\in [-5,5]$ and $\delta_{\rm Z}=0.0005$. Black dashed lines highlight the boundaries of four regions for the four types of FM background of a skyrmion.}
\end{figure}

\subsection{Size}
\label{subsec:Size}
The size of a CP$^{3}$ skyrmion is characterized by the optimal value of the $\lambda_0$ parameter in the skyrmion ansatz. As discussed in Sec.~\ref{subsec:Skyrmion-Ansatz}, it is determined by the competition between the Coulomb energy $E_{\rm C}$ and the excess anisotropic energy $E'_{\rm A}$. 
To be more precise, according to Eq.~(\ref{eq:lambda_dep}) the optimal value of the $\lambda_0$ parameter in the skyrmion ansatz is obtained by solving $\partial E_{\rm sk}\left(\lambda_{0}\right)/\partial\lambda_{0}=0$, which gives 
\begin{equation}
\lambda_{0}=\left(\frac{E_{\rm C}[\tilde{Z}_{\rm sk}]}{2E_{\rm A}'[\tilde{Z}_{\rm sk}]}\right)^{\frac{1}{3}}, \label{eq:lambda_0}
\end{equation}
where $\tilde{Z}_{\rm sk}$ is the $\CP{3}$-field for an optimal skyrmion, but $\lambda_0$ has been set to $1$. It has the same meaning as in Eq.~(\ref{eq:lambda_dep}) in Sec.~\ref{subsec:Radial-deformation}. 

We define the skyrmion size $R$ (in unit of $l_B$) by averaging $r$ over the topological charge density $\rho_{\rm topo}(\br)$
\begin{equation}
R=\int_0^{\infty} r\rho_{\rm topo}(\br)d^{2}r ,
\end{equation}
because $\rho_{\rm topo}(\br)$ can be viewed as the excess charge density induced by the texture in a CP$^{3}$ skyrmion and is thus a good measure for the spatial extent of the texture. For a radially deformed skyrmion, we have 
\begin{equation}\label{eq:Size_propto_lambda}
R=\alpha\left(\kappa\right)\lambda_{0}
\end{equation}
with an $r$-dependent size parameter $\lambda\left(r\right)=\lambda_{0}\exp\left(-r^{2}/\kappa\lambda_{0}^{2}\right)$ and topological charge density  
\begin{equation}
\rho_{{\rm topo}}\left(r\right)=\frac{\lambda^{2}(r)}{\pi\left[r^{2}+\lambda^{2}(r)\right]^{2}}\left(1+\frac{2r^{2}}{\kappa\lambda_{0}^{2}}\right). \label{eq:rho-topo-deformed}
\end{equation}
The coefficient $\alpha(\kappa)$ is a monotonically increasing function of the radial deformation parameter $\kappa$. Its actual form depends on the radial deformation and is \emph{not} related to the energy minimization. In the scale-invariant limit $\kappa\rightarrow \infty$ we have $\alpha(\kappa)\rightarrow \pi/2$. Away from this limit, $\alpha\left(\kappa\right)$ does not change much: for instance $\alpha(10)\sim 1.0$. 

\begin{figure}[t]
\centering
\includegraphics[width=0.8\columnwidth]{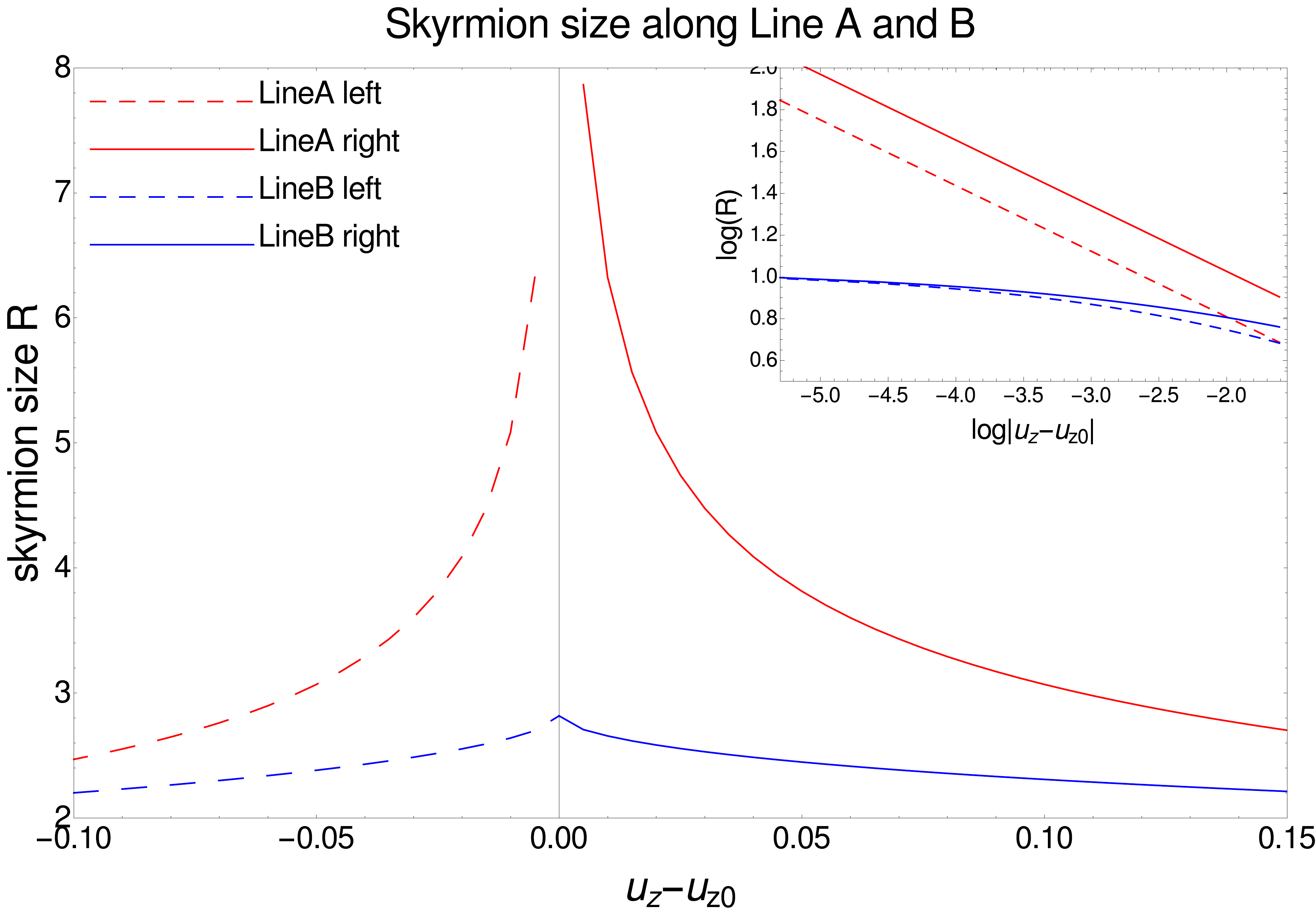}
\caption{\label{fig:SkyrmSize} Skyrmion size (in unit of magnetic length $l_B$) as a function of $u_{\rm z}$, along the lines A and B in Fig. \ref{fig:Skyrmion_Size_All}. }
\end{figure}

Fig.~\ref{fig:Skyrmion_Size_All} displays the size $R$ of optimal skyrmions obtained in Sec.~\ref{subsec:Minimization-results} ($u_{\perp},u_{\rm z}\in[-5,5]$, $\delta_{\rm Z}=0.0005$). 
One notices generally an increase in the Skyrmion size for $(u_{\perp},u_{\rm z})$ close to a transition between different ferromagnetic backgrounds. These transitions are indicated by dotted black lines in Fig.~\ref{fig:Skyrmion_Size_All}. In order to investigate in more detail the increase in skyrmion size,  we plot, in Fig.~\ref{fig:SkyrmSize}, the skyrmion size $R$ as a function of $u_{\rm z}-u_{\rm z0}$ along line A and line B. Here, $u_{\rm z0}$ denotes the value at the border between two regions of different FM background, which is $u_{\rm z}=u_{\perp}=-1.75$ for line A separating the unentangled easy-axis from the unentangled easy-plane FM and $u_{\rm z}=1/2$ for line B separating the unentangled from the entangled easy-axis FM. 

One sees in Fig.~\ref{fig:SkyrmSize} that along line A, the skyrmion size diverges (red curves) when approaching the transition at $u_{\rm z}=u_{\perp}$. 
This is expected because, at the transition, the pseudospin texture costs the same amount of anisotropic energy as the state with uniform pseudospin magnetization, giving a vanishing excess anisotropic energy $E'[\check{Z}_{\rm sk}] = E[\check{Z}_{\rm sk}] - E_{\rm A}[F]$ and thus a divergent $\lambda_0$. Physically this means that the pseudospin $\SU{2}$ symmetry is restored when $u_{\perp}=u_{\rm z}$, and the Coulomb energy of the skyrmion tends to inflate the skyrmion to infinite size.  
In the vicinity of the point $u_{\perp}=u_{\rm z}=-1.75$, $E'[\check{Z}_{\rm sk}]$ is proportional to $|u_{\perp}-u_{\rm z}|$. 
Consequently, $E'[\check{Z}_{\rm sk}] \propto |u_{\rm z}-u_{\rm z0}|$ with $u_{\rm z0}=-1.75$, then from Eq.~(\ref{eq:lambda_0}) we know $\lambda_0 \propto |u_{\rm z}-u_{\rm z0}|^{-1/3}$. Therefore along line A, the skyrmion size $R$ scales as $R \propto |u_{\rm z}-u_{\rm z0}|^{-1/3}$. 
The exponent $-0.32$ is extracted from the red lines of the log-log plot in the inset of Fig.~\ref{fig:SkyrmSize}, in agreement with the above scaling argument.
The divergence of the skyrmion size at the line $u_{\rm z} = u_\perp$ reflects the underlying transition between an easy-axis and an easy-plane pseudospin FM background, as long as both are unentangled. 

The blue curves in Fig.~\ref{fig:SkyrmSize} show the evolution of the skyrmion size along line B cutting the border $u_{\rm z}=1/2$, where one also notices an enhanced albeit non-divergent size.  
In contrast to the transition at $u_{\rm z}=u_{\perp}$, there is no symmetry restoration that would lead to a divergent skyrmion size, i.e. the anisotropy energy associated with the pseudospin remains finite here. This yields a ``truncated power law'',
\begin{equation}
\lambda \sim (|u_{\rm z}-1/2| + C)^{\gamma}, 
\end{equation}
that is now cut off by a non-zero constant constant $C$. As one may expect from our simplified scaling analysis, the exponent $\gamma$ is again close to $-1/3$. Even if there is no fully developed divergence in the skyrmion size, its increase unveils again a transition between different underlying FM background states.

The effect of disorder on skyrmions becomes significant when the length scale of the disorder potential is comparable to the skyrmion size. In such conditions, the charge-carrying skyrmion is trapped to the disorder potential that is coupled to electric charge. 
Moreover, if there is also pseudospin disorder in the hosting system, then a $\CP{3}$-pseudospin skyrmion is frustrated by the two types of disorders. In the experiments where one can control the parameters $u_{\perp}$ and $u_{\rm z}$, the relation between the skyrmion size $R$ and the disorder length scale can be revealed in transport measurements. 

\begin{figure}[t]
\includegraphics[width=0.8\columnwidth]{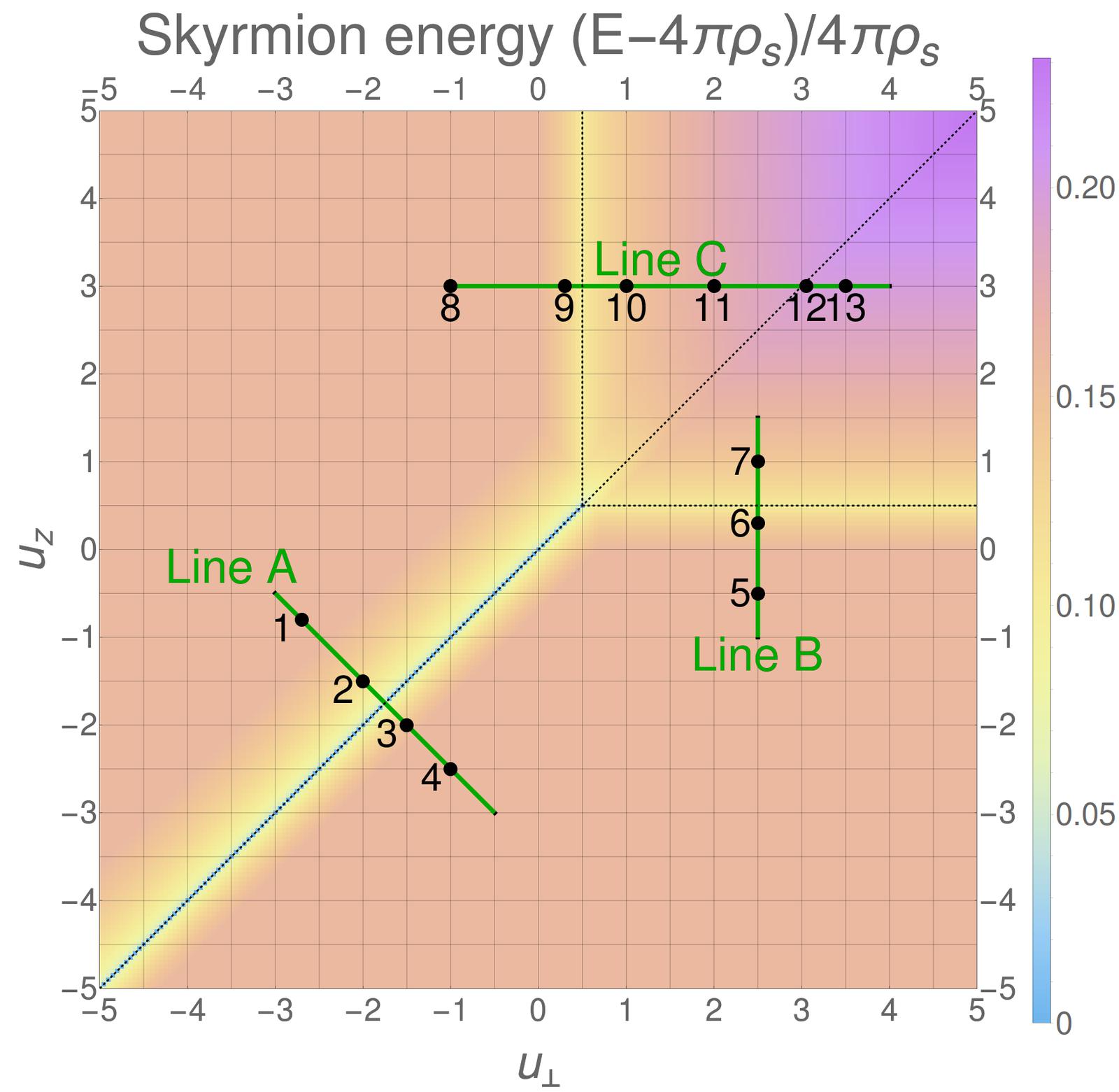}
\caption{\label{fig:Skyrmion_Energy_All} Energy of optimal skyrmions at $u_{\perp},u_{\rm z}\in [-5,5]$ and $\delta_{\rm Z}=0.0005$. The dimensionless value $(E-4\pi\rho_{\rm s})/4\pi\rho_{\rm s}$ is plotted in the figure, with the numerical value of $4\pi\rho_{\rm s}=3.9374\times(e^2/4\pi\epsilon l_B)$. Black dashed lines highlight the boundaries of four regions for the four types of FM background of a skyrmion. }
\end{figure}

\subsection{Energy}
\label{subsec:Energy}
As discussed in Sec.~\ref{subsec:Skyrmion-Ansatz} and Sec.~\ref{subsec:Radial-deformation}, all scale-invariant $\CP{3}$-skyrmions carrying topological charge ${\cal Q}=1$ have size-independent energy $E^{(0)}_{\rm NLSM}=4\pi\rho_s$, and size-dependent energy $E_{\rm C}+E_{\rm A}$. 
The skyrmion size is determined by the competition between the Coulomb energy $E_{\rm C}$ and the anisotropic energy $E_{\rm A}$. 
The radius deformation of skyrmion is introduced to make $E_{\rm A}$ finite, at a cost of slight increase of $E_{\rm NLSM}$ and $E_{\rm C}$. 
After minimization of the skyrmion energy at $\delta_{\rm Z}=0.0005$ and $u_{\perp},u_{\rm z}\in[-5,5]$, we find that the ratio $(E-4\pi\rho_s)/4\pi\rho_s$ shown in Fig.~\ref{fig:Skyrmion_Energy_All} is between $0.15$ and $0.20$, whereas the ratio $E_{\rm NLSM}/4\pi\rho_s$ is always close to $1$. Thus the Coulomb energy and anisotropic energy have significant contribution to the total energy of a $\CP{3}$-skyrmion described by our model. Despite of such contribution, a $\CP{3}$-skyrmion still has lower energy compared to the quasiparticle with a single spin/pseudospin flip on top of the QHFM state at quarter filling of the $N=0$ LL. 

Figure~\ref{fig:Skyrmion_Energy_All} shows the value $E[Z_{\rm sk}]/4\pi\rho_s-1$, i.e. the relative energy difference between the optimal skyrmion $\check{Z}_{\rm sk}$, at different $u_{\perp}$ and $u_{\rm z}$, and the scale-invariant skyrmion, with energy $4\pi\rho_s$.  
We observe that the energy of an optimal skyrmion does not depend on $u_{\perp}$ and $u_{\rm z}$ except when they are close to the lines $u_{\perp}=1/2$ or $u_{\rm z}=1/2$, as well as close to the line $u_{\perp}=u_{\rm z}\leq 1/2$. As shown in the skyrmion phase diagram, Fig.~\ref{fig:Skyrmion_Type_Diagram}, one obtains pseudospin skyrmions in the vicinity of the latter line. Their energy naturally depends on $u_{\perp}$ and $u_z$ in contrast to that of spin skyrmions that are encountered away from this line and that only depend on $\delta_{\rm Z}$, which is constant here. The energy of the spin skyrmions thus remains constant, as seen in the light orange parts of Fig.~\ref{fig:Skyrmion_Energy_All} that match the (light red) parts of the skyrmion phase diagram (Fig.~\ref{fig:Skyrmion_Type_Diagram}) where one obtains pure spin skyrmions. 

Together with the plot of skyrmion size in Fig.~\ref{fig:Skyrmion_Size_All}, we also observed that, for different anisotropy parameter $u_{\perp}$ and $u_{\rm z}$, an \emph{optimal} skyrmion has a lower \emph{optimal} energy when its \emph{optimal} size is larger. The lowest value of $E[Z_{\rm sk}]$ is $4\pi\rho_s$, which is achieved by the optimal skyrmion $Z_{\rm sk}$ obtained at $u_{\rm z}=u_{\perp}\le 1/2$. At this condition, the size of the optimal skyrmion diverges and its Coulomb energy vanishes, because the skyrmion energy has a pseudospin $\SU{2}$ symmetry. 

\subsection{Magnetic-field scaling of skyrmion size and energy}

\begin{figure*}[t]
\includegraphics[width=0.95\columnwidth]{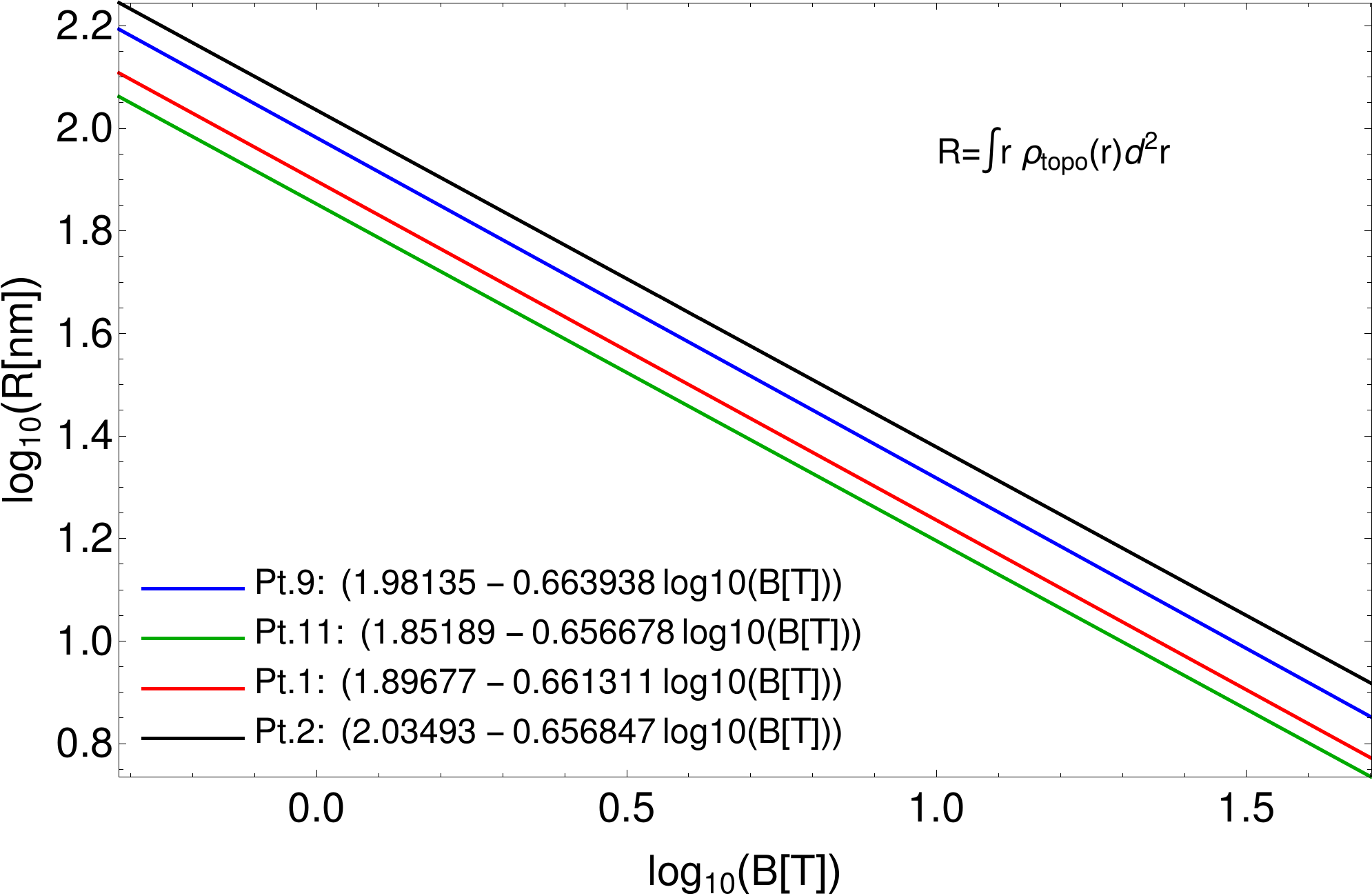}
\includegraphics[width=0.95\columnwidth]{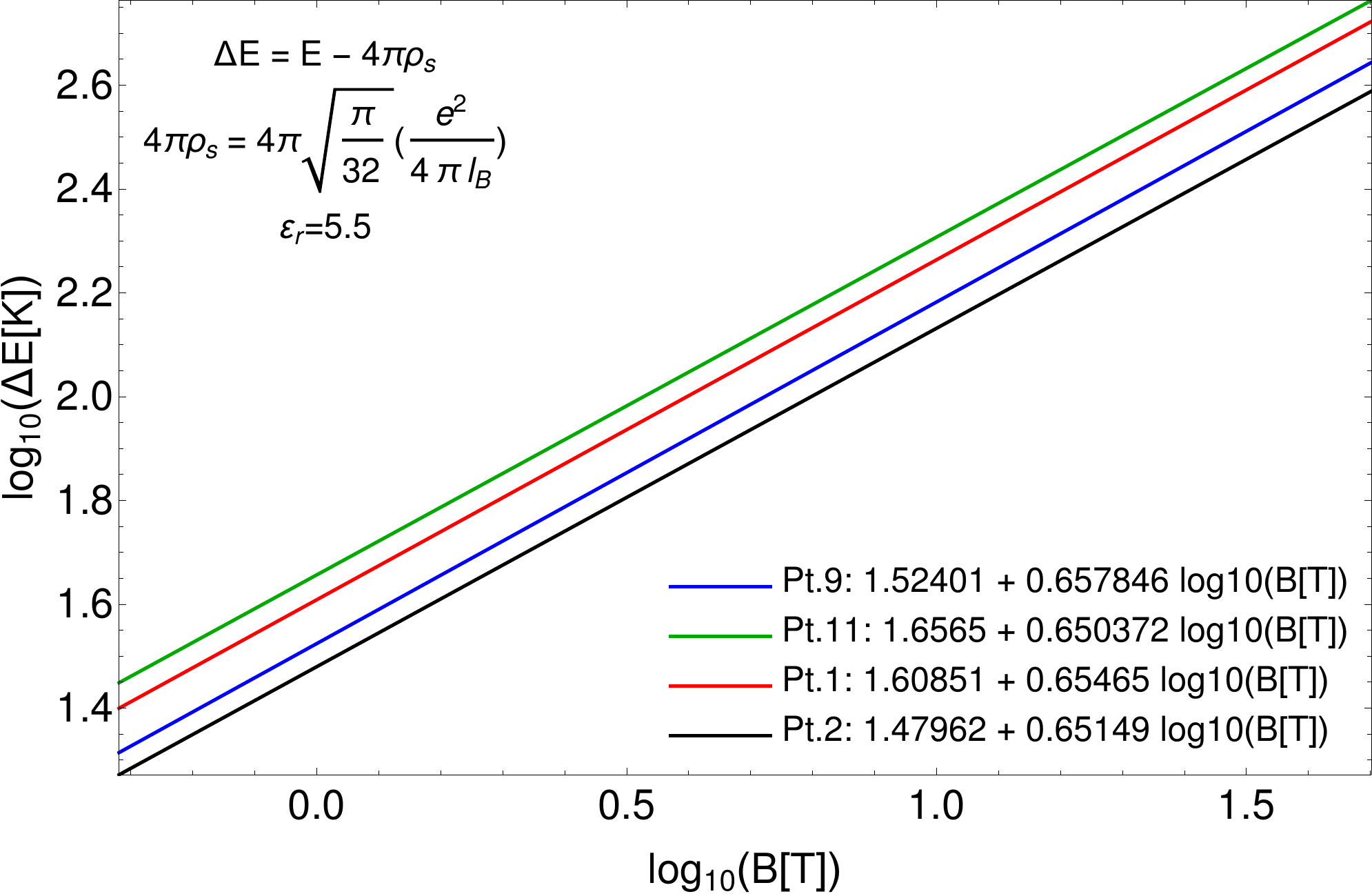}
\caption{\label{fig:Energy_size_B} Size (upper figure) and energy (lower figure) of optimal skyrmions at $\delta_{\rm Z}\in[0.00005,0.00075]$. Green, red, blue, black lines represent skyrmions at point 11, point 1, point 9, point 2 in the skyrmion type diagram Fig.~\ref{fig:Skyrmion_Type_Diagram} respectively.}
\end{figure*}

The correlation between skyrmion energy and skyrmion size still holds when we vary the perpendicular magnetic field $B_{\perp}$, which we choose here to be identical to the total magnetic field, i.e. we discard a tilt of field described in the following section. This is shown in Fig.~\ref{fig:Energy_size_B}, where we plot in a log-log scale the numerically obtained size and energy of four types of optimal skyrmions under $B\le 50\mbox{T}$. Different types of optimal skyrmion have different value of energy and size, but the shape of the lines are similar, which root from same scaling relations. Indeed Eq.~(\ref{eq:lambda_0}) implies that the skyrmion size (in unit of $l_B$) scales as $\delta_{\rm Z}^{-1/3}$, and one obtains from Eq.~(\ref{eq:delta_Z-B}) then gives 
\begin{equation}\label{scalingRB}
R\sim B^{-2/3}
\end{equation}
in natural units that are independent of the magnetic field. One notices that the numerically obtained exponents in Fig.~\ref{fig:Energy_size_B}(a) agree to great accuracy (within 2\%) with that obtained from our simple scaling analysis even if the latter does not take into account the radial deformation in Eq.~(\ref{eq:lambda_r}). 

Similarly, insert Eq.~(\ref{eq:lambda_0}) to Eq.~(\ref{eq:lambda_dep}), we obtain the scaling of the skyrmion energy $E-E^0_{\rm NLSM}\sim  \delta_{\rm Z}^{1/3}$ or, in natural $B$-independent units
\begin{equation}\label{scalingEB}
E-E^{(0)}_{\rm NLSM}\sim  B^{2/3},
\end{equation}
which again agrees well (within 2\%) with the numerically obtained scaling of the skyrmion energy plotted in Fig.~\ref{fig:Energy_size_B}(b).

\section{Modification of the anisotropy energy}
\label{sec:Modification-of-the-anisotropy-energy}
In the previous sections, we discussed the skyrmion type diagram, which provides a global view of the possible types of skyrmions at quarter filling of the $N=0$ LL of graphene monolayer. In this section we show how this diagram is changed by modifications of the anisotropic energy.

We discuss three modifications to the anisotropic energy, each corresponding to one of the three contributions (namely the in-plane pseudospin contribution, $z$-component pseudospin contribution and spin contribution). The common way to modify the anisotropic energy is by sample tilting, where the sample is tilted away from the upright position while increasing the total magnetic field $B_{\rm T}$, so that the strength of the perpendicular component $B_{\perp}$ is kept to a constant value. The electron spin is coupled to the total applied magnetic field, while the pseudospin contribution to the anisotropic energy is proportional to the component perpendicular to the sample. Therefore, sample tilting guarantees that the pseudospin contribution to the anisotropic energy is unaltered, but the electron spin contribution is increasing. 
We further discuss terms that breaks the U$(1)$ symmetry between the $x$- and $y$-components of the pseudospin, whose contribution to the anisotropic energy is changed accordingly. 
At last we discuss the pseudospin analogue of the Zeeman coupling, i.e. the term proportional to $\int d^2 r M_{\rm Pz}$, which describes the sublattice asymmetry and that could eventually be induced by substrates that are roughly commensurate with the carbon spacing in graphene (e.g. Boron-Nitride substrates).

\subsection{Sample tilting}
\label{subsec:Tilting-sample}

The effect of sample tilting by an angle $\varphi$ is the rescaling of the parameters $u_\perp$, $u_{\rm z}$ and $\delta_{\rm Z}$ in the anisotropic energy. If we maintain our definition of the effective parameters $u_{\perp,\rm{z}}$ in terms of the perpendicular component $B_{\perp}$ of the magnetic field, $u_{\perp,\rm{z}}=U_{\perp,\rm{z}}/g\mu_{\rm B} B_{\perp}$, and use the Zeeman coupling, sensitive to the total magnetic field $B_{\rm T}$, as the overall energy scale, one obtains from Eq.~(\ref{eq:EnAnisotropic})
\begin{eqnarray}
E_{\rm A}[Z] &=& \frac{\Delta_{\rm Z}(B_{\rm T})}{2}\int \frac{d^{2}r}{2\pi l_{B}^{2}}  \times\nn\\
&&\left[ u_{\perp}\cos\varphi\left( M_{\rm Px}^{2}+M_{\rm Py}^{2}\right)+u_{\rm z}\cos\varphi M_{\rm Pz}^{2}-M_{\rm Sz}\right],\nn\\\label{eq:EnAnisotropicTilt}
\end{eqnarray}
where $\varphi$ is the tilting angle from an upright direction, $B_{\perp}=B_{\rm T}\cos\varphi$, and $M_{\rm P}$, $M_{\rm S}$ are computed by Eq.~(\ref{eq:Ti_Z}), Eq.~(\ref{eq:Si_Z}) respectively.  
Such a constant $B_{\perp}$ guarantees a constant length scale $l_B$ and a constant energy scale $e^2/4\pi\epsilon l_B$. After the rescaling of length and energy to unit $l_B$ and $e^2/4\pi\epsilon l_B$, $\Delta_{\rm Z}(B_{\rm T})$ is replaced by $\delta_{\rm Z 0}/\cos\varphi$, where $\delta_{\rm Z 0}$ is computed by Eq.~(\ref{eq:delta_Z-B}) with $B_{\perp}=B_{\rm T}$, i.e. in the absence of sample tilting. At this stage, we see that the sample tilting by an angle $\varphi$ changes the anisotropic energy as if we simply replace $\delta_{\rm Z 0},u_{\perp}$, and $u_{\rm z}$ in the model without tilting by $\delta_{\rm Z 0}/\cos\varphi,u_{\perp}\cos\varphi$, and $u_{\rm z}\cos\varphi$.

\begin{figure}[t]
\includegraphics[width=0.9\columnwidth]{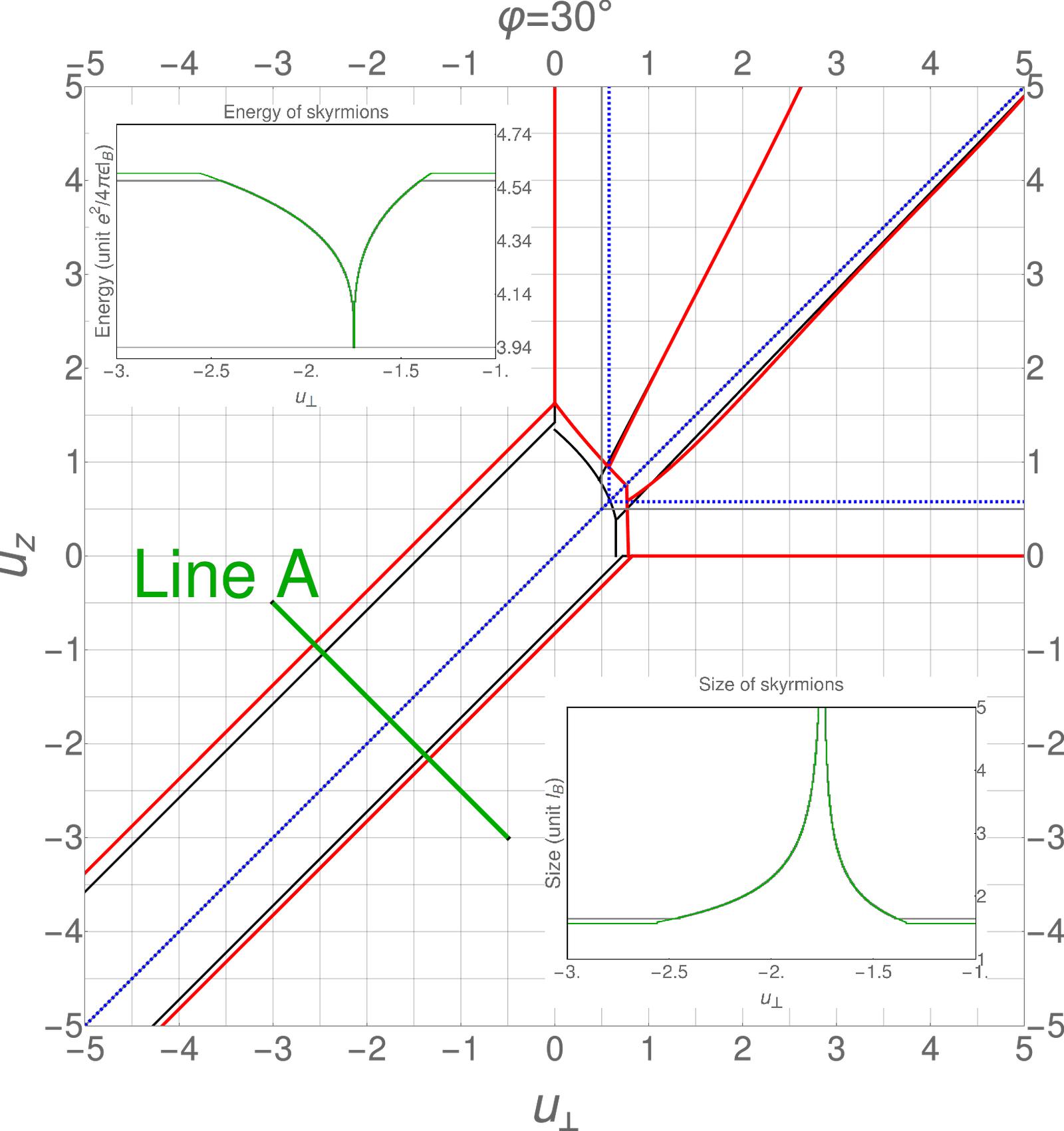}
\caption{\label{fig:Skyrmion-type-tilting} 
Change of the skyrmion type diagram for sample tilting by angle $\varphi=\pi/6$. 
Shifted borders are marked by blue dashed lines for different types of FM backgrounds, and red for different types of skyrmions, in contrast to the grey and black lines which mark the borders in Fig.~\ref{fig:Skyrmion_Type_Diagram} with $\delta u_{\perp}=\delta u_{\rm z}=\varphi=0$.
The insets show the evolution of the skyrmion energy (still in units of $e^2/4\pi \epsilon l_B$) and size across line A.}
\end{figure}

To illustrate the effect of the tilt, we have plotted the evolution of the phase diagram in Fig.~\ref{fig:Skyrmion-type-tilting} for a value of $\varphi=30^\circ$, as well as that for the skyrmion energy and size along line A in the insets. One first notices that the tilt shifts the borders between the unentangled and entangled types of optimal FM states $Z(\br)=F$. Generally the unentangled states are slightly favoured. This can easily be understood as the consequence of the relative increase of the Zeeman energy with respect to the pseudospin-symmetry-breaking terms -- the Zeeman coupling favours indeed a full spin polarization and thus a maximal value of $\cos\alpha$, which precisely means that the unentangled FM states are favoured. 

Furthermore, the borders (red lines in Fig.~\ref{fig:Skyrmion-type-tilting} for $\varphi=30^\circ$ as compared to black lines for $\varphi-0$) between the pseudospin skyrmions in the vicinity of $u_{\rm z}=u_{\perp}\leq 1/2$ are shifted. Indeed, the pseudospin skyrmions are favoured over a larger range as one may expect from the relative increase of the Zeeman energy that renders larger the energy cost to create a spin skyrmion. This is clearly seen in the inset of Fig.~\ref{fig:Skyrmion-type-tilting} -- while the energy of the pseudospin skyrmion, in the tilt-independent units of $e^2/4\pi \epsilon l_B$, remains the same as for $\varphi=0$, the $u_{\perp, \rm z}$-independent energy of the spin skyrmion is increased and its size reduced such that the region of stable pseudospin skyrmions is increased. 

Notice finally that the scaling relations (\ref{scalingRB}) and (\ref{scalingEB}) are also affected by the sample tilting. The scaling of the size of pseudospin skyrmions still obeys the $B_\perp^{-2/3}$ law because the pseudospin anisotropy depends only on $B_\perp$. We can also express such scaling in the total magnetic field and the tilting angle, namely 
\begin{equation}
R_{\rm p}\sim B_{\rm T}^{-2/3}(\cos\varphi)^{-2/3}.
\end{equation}
By contrast, the spin skyrmion follows a scaling relation
\begin{equation}
R_{\rm s} \sim B_\perp^{-2/3}(\cos\varphi)^{1/3} = B_{\rm T}^{-2/3}(\cos\varphi)^{-1/3},
\end{equation}
The exponents on $\cos\varphi$ and the magnetic field $B_\perp$ or $B_{\rm T}$ are different because the Coulomb energy $E_{\rm C}$ depends on $l_B^{-1}\sim B_\perp^{-1/2}$ while the Zeeman coupling depends on $B_{\rm T}=B_\perp / \cos\varphi$. 
If one perform the sample tilting at constant total magnetic field $B_{\rm T}$, spin and pseudospin skyrmions would show a different scaling with respect to the tilt angle.

\subsection{Anisotropy of substrate}
\label{subsec:Anisotropy-of-substrate}

\begin{figure}[t]
\includegraphics[width=0.9\columnwidth]{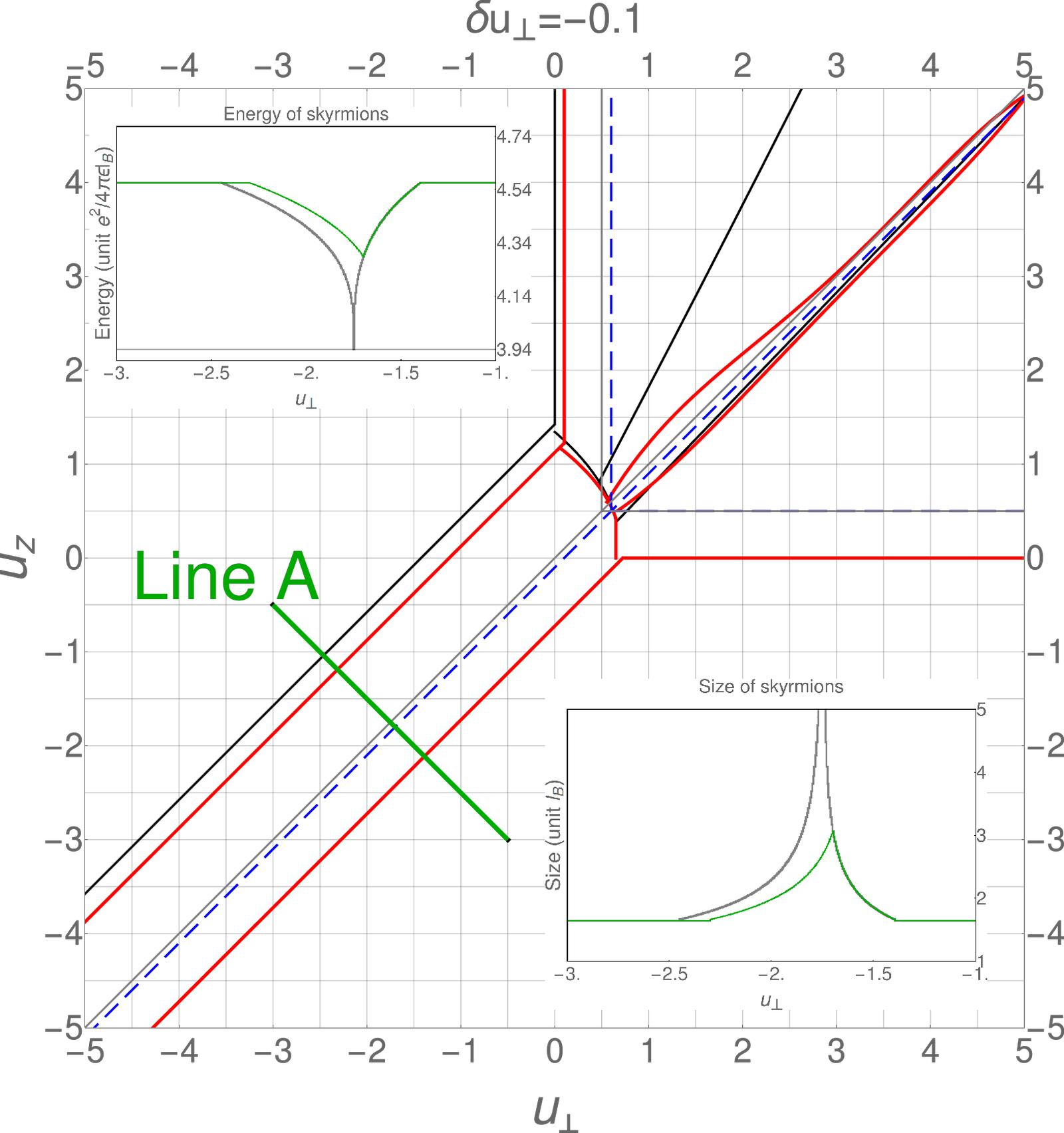}
\caption{\label{fig:Skyrmion-type-substrate-anisotropy} Change of the skyrmion type diagram for $\delta u_{\perp}=-0.1$.  The color code is the same as in Fig.~\ref{fig:Skyrmion-type-tilting}, in the absence of tilt ($\varphi=0$).}
\end{figure}

Our model for the anisotropic energy $E_{\rm A}$ in Eq.~(\ref{eq:EnAnisotropic}) [also Eq.~(\ref{eq:EnAnisotropicTilt})] has a U$(1)$ symmetry for the $x$- and $y$-component of the pseudospin. 
Formally, this symmetry can be broken with the help of the parameter $\delta u_{\perp}$, which yields the modified energy
\begin{eqnarray}
E_{\rm A}[Z] &=& \frac{\Delta_{\rm Z}(B_{\rm T})}{2}\int\frac{d^{2}r}{2\pi l_{B}^{2}}\left[(u_{\perp}
+\delta u_{\perp})\cos\varphi M_{\rm Px}^{2}\right.
\nn\\
&& \left. +(u_{\perp}-\delta u_{\perp})\cos\varphi M_{\rm Py}^{2} + u_{\rm z}\cos\varphi M_{\rm Pz}^{2}-M_{\rm Sz}\right],\nn\\\label{eq:EnAnisotropicSubstrateAniso}
\end{eqnarray}
where we have also considered sample tilting. It is tempting to attribute this symmetry breaking to an $xy$-anisotropy in the graphene sheet, e.g. when one applies uni-axial strain to it that breaks the $120^{\circ}$ rotation symmetry of the lattice. However, more microscopic calculations beyond the scope of our paper, would be required to corroborate this picture. 

The main effect of this symmetry breaking is to favour, in the case of an easy-plane pseudospin FM, an orientation of the pseudospin magnetization along the $x$-axis ($\phi_{\rm P}=0$) for $\delta u<0$ or along the $y$-axis ($\phi_{\rm P}=\pi/2$) when $\delta u>0$, while no particular direction is favoured in the isotropic case ($\delta u=0$). The relevant energy scale for the in-plane anisotropy is therefore reduced, and we can use the above model (\ref{eq:EnAnisotropicTilt}) also to treat the present case if one replaces the model parameters $(u_{\perp},u_{\rm z})$ by $(u_{\perp}-|\delta u_{\perp}|,u_{\rm z})$. 

The evolution of the phase diagram is shown in Fig.~\ref{fig:Skyrmion-type-substrate-anisotropy} for a value $\delta u_{\perp}= -0.1$. One first notices that the transition lines towards the easy-plane FM states is modified, whereas the easy-axis FM states are naturally not affected in energy by the change in $u_{\perp}$. Indeed, the transition line between the unentangled easy-plane and the easy-axis FM states is shifted downwards -- the transition occurs at $u_{\rm z}=u_\perp + \delta u_\perp$ instead of $u_{\rm z}=u_{\perp}$ as in the isotropic case. Furthermore, the Zeeman energy is again effectively increased with respect to the easy-plane FM energy scale (for positive values of $u_\perp$) such that the unentangled easy-plane FM states are stabilized, whence the right shift of the transition between the unentangled and entangled easy-plane FM states in Fig.~\ref{fig:Skyrmion-type-substrate-anisotropy}.

Similarly one may understand the borders of the regions for the different skyrmion types. While the transition between the spin and the pseudospin skyrmion is unaffected in the case of an easy-axis FM background characterized by the unaltered energy scale $u_{\rm z}$, that between the spin an pseudospin skyrmions in an easy-plane FM background is shifted downwards (red line in Fig.~\ref{fig:Skyrmion-type-substrate-anisotropy}. This can be understood from the evolution of the skyrmion size shown in the lower inset (cut along line A). When approaching the transition between the two different underlying FM states from the right hand side (by lowering the value of $u_\perp$, the skyrmion size follows first the same evolution as in the isotropic case until the maximum is reached at the value of the underlying transition. Notice, however, that due to the in-plane anisotropy the pseudospin $\SU{2}$ symmetry is no longer restored so that the skyrmion size does not diverge. Left to the transition, the skyrmion size is therefore generically smaller than in the isotropic case, and the skyrmion energy higher (see upper inset of Fig.~\ref{fig:Skyrmion-type-substrate-anisotropy}). The energy of the spin skyrmion is therefore reached at smaller values of $|u_\perp|$ such that the region where one encounters pseudospin skyrmions becomes smaller. Furthermore, the transition between the region where one finds entanglement skyrmions and that of deflated pseudospin skyrmions in the region $u_{\rm z}>u_\perp$ is strongly affected by the in-plane anisotropy. This can be attributed to the very small energy difference and thus delicate competition between entanglement and deflated pseudospin skyrmions that are also characterized by a certain degree of entanglement, as discussed in Sec.~\ref{subsec:deflated-pseudospin-skyrmion}.

\subsection{Pseudospin analogue of Zeeman coupling}
\label{subsec:Pseudospin-Zeeman-coupling}

\begin{figure}[t]
\includegraphics[width=0.9\columnwidth]{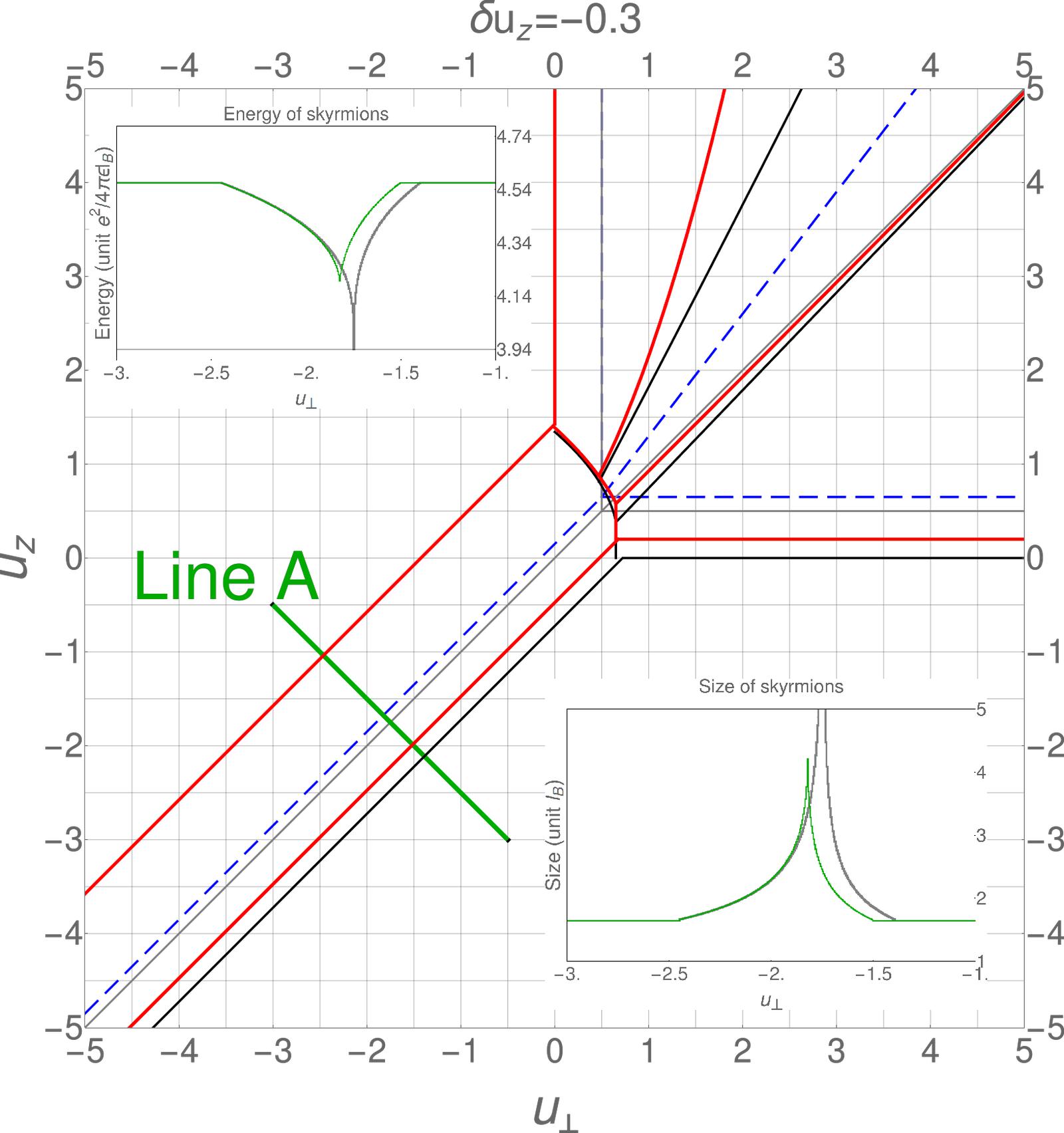}
\caption{\label{fig:Skyrmion-type-pseudospin-Zeeman} Change of the skyrmion type diagram for $\delta u_{\rm z}=-0.3$. The color code is the same as in Fig.~\ref{fig:Skyrmion-type-tilting}, in the absence of tilt ($\varphi=0$). }
\end{figure}

The anisotropic energy Eq.~(\ref{eq:EnAnisotropic}) can also be modified by adding the term $\Delta_{\rm P}\int d^2 r M_{\rm Pz}$, which is a pseudospin analogue of the Zeeman coupling and describes the sublattice asymmetry. Omitting the $\delta u_{\perp}$ term discussed in the previous section but including the sample tilting, we have 
\begin{eqnarray}
E_{\rm A}[Z] &=& \frac{\Delta_{\rm Z}(B_{\rm T})}{2}\int\frac{d^{2}r}{2\pi l_{B}^{2}}
\left[ u_{\perp}\cos\varphi\left( M_{\rm Px}^{2}+M_{\rm Py}^{2}\right)\right.
\nn\\
&& \left.+ u_{\rm z}\cos\varphi M_{\rm Pz}^{2} + \delta u_{\rm z}\cos\varphi M_{\rm Pz} -M_{\rm Sz}\right],\nn\\\label{eq:EnAnisotropicPspinZeeman}
\end{eqnarray}
where $\delta u_{\rm z}=\Delta_{\rm P}\cos\varphi/[\Delta_{\rm Z}(B_{\rm T})/2]$ is the relative energy scale of the pseudospin analogue of the Zeeman coupling. 
The original $\mathbb{Z}_2$ symmetry (equivalence of a pseudospin orientation along the $z$ or $-z$-axis) is now broken, and pseudospin orientation along $z$ are favoured for $\delta u_{\rm z}<0$ and along $-z$ for $\delta u_{\rm z}>0$. 

The $\delta u_{\rm z}$ term modifies the optimal FM states $Z(\br)=F$ for the anisotropic energy in a complementary manner as $\delta u_{\perp}$ -- instead of the easy-plane FM states, the easy-axis FM states are affected by this type of anisotropy. This can indeed be seen in the phase diagram shown in Fig.~\ref{fig:Skyrmion-type-pseudospin-Zeeman} for $\delta u_{\rm z}=-0.3$. First, it shifts the border between the regions for unentangled easy-axis FM and for unentangled easy-plane FM, as well as the border between the regions for unentangled easy-axis FM and entangled easy-axis FM. Both are shifted upwards by $2\delta u_{\rm z}\cos\varphi$. 
Second, the slope of the border between the regions for entangled easy-axis FM and for entangled easy-plane FM, is increased from $1$ to $1+|\delta u_{\rm z}|\cos\varphi$. Moreover, at the upper-left side of the sloped borders separating the easy-axis FM and easy-plane FM regions, the $(u_\perp,u_{\rm z})$ points close to the border corresponds to FM spinor $F$ with optimal value of $\theta_{\rm P}$ less than $\pi/2$. As $(u_\perp,u_{\rm z})$ moves away from these borders (to the upper-left side), $\theta_{\rm P}$ approaches to $\pi/2$. These modifications can be understood by rewriting $E_{\rm A}[Z]$ as 
\begin{eqnarray}
E_{\rm A}[Z] &=& \frac{\Delta_{\rm Z}(B_{\rm T})}{2}\int\frac{d^{2}r}{2\pi l_{B}^{2}}\times\nonumber\\
&& \left[u_{\perp}\cos\varphi\left(M_{{\rm Px}}^{2}+M_{{\rm Py}}^{2}\right)+\left(u_{{\rm z}}+\frac{\delta u_{{\rm z}}}{2}\right)\cos\varphi M_{{\rm Pz}}^{2}\right.\nonumber\\
&& \left.-\frac{\delta u_{{\rm z}}}{2}\cos\varphi(M_{{\rm Pz}}-1)^{2}-M_{{\rm Sz}}+\frac{\delta u_{{\rm z}}}{2}\cos\varphi\right]
\label{eq:EnAnisotropicPspinZeeman1}
\end{eqnarray}
The analysis is similar to those for the the $\delta u_{\perp}$ term for the substrate anisotropy in the previous subsection. The upward shift of the transition between the easy-axis and easy-plane FM states is a consequence of the lowered energy of the former (for $u_{\rm z}<0$) that favours easy-axis states. Similarly to the easy-plane anisotropy discussed above, there is no restoration of the $\SU{2}$ pseudospin symmetry at the transition line $u_\perp = u_{\rm z} + \delta u_{\rm z}/2$ because of the remaining term 
$\delta u_{\rm z}\cos \varphi (M_{\rm Pz}-1)^2/2$ in Eq.~(\ref{eq:EnAnisotropicPspinZeeman1}. As a consequence, the skyrmion size does not diverge at this value and its energy does not reach the minimal (scale-invariant) value $4\pi \rho_{\rm s}$ (see insets of Fig.~\ref{fig:Skyrmion-type-pseudospin-Zeeman}).

\section{Summary and discussions}
\label{sec:Summary-and-discussions}

In conclusion, our analysis of the quarter-filled $N=0$ LL of graphene shows that the anisotropic terms in the energy functional, which break the spin and valley pseudospin symmetry of the leading Coulomb energy, lead to four types of ferromagnetic ground state and various types of $\CP{3}$-skyrmions. A skyrmion is embedded in the ferromagnetic background $F$, and the center spinor $C$ gives rise to different types of skyrmions because the orthogonality condition between the spinors, $F^{\dagger}C=0$, does not entirely fix the skyrmion type. The favored skyrmion type is then obtained by a minimization of the anisotropic energy terms. 
When the electron spin and pseudospin contribution to the anisotropic energy varies, a spin skyrmion can change to a pseudospin skyrmion while the ferromagnetic background remains unaltered. This is strikingly different from the two-fold degenerate LL split by electron spin, where a skyrmion simply disappears when the magnetic field, and hence the Zeeman coupling increases to a critical value. A large spin skyrmion requires weak Zeeman coupling and hence a weak applied magnetic field. 
However, in our four-component system where $\CP{3}$-skyrmions are formed, one can tune the model parameters $u_{\perp},u_{\rm z}$ to approximately restore the $\SU{2}$ symmetry of the valley pseudospin, so as to have a large pseudospin skyrmion, even when the anisotropic energy for both spin and valley pseudospin is high. By analogy, skyrmion also appears in graphene monolayer at $\nu=0$ when the valley pseudospin $\SU{2}$ symmetry or the $\SO{5}$ symmetry\cite{SO5} is approximately restored. 

In addition to these skyrmion types, one encounters more exotic skyrmions that are formed due to spin-pseudospin entanglement. A direct transposition of the concept of a skyrmion that entirely covers the spin or pseudospin Bloch sphere yields the \textit{entanglement} skyrmion, which covers the third \textit{entanglement} Bloch sphere. This Bloch sphere is required, at $\nu=\pm 1$, to fully characterize the skyrmion and FM states i.e. to account for two different angles $\alpha$ and $\beta$ that naturally arise in our $\CP{3}$ description of $\SU{4}$ states. In the case of easy-axis FM states, entanglement manifests itself, rather counter-intuitively, in the form of locally anti-ferromagnetic or anti-ferrimagnetic patterns. Moreover, we have identified another type of skyrmion, a deflated pseudospin skyrmions with partial entanglement. While all directions in the pseudospin Bloch sphere are explored within this skyrmion type, similarly to an unentangled pseudospin skyrmion, the magnitude of the pseudospin magnetization is reduced in most places. This provides the deflated image of the representation on the Bloch sphere, accompanied by an exploration of parts of the spin Bloch sphere. In lattice-resolved images, deflated pseudospin skyrmions are manifest by a lower spin-pseudospin contrast and by local deviations of the spin magnetization from the $z$-axis. 

The lattice-resolved images we obtain in our study can in principle be used as a guide in a possible experimental identification of the various skyrmions in graphene, e.g. in spin-resolved STM/STS measurements. While the varying spin magnetization can be probed by a combination of magnetic tips, not only in the $z$-direction but also in the $x$- and $y$-directions, the $z$-component of the pseudospin magnetization can directly be read of from the relative electronic occupation of the different sublattices in graphene. In contrast to the spin magnetization, the $x$- and $y$-components of the pseudospin magnetization are unfortunately not accessible in this type of experiments that do not provide insight into the combination $\phi_{\rm P} - \beta$ of the parameters in our parametrization of $\CP{3}$ fields. 

Our analysis also shows that the model parameters $u_{\perp}$ and $u_{\rm z}$ that correspond to large optimal $\CP{3}$-skyrmions reside in the vicinity of the borders between two different types of ferromagnetic backgrounds. Indeed, the transitions are characterized by either (i) a partial symmetry restoration, e.g. of the $\SU{2}$ pseudospin symmetry at $u_{\rm z}=u_\perp\leq 1/2$, or (ii) a tendency of such a symmetry restoration, e.g. at $u_{\rm z}=1/2$ for $u_\perp>1/2$ or $u_\perp=1/2$ for $u_{\rm z}\geq 1/2$ between unentangled and entangle FM backgrounds. In general, the energy to create a skyrmion is reduced due to a reduction of the anisotropy energy. In the case (i), the partial symmetry restoration yields a divergent skyrmion size described by a power law $\lambda \sim |u_{\rm z}-u_\perp|^{\gamma}$, with a critical exponent $\gamma$ close to $-1/3$, as expected from simple scaling arguments. At the same time, the skyrmion size -- in $B$-independent units -- scales in the vicinity of the transition as $R\sim B^{-2/3}$ with the $B$-field and its energy as $E-E_{\rm NLSM}^{(0)}\sim B^{2/3}$ when compared to the energy $E_{\rm NLSM}^{(0)}=4\pi \rho_{\rm s}$ of the non-linear sigma model. Right at the transition, the skyrmion energy retrieves its scale-invariant value $E_{\rm NLSM}^{(0)}=4\pi \rho_{\rm s}$.
If there is only a tendency towards symmetry restoration [case (ii)], either in the above mentioned transition between unentangled and entangled FM backgrounds or in the presence of an anisotropy in the substrate that couples to the pseudospin, the divergence in the skyrmion size is truncated and follows approximately the law $\lambda \sim (|u-u_c|+C)^{\gamma}$, where $C$ is some positive constant and $u$ is a generic coupling constant with critical value $u_c$. Again we find that the exponent $\gamma$ is close to $-1/3$.

Several terms are not included in our model. 
The $\SU{4}$ invariance of the non-linear sigma model energy $E_{\rm NLSM}[Z]$ can be broken by a term of the form\cite{SO5,Moon1995} $2\rho_s'\int d^2r {\rm Tr}[\Gamma \bD Z(\br)\cdot \bD Z^{\dagger}(\br)]$, where $\Gamma$ is a superposition of the tensor product of Pauli matrices, and the coefficient $2\rho_s'$ is of the same order as the input parameters $u_{\perp}$, $u_{\rm z}$ for the anisotropic energy in our model. 
At the fourth order of the gradient expansion, which may be relevant for small skyrmion sizes, 
we may have\cite{Apel1998} $-(3\rho_s/8)\int d^2r [\Delta^2 (Z^{\dagger}\boldsymbol{\sigma}Z)]^2$, where $\Delta=\partial_x^2 +\partial_y^2$ is the Laplacian. The minus sign of such a term is on the same order as the Coulomb energy $E_{\rm C}[Z]$ in the gradient expansion, and it implies that our model has overestimated\cite{Abolfath1997} the energy of a skyrmion, or equivalently, underestimated the skyrmion size. Generally speaking, the non-linear sigma model of skyrmions does not take account for the reduction of magnetization in the core of the texture caused by quantum fluctuations.\cite{Sondhi1993,Abolfath1997}

\section*{Acknowledgement}
We most gratefully acknowledge fruitful discussions and former collaboration with Achim Rosch. Furthermore, we thank Markus Morgenstern for helpful experimental insights. YL is funded by a scholarship from the China Scholarship Council.

\appendix

\section{Visualization of a CP$^{3}$-skyrmion on honeycomb lattice}
\label{subsec:Visualization-CP3-skyrmion-on-honeycomb-lattice}
In Sec.~\ref{sec:4-types-skyrmion} we represent the $\CP{3}$ textures $Z_{\rm skyr}(\br)$ in lattice-scale profiles of the electron density $\rho_{\rm Total}(\br)$ and the $z$-component of the spin magnetization $M_{\rm Sz}(\br)$ for the $\CP{3}$-field $Z_{\rm skyr}(\br)$. These profiles are computed with the help of Eqs.~(\ref{eq:Si_Z}) and (\ref{eq:Ti_Z}), where the components of the $\CP{3}$-field are convoluted with a form factor 
\begin{equation}
f_{\lambda}(\br)=\sum_{\br_{j}^{\lambda}}g(\br-\br_{j}^{\lambda})
\end{equation}
that mimics the amplitude square of the atomic wave function. Here, $\br_{j}^{\lambda}$ is the position on the sublattice $\lambda$ in the $j$-th unit cell. For illustration purposes, we have chosen a Gaussian profile for $g(\br)$ that is normalized as $g(\br=0)=1$. Moreover, if we neglect the overlap between the atomic wave functions centred around different lattice site in the function $g(\br-\br_{j}^{\lambda})$, the expressions for $\rho_{\rm Total}(\br)$ and $M_{\rm Sz}(\br)$ simply read
\begin{eqnarray}
\rho_{\rm Total}(\br) &= \sum_{\br_{j}^{\lambda}}\rho_{\rm Total}(\br_{j}^{\lambda})g(\br-\br_{j}^{\lambda})\\
M_{\rm Sz}(\br) &= \sum_{\br_{j}^{\lambda}}M_{\rm Sz}(\br_{j}^{\lambda})g(\br-\br_{j}^{\lambda})
\end{eqnarray}
The electronic occupation of the sublattices A and B in the $j$-th unit cell is given by $\rho_{\rm Total}(\br_{j}^{A,B})$, while the $z$-component of the spin magnetization is $S_{\rm z}(\br_{j}^{A,B})$, with
\begin{eqnarray}
\rho_{\rm Total}(\br_{j}^{A,B}) &= Z^{\dagger}(\br_{j}^{A,B})\left[\frac{\sigma_{z}\pm1}{2}\otimes1\right]Z(\br_{j}^{A,B}),\\
S_{\rm z}(\br_{j}^{A,B}) &= Z^{\dagger}(\br_{j}^{A,B})\left[\frac{\sigma_{z}\pm1}{2}\otimes\sigma_{z}\right]Z(\br_{j}^{A,B}).
\end{eqnarray}
In the above expression, the $+$-sign in the projector is chosen for a site on the $A$-sublattice and $-$ for a site on the $B$-sublattice.

\section{Second homotopy group for ${\rm CP}^3$}
\label{subsec:second-homotopy}
The second homotopy group\cite{TopologyTextbooks} $\pi_2({\cal M})$ of a topological manifold ${\cal M}$ is defined as the equivalence class of the continuous mapping $f:S^2 \rightarrow {\cal M}$, where two mappings $f_0$ and $f_1$ are equivalent means that there exists a continuous mapping $F:[0,1]\times S^2\rightarrow {\cal M}$ such that $F(0,\cdot)=f_0(\cdot)$ and $F(1,\cdot)=f_1(\cdot)$. The multiplication in the group is the homotopic composition of the continuous mappings. 

The second homotopy group $\pi_2(S^2)=\mathbb{Z}$ of the 2-sphere $S^2$ can be understood as ``wrapping a 2-sphere on another 2-sphere''. The winding number (Brouwer degree) of the mapping $f:S^2 \rightarrow S^2$ classifies all possible ways of the ``wrapping''. Since ${\rm CP}^1\cong S^2$, we have $\pi_2( {\rm CP}^1 )=\mathbb{Z}$. It can also be computed\cite{Hasebe2002,TopologyTextbooks} by the following relations between a simply connected Lie group ${\cal G}$, one of its subgroups ${\cal H}$ and the coset space ${\cal G}/{\cal H}$ : 
\begin{equation}
\pi_2({\cal G}/{\cal H}) = \pi_1({\cal H}),
\end{equation}
where $\pi_1({\cal M})$ is the fundamental group of a manifold defined by replacing $S^2$ in the definition of $\pi_2({\cal M})$ by $S^1$. The manifold $\CP{1}$ is isomorphic to $\SU{2}/{\rm U}(1)$, therefore 
\begin{equation}
\pi_2({\rm CP}^1) = \pi_2(\SU{2}/{\rm U}(1)) = \pi_1({\rm U}(1)),
\end{equation}
where the last equation $\pi_1({\rm U}(1))=\mathbb{Z}$ follows from the classification of mappings from a closed path to another closed path. The manifold ${\rm CP}^3$ is isomorphic to 
\begin{equation}\label{eq:group}
{\rm U}(4)/[{\rm U}(3)\times {\rm U}(1)] \cong \SU{4}/[\SU{3}\times {\rm U}(1)], 
\end{equation}
therefore 
\begin{equation}
\pi_2({\rm CP}^3) = \pi_2(\SU{4}/[\SU{3}\times {\rm U}(1)]) = \pi_1(\SU{3}\times {\rm U}(1)).\nn
\end{equation}
Using the fact that the homotopy group for product manifold factorizes, i.e. 
\begin{equation}
\pi_k({\cal G}\times{\cal H}) = \pi_k({\cal G})\times\pi_k({\cal H}),
\end{equation}
and the fact that any simple Lie group ${\cal G}$ has $\pi_1({\cal G})=0$, we obtain 
\begin{equation}
\pi_2({\rm CP}^3) = \pi_1(\SU{3})\times \pi_1({\rm U}(1)) = \mathbb{Z}.
\end{equation}
This result can be generalized\cite{dadda1978} to compute $\pi_2({\rm CP}^{\Nmath-1})$ for integer $\Nmath>1$, because 
\begin{eqnarray}\label{dadda}
\pi_2({\rm CP}^{\Nmath-1}) & = & \pi_2(\SU{\Nmath}/[\SU{\Nmath-1}\times {\rm U}(1)]) \nn\\
& = & \pi_1(\SU{\Nmath-1}\times {\rm U}(1))\nn\\
& = & \pi_1(\SU{\Nmath-1})\times \pi_1({\rm U}(1))\nn\\
& = & \mathbb{Z}
\end{eqnarray}
This argument shows that any $\Nmath$-component QH system or, more generally, any $\Nmath$-component ferromagnet in two spatial dimensions with one filled component can principally host topological textures in the form of skyrmions that are characterized by an integer topological charge $\Qmath\in \mathbb{Z}$. The specificity of the quantum Hall ferromagnet is that this topological charge is directly related to an electric charge. 

\section{CP$^{3}$-field and non-linear sigma model}
\label{subsec:CP3-spinor-and-NLSM}
The $\CP{3}$ space is a collection of normalized, four-component complex vectors $Z=(z_{1},z_{2},z_{3},z_{4})^{T}$. In addition, two vectors $Z$ and $e^{i\varphi}Z$ are equivalent for arbitrary $\varphi\in\mathbb{R}$, in the sense that they correspond to the same matrix $Q=2ZZ^{\dagger}-1$, which plays the role of order parameter in the theory of QHFM in Ref.~[\onlinecite{Yang2006}]. To be more precise, $Z$ should be understood as a representative of the equivalent class $[Z]\in V/\sim$, where $V$ denotes the set of normalized $\mathbb{C}^{4}$ vectors and $\sim$ denotes the equivalence relation $W\sim e^{i\varphi}W$ for $W\in V$ and $\varphi\in\mathbb{R}$. In this paper, we call the normalized $\mathbb{C}^{4}$ vector $Z$ a ``$\CP{3}$-spinor'' if it represents an element in the $\CP{3}$ space. The name ``spinor'' here is inherited from the physics literature without mathematical rigour.

The $\CP{3}$ space can also be understood as a coset space (See Eq.~(\ref{eq:group}) in App.~\ref{subsec:second-homotopy})
\begin{equation}
\CP{3} \cong {\rm U}(4)/[{\rm U}(3)\times {\rm U}(1)] \cong \SU{4}/[\SU{3}\times {\rm U}(1)],
\end{equation}
which means that the integer filling of one of the four Landau sublevels is distinguished by removing the $\U{3}$ (or $\SU{3}$) rotations among three empty sublevels from the entire rotation group $\U{4}$ (or $\SU{4}$) among all four sublevels, followed by a removal of a global phase factor $\U{1}$. 

The parametrization of a CP$^{3}$-spinor is discussed in Sec.~\ref{subsec:QHFM-ground-states}. There we construct the CP$^{3}$-spinor from six angles $\theta_{\rm S}$, $\phi_{\rm S}$, $\theta_{\rm P}$, $\phi_{\rm P}$, $\alpha$, $\beta$, and visualized them via a triplet of Bloch spheres. The inverse problem -- to obtain the six angles parametrizing a given CP$^{3}$-spinor up to an overall phase factor $e^{i\phi}$ -- is equally important for the complete understanding of the parametrization Eq.~(\ref{eq:parametrizationZ}). 
Given a CP$^{3}$ -spinor $Y$, we have access to the parameter $\alpha$ by computing the magnitude of spin and pseudospin magnetization via Eqs.~(\ref{eq:Si_F}) and (\ref{eq:Ti_F}). There are two possible values: $\alpha=\cos^{-1}|\bM_{\rm S/P}|$ and $\alpha=\pi-\cos^{-1}|\bM_{\rm S/P}|$. They correspond to the same CP$^{3}$-spinor because of the following equivalence between two CP$^{3}$-spinors parametrized in Eq.~(\ref{eq:parametrizationZ}):
\begin{eqnarray}
& e^{i(\phi_{\rm S}+\phi_{\rm P}-\beta)} Y(\theta_{\rm S},\phi_{\rm S},\theta_{\rm P},\phi_{\rm P},\alpha,\beta)= \nonumber\\
& Y(\pi-\theta_{\rm S},\pi+\phi_{\rm S},\pi-\theta_{\rm P},\pi+\phi_{\rm P},\pi-\alpha,\beta')\nonumber\\
& \beta' = -\beta+2\phi_{\rm S}+2\phi_{\rm P}.\label{eq:parametrization_redundancy}
\end{eqnarray}
The effect of changing $\alpha$ into $\pi-\alpha$ while keeping the same CP$^{3}$-spinor is to reverse the direction of the spin and pseudospin magnetization. It reveals the fact that our parametrization is redundant for $\pi/2\le \alpha\le\pi$. In the discussion of QHFM states, we can always restrict $\alpha$ in the range $[0,\pi/2]$. Under this restriction, we have $\cos\alpha\ge0$ and $\sin\alpha\ge0$, in agreement with the Schmidt decomposition theorem, which claims that we have non-negative real numbers as the coefficients in front of the direct-product-state basis obtained from the decomposition. 
Notice, however, that this restriction is not the unique choice to avoid redundancies in the parametrization. Instead one can also choose the restriction $\theta_{\rm S}$ or $\theta_{\rm P}$ in $[0,\pi/2]$, such that $\alpha$ and $\beta$ span the full Bloch sphere, as in the discussion of entanglement skyrmions in Sec.~\ref{subsec:entanglement-skyrmion}.

The CP$^{3}$-non-linear sigma model (CP$^{3}$-NLSM) has the following form of energy:
\begin{equation}
{\cal E}[Z]=\int d^2r \bD Z^{\dagger}(\br)\cdot \bD Z(\br),\label{eq:NLSM-general}
\end{equation}
where we defined $\bD Z = \nabla Z(\br) - [Z^{\dagger}(\br)\nabla Z(\br)] Z(\br)$. The finite-energy configurations of $Z(\br)$ satisfies the boundary condition at spatial infinity:
\begin{equation}
\lim_{\left|\br\right|\rightarrow\infty}Z(\br) = e^{i g(\varphi)} Z_{\infty} ,
\end{equation}
where $g(\varphi)$ is a function of the polar angle $\varphi$ of coordinate plane, and $Z_{\infty}$ is a constant CP$^{3}$-spinor. Under such boundary condition, the base manifold of field $Z(\br)$ can be extended to the Riemann sphere $S^{2}=\mathbb{R}^{2}\cup\{\infty\}$ via Riemann stereographic projection,\cite{TopologyTextbooks} and the value of the CP$^{3}$-spinor field at spatial infinity is $Z(\infty)=Z_{0}$. After the extension, each field $Z(\br)$ can be considered as a mapping $S^{2}\rightarrow CP^{3}$ and can be classified by the degree of mapping
\begin{eqnarray}
{\cal Q}[Z] &= \int\left(-\frac{i}{2\pi}\right)[\bD Z(\br)^{\dagger} \times \bD Z(\br)]_z d^{2}r\nonumber\\
&= \int\rho_{\rm topo}(\br)d^{2}r
\end{eqnarray}
because the second homotopy group $\pi_{2}({\rm CP}^{3})=\mathbb{Z}$ of the CP$^{3}$ space (as target space of field $Z(\br)$) is non-trivial. ${\cal Q}$ is called \emph{topological charge} in the physics literature, and correspondingly, $\rho_{\rm topo}(\br)$ is called \emph{topological charge density}. It is impossible to continuously deform a field $Z_{a}(\br)$ into another configuration $Z_{b}(\br)$ of different topological charge ${\cal Q}[Z_{b}]\neq{\cal Q}[Z_{a}]$. Therefore, our variational analysis on ${\cal E}[Z]$ (Eq.~(\ref{eq:NLSM-general})) is limited within a subspace of field configurations
$${\cal C}_{M}=\left\{ Z(\br)\in {\rm CP}^{3}\mbox{-field}\,|\,{\cal Q}[Z]=M\in\mathbb{Z}\right\} .$$ 

\section{Energy-minimizing solutions in each topological sector}
\label{subsec:Energy-minimizing-solutions-in-each-topological-sector}
The minimal-energy configuration $Z_{[{\cal Q}]}(\br)$ for ${\cal E}[Z]$ in each ${\cal C}_{{\cal Q}}$ serves as the starting point of our analysis of CP$^{3}$-skyrmions. We observe an inequality
\begin{eqnarray}
\int\delta_{ij}(D_{i}Z\pm i\epsilon_{ik}D_{k}Z)^{\dagger}(D_{j}Z\pm i\epsilon_{jl}D_{l}Z)\nonumber\\
=2{\cal E}[Z]\pm4\pi{\cal Q}[Z]\ge 0
\end{eqnarray}
or, equivalently, 
\begin{equation}
{\cal E}[Z]\ge2\pi\left|{\cal Q}[Z]\right|.
\end{equation}
It means that for each ${\cal C}_{{\cal Q}}$, the lower bound of ${\cal E}[Z\in {\cal C}_{{\cal Q}}]$ is $2\pi\left|{\cal Q}[Z]\right|$. Comparing Eq.~(\ref{eq:NLSM-general}) to Eq.~(\ref{eq:EnNLSM}) to restore the units of energy, we find that the right hand side in the above inequality gives $4\pi\rho_{\rm s}$ as the lower bound of $E_{\rm NLSM}$ for charge $|{\cal Q}|=1$ skyrmions.  
Moreover, the minimal-energy configuration which saturates the inequality is the solution of the following first order partial differential equations, which are called the Bogomol'nyi-Prasad-Sommerfield (BPS) equations:
\begin{eqnarray}
D_{i}Z+i\epsilon_{ik}D_{k}Z &= 0\;\mbox{ for }{\cal Q}[Z]<0,\label{eq:BPS_minus}\\
D_{i}Z-i\epsilon_{ik}D_{k}Z &= 0\;\mbox{ for }{\cal Q}[Z]>0.\label{eq:BPS_plus}
\end{eqnarray}
Inserting $Z=W/\sqrt{W^{\dagger}W}$ with the \emph{unormalized} complex vector $W\in\mathbb{C}^{4}$ and using the complex coordinate $z=x+iy$, the above equations are rewritten as
\begin{eqnarray}
\partial W &= 0\;\mbox{ for }{\cal Q}<0\\
\bar{\partial}W &= 0\;\mbox{ for }{\cal Q}>0
\end{eqnarray}
where $\partial=(\partial_{x}-i\partial_{y})/2$ and $\bar{\partial}=(\partial_{x}+i\partial_{y})/2$.

The topological charge ${\cal Q}$ can also be expressed in $W$ with the help of Stoke's theorem:
\begin{equation}\label{eq:topoU1charge}
{\cal Q}=\frac{1}{2\pi i}\oint_{{\cal C}}\frac{W^{\dagger}\boldsymbol{\nabla}W}{W^{\dagger}W}\cdot d\boldsymbol{l},
\end{equation}
where the contour ${\cal C}=\partial {\cal D}$ encloses a simply connected region ${\cal D}$, and ${\cal D}$ maximally covers the punctured complex plane where $W$ and $\boldsymbol{\nabla}W$ are well defined. 
The generic solution of BPS equation Eq.~(\ref{eq:BPS_plus}) for ${\cal Q}>0$ (${\cal Q}<0$) is a four-component vector field $(w_1(z),w_2(z),w_3(z),w_4(z))$, in which the components $w_i(z)$ are meromorphic (anti-meromorphic) functions. 

In the main text we are interested in the solution of BPS equation with topological charge ${\cal Q}=1$. The generic form is 
\begin{equation}
W_{[{\cal Q}=1]}(z) = \left[ f_1 z - c_1, f_2 z - c_2, f_3 z - c_3, f_4 z - c_4 \right].
\end{equation}
However, to demonstrate the use of the above formula in Eq.~(\ref{eq:topoU1charge}), let us compute ${\cal Q}$ for the following complex vector field
\begin{equation}
W(z)=\left[1,\frac{z-a_1}{z-b_1},z-a_2,z-a_3\right],\label{eq:W-example}
\end{equation}
where the topological charge cannot be simply read off. 
The contour ${\cal C}$ consists three parts -- an anticlockwise circle ${\cal C}_1=\{z:|z|=R\rightarrow\infty\}$, a clockwise circle ${\cal C}_2=\{z:|z-b_1|=R\rightarrow 0\}$, and a pair of straight lines $l_1,l_2$ of opposite directions connecting the two circles. We have
\begin{eqnarray}
{\cal Q} & = & \frac{1}{2\pi i}\left[ \oint_{{\cal C}_1} +\oint_{{\cal C}_2} +\left( \oint_{l_1} +\oint_{l_2} \right) \right] \frac{W^{\dagger}\boldsymbol{\nabla}W}{W^{\dagger}W}\cdot d\boldsymbol{l} \nn\\
& = & \frac{1}{2\pi i}\left[ \left(0 +0 +\frac{2\pi i}{2} +\frac{2\pi i}{2}\right) - \left(0 - 2\pi i +0 +0\right) + \left(0\right) \right]\nn\\
& = & 2,\nn
\end{eqnarray}
where each term in the round bracket corresponds to the contribution from the components of $W(z)$ in Eq.~(\ref{eq:W-example}) accordingly. It is easy to verify the above result for the topological charge of $W(z)$ in Eq.~(\ref{eq:W-example}) by directly computation with Eq.~(\ref{eq:TopologicalDensity}) in the main text. 

\section{$\kappa$-dependence of energy for deformed skyrmions}
\label{subsec:kappa-dependence-of-energy-for-deformed-skyrmions}
\begin{figure}[t]
\includegraphics[width=0.7\columnwidth]{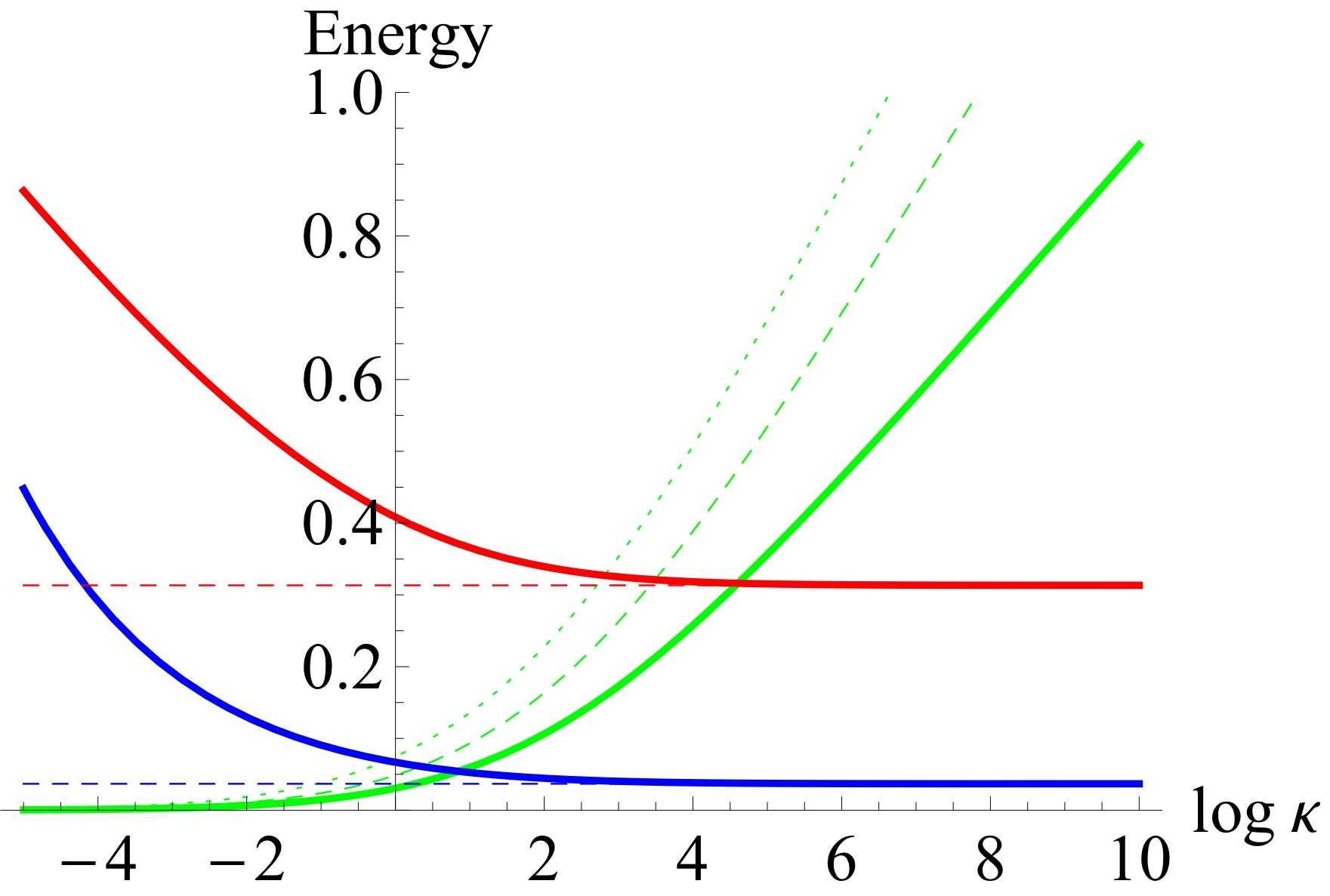}
\caption{\label{fig:kappa_dependence}$\kappa$-dependence of the NLSM energy $E_{\rm NLSM}$ (red line), the Coulomb interaction energy $E_{\rm C}$ (blue line) and the anisotropic energy $E_{\rm A}$ (green lines).}
\end{figure}

In Sec.~\ref{subsec:Radial-deformation} we discussed the radial deformed skyrmion $\check{Z}_{\rm sk}$. It has the form $$\check{Z}_{\rm sk}=\check{W}_{\rm [1]}/\sqrt{\check{W}_{\rm [1]}^{\dagger}\check{W}_{\rm [1]}}$$ with $\check{W}_{\rm [1]}(x,y) = (x+iy)F-\lambda(r)C$ and $\lambda(r) = \lambda_{0}\exp(-r^{2}/\kappa\lambda_{0}^2)$. 
In the limit $\kappa\rightarrow+\infty$, the three components of the energy $E[\check{Z}_{\rm sk}]$ have the following behavior
\begin{eqnarray}
E_{\rm NSLM}[\check{Z}_{\rm sk}] &\rightarrow &  4\pi\sqrt{\frac{\pi}{32}} \frac{e^{2}}{4\pi\epsilon l_{B}},\\
E_{\rm C}[\check{Z}_{\rm sk}] &\rightarrow & \frac{3\pi}{64} \frac{e^{2}}{4\pi\epsilon l_{B}} \left(\frac{\lambda_0}{l_B}\right)^{-1},\\
E_{\rm A}[\check{Z}_{\rm sk}] &\sim & \left( A_{0}+A_{1}\left|\Delta_{\rm Z}\right|\log\kappa\right)\left(\frac{\lambda_0}{l_B}\right)^{-2}.
\end{eqnarray}
As a function of $\kappa$, $E_{\rm NSLM}$ and $E_{\rm C}$ decrease monotonically. Meanwhile, $E_{\rm A}$ is a monotonic increase function of $\kappa$. Fig.~\ref{fig:kappa_dependence} displays their $\kappa$ dependence. Based on this analysis, the radial-deformation ansatz $\check{Z}_{\rm sk}$ gives a finite anisotropic energy $E_{\rm A}$ and a lowered total energy.

\section{Visualization of a CP$^{3}$-skyrmion on Bloch spheres}
\label{subsec:Visualization-CP3-skyrmion-on-Bloch-spheres}
\begin{figure}[t]
\includegraphics[width=\columnwidth]{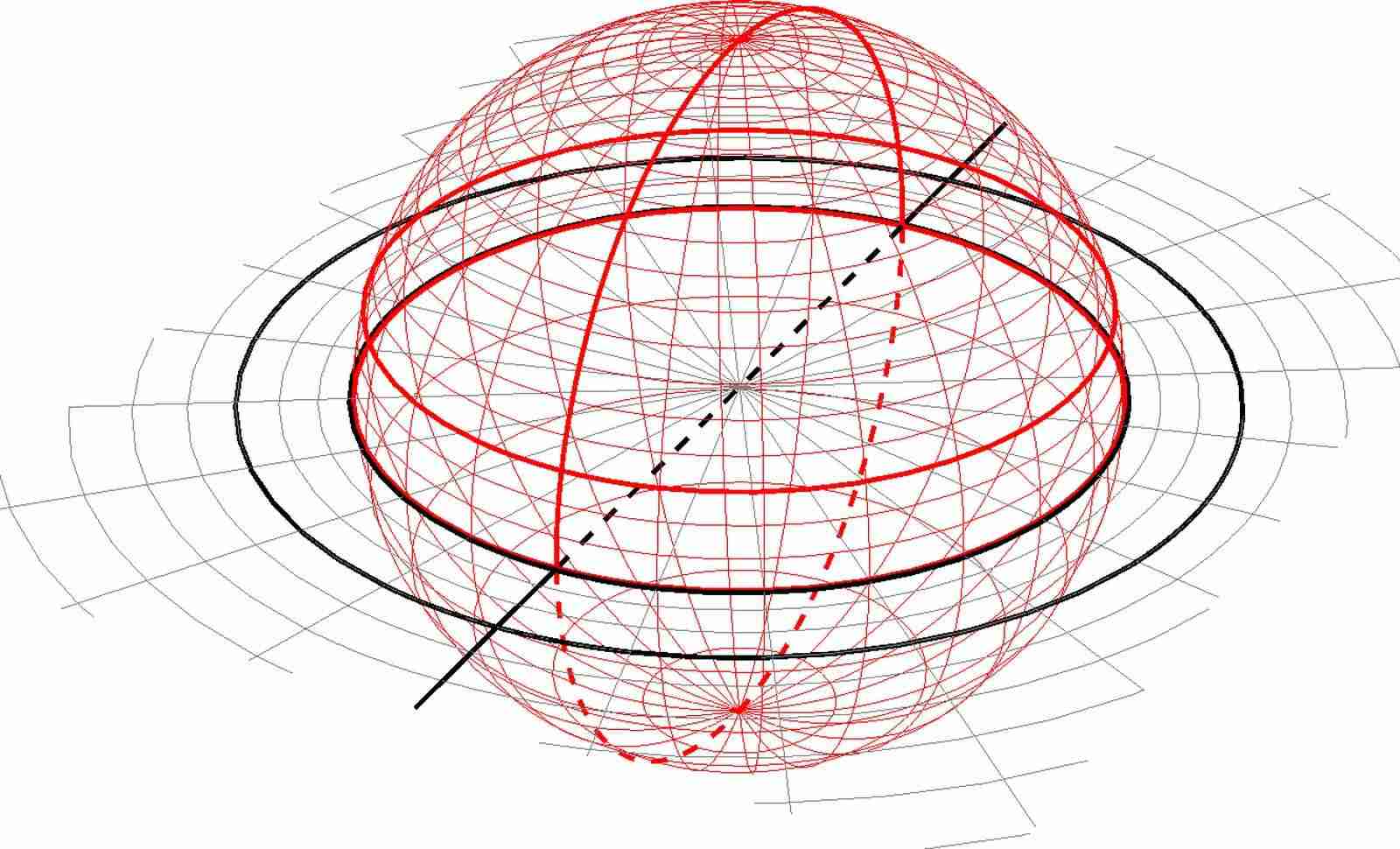}
\caption{\label{fig:Riemann_Sphere}The Riemann sphere and the stereographic projection. Lines on the $xy$-plane (where the CP$^{3}$-skyrmions live in) are colored in gray and lines on the Riemann sphere is colored in red.}
\end{figure}

The $\CP{1}$-skyrmion (or O(3)-skyrmion) can be visualized by the image of the Riemann sphere on the Bloch sphere. The Riemann sphere is obtained by the stereographic projection between the Cartesian coordinates $(X,Y)$ on the plane and the Cartesian coordinates $(x,y,z)$ on the Riemann sphere:
\begin{eqnarray}
(X,Y) &= (\frac{x}{1-z},\frac{y}{1-z})\\
(x,y,z) &= \frac{(2X,2Y,-1+X^{2}+Y^{2})}{1+X^{2}+Y^{2}}
\end{eqnarray}
It is illustrated in Fig.~\ref{fig:Riemann_Sphere}. The ``hedgehog'' skyrmion is in fact an identity mapping from the Riemann sphere to the Bloch sphere. 

As discussed in Sec.~\ref{subsec:QHFM-ground-states} and Appendix.~\ref{subsec:CP3-spinor-and-NLSM}, a CP$^{3}$-spinor can be parametrized by six angles $\theta_{\rm S}$, $\phi_{\rm S}$, $\theta_{\rm P}$, $\phi_{\rm P}$, $\alpha$, $\beta$, and therefore can be visualized by three Bloch spheres, namely the spin Bloch sphere (for $\theta_{\rm S}$ and $\phi_{\rm S}$), the pseudospin Bloch sphere (for $\theta_{\rm P}$ and $\phi_{\rm P}$) and the entanglement Bloch sphere (encodes $\alpha$ and $\beta$). Hence a CP$^{3}$-skyrmion can be visualized by plotting the image of the Riemann sphere on the three Bloch spheres via the parametrization of the CP$^{3}$-spinor.

In practice, we draw the spin magnetization and pseudospin magnetization
\begin{eqnarray}
\bM_{\rm S}(r) &= Z^{\dagger}(r)(1\otimes\boldsymbol{\sigma})Z(r)\nonumber\\
\bM_{\rm P}(r) &= Z^{\dagger}(r)(\boldsymbol{\sigma}\otimes1)Z(r)\nonumber
\end{eqnarray}
at points specified by longitudinal and latitudinal lines on the Riemann sphere. These lines correspond to rays of constant angle $\theta$ and circles of constant radius $r$ circles on the $xy$-plane where the skyrmion lives. In general, at a given point $r_0$ on the $xy$-plane, $\bM_{\rm S}(r_0)$ and $\bM_{\rm P}(r_0)$ may have a magnitude less than $1$, because $\alpha(r_0)$ can be different from $0$ or $\pi$. As a consequence on the visualization, the spin and pseudospin magnetization lay \emph{inside} the corresponding Bloch spheres. This is the case for the CP$^{3}$-entanglement skyrmion and the deflated CP$^{3}$-pseudospin skyrmion.


\end{document}